\documentclass{article}

 \usepackage[preprint]{neurips_2025}

\usepackage[utf8]{inputenc} 
\usepackage[T1]{fontenc}    
\usepackage{hyperref}       
\usepackage{url}            
\usepackage{booktabs}       
\usepackage{amsfonts}       
\usepackage{nicefrac}       
\usepackage{microtype}      
\usepackage{xcolor}         

\usepackage{enumitem}

\usepackage{amsmath}
\usepackage{amssymb}
\usepackage{mathtools}
\usepackage{amsthm}
\usepackage[bb=px]{mathalpha}
\newcommand{\ket}[1]{\left|#1\right>}

\usepackage[capitalize,noabbrev]{cleveref}

\definecolor{codegreen}{rgb}{0,0.6,0}
\definecolor{codegray}{rgb}{0.5,0.5,0.5}
\definecolor{codepurple}{rgb}{0.58,0,0.82}
\definecolor{backcolour}{rgb}{0.97,0.97,0.97}
\usepackage{listings}
\lstdefinestyle{markdownstyle}{
    basicstyle=\ttfamily\tiny,
    backgroundcolor=\color{backcolour},
    xleftmargin=0.05\textwidth,
    xrightmargin=0.05\textwidth,
    breakindent=0\dimen0,
    columns=flexible,
    showspaces=false,
    showstringspaces=false,
    breaklines=true,
    breakatwhitespace=true,
    breakautoindent=true,
}

\theoremstyle{plain}

\theoremstyle{definition}

\theoremstyle{remark}

\newcommand{\bra}[1]{\left\langle #1 \right|}

\usepackage{subfig}
\usepackage[textsize=tiny]{todonotes}

\title{Newton to Einstein: Axiom-Based Discovery via Game Design}

\author{%
	Pingchuan Ma \\
	MIT CSAIL
	\And
	Benjamin Jones \\
	MIT CSAIL
	\And 
	Tsun-Hsuan Wang \\
	MIT CSAIL
	\And 
	Minghao Guo \\
	MIT CSAIL
	\And
	Michal Piotr Lipiec \\
	MIT CSAIL
	\And
	Chuang Gan \\
	UMass Amherst
	\And
	Wojciech Matusik \\
	MIT CSAIL
}

\begin{document}

\maketitle

\begin{abstract}

This position paper argues that machine learning for scientific discovery should shift from inductive pattern recognition to axiom-based reasoning. We propose a game design framework in which scientific inquiry is recast as a rule-evolving system: agents operate within environments governed by axioms and modify them to explain outlier observations. Unlike conventional ML approaches that operate within fixed assumptions, our method enables the discovery of new theoretical structures through systematic rule adaptation. We demonstrate the feasibility of this approach through preliminary experiments in logic-based games, showing that agents can evolve axioms that solve previously unsolvable problems. This framework offers a foundation for building machine learning systems capable of creative, interpretable, and theory-driven discovery.

\end{abstract}
\section{Introduction}
\label{sec:intro}

Throughout history, the most transformative scientific breakthroughs have emerged not from accumulating more data, but from fundamentally revising foundational principles. Einstein's relativity theory built upon Newton's laws while transcending their limitations. Darwin's evolution unified disparate biological observations through a new theoretical framework. These paradigm shifts demonstrate that \textbf{scientific discovery should shift from inductive pattern recognition to axiom-based reasoning}.

\textbf{We advocate for using axioms as the primary representation in automatic scientific discovery}, where explicit reasoning with established principles can efficiently generate new theoretical insights. This position challenges the current emphasis on computational induction exemplified by AlphaFold \citep{jumper2021highly, roy2024alphafold3} and symbolic regression \citep{udrescu2020ai, schmidt2009distilling, cranmer2020discovering}, which require extensive datasets and computational resources to derive principles from scratch.

\begin{figure}[t]
    \centering
    \subfloat[\textbf{Technology Breakthroughs.} ]{
        \includegraphics[width=0.35\linewidth]{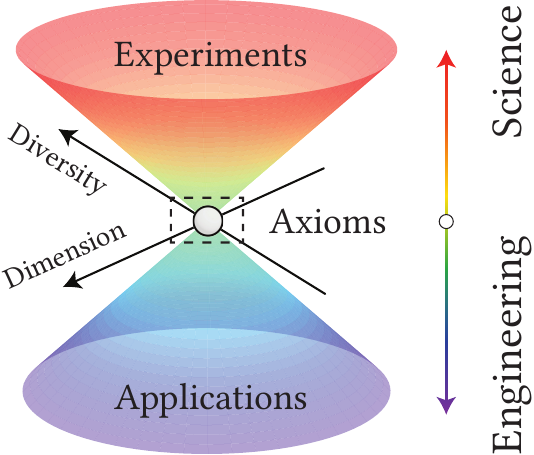}
    \label{fig:hourglass}
    } \hspace{0.1\textwidth}
    \subfloat[\textbf{Axiom Set Update}.]{
   \includegraphics[width=0.35\linewidth]{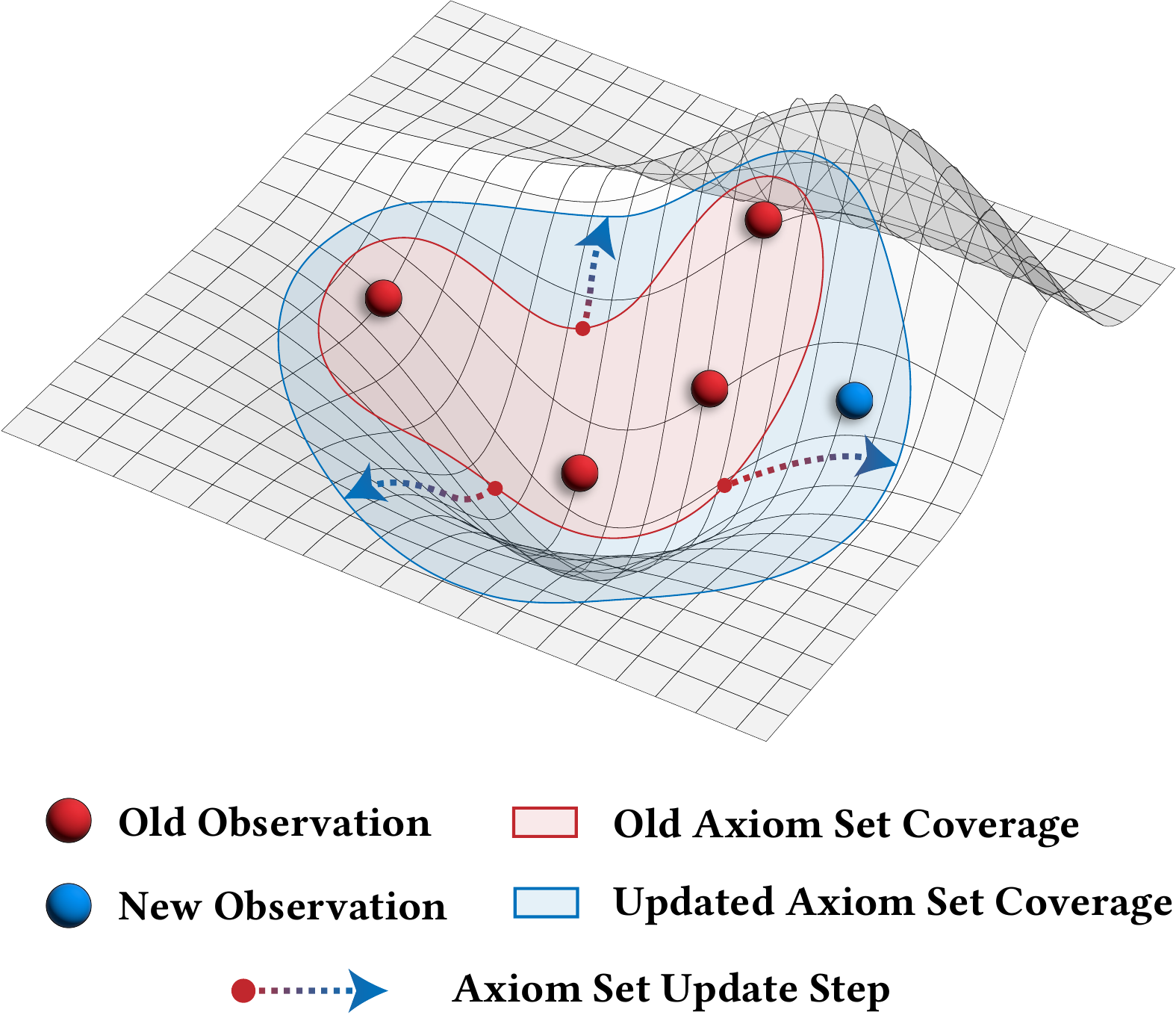}
    \label{fig:manifold}
    }
    \caption{\textbf{Axioms in scientific discovery}. (a) Scientists build simple models based on axioms that explain experimental observations. These fundamental principles, when combined, derive numerous predictions and generate new knowledge, often leading to paradigm shifts. (b) When observations emerge that cannot be explained, scientists modify and extend current axiom sets rather than develop entirely new theories. This approach expands explanatory power while maintaining validity of previously explained phenomena.}
    \label{fig:introduc}
\end{figure}

Our position stems from three key observations. First, many scientific fields possess robust foundational axioms, providing a solid base for development without starting from scratch. Second, reasoning from existing axioms to explain outlier observations is computationally more efficient than large-scale inductive training. Third, this approach enables iterative, continual updating as new evidence emerges.

The traditional scientific discovery process faces significant challenges when existing theories fail to explain new observations. Manual experimental design requires extensive trial and error, consuming substantial resources, while developing new theories demands considerable expertise. An automated scientific discovery pipeline could systematize both theoretical development and experimental validation \citep{king2009automation}, accelerating scientific progress.

Technology breakthroughs follow two interconnected branches: science and engineering (Fig.~\ref{fig:hourglass}). In science, fundamental discoveries emerge through inductive reasoning \citep{haig1995grounded}, where researchers gather empirical data through experiments. Kepler's planetary laws, derived from astronomical observations, enabled Newton to formulate universal gravitation and motion principles. These axioms then serve as building blocks for engineering applications through deductive reasoning. Newton's laws guide robotics motion planning \citep{featherstone2014rigid, siciliano2008springer, spong2020robot}, while molecular structures discovered in chemistry foundation drug discovery processes \citep{ferreira2015molecular, drews2000drug}.

To demonstrate feasibility, we present a framework translating scientific discovery into game design with three components: (1) rule/axiom set, (2) simulator deriving facts through gameplay, and (3) goal comparison mechanism. Game rules parallel scientific axioms, and when facing unwinnable states, our system evolves rules while maintaining proximity to original axioms.

We implement this using logic programming \citep{lloyd2012foundations}, providing formal language for expressing game mechanics and evolution processes. The simulator evaluates rule modification effectiveness by playing each variant. Our experiments demonstrate the system's ability to evolve valid axiom sets that transform unwinnable games into solvable ones. The framework automatically generates detailed evolution trajectories, capturing step-by-step axiom modifications for training future automated discovery systems.

In summary, our contributions are:
\begin{itemize}[align=right,itemindent=0em,labelsep=2pt,labelwidth=1em,leftmargin=*,itemsep=0em]
\item Proposing a shift from induction models to efficient axiom-based reasoning for automatic discovery.
\item Comprehensive analysis of historical scientific breakthroughs through axiom reasoning.
\item Implementation of a computational framework translating scientific discovery into game design with automated reasoning capabilities.
\item Development of a systematic data generation pipeline capturing detailed reasoning processes for automated discovery systems.
\end{itemize}

\section{Motivational Example}
\label{sec:motivation}

To illustrate axiom reasoning in scientific discovery, we examine the historical transformation from Newtonian mechanics to general relativity—one of physics' most significant paradigm shifts. This transformation exemplifies our core argument: rather than abandoning existing theories when faced with anomalies, scientists systematically modify foundational axioms to accommodate new observations while preserving established principles.

\textbf{The Challenge:} By the early 1900s, several observations challenged Newtonian mechanics. Mercury's perihelion precession deviated from predictions, the Michelson-Morley experiment failed to detect Earth's motion through the ether, and high-velocity particle behavior contradicted classical expectations. Rather than collecting more data or developing entirely new theories, Einstein employed axiom reasoning—systematically modifying Newton's foundational assumptions.

Consider the evolution from Newton's Second Law to Einstein's Geodesic Equation:
\begin{align}
\textbf{Newton:}\ &&\vec{F}=m\frac{d\vec{v}}{dt},&&\text{where}\ v^i=\frac{dx^i}{dt} \label{eq:newton} \\
\textbf{Einstein:}\ &&\nabla_{\tau}v=0\ \text{for}\ \vec{F}=0,&&\text{where}\ v^{\mu}=\frac{dx^{\mu}}{d\tau} \label{eq:einstein}
\end{align}

This transformation demonstrates three key principles of axiom reasoning:
\begin{enumerate}[align=right,itemindent=0em,labelsep=2pt,labelwidth=1em,leftmargin=*,itemsep=0em]
\item \textbf{Systematic generalization:} Newton's absolute time $t$ and 3D coordinates $x^i$ generalize to proper time $\tau$ and 4D spacetime coordinates $x^{\mu}$.
\item \textbf{Preservation of core concepts:} The fundamental notion of force-free motion remains intact, expressed through the covariant derivative $\nabla_{\tau}$.
\item \textbf{Minimal modification:} The changes are precisely targeted to address specific anomalies while maintaining consistency with established phenomena.
\end{enumerate}

\textbf{Connection to Our Framework:} This historical example parallels our game design approach: when encountering unwinnable states (like unexplained observations), we modify rules (axioms) while maintaining system integrity. Like Einstein's work, we seek minimal modifications that resolve challenges while preserving fundamental structure. This example, detailed in \Cref{sec:newtonian_general}, shows how axiom reasoning enables breakthroughs through systematic modification rather than replacement.
\section{General Scientific Discovery}
\label{sec:method}

\begin{figure}[t]
    \centering
    \includegraphics[width=0.7\linewidth]{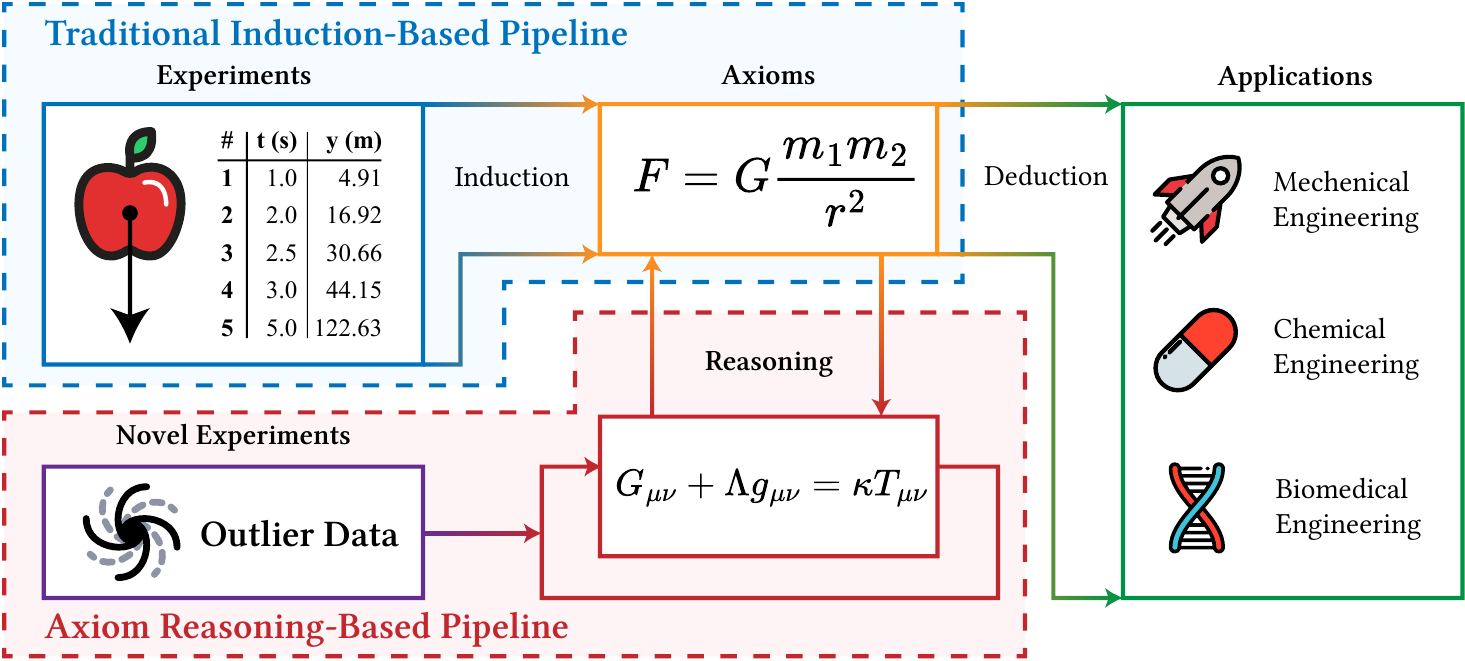}
    \caption{\textbf{Illustration of Two Pipelines for Scientific Discovery}. Traditional pipeline derives axioms through statistical inference from extensive experimental data, while our proposed pipeline modifies existing axioms through iterative reasoning when encountering new observations, leveraging language models to reduce data requirements.}
    \vspace{-2mm}
    \label{fig:pipeline}
\end{figure}

In this section, we examine the contrasting approaches to scientific discovery. We visualize the pipeline and the comparison in Fig.~\ref{fig:pipeline}.


\subsection{Induction-Based Pipeline}

The traditional approach follows three resource-intensive steps:
\textit{(1)} \textbf{Experiment}. Researchers design repeatable experiments with variable inputs hypothesized to determine outputs, requiring controlled environments and systematic parameter variation. \textit{(2)} \textbf{Observation}. Multiple experimental iterations accumulate comprehensive datasets capturing the full range of input-output relationships across all conditions. \textit{(3)} \textbf{Induction}. Scientists identify patterns to formulate axiom sets satisfying three criteria: completeness (explaining all observed data), compactness (minimal number of axioms), and interpretability (meaningful principles).

This process can be formalized mathematically. Given a dataset with inputs $\mathbf{x}$ and outputs $g(\mathbf{x},\epsilon)$, where $g(\cdot)$ maps inputs to outputs and $\epsilon$ represents noise or unmeasured effects, the goal is discovering a finite axiom set $\{\mathbf{A}\}$ that explains all observations. This requires the dataset $\langle\mathbf{x},g(\mathbf{x},\epsilon)\rangle$ to lie within the space spanned by the axioms: $\langle\mathbf{x},g(\mathbf{x},\epsilon)\rangle\in\text{span}(\{\mathbf{A}\})$.

Incorporating the preference for compact axiom sets through minimizing the $0$-norm $\mathcal{L}_\text{compact}=||\{\mathbf{A}\}||_0$, the optimization problem becomes:
\begin{align}
    \{\mathbf{A}\}={\arg\min}_k||\{\mathbf{A}\}_k||_0, \quad
    \text{s.t.} \quad \langle\mathbf{x},g(\mathbf{x},\epsilon)\rangle\in\text{span}(\{\mathbf{A}\}_k)
\end{align}
where $k$ indexes all possible axiom combinations. This formulation captures both complete observational coverage and parsimony in scientific explanations.

\subsection{Axiom Reasoning-Based Pipeline}

Our approach builds upon existing theories rather than starting from scratch, modifying established axioms to accommodate new observations. This is especially efficient in fields with strong foundations, preserving valuable aspects of current theories while enabling targeted improvements. The axiom reasoning-based process consists of three key stages:

\begin{enumerate}[align=right,itemindent=0em,labelsep=2pt,labelwidth=1em,leftmargin=*,itemsep=0em]
\item \textbf{Outlier Detection}. During exploration, scientists encounter observations beyond current axiom coverage. These outliers highlight theory limitations and guide targeted experiments to study anomalies systematically.
\item \textbf{Axiom Modification}. Scientists modify existing axioms to accommodate outliers while preserving established explanations. This process relies on scientific intuition, minimizing deviation from original axioms while maximizing explanatory power.
\item \textbf{Iterative Refinement}. Each outlier contributes to theoretical evolution through gradual refinement rather than replacement, maintaining consistency while efficiently incorporating new information.
\end{enumerate}

\textbf{Mathematical Formulation:} Let $\hat{\mathbf{x}}, g(\hat{\mathbf{x}},\epsilon)$ represent new observations and $\{\mathbf{A}\}^n$ the axiom set at iteration $n$. When $\langle\hat{\mathbf{x}}, g(\hat{\mathbf{x}},\epsilon)\rangle \notin \text{span}(\{\mathbf{A}\}^n)$, we seek updated axioms encompassing both existing and new observations:
\begin{align}
    \{\langle\mathbf{x}, g(\mathbf{x},\epsilon)\rangle, \langle\hat{\mathbf{x}}, g(\hat{\mathbf{x}},\epsilon)\rangle\} \in \text{span}(\{\mathbf{A}\}^{n+1})
\end{align}

To ensure minimal deviation, we introduce locality regularization $\mathcal{L}_{\text{local}} = \|\{\mathbf{A}\}^{n+1} - \{\mathbf{A}\}^n\|$, where distance is defined semantically. The complete optimization combines compactness and locality:
\begin{align}
    \{\mathbf{A}\} = \arg\min_k \left( \|\{\mathbf{A}\}_k\|_0 + \alpha \|\{\mathbf{A}\}^{n+1}_k - \{\mathbf{A}\}^n\| \right)
\end{align}
subject to the spanning constraint above, where $\alpha$ balances compactness and stability. Starting from an initial axiom set ($n=0$), typically established through traditional scientific methods, this formulation enables an automated, iterative process for scientific axiom reasoning.

\subsection{From Scientific Discovery to Game Design}

Games provide structured environments that share fundamental characteristics with scientific systems, making them ideal testbeds for axiom reasoning:

\begin{enumerate}[align=right,itemindent=0em,labelsep=2pt,labelwidth=1em,leftmargin=*,itemsep=0em]
\item \textbf{Repeatable runs}. Games enable systematic data collection through repeatable executions.
\item \textbf{Rule-based system}. Games operate on explicit  rules that directly parallel scientific axioms.
\item \textbf{Deductive generalization}. Finite rules generate infinite valid states, similar to how physical laws describe natural phenomena.
\item \textbf{Goal composition}. Clear objectives guide exploration and provide evaluation metrics, paralleling scientific hypothesis testing.
\end{enumerate}

\textbf{Key Advantages:} Games offer tractability--we can precisely define success criteria, automatically verify outcomes, and systematically explore rule modifications. This enables studying axiom-based reasoning in controlled environments with immediate feedback on rule changes, making them ideal for developing automated reasoning strategies.

\textbf{Framework Components:} We decompose game systems into four essential elements that directly parallel scientific discovery:

\begin{enumerate}[align=right,itemindent=0em,labelsep=2pt,labelwidth=1em,leftmargin=*,itemsep=0em]
    \item \textbf{Game rules} serve as fundamental axioms governing system behavior and defining valid transitions.
    \item \textbf{Initial setups} specify starting conditions, analogous to experimental configurations.
    \item \textbf{Simulator} executes rules to generate trajectories, comparable to natural process unfolding.
    \item \textbf{Goal} defines desired outcomes that guide exploration, similar to research objectives.
\end{enumerate}

\textbf{Optimization Framework:} Our approach maintains simulator mechanics and goals while focusing on rule optimization, mirroring how scientific theories evolve while observations remain constant. The game solver finds optimal trajectories within current rules, like natural phenomena following least-action paths. We frame this as an optimization task: modifying rules to achieve previously impossible outcomes while minimizing solution path length and incorporating regularization terms for rule compactness and deviation from originals. This formulation captures the balance between improving explanatory power, maintaining elegance, and preserving stability in scientific discovery.

\subsection{Planning Domain Definition Language}

We formalize game axioms using the Planning Domain Definition Language (PDDL), a standard language for classical AI planning~\cite{gel98, aeronautiques1998pddl}. PDDL provides an ideal formalism for our axiom reasoning framework due to its explicit separation of rules from problem instances and its mature ecosystem of automated reasoning tools.
\paragraph{PDDL Structure} The language represents systems through states (defined by predicates) and actions (rules for state transitions), where actions execute when specific preconditions are satisfied. PDDL separates problems into two components: \textit{(1)} a domain file specifying predicates and actions (analogous to our game rules/axioms), and \textit{(2)} a problem file defining initial and goal states (analogous to our experimental setups and objectives).
\paragraph{Framework Alignment} This structure aligns with our framework while enabling efficient AI planners as simulators. The explicit rule representation facilitates axiom modification, while the domain-problem separation allows testing across multiple scenarios~\cite{malte2006fastdownward}.
\paragraph{Advantages for Axiom Reasoning} PDDL offers several key benefits: \textit{(1)} \textbf{Explicit axiom representation} enables direct manipulation of game rules, \textit{(2)} \textbf{Formal semantics} ensure consistent interpretation across modifications, \textit{(3)} \textbf{Automated verification} through planners validates rule set completeness, and \textit{(4)} \textbf{ Established toolchain} provides efficient simulators for trajectory generation.
\paragraph{Example Application} Consider a blocks world domain where the original axiom "move block A onto block B" requires A to be clear (nothing on top). Our framework can systematically modify this rule to "move block A from middle of stack," creating new solution possibilities while maintaining formal consistency through PDDL's structured representation.

\begin{figure*}[t]
    \centering
    \includegraphics[width=\linewidth]{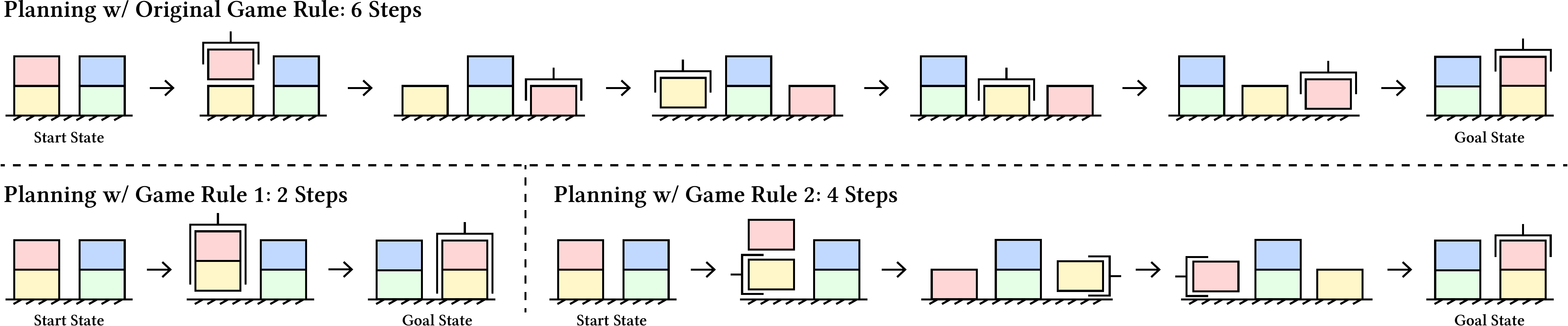}
    \vspace{-4mm}
    \caption{\textbf{Illustration of Game Design}. The top row demonstrates a conventional blocks world planning trajectory under original rules, requiring six steps to reach the goal state. Our approach proposes two elegant rule updates that simplify the solution: (1) enabling simultaneous lifting of multiple blocks (bottom left; two steps required), and (2) allowing blocks to be extracted from the middle of a stack (bottom right; four steps required).}
    \vspace{-4mm}
    \label{fig:game}
\end{figure*}

\subsection{Distance Function}

Measuring distance between axiom sets requires understanding semantic relationships and domain knowledge--a core challenge in our axiom reasoning pipeline. We propose and evaluate several approaches for implementing this distance function, where the metric choice influences how rule modifications are evaluated, similar to how scientific communities assess theoretical changes.
\begin{enumerate}[align=right,itemindent=0em,labelsep=2pt,labelwidth=1em,leftmargin=*,itemsep=0em]
\item \textbf{Levenshtein Distance} Levenshtein distance measures string similarity by counting minimum single-character edits. While computationally simple and readily applicable to PDDL-formatted rules, it suffers from critical limitations. For example, changing \texttt{(and (clear ?x) (clear ?y))} to \texttt{(or (clear ?x) (clear ?y))} requires only three character edits but fundamentally alters rule semantics from requiring both conditions to accepting either. This syntactic-semantic disconnect makes Levenshtein inadequate for meaningful axiom evaluation.
\item \textbf{Semantic Relative Distance} We propose a semantics-aware approach addressing two key requirements: \textit{(1)} accounting for rule implications through deductive expansions, and \textit{(2)} prioritizing relative rankings over absolute values. Our method employs language models to perform atomic comparisons between rule sets based on logical implications and behavioral outcomes.
\end{enumerate}
\paragraph{Implementation} Given two rule sets $\mathbf{R}_1$ and $\mathbf{R}_2$ and reference set $\widehat{\mathbf{R}}$, we query a language model: \textit{"Which rule set is semantically closer to the reference: $\mathbf{R}_1$ or $\mathbf{R}_2$?"} The model analyzes rule structures, preconditions, and effects to determine relative proximity. This atomic comparison integrates into standard sorting algorithms (merge sort, quicksort) achieving $O(n\log n)$ complexity for ordering multiple rule sets.
\paragraph{Example Application} Consider reference rule \texttt{(move ?x ?y)}, and modifications: $\mathbf{R}_1$ adds precondition \texttt{(adjacent ?x ?y)} while $\mathbf{R}_2$ removes precondition \texttt{(clear ?y)}. The language model evaluates that $\mathbf{R}_1$ represents a smaller semantic change (adding constraint vs. removing fundamental requirement), ranking it closer to the reference.
\paragraph{Hybrid Approach} For practical implementation, we combine both metrics: Levenshtein distance provides fast initial filtering, while semantic distance handles final ranking among candidates. This hybrid approach balances computational efficiency with semantic accuracy, enabling scalable axiom modification evaluation in our framework.

\subsection{Axiom Search Algorithm}

\paragraph{Search Algorithms with Language Models}
Scientific discovery requires both semantic reasoning and creative thinking. Our system leverages LLMs to generate axiom modifications, with classical search algorithms organizing and evaluating these proposals. All algorithms use LLMs by prompting with current axioms, failures, and desired outcomes to suggest specific PDDL rule changes. We implemented the following search strategy variants.
\begin{itemize}[align=right,itemindent=0em,labelsep=2pt,labelwidth=1em,leftmargin=*,itemsep=0em]
\item \textbf{Breadth-First Search} Systematically explores single-rule modifications before multiple-rule changes. Guarantees minimal solutions but scales poorly beyond depth 3-4.
\item \textbf{Monte Carlo Tree Search} Uses UCB1 selection and random rollouts to balance exploitation and exploration. Effective for complex domains requiring creative modification sequences.
\item \textbf{Genetic Algorithms} Uniquely employs LLMs for both crossover and mutation operations. Suited for open-ended exploration.
\item \textbf{Beam Search} Maintains top-k candidates based on success rate and semantic distance, offering efficient focused exploration.
\end{itemize}

\paragraph{Algorithm Selection}
Choice depends on domain characteristics: BFS for minimal changes in simple domains, MCTS for complex creative solutions, genetic algorithms for open-ended exploration, and beam search for resource-constrained applications.

\section{Preliminary Experiments}
\label{sec:exp}

To demonstrate the practical feasibility of axiom-based reasoning for game design, we developed an integrated experimental platform combining three key components: \textit{(1)} a semantic relative distance function powered by language models to evaluate axiom modifications, \textit{(2)} various language model-driven search algorithms for exploring rule modifications, and \textit{(3)} a comprehensive suite of 12 classic logical games (e.g., blocks world~\citep{russell2003artificial}) as test environments\footnote{We will release the codebase and the generated dataset.}.

\subsection{Experimental Platform}

Our modular platform enables experimentation with different language models, search strategies, and game domains. The architecture combines semantic distance functions for evaluating modifications, search algorithms for exploring rule changes, and a game suite for validation.

\subsection{Blocks World Case Study}

We demonstrate our approach using the blocks world domain with OpenAI's \texttt{gpt-4o-mini-2024-07-18} as the backbone language model~\citep{hurst2024gpt}. Our experimental configuration employed semantic relative distance as the distance metric and beam search with width 8 as the search algorithm. To ensure statistical stability in the reasoning process, we repeated the decoding process 16 times. The original blocks world rules require a 6-step planning trajectory to reach the specified goal state. We challenged our system to discover rule modifications that would enable solutions within 4 steps, creating a concrete optimization target that mirrors how scientific theories evolve to explain phenomena more efficiently.

\paragraph{Discovered Modifications} Our pipeline successfully identified two distinct categories of effective rule modifications:
\begin{enumerate}[align=right,itemindent=0em,labelsep=2pt,labelwidth=1em,leftmargin=*,itemsep=0em]
\item \textbf{Simultaneous manipulation:} Allowing simultaneous manipulation of multiple blocks dramatically reduces the solution to just 2 steps while maintaining physical plausibility. This modification demonstrates how our system can identify non-obvious extensions to existing rules that preserve the domain's fundamental constraints.
\item \textbf{Stack extraction:} Enabling direct extraction of blocks from the middle of stacks achieves a 4-step solution by introducing new manipulation capabilities. This modification shows the system's ability to identify targeted rule relaxations that address specific solution bottlenecks.
\end{enumerate}
\textbf{Significance:} Both modifications show how our system finds meaningful rule changes that improve efficiency while preserving core puzzle mechanics. The changes extend existing capabilities coherently rather than adding arbitrary mechanics, validating our hypothesis that systematic axiom modification can enhance performance while maintaining essential characteristics.

\textbf{Implications for Scientific Discovery:} This case study illustrates how our framework could assist in scientific contexts where existing theories need modification to accommodate new observations or requirements. The process mirrors how Einstein modified Newtonian mechanics—preserving core principles while extending capabilities to handle previously intractable scenarios. Additional experimental details and results from the complete 12-game suite are provided in the appendix.


\section{Related Work}
\label{sec:related}

The integration of artificial intelligence into scientific discovery has enhanced efficiency by automating hypothesis generation, experiment design, and data analysis. AI-driven approaches accelerate both inductive learning, which extracts governing principles from experimental data, and deductive reasoning, which validates principles against established axioms.

Recent advances in differentiable simulation have enabled automated identification of fundamental laws directly from data~\citep{du2021diffpd, ma2023learning, ma2021risp, ma2024llm}, while molecular design has leveraged machine learning for structure prediction and optimization~\citep{jin2018junction, zhou2019optimization, schneider2018automating, jumper2021highly, roy2024alphafold3}. Symbolic regression methods such as AI Feynman~\citep{udrescu2020ai} aim to recover equations from data, but do not support dynamic axiom evolution based on semantic reasoning or LLM-driven proposal generation. These approaches typically require extensive datasets and focus on parameter identification rather than axiom modification.

Our work relates to several distinct research directions. Classical AI planning using PDDL~\citep{gel98, aeronautiques1998pddl} provides our formal foundation, though traditional planning assumes fixed rule sets rather than modifying axioms themselves. Automated theorem proving seeks to discover mathematical theorems~\citep{trinh2024solving} but works within formal systems rather than accommodating new observations through axiom modification.

Molecular grammar approaches encode domain constraints in explicit production rules~\citep{kajino2019molecular, guo2022data, guo2023hierarchical, sun2024representing}, offering compact, interpretable representations that align with axiomatic reasoning paradigms. These grammars demonstrate the value of structure, but lack the iterative, semantically-informed axiom refinement we enable. Whether manually designed or learned from data, they represent static rule sets rather than dynamic modification frameworks.

While games have traditionally served as AI testbeds with fixed rules, we treat them as systems with modifiable axioms. Language models' semantic reasoning capabilities~\citep{hurst2024gpt} enable axiom modification proposals, extending beyond traditional NLP applications.

Unlike traditional AI4Science approaches that rely on inductive learning from large datasets, our framework modifies established axioms systematically. This provides computational efficiency, interpretability through explicit rules, and alignment with historical scientific processes, while our game implementation offers a controlled testbed for axiom reasoning.


\begin{table}
    \tiny
    \centering
    \begin{tabular}{|c|c|c|c|}\hline
           \hyperref[sec:flat_to_sphere]{Flat $\to$ Spherical Earth}&  \hyperref[sec:ml]{Neural Networks $\to$ Transformers}&  \hyperref[sec:ray_to_wave]{Ray $\to$ Wave Optics}&  \hyperref[sec:wave_to_quantum]{Wave $\to$ Quantum Optics}\\\hline
          \hyperref[sec:Newtonian_to_Hamiltonian]{Newtonian $\to$ Hamiltonian Mechanics} & \hyperref[sec:Hamiltonian_to_quantum]{Hamiltonian $\to$ Quantum Mechanics} &  \hyperref[sec:genetics_to_molecular]{Classical $\to$ Molecular Biology}&  \hyperref[sec:Euclidean_to_Hyperbolic]{Euclidean $\to$ Hyperbolic Geometry}\\\hline
           \hyperref[sec:heliocentric_to_cosmological]{Heliocentric $\to$ Cosmological Model}&  \hyperref[sec:fourier_to_wavelet]{Fourier analysis $\to$ Wavelet Theory}&  \hyperref[sec:photography_digital]{Analog $\to$ Digital Imaging}& \hyperref[sec:conventional_metamaterials]{Metals $\to$ Metamaterials} \ref{sec:wave_to_quantum}\\\hline
  \hyperref[sec:chemistry_to_bohr]{Classical Chemistry $\to$ Bohr Theory}& \hyperref[sec:geocentric_to_heliocentric]{Geocentric $\to$ Heliocentric Model}& \hyperref[sec:newtonian_general]{Newtonian Physics $\to$ Relativity}& \hyperref[sec:classical_to_statistical]{Classical $\to$ Statistical Thermodynamics}\\\hline
  \hyperref[sec:analog_to_digital]{Analog $\to$ Digital Electronics}& \hyperref[sec:digital_to_quantum]{Digital $\to$ Quantum Computing}& & \\\hline
    \end{tabular}
    \caption{Summary of transformations presented in the appendix. In each case, we demonstrate gradual changes of axioms that lead from one theory to another.}
    \label{tab:transf_links}
    \vspace{-5mm}
\end{table}

\section{From Game Design to Scientific Discovery}
\label{sec:science_application}

Our game framework extends naturally to scientific discovery, viewing scientific theories as rule sets governing phenomena, with experiments serving as validation cases that verify or challenge these rules. Consider the historical transition from flat Earth to spherical Earth models, detailed in Appendix~\ref{sec:flat_to_sphere}. This transformation exemplifies our framework's core components in action. The initial ``rule'' represents the assumed Earth geometry--flat versus spherical. The experimental setup involves strategically placed sticks casting shadows at different latitudes, designed to discriminate between competing geometric models. The simulator employs ray geometry to predict shadow angles under each geometric assumption. The ``goal'' requires accurate prediction of observed shadow patterns. Critically, the flat Earth model fails to achieve this goal - predicted and observed angles diverge systematically. This failure triggers our axiom modification process: updating the geometric rule from flat to spherical Earth. The modified theory successfully explains the observations while maintaining theoretical continuity--the spherical model approaches the flat model in the limit of infinite radius ($R \to \infty$), preserving local accuracy while enabling global explanatory power. This example validates three key aspects of our approach. First, it demonstrates how established theories (flat Earth) can systematically evolve to accommodate anomalous observations rather than requiring wholesale abandonment. Second, it shows how our framework naturally captures the scientific preference for minimal modifications that preserve existing explanatory success while extending to new phenomena. Third, it illustrates how the game analogy makes explicit the typically implicit process of theory revision in response to empirical challenges. The general pattern - initial rule inadequacy, systematic modification, improved explanatory coverage - applies broadly across scientific domains. Table~\ref{tab:transf_links} catalogs additional historical transformations following this pattern, from Newtonian to relativistic mechanics, classical to quantum theory, and geocentric to heliocentric models. Each transformation demonstrates how our game-theoretic framework captures fundamental aspects of scientific progress through axiom reasoning.

\section{Alternative Views}
\label{sec:alternative}
While we advocate for axiom-based reasoning, several alternative perspectives merit consideration.
\paragraph{Pure Data-Driven Discovery} Modern machine learning's pattern recognition capabilities might suggest that axiom-based reasoning is obsolete. Deep learning models can discover complex patterns without explicit axioms, yet they often lack interpretability and theoretical grounding—qualities that become crucial when data is limited or causal understanding is essential.
\paragraph{Hybrid Approaches Over Pure Axiom Reasoning} 
Consider neural architecture evolution from fully connected networks ($Wx$) to attention mechanisms ($QK^TV$), described in Appendix~\ref{sec:ml}. While we frame this as axiom modification, it equally represents empirical optimization driven by experimental results on large datasets. The success of these architectures emerged from combining theoretical insights with empirical validation rather than pure axiom reasoning. This suggests that effective scientific discovery may require integrating both theoretical frameworks and data-driven exploration, rather than relying solely on either approach.
\paragraph{Discontinuous Innovation}
Some breakthrough discoveries arise from discontinuous thinking beyond systematic axiom modification, such as penicillin's accidental discovery or quantum mechanics' conceptual revolution. These examples challenge the assumption that scientific progress always follows gradual evolution. While our framework demonstrates the value of axiom-based reasoning, a comprehensive approach to automated scientific discovery may need to combine multiple strategies, including data-driven methods and mechanisms for discontinuous innovation.

\section{Limitations and Future Work}
\label{sec:con}

\subsection{Current Limitations}

\textit{(1)} \textbf{Domain Scope:} While effective with established axioms, the approach struggles in emerging fields like computational biology where theoretical foundations are unclear. \textit{(2)} \textbf{Discontinuous Innovation:} The framework cannot capture paradigmatic shifts requiring complete axiom replacement rather than modification, as discussed in Section~\ref{sec:alternative}. \textit{(3)} \textbf{Semantic Distance Metrics:} Current language model-based distance functions may inadequately capture complex theoretical relationships requiring domain expertise. \textit{(4)} \textbf{Scalability Challenges:} Game environments simplify real scientific complexity; scaling to domains with numerous interacting axioms and multi-scale phenomena remains challenging.

\subsection{Future Research Directions}

\textit{(1)} \textbf{Hybrid Reasoning Systems:} Integrate discontinuous innovation mechanisms through neural-symbolic architectures that can propose entirely new axiom categories, not just modifications of existing ones. \textit{(2)} \textbf{Domain-Specific Extensions:} Develop specialized implementations for specific scientific fields, incorporating domain-appropriate distance metrics and constraint systems. Physics, chemistry, and biology each require tailored approaches to axiom representation and modification. \textit{(3)} \textbf{Human-AI Collaboration:} Design interactive frameworks where domain experts guide axiom modification processes, combining human intuition with systematic computational exploration. \textit{(4)} \textbf{Empirical Validation:} Test the framework on historical scientific discoveries with known outcomes, systematically evaluating whether our approach can reproduce major theoretical transitions under controlled conditions.

\subsection{Broader Impact}

This work contributes to the growing intersection of AI and scientific methodology, offering a systematic approach to theory evolution that complements existing data-driven discovery methods. As scientific domains increasingly rely on computational assistance, frameworks emphasizing interpretable, knowledge-building approaches may prove essential for maintaining scientific understanding alongside accelerated discovery.

\section*{Impact Statement}

This paper presents work whose goal is to advance the field of 
Machine Learning. There are many potential societal consequences 
of our work, but none we necessitate highlighting here.

\nocite{langley00}

\bibliographystyle{unsrtnat}
\small{
\bibliography{reference}

\begin{thebibliography}{31}
\providecommand{\natexlab}[1]{#1}
\providecommand{\url}[1]{\texttt{#1}}
\expandafter\ifx\csname urlstyle\endcsname\relax
  \providecommand{\doi}[1]{doi: #1}\else
  \providecommand{\doi}{doi: \begingroup \urlstyle{rm}\Url}\fi

\bibitem[Jumper et~al.(2021)Jumper, Evans, Pritzel, Green, Figurnov,
  Ronneberger, Tunyasuvunakool, Bates, {\v{Z}}{\'\i}dek, Potapenko,
  et~al.]{jumper2021highly}
John Jumper, Richard Evans, Alexander Pritzel, Tim Green, Michael Figurnov,
  Olaf Ronneberger, Kathryn Tunyasuvunakool, Russ Bates, Augustin
  {\v{Z}}{\'\i}dek, Anna Potapenko, et~al.
\newblock Highly accurate protein structure prediction with alphafold.
\newblock \emph{nature}, 596\penalty0 (7873):\penalty0 583--589, 2021.

\bibitem[Roy and Al-Hashimi(2024)]{roy2024alphafold3}
Rohit Roy and Hashim~M Al-Hashimi.
\newblock Alphafold3 takes a step toward decoding molecular behavior and
  biological computation.
\newblock \emph{Nature Structural \& Molecular Biology}, 31\penalty0
  (7):\penalty0 997--1000, 2024.

\bibitem[Udrescu and Tegmark(2020)]{udrescu2020ai}
Silviu-Marian Udrescu and Max Tegmark.
\newblock Ai feynman: A physics-inspired method for symbolic regression.
\newblock \emph{Science Advances}, 6\penalty0 (16):\penalty0 eaay2631, 2020.

\bibitem[Schmidt and Lipson(2009)]{schmidt2009distilling}
Michael Schmidt and Hod Lipson.
\newblock Distilling free-form natural laws from experimental data.
\newblock \emph{science}, 324\penalty0 (5923):\penalty0 81--85, 2009.

\bibitem[Cranmer et~al.(2020)Cranmer, Sanchez~Gonzalez, Battaglia, Xu, Cranmer,
  Spergel, and Ho]{cranmer2020discovering}
Miles Cranmer, Alvaro Sanchez~Gonzalez, Peter Battaglia, Rui Xu, Kyle Cranmer,
  David Spergel, and Shirley Ho.
\newblock Discovering symbolic models from deep learning with inductive biases.
\newblock \emph{Advances in neural information processing systems},
  33:\penalty0 17429--17442, 2020.

\bibitem[King et~al.(2009)King, Rowland, Oliver, Young, Aubrey, Byrne, Liakata,
  Markham, Pir, Soldatova, et~al.]{king2009automation}
Ross~D King, Jem Rowland, Stephen~G Oliver, Michael Young, Wayne Aubrey, Emma
  Byrne, Maria Liakata, Magdalena Markham, Pinar Pir, Larisa~N Soldatova,
  et~al.
\newblock The automation of science.
\newblock \emph{Science}, 324\penalty0 (5923):\penalty0 85--89, 2009.

\bibitem[Haig(1995)]{haig1995grounded}
Brian~D Haig.
\newblock Grounded theory as scientific method.
\newblock \emph{Philosophy of education}, 28\penalty0 (1):\penalty0 1--11,
  1995.

\bibitem[Featherstone(2014)]{featherstone2014rigid}
Roy Featherstone.
\newblock \emph{Rigid body dynamics algorithms}.
\newblock Springer, 2014.

\bibitem[Siciliano et~al.(2008)Siciliano, Khatib, and
  Kr{\"o}ger]{siciliano2008springer}
Bruno Siciliano, Oussama Khatib, and Torsten Kr{\"o}ger.
\newblock \emph{Springer handbook of robotics}, volume 200.
\newblock Springer, 2008.

\bibitem[Spong et~al.(2020)Spong, Hutchinson, and Vidyasagar]{spong2020robot}
Mark~W Spong, Seth Hutchinson, and Mathukumalli Vidyasagar.
\newblock \emph{Robot modeling and control}.
\newblock John Wiley \& Sons, 2020.

\bibitem[Ferreira et~al.(2015)Ferreira, Dos~Santos, Oliva, and
  Andricopulo]{ferreira2015molecular}
Leonardo~G Ferreira, Ricardo~N Dos~Santos, Glaucius Oliva, and Adriano~D
  Andricopulo.
\newblock Molecular docking and structure-based drug design strategies.
\newblock \emph{Molecules}, 20\penalty0 (7):\penalty0 13384--13421, 2015.

\bibitem[Drews(2000)]{drews2000drug}
Jurgen Drews.
\newblock Drug discovery: a historical perspective.
\newblock \emph{science}, 287\penalty0 (5460):\penalty0 1960--1964, 2000.

\bibitem[Lloyd(2012)]{lloyd2012foundations}
John~W Lloyd.
\newblock \emph{Foundations of logic programming}.
\newblock Springer Science \& Business Media, 2012.

\bibitem[Gelfond and Lifschitz(1998)]{gel98}
Michael Gelfond and Vladimir Lifschitz.
\newblock Action languages.
\newblock \emph{Electronic Transactions on Artificial Intelligence},
  3:\penalty0 195--210, 1998.
\newblock URL \url{http://www.cs.utexas.edu/users/ai-lab?gel98}.

\bibitem[Aeronautiques et~al.(1998)Aeronautiques, Howe, Knoblock, McDermott,
  Ram, Veloso, Weld, Sri, Barrett, Christianson, et~al.]{aeronautiques1998pddl}
Constructions Aeronautiques, Adele Howe, Craig Knoblock, ISI~Drew McDermott,
  Ashwin Ram, Manuela Veloso, Daniel Weld, David~Wilkins Sri, Anthony Barrett,
  Dave Christianson, et~al.
\newblock Pddl| the planning domain definition language.
\newblock \emph{Technical Report, Tech. Rep.}, 1998.

\bibitem[Helmert(2006)]{malte2006fastdownward}
Malte Helmert.
\newblock The fast downward planning system.
\newblock \emph{J. Artif. Int. Res.}, 26\penalty0 (1):\penalty0 191–246, July
  2006.
\newblock ISSN 1076-9757.

\bibitem[Russell and Norvig(2003)]{russell2003artificial}
SJ~Russell and P~Norvig.
\newblock Artificial intlligence: A modern approach (2nd).
\newblock \emph{Artificial Intelligence and Machine Learning Book}, 2003.

\bibitem[Hurst et~al.(2024)Hurst, Lerer, Goucher, Perelman, Ramesh, Clark,
  Ostrow, Welihinda, Hayes, Radford, et~al.]{hurst2024gpt}
Aaron Hurst, Adam Lerer, Adam~P Goucher, Adam Perelman, Aditya Ramesh, Aidan
  Clark, AJ~Ostrow, Akila Welihinda, Alan Hayes, Alec Radford, et~al.
\newblock Gpt-4o system card.
\newblock \emph{arXiv preprint arXiv:2410.21276}, 2024.

\bibitem[Du et~al.(2021)Du, Wu, Ma, Wah, Spielberg, Rus, and
  Matusik]{du2021diffpd}
Tao Du, Kui Wu, Pingchuan Ma, Sebastien Wah, Andrew Spielberg, Daniela Rus, and
  Wojciech Matusik.
\newblock Diffpd: Differentiable projective dynamics.
\newblock \emph{ACM Transactions on Graphics (TOG)}, 41\penalty0 (2):\penalty0
  1--21, 2021.

\bibitem[Ma et~al.(2023)Ma, Chen, Deng, Tenenbaum, Du, Gan, and
  Matusik]{ma2023learning}
Pingchuan Ma, Peter~Yichen Chen, Bolei Deng, Joshua~B Tenenbaum, Tao Du, Chuang
  Gan, and Wojciech Matusik.
\newblock Learning neural constitutive laws from motion observations for
  generalizable pde dynamics.
\newblock In \emph{International Conference on Machine Learning}. PMLR, 2023.

\bibitem[Ma et~al.(2021)Ma, Du, Tenenbaum, Matusik, and Gan]{ma2021risp}
Pingchuan Ma, Tao Du, Joshua~B Tenenbaum, Wojciech Matusik, and Chuang Gan.
\newblock Risp: Rendering-invariant state predictor with differentiable
  simulation and rendering for cross-domain parameter estimation.
\newblock In \emph{International Conference on Learning Representations}, 2021.

\bibitem[Ma et~al.(2024)Ma, Wang, Guo, Sun, Tenenbaum, Rus, Gan, and
  Matusik]{ma2024llm}
Pingchuan Ma, Tsun-Hsuan Wang, Minghao Guo, Zhiqing Sun, Joshua~B Tenenbaum,
  Daniela Rus, Chuang Gan, and Wojciech Matusik.
\newblock Llm and simulation as bilevel optimizers: A new paradigm to advance
  physical scientific discovery.
\newblock \emph{arXiv preprint arXiv:2405.09783}, 2024.

\bibitem[Jin et~al.(2018)Jin, Barzilay, and Jaakkola]{jin2018junction}
Wengong Jin, Regina Barzilay, and Tommi Jaakkola.
\newblock Junction tree variational autoencoder for molecular graph generation.
\newblock In \emph{International conference on machine learning}, pages
  2323--2332. PMLR, 2018.

\bibitem[Zhou et~al.(2019)Zhou, Kearnes, Li, Zare, and
  Riley]{zhou2019optimization}
Zhenpeng Zhou, Steven Kearnes, Li~Li, Richard~N Zare, and Patrick Riley.
\newblock Optimization of molecules via deep reinforcement learning.
\newblock \emph{Scientific reports}, 9\penalty0 (1):\penalty0 10752, 2019.

\bibitem[Schneider(2018)]{schneider2018automating}
Gisbert Schneider.
\newblock Automating drug discovery.
\newblock \emph{Nature reviews drug discovery}, 17\penalty0 (2):\penalty0
  97--113, 2018.

\bibitem[Trinh et~al.(2024)Trinh, Wu, Le, He, and Luong]{trinh2024solving}
Trieu~H Trinh, Yuhuai Wu, Quoc~V Le, He~He, and Thang Luong.
\newblock Solving olympiad geometry without human demonstrations.
\newblock \emph{Nature}, 625\penalty0 (7995):\penalty0 476--482, 2024.

\bibitem[Kajino(2019)]{kajino2019molecular}
Hiroshi Kajino.
\newblock Molecular hypergraph grammar with its application to molecular
  optimization.
\newblock In \emph{International Conference on Machine Learning}, pages
  3183--3191. PMLR, 2019.

\bibitem[Guo et~al.(2022)Guo, Thost, Li, Das, Chen, and Matusik]{guo2022data}
Minghao Guo, Veronika Thost, Beichen Li, Payel Das, Jie Chen, and Wojciech
  Matusik.
\newblock Data-efficient graph grammar learning for molecular generation.
\newblock \emph{arXiv preprint arXiv:2203.08031}, 2022.

\bibitem[Guo et~al.(2023)Guo, Thost, Song, Balachandran, Das, Chen, and
  Matusik]{guo2023hierarchical}
Minghao Guo, Veronika Thost, Samuel~W Song, Adithya Balachandran, Payel Das,
  Jie Chen, and Wojciech Matusik.
\newblock Hierarchical grammar-induced geometry for data-efficient molecular
  property prediction.
\newblock In \emph{International Conference on Machine Learning}, pages
  12055--12076. PMLR, 2023.

\bibitem[Sun et~al.(2024)Sun, Guo, Yuan, Thost, Owens, Grosz, Selvan, Zhou,
  Mohiuddin, Pedretti, et~al.]{sun2024representing}
Michael Sun, Minghao Guo, Weize Yuan, Veronika Thost, Crystal~Elaine Owens,
  Aristotle~Franklin Grosz, Sharvaa Selvan, Katelyn Zhou, Hassan Mohiuddin,
  Benjamin~J Pedretti, et~al.
\newblock Representing molecules as random walks over interpretable grammars.
\newblock \emph{arXiv preprint arXiv:2403.08147}, 2024.

\bibitem[Langley(2000)]{langley00}
P.~Langley.
\newblock Crafting papers on machine learning.
\newblock In Pat Langley, editor, \emph{Proceedings of the 17th International
  Conference on Machine Learning (ICML 2000)}, pages 1207--1216, Stanford, CA,
  2000. Morgan Kaufmann.

\end{thebibliography}
}



\newpage

\appendix


\section*{Appendix: Table of Contents}
\contentsline {section}{\numberline {A}Implementation Details}{18}{appendix.A}%
\contentsline {subsection}{\numberline {A.1}Implemented Games}{18}{subsection.A.1}%
\contentsline {subsubsection}{\numberline {A.1.1}Blocks World}{18}{subsubsection.A.1.1}%
\contentsline {subsubsection}{\numberline {A.1.2}Briefcase}{18}{subsubsection.A.1.2}%
\contentsline {subsubsection}{\numberline {A.1.3}Bulldozer}{18}{subsubsection.A.1.3}%
\contentsline {subsubsection}{\numberline {A.1.4}Casino}{19}{subsubsection.A.1.4}%
\contentsline {subsubsection}{\numberline {A.1.5}Depot}{20}{subsubsection.A.1.5}%
\contentsline {subsubsection}{\numberline {A.1.6}Ferry}{20}{subsubsection.A.1.6}%
\contentsline {subsubsection}{\numberline {A.1.7}Gripper}{21}{subsubsection.A.1.7}%
\contentsline {subsubsection}{\numberline {A.1.8}Hanoi}{22}{subsubsection.A.1.8}%
\contentsline {subsubsection}{\numberline {A.1.9}Logistics}{22}{subsubsection.A.1.9}%
\contentsline {subsubsection}{\numberline {A.1.10}Maze}{23}{subsubsection.A.1.10}%
\contentsline {subsubsection}{\numberline {A.1.11}Miconic}{23}{subsubsection.A.1.11}%
\contentsline {subsubsection}{\numberline {A.1.12}Monkey}{24}{subsubsection.A.1.12}%
\contentsline {subsection}{\numberline {A.2}Discussion on the Game Proposal}{25}{subsection.A.2}%
\contentsline {paragraph}{Symbolic Reasoning Games}{25}{section*.22}%
\contentsline {paragraph}{Observation-Driven Games}{25}{section*.23}%
\contentsline {paragraph}{Multi-Scale Reasoning Games}{25}{section*.24}%
\contentsline {paragraph}{Constraint Satisfaction Games}{25}{section*.25}%
\contentsline {paragraph}{Theory Unification Games}{25}{section*.26}%
\contentsline {subsection}{\numberline {A.3}Empirical Study of Different Searching Algorithms}{26}{subsection.A.3}%
\contentsline {paragraph}{Time Consumption Analysis}{26}{section*.27}%
\contentsline {paragraph}{Performance Comparison}{26}{section*.28}%
\contentsline {paragraph}{Algorithm-Specific Characteristics}{26}{section*.29}%
\contentsline {paragraph}{Resource-Performance Trade-offs}{26}{section*.30}%
\contentsline {paragraph}{Practical Recommendations}{26}{section*.31}%
\contentsline {section}{\numberline {B}General Introduction to Axiom Transformations}{26}{appendix.B}%
\contentsline {section}{\numberline {C}Illustrative Example: Flat Earth to Spherical Earth}{27}{appendix.C}%
\contentsline {subsection}{\numberline {C.1}Game Abstraction for Axiom-Based Discovery}{27}{subsection.C.1}%
\contentsline {subsection}{\numberline {C.2}Game Instances: Flat Earth and Spherical Earth Models}{27}{subsection.C.2}%
\contentsline {paragraph}{1. Rule Set (\(\mathcal {R}\))}{28}{section*.32}%
\contentsline {paragraph}{2. Initial Conditions (\(\mathcal {I}\))}{28}{section*.33}%
\contentsline {paragraph}{3. Simulator (\(\mathcal {S}\))}{28}{section*.34}%
\contentsline {paragraph}{4. Goal (\(\mathcal {T}\))}{28}{section*.35}%
\contentsline {subsection}{\numberline {C.3}Prediction Using the Flat Earth Assumption: $\mathcal {S}(\mathcal {R}_{\text {Flat}}, \mathcal {I})$}{28}{subsection.C.3}%
\contentsline {subsection}{\numberline {C.4}Symbolic Axiom Evolution: Flat Earth to Spherical Earth}{29}{subsection.C.4}%
\contentsline {paragraph}{Initial Axiom Set (\(\mathcal {R}_{\text {Flat}}\)): Flat Earth}{29}{section*.36}%
\contentsline {paragraph}{Rewrite Step 1: Introduce Spherical Topology. \newline }{29}{section*.37}%
\contentsline {paragraph}{Final Axiom Set (\(\mathcal {R}_1\)): Spherical Earth}{29}{section*.38}%
\contentsline {paragraph}{Summary of Evolution (compact form):}{29}{section*.39}%
\contentsline {subsection}{\numberline {C.5}Prediction Using the Spherical Earth Assumption: $\mathcal {S}(\mathcal {R}_{\text {Sphere}}, \mathcal {I})$}{29}{subsection.C.5}%
\contentsline {subsection}{\numberline {C.6}Note on the Following Sections}{30}{subsection.C.6}%
\contentsline {section}{\numberline {D}From Neural Networks to Transformers}{30}{appendix.D}%
\contentsline {section}{\numberline {E}From Ray Optics to Wave Optics}{31}{appendix.E}%
\contentsline {subsection}{\numberline {E.1}Ray Optics}{31}{subsection.E.1}%
\contentsline {subsubsection}{\numberline {E.1.1}Axioms}{31}{subsubsection.E.1.1}%
\contentsline {subsubsection}{\numberline {E.1.2}Completeness}{31}{subsubsection.E.1.2}%
\contentsline {subsubsection}{\numberline {E.1.3}Independence}{31}{subsubsection.E.1.3}%
\contentsline {subsection}{\numberline {E.2}Wave Optics}{32}{subsection.E.2}%
\contentsline {subsubsection}{\numberline {E.2.1}Axioms}{32}{subsubsection.E.2.1}%
\contentsline {subsubsection}{\numberline {E.2.2}Completeness}{32}{subsubsection.E.2.2}%
\contentsline {subsubsection}{\numberline {E.2.3}Independence}{32}{subsubsection.E.2.3}%
\contentsline {subsection}{\numberline {E.3}Transformation}{32}{subsection.E.3}%
\contentsline {subsubsection}{\numberline {E.3.1}Axiom 1}{32}{subsubsection.E.3.1}%
\contentsline {subsubsection}{\numberline {E.3.2}Axiom 2}{34}{subsubsection.E.3.2}%
\contentsline {subsubsection}{\numberline {E.3.3}Axiom 3}{36}{subsubsection.E.3.3}%
\contentsline {section}{\numberline {F}From Wave Optics to Quantum Optics}{38}{appendix.F}%
\contentsline {subsection}{\numberline {F.1}Wave Optics}{38}{subsection.F.1}%
\contentsline {subsection}{\numberline {F.2}Quantum Optics}{38}{subsection.F.2}%
\contentsline {subsubsection}{\numberline {F.2.1}Axioms}{38}{subsubsection.F.2.1}%
\contentsline {subsubsection}{\numberline {F.2.2}Completeness}{39}{subsubsection.F.2.2}%
\contentsline {subsubsection}{\numberline {F.2.3}Independence}{39}{subsubsection.F.2.3}%
\contentsline {subsection}{\numberline {F.3}Transformations}{39}{subsection.F.3}%
\contentsline {subsubsection}{\numberline {F.3.1}Axiom 1}{39}{subsubsection.F.3.1}%
\contentsline {paragraph}{Initial Set (Wave Equation).}{39}{section*.40}%
\contentsline {paragraph}{Rewrite Step 1: Mode Expansion.}{39}{section*.41}%
\contentsline {paragraph}{Rewrite Step 2: Promote Amplitudes to Operators.}{39}{section*.42}%
\contentsline {paragraph}{Rewrite Step 3: Canonical Commutation.}{40}{section*.43}%
\contentsline {paragraph}{Final Set (Quantum Optics Axiom 1: Field Quantization).}{40}{section*.44}%
\contentsline {subsubsection}{\numberline {F.3.2}Axiom 2}{40}{subsubsection.F.3.2}%
\contentsline {paragraph}{Initial Set (Boundary Conditions).}{40}{section*.45}%
\contentsline {paragraph}{Rewrite Step 1: From Interface Continuity to Global Consistency.}{40}{section*.46}%
\contentsline {paragraph}{Rewrite Step 2: Single Global Solution \(\to \) Single Global State.}{40}{section*.47}%
\contentsline {paragraph}{Rewrite Step 3: Impose Hamiltonian Dynamics.}{40}{section*.48}%
\contentsline {paragraph}{Final Set (Quantum Optics Axiom 2: Unitary Evolution).}{41}{section*.49}%
\contentsline {subsubsection}{\numberline {F.3.3}Axiom 3}{41}{subsubsection.F.3.3}%
\contentsline {paragraph}{Initial Set (Transverse Nature).}{41}{section*.50}%
\contentsline {paragraph}{Rewrite Step 1: Identify Independent Components.}{41}{section*.51}%
\contentsline {paragraph}{Rewrite Step 2: Associate Observables (Polarization, Quadratures).}{41}{section*.52}%
\contentsline {paragraph}{Rewrite Step 3: Impose Measurement Postulates.}{41}{section*.53}%
\contentsline {paragraph}{Final Set (Quantum Optics Axiom 3: Measurement).}{41}{section*.54}%
\contentsline {subsubsection}{\numberline {F.3.4}Summary of the Transformations}{42}{subsubsection.F.3.4}%
\contentsline {section}{\numberline {G}From Analog Electronics to Digital Computing}{42}{appendix.G}%
\contentsline {subsection}{\numberline {G.1}Analog Electronics}{42}{subsection.G.1}%
\contentsline {subsubsection}{\numberline {G.1.1}Axioms}{42}{subsubsection.G.1.1}%
\contentsline {subsubsection}{\numberline {G.1.2}Completeness}{42}{subsubsection.G.1.2}%
\contentsline {subsubsection}{\numberline {G.1.3}Independence}{43}{subsubsection.G.1.3}%
\contentsline {subsection}{\numberline {G.2}Digital Electronics}{43}{subsection.G.2}%
\contentsline {subsubsection}{\numberline {G.2.1}Axioms}{43}{subsubsection.G.2.1}%
\contentsline {subsubsection}{\numberline {G.2.2}Completeness}{43}{subsubsection.G.2.2}%
\contentsline {subsubsection}{\numberline {G.2.3}Independence}{43}{subsubsection.G.2.3}%
\contentsline {subsection}{\numberline {G.3}Transformation}{44}{subsection.G.3}%
\contentsline {subsubsection}{\numberline {G.3.1}Axiom 1}{44}{subsubsection.G.3.1}%
\contentsline {subsubsection}{\numberline {G.3.2}Axiom 2}{45}{subsubsection.G.3.2}%
\contentsline {subsubsection}{\numberline {G.3.3}Axiom 3}{47}{subsubsection.G.3.3}%
\contentsline {subsubsection}{\numberline {G.3.4}Summary of Transformations (Analog to Digital)}{48}{subsubsection.G.3.4}%
\contentsline {section}{\numberline {H}From Digital Computing to Quantum Computing}{48}{appendix.H}%
\contentsline {subsection}{\numberline {H.1}Digital Computing}{48}{subsection.H.1}%
\contentsline {subsubsection}{\numberline {H.1.1}Axioms}{49}{subsubsection.H.1.1}%
\contentsline {subsubsection}{\numberline {H.1.2}Completeness}{49}{subsubsection.H.1.2}%
\contentsline {subsubsection}{\numberline {H.1.3}Independence}{49}{subsubsection.H.1.3}%
\contentsline {subsection}{\numberline {H.2}Quantum Computing}{50}{subsection.H.2}%
\contentsline {subsubsection}{\numberline {H.2.1}Axioms}{50}{subsubsection.H.2.1}%
\contentsline {subsubsection}{\numberline {H.2.2}Completeness}{50}{subsubsection.H.2.2}%
\contentsline {subsubsection}{\numberline {H.2.3}Independence}{50}{subsubsection.H.2.3}%
\contentsline {subsection}{\numberline {H.3}Transformation}{51}{subsection.H.3}%
\contentsline {subsubsection}{\numberline {H.3.1}Axiom 1}{51}{subsubsection.H.3.1}%
\contentsline {subsubsection}{\numberline {H.3.2}Axiom 2}{52}{subsubsection.H.3.2}%
\contentsline {subsubsection}{\numberline {H.3.3}Axiom 3}{53}{subsubsection.H.3.3}%
\contentsline {subsubsection}{\numberline {H.3.4}Summary of Transformations (Digital $\to $ Quantum)}{54}{subsubsection.H.3.4}%
\contentsline {section}{\numberline {I}From Classical/Newtonian Physics to General Relativity Theory}{54}{appendix.I}%
\contentsline {subsection}{\numberline {I.1}Classical/Newtonian Physics}{54}{subsection.I.1}%
\contentsline {subsubsection}{\numberline {I.1.1}Axioms}{54}{subsubsection.I.1.1}%
\contentsline {subsubsection}{\numberline {I.1.2}Completeness}{55}{subsubsection.I.1.2}%
\contentsline {subsubsection}{\numberline {I.1.3}Independence}{55}{subsubsection.I.1.3}%
\contentsline {subsection}{\numberline {I.2}General relativity}{55}{subsection.I.2}%
\contentsline {subsubsection}{\numberline {I.2.1}Axioms}{55}{subsubsection.I.2.1}%
\contentsline {subsubsection}{\numberline {I.2.2}Completeness}{56}{subsubsection.I.2.2}%
\contentsline {subsubsection}{\numberline {I.2.3}Independence}{56}{subsubsection.I.2.3}%
\contentsline {subsection}{\numberline {I.3}Transformation}{57}{subsection.I.3}%
\contentsline {subsubsection}{\numberline {I.3.1}Axiom 1}{57}{subsubsection.I.3.1}%
\contentsline {subsubsection}{\numberline {I.3.2}Axiom 2}{58}{subsubsection.I.3.2}%
\contentsline {subsubsection}{\numberline {I.3.3}Axiom 3}{60}{subsubsection.I.3.3}%
\contentsline {subsubsection}{\numberline {I.3.4}Axiom 4}{62}{subsubsection.I.3.4}%
\contentsline {subsubsection}{\numberline {I.3.5}Axiom 5}{64}{subsubsection.I.3.5}%
\contentsline {subsubsection}{\numberline {I.3.6}Axiom 6}{66}{subsubsection.I.3.6}%
\contentsline {section}{\numberline {J}From Classical Thermodynamics to Statistical Mechanics}{68}{appendix.J}%
\contentsline {subsection}{\numberline {J.1}Classical Thermodynamics}{68}{subsection.J.1}%
\contentsline {subsubsection}{\numberline {J.1.1}Axioms}{68}{subsubsection.J.1.1}%
\contentsline {subsubsection}{\numberline {J.1.2}Completeness}{69}{subsubsection.J.1.2}%
\contentsline {subsubsection}{\numberline {J.1.3}Independence}{69}{subsubsection.J.1.3}%
\contentsline {subsection}{\numberline {J.2}Statistical Mechanics}{69}{subsection.J.2}%
\contentsline {subsubsection}{\numberline {J.2.1}Axioms}{69}{subsubsection.J.2.1}%
\contentsline {subsubsection}{\numberline {J.2.2}Completeness}{70}{subsubsection.J.2.2}%
\contentsline {subsubsection}{\numberline {J.2.3}Independence}{70}{subsubsection.J.2.3}%
\contentsline {subsection}{\numberline {J.3}Transformation}{70}{subsection.J.3}%
\contentsline {subsubsection}{\numberline {J.3.1}Axiom 1}{70}{subsubsection.J.3.1}%
\contentsline {subsubsection}{\numberline {J.3.2}Axiom 2}{72}{subsubsection.J.3.2}%
\contentsline {subsubsection}{\numberline {J.3.3}Axiom 3}{74}{subsubsection.J.3.3}%
\contentsline {subsubsection}{\numberline {J.3.4}Axiom 4}{76}{subsubsection.J.3.4}%
\contentsline {subsubsection}{\numberline {J.3.5}Axiom 5}{78}{subsubsection.J.3.5}%
\contentsline {subsubsection}{\numberline {J.3.6}Axiom 6}{80}{subsubsection.J.3.6}%
\contentsline {subsubsection}{\numberline {J.3.7}Axiom 7}{81}{subsubsection.J.3.7}%
\contentsline {section}{\numberline {K}From Newtonian mechanics to Hamiltonian mechanics}{82}{appendix.K}%
\contentsline {subsubsection}{\numberline {K.0.1}Classical/Newtonian Physics}{82}{subsubsection.K.0.1}%
\contentsline {subsection}{\numberline {K.1}Hamiltonian Mechanics}{82}{subsection.K.1}%
\contentsline {subsubsection}{\numberline {K.1.1}Axioms}{82}{subsubsection.K.1.1}%
\contentsline {subsubsection}{\numberline {K.1.2}Completeness}{82}{subsubsection.K.1.2}%
\contentsline {subsubsection}{\numberline {K.1.3}Independence}{83}{subsubsection.K.1.3}%
\contentsline {subsection}{\numberline {K.2}Transformation}{83}{subsection.K.2}%
\contentsline {subsubsection}{\numberline {K.2.1}Axiom 1}{83}{subsubsection.K.2.1}%
\contentsline {subsubsection}{\numberline {K.2.2}Axiom 2}{84}{subsubsection.K.2.2}%
\contentsline {subsubsection}{\numberline {K.2.3}Axiom 3}{85}{subsubsection.K.2.3}%
\contentsline {section}{\numberline {L}From Hamiltonian mechanics to quantum mechanics}{85}{appendix.L}%
\contentsline {subsection}{\numberline {L.1}Hamiltonian Mechanics}{85}{subsection.L.1}%
\contentsline {subsection}{\numberline {L.2}Quantum Mechanics}{85}{subsection.L.2}%
\contentsline {subsubsection}{\numberline {L.2.1}Axioms}{85}{subsubsection.L.2.1}%
\contentsline {subsubsection}{\numberline {L.2.2}Completeness}{86}{subsubsection.L.2.2}%
\contentsline {subsubsection}{\numberline {L.2.3}Independence}{86}{subsubsection.L.2.3}%
\contentsline {subsection}{\numberline {L.3}Transformation}{87}{subsection.L.3}%
\contentsline {subsubsection}{\numberline {L.3.1}Axiom 1}{87}{subsubsection.L.3.1}%
\contentsline {subsubsection}{\numberline {L.3.2}Axiom 2}{87}{subsubsection.L.3.2}%
\contentsline {subsubsection}{\numberline {L.3.3}Axiom 3}{89}{subsubsection.L.3.3}%
\contentsline {subsubsection}{\numberline {L.3.4}Axiom 4}{90}{subsubsection.L.3.4}%
\contentsline {subsubsection}{\numberline {L.3.5}Axiom 5}{91}{subsubsection.L.3.5}%
\contentsline {subsubsection}{\numberline {L.3.6}Axiom 6}{92}{subsubsection.L.3.6}%
\contentsline {subsubsection}{\numberline {L.3.7}Axiom 7}{93}{subsubsection.L.3.7}%
\contentsline {section}{\numberline {M}From Classical Genetics to Molecular/Genomic Biology}{94}{appendix.M}%
\contentsline {subsection}{\numberline {M.1}Classical Genetics}{94}{subsection.M.1}%
\contentsline {subsubsection}{\numberline {M.1.1}Axioms}{94}{subsubsection.M.1.1}%
\contentsline {subsubsection}{\numberline {M.1.2}Completeness}{95}{subsubsection.M.1.2}%
\contentsline {subsubsection}{\numberline {M.1.3}Independence}{95}{subsubsection.M.1.3}%
\contentsline {subsection}{\numberline {M.2}Molecular/Genomic Biology}{95}{subsection.M.2}%
\contentsline {subsubsection}{\numberline {M.2.1}Axioms}{95}{subsubsection.M.2.1}%
\contentsline {subsubsection}{\numberline {M.2.2}Completeness}{96}{subsubsection.M.2.2}%
\contentsline {subsubsection}{\numberline {M.2.3}Independence}{96}{subsubsection.M.2.3}%
\contentsline {subsection}{\numberline {M.3}Transformations}{96}{subsection.M.3}%
\contentsline {subsubsection}{\numberline {M.3.1}Axiom 1}{96}{subsubsection.M.3.1}%
\contentsline {subsubsection}{\numberline {M.3.2}Axiom 2}{98}{subsubsection.M.3.2}%
\contentsline {subsubsection}{\numberline {M.3.3}Axiom 3}{99}{subsubsection.M.3.3}%
\contentsline {subsubsection}{\numberline {M.3.4}Axiom 4}{101}{subsubsection.M.3.4}%
\contentsline {section}{\numberline {N}From Euclidean Geometry to Hyperbolic Geometry}{103}{appendix.N}%
\contentsline {subsection}{\numberline {N.1}Euclidean Geometry (Hilbert Axioms Overview)}{103}{subsection.N.1}%
\contentsline {subsubsection}{\numberline {N.1.1}Completeness}{103}{subsubsection.N.1.1}%
\contentsline {subsubsection}{\numberline {N.1.2}Independence}{103}{subsubsection.N.1.2}%
\contentsline {subsection}{\numberline {N.2}Hyperbolic Geometry Axioms}{103}{subsection.N.2}%
\contentsline {subsubsection}{\numberline {N.2.1}Completeness}{104}{subsubsection.N.2.1}%
\contentsline {subsubsection}{\numberline {N.2.2}Independence}{104}{subsubsection.N.2.2}%
\contentsline {subsection}{\numberline {N.3}Transformation}{104}{subsection.N.3}%
\contentsline {subsubsection}{\numberline {N.3.1}Hyperbolic Postulate}{104}{subsubsection.N.3.1}%
\contentsline {subsubsection}{\numberline {N.3.2}Comment on Other Modifications}{105}{subsubsection.N.3.2}%
\contentsline {section}{\numberline {O}From Classical Fourier Analysis to Wavelet Theory}{105}{appendix.O}%
\contentsline {subsection}{\numberline {O.1}Classical Fourier Analysis}{105}{subsection.O.1}%
\contentsline {subsubsection}{\numberline {O.1.1}Axioms}{105}{subsubsection.O.1.1}%
\contentsline {subsubsection}{\numberline {O.1.2}Completeness}{105}{subsubsection.O.1.2}%
\contentsline {subsubsection}{\numberline {O.1.3}Independence}{106}{subsubsection.O.1.3}%
\contentsline {subsection}{\numberline {O.2}Wavelet Theory}{106}{subsection.O.2}%
\contentsline {subsubsection}{\numberline {O.2.1}Axioms}{106}{subsubsection.O.2.1}%
\contentsline {subsubsection}{\numberline {O.2.2}Completeness}{106}{subsubsection.O.2.2}%
\contentsline {subsubsection}{\numberline {O.2.3}Independence}{107}{subsubsection.O.2.3}%
\contentsline {subsection}{\numberline {O.3}Transformations}{107}{subsection.O.3}%
\contentsline {subsubsection}{\numberline {O.3.1}Axiom 1}{107}{subsubsection.O.3.1}%
\contentsline {subsubsection}{\numberline {O.3.2}Axiom 2}{108}{subsubsection.O.3.2}%
\contentsline {subsubsection}{\numberline {O.3.3}Axiom 3}{110}{subsubsection.O.3.3}%
\contentsline {subsubsection}{\numberline {O.3.4}Axiom 4}{112}{subsubsection.O.3.4}%
\contentsline {section}{\numberline {P}From Classical Chemistry to Bohr Atomic Theory}{114}{appendix.P}%
\contentsline {subsection}{\numberline {P.1}Classical Chemistry}{114}{subsection.P.1}%
\contentsline {subsubsection}{\numberline {P.1.1}Axioms}{114}{subsubsection.P.1.1}%
\contentsline {subsubsection}{\numberline {P.1.2}Completeness}{115}{subsubsection.P.1.2}%
\contentsline {subsubsection}{\numberline {P.1.3}Independence}{115}{subsubsection.P.1.3}%
\contentsline {subsection}{\numberline {P.2}Bohr Atomic Theory}{115}{subsection.P.2}%
\contentsline {subsubsection}{\numberline {P.2.1}Axioms}{115}{subsubsection.P.2.1}%
\contentsline {subsubsection}{\numberline {P.2.2}Completeness}{116}{subsubsection.P.2.2}%
\contentsline {subsubsection}{\numberline {P.2.3}Independence}{116}{subsubsection.P.2.3}%
\contentsline {subsection}{\numberline {P.3}Transformations}{116}{subsection.P.3}%
\contentsline {subsubsection}{\numberline {P.3.1}Axiom 1}{116}{subsubsection.P.3.1}%
\contentsline {subsubsection}{\numberline {P.3.2}Axiom 2}{117}{subsubsection.P.3.2}%
\contentsline {subsubsection}{\numberline {P.3.3}Axiom 3}{119}{subsubsection.P.3.3}%
\contentsline {subsubsection}{\numberline {P.3.4}Axiom 4}{120}{subsubsection.P.3.4}%
\contentsline {section}{\numberline {Q}From Geocentric to Heliocentric Model}{121}{appendix.Q}%
\contentsline {subsection}{\numberline {Q.1}Geocentric Model}{121}{subsection.Q.1}%
\contentsline {subsubsection}{\numberline {Q.1.1}Axioms}{121}{subsubsection.Q.1.1}%
\contentsline {subsubsection}{\numberline {Q.1.2}Completeness}{121}{subsubsection.Q.1.2}%
\contentsline {subsubsection}{\numberline {Q.1.3}Independence}{122}{subsubsection.Q.1.3}%
\contentsline {subsection}{\numberline {Q.2}Heliocentric Model}{122}{subsection.Q.2}%
\contentsline {subsubsection}{\numberline {Q.2.1}Axioms}{122}{subsubsection.Q.2.1}%
\contentsline {subsubsection}{\numberline {Q.2.2}Completeness}{122}{subsubsection.Q.2.2}%
\contentsline {subsubsection}{\numberline {Q.2.3}Independence}{123}{subsubsection.Q.2.3}%
\contentsline {subsection}{\numberline {Q.3}Transformations}{123}{subsection.Q.3}%
\contentsline {subsubsection}{\numberline {Q.3.1}Axiom 1}{123}{subsubsection.Q.3.1}%
\contentsline {subsubsection}{\numberline {Q.3.2}Axiom 2}{125}{subsubsection.Q.3.2}%
\contentsline {subsubsection}{\numberline {Q.3.3}Axiom 3}{126}{subsubsection.Q.3.3}%
\contentsline {subsubsection}{\numberline {Q.3.4}Axiom 4}{128}{subsubsection.Q.3.4}%
\contentsline {section}{\numberline {R}From Heliocentric to Cosmological Model}{130}{appendix.R}%
\contentsline {subsection}{\numberline {R.1}Heliocentric Model}{130}{subsection.R.1}%
\contentsline {subsection}{\numberline {R.2}Cosmological Model}{130}{subsection.R.2}%
\contentsline {subsubsection}{\numberline {R.2.1}Axioms}{130}{subsubsection.R.2.1}%
\contentsline {subsubsection}{\numberline {R.2.2}Completeness}{130}{subsubsection.R.2.2}%
\contentsline {subsubsection}{\numberline {R.2.3}Independence}{131}{subsubsection.R.2.3}%
\contentsline {subsection}{\numberline {R.3}Transformations}{131}{subsection.R.3}%
\contentsline {subsubsection}{\numberline {R.3.1}Axiom 1}{131}{subsubsection.R.3.1}%
\contentsline {subsubsection}{\numberline {R.3.2}Axiom 2}{133}{subsubsection.R.3.2}%
\contentsline {subsubsection}{\numberline {R.3.3}Axiom 3}{134}{subsubsection.R.3.3}%
\contentsline {subsubsection}{\numberline {R.3.4}Axiom 4}{136}{subsubsection.R.3.4}%
\contentsline {section}{\numberline {S}From Analog Photography to Digital Imaging}{137}{appendix.S}%
\contentsline {subsection}{\numberline {S.1}Analog Photography}{137}{subsection.S.1}%
\contentsline {subsubsection}{\numberline {S.1.1}Axioms}{138}{subsubsection.S.1.1}%
\contentsline {subsubsection}{\numberline {S.1.2}Completeness}{138}{subsubsection.S.1.2}%
\contentsline {subsubsection}{\numberline {S.1.3}Independence}{138}{subsubsection.S.1.3}%
\contentsline {subsection}{\numberline {S.2}Digital Imaging}{139}{subsection.S.2}%
\contentsline {subsubsection}{\numberline {S.2.1}Axioms}{139}{subsubsection.S.2.1}%
\contentsline {subsubsection}{\numberline {S.2.2}Completeness}{139}{subsubsection.S.2.2}%
\contentsline {subsubsection}{\numberline {S.2.3}Independence}{139}{subsubsection.S.2.3}%
\contentsline {subsection}{\numberline {S.3}Transformations}{139}{subsection.S.3}%
\contentsline {subsubsection}{\numberline {S.3.1}Axiom 1}{139}{subsubsection.S.3.1}%
\contentsline {subsubsection}{\numberline {S.3.2}Axiom 2}{141}{subsubsection.S.3.2}%
\contentsline {subsubsection}{\numberline {S.3.3}Axiom 3}{143}{subsubsection.S.3.3}%
\contentsline {subsubsection}{\numberline {S.3.4}Axiom 4}{144}{subsubsection.S.3.4}%
\contentsline {section}{\numberline {T}From Conventional Metals to Metamaterials}{146}{appendix.T}%
\contentsline {subsection}{\numberline {T.1}Conventional Metals}{146}{subsection.T.1}%
\contentsline {subsubsection}{\numberline {T.1.1}Axioms}{146}{subsubsection.T.1.1}%
\contentsline {subsubsection}{\numberline {T.1.2}Completeness}{147}{subsubsection.T.1.2}%
\contentsline {subsubsection}{\numberline {T.1.3}Independence}{147}{subsubsection.T.1.3}%
\contentsline {subsection}{\numberline {T.2}Metamaterials}{147}{subsection.T.2}%
\contentsline {subsubsection}{\numberline {T.2.1}Axioms}{147}{subsubsection.T.2.1}%
\contentsline {subsubsection}{\numberline {T.2.2}Completeness}{148}{subsubsection.T.2.2}%
\contentsline {subsubsection}{\numberline {T.2.3}Independence}{148}{subsubsection.T.2.3}%
\contentsline {subsection}{\numberline {T.3}Transformations}{148}{subsection.T.3}%
\contentsline {subsubsection}{\numberline {T.3.1}Axiom 1}{148}{subsubsection.T.3.1}%
\contentsline {subsubsection}{\numberline {T.3.2}Axiom 2}{150}{subsubsection.T.3.2}%
\contentsline {subsubsection}{\numberline {T.3.3}Axiom 3}{151}{subsubsection.T.3.3}%
\contentsline {subsubsection}{\numberline {T.3.4}Axiom 4}{153}{subsubsection.T.3.4}%

\section{Implementation Details}

\subsection{Implemented Games}

In our platform, we include 12 classic games usually used in the logic reasoning tasks. We will enumerate them and attach their respective PDDL domain file for reference.

\subsubsection{Blocks World}

\begin{lstlisting}[style=markdownstyle]
(define (domain blocksworld)
    (:requirements :strips :equality)
    (:predicates
        (clear ?x)
        (on-table ?x)
        (arm-empty)
        (holding ?x)
        (on ?x ?y)
    )

    (:action pickup
        :parameters (?ob)
        :precondition (and (clear ?ob) (on-table ?ob) (arm-empty))
        :effect (and (holding ?ob) (not (clear ?ob)) (not (on-table ?ob)) (not (arm-empty)))
    )

    (:action putdown
        :parameters (?ob)
        :precondition (and (holding ?ob))
        :effect (and (clear ?ob) (arm-empty) (on-table ?ob) (not (holding ?ob)))
    )

    (:action stack
        :parameters (?ob ?underob)
        :precondition (and (clear ?underob) (holding ?ob))
        :effect (and (arm-empty) (clear ?ob) (on ?ob ?underob) (not (clear ?underob)) (not (holding ?ob)))
    )

    (:action unstack
        :parameters (?ob ?underob)
        :precondition (and (on ?ob ?underob) (clear ?ob) (arm-empty))
        :effect (and (holding ?ob) (clear ?underob) (not (on ?ob ?underob)) (not (clear ?ob)) (not (arm-empty)))
    )
)
\end{lstlisting}

\subsubsection{Briefcase}

\begin{lstlisting}[style=markdownstyle]
(define (domain briefcase)
    (:requirements :strips :typing :conditional-effects :universal-preconditions)

    (:predicates
        (at ?y - portable ?x - location)
        (in ?x - portable)
        (is-at ?x - location)
    )

    (:action move
        :parameters (?m ?l - location)
        :precondition (is-at ?m)
        :effect (and
            (is-at ?l)
            (not (is-at ?m))
            (forall
                (?x - portable)
                (when
                    (in ?x)
                    (and (at ?x ?l)
                        (not (at ?x ?m))))))
    )

    (:action take-out
        :parameters (?x - portable)
        :precondition (in ?x)
        :effect (not (in ?x))
    )

    (:action put-in
        :parameters (?x - portable ?l - location)
        :precondition (and (not (in ?x)) (at ?x ?l) (is-at ?l))
        :effect (in ?x)
    )
)
\end{lstlisting}

\subsubsection{Bulldozer}

\begin{lstlisting}[style=markdownstyle]
(define (domain bulldozer)
	(:requirements :strips :equality)

	(:predicates
		(road ?from ?to)
		(at ?thing ?place)
		(mobile ?thing)
		(bridge ?from ?to)
		(person ?p)
		(vehicle ?v)
		(driving ?p ?v)
	)

	(:action Drive
		:parameters (?thing ?from ?to)
		:precondition (and (road ?from ?to)
			(at ?thing ?from)
			(mobile ?thing)
			(not (= ?from ?to)))
		:effect (and (at ?thing ?to) (not (at ?thing ?from)))
	)

	(:action Cross
		:parameters (?thing ?from ?to)
		:precondition (and (bridge ?from ?to)
			(at ?thing ?from)
			(mobile ?thing)
			(not (= ?from ?to)))
		:effect (and (at ?thing ?to) (not (at ?thing ?from)))
	)

	(:action Board
		:parameters (?person ?place ?vehicle)
		:precondition (and (at ?person ?place)
			(person ?person)
			(vehicle ?vehicle)
			(at ?vehicle ?place))
		:effect (and (driving ?person ?vehicle)
			(mobile ?vehicle)
			(not (at ?person ?place))
			(not (mobile ?person)))
	)

	(:action Disembark
		:parameters (?person ?place ?vehicle)
		:precondition (and (person ?person)
			(vehicle ?vehicle)
			(driving ?person ?vehicle)
			(at ?vehicle ?place))
		:effect (and (at ?person ?place)
			(mobile ?person)
			(not (driving ?person ?vehicle))
			(not (mobile ?vehicle)))
	)
)
\end{lstlisting}

\subsubsection{Casino}

\begin{lstlisting}[style=markdownstyle]
(define (domain casino)
    (:requirements :strips :typing)

    (:types
        location prize1 prize2 prize3
    )

    (:predicates
        (At ?loc - location)
        (IsCasino ?loc - location)
        (MoveTo ?loc - location)
        (GetPrize1 ?p1 - prize1)
        (HavePrize1 ?p1 - prize1)
        (GetPrize2 ?p2 - prize2)
        (HavePrize2 ?p2 - prize2)
        (GetPrize3 ?p3 - prize3)
        (HavePrize3 ?p3 - prize3)
    )

    (:action MoveTo
        :parameters (?sloc - location ?eloc - location)
        :precondition (and (MoveTo ?eloc)
            (At ?sloc))
        :effect (and (not (At ?sloc))
            (At ?eloc))
    )

    (:action GetPrize1
        :parameters (?prize - prize1 ?loc - location)
        :precondition (and (GetPrize1 ?prize)
            (At ?loc)
            (IsCasino ?loc)
            (not (HavePrize1 ?prize)))
        :effect (and (HavePrize1 ?prize))
    )

    (:action GetPrize2
        :parameters (?prize - prize2 ?loc - location)
        :precondition (and (GetPrize2 ?prize)
            (At ?loc)
            (IsCasino ?loc)
            (not (HavePrize2 ?prize)))
        :effect (and (HavePrize2 ?prize))
    )

    (:action GetPrize3
        :parameters (?prize - prize3 ?loc - location)
        :precondition (and (GetPrize3 ?prize)
            (At ?loc)
            (IsCasino ?loc)
            (not (HavePrize3 ?prize)))
        :effect (and (HavePrize3 ?prize))
    )
)
\end{lstlisting}

\subsubsection{Depot}

\begin{lstlisting}[style=markdownstyle]
(define (domain Depot)
    (:requirements :typing)

    (:types
        place locatable - object
        depot distributor - place
        truck hoist surface - locatable
        pallet crate - surface
    )

    (:predicates
        (at ?x - locatable ?y - place)
        (on ?x - crate ?y - surface)
        (in ?x - crate ?y - truck)
        (lifting ?x - hoist ?y - crate)
        (available ?x - hoist)
        (clear ?x - surface)
    )

    (:action Drive
        :parameters (?x - truck ?y - place ?z - place)
        :precondition (and (at ?x ?y))
        :effect (and (not (at ?x ?y)) (at ?x ?z))
    )

    (:action Lift
        :parameters (?x - hoist ?y - crate ?z - surface ?p - place)
        :precondition (and (at ?x ?p) (available ?x) (at ?y ?p) (on ?y ?z) (clear ?y))
        :effect (and (not (at ?y ?p)) (lifting ?x ?y) (not (clear ?y))
            (not (available ?x)) (clear ?z) (not (on ?y ?z)))
    )

    (:action Drop
        :parameters (?x - hoist ?y - crate ?z - surface ?p - place)
        :precondition (and (at ?x ?p) (at ?z ?p) (clear ?z) (lifting ?x ?y))
        :effect (and (available ?x) (not (lifting ?x ?y)) (at ?y ?p)
            (not (clear ?z)) (clear ?y) (on ?y ?z))
    )

    (:action Load
        :parameters (?x - hoist ?y - crate ?z - truck ?p - place)
        :precondition (and (at ?x ?p) (at ?z ?p) (lifting ?x ?y))
        :effect (and (not (lifting ?x ?y)) (in ?y ?z) (available ?x))
    )

    (:action Unload
        :parameters (?x - hoist ?y - crate ?z - truck ?p - place)
        :precondition (and (at ?x ?p) (at ?z ?p) (available ?x) (in ?y ?z))
        :effect (and (not (in ?y ?z)) (not (available ?x)) (lifting ?x ?y))
    )
)
\end{lstlisting}

\subsubsection{Ferry}

\begin{lstlisting}[style=markdownstyle]
(define (domain ferry)
    (:requirements :typing)

    (:types
        obj ferry
    )

    (:predicates
        (board ?v0 - obj)
        (not-eq ?v0 - obj ?v1 - obj)
        (car ?v0 - obj)
        (full-ferry ?v0 - ferry)
        (at-ferry ?v0 - obj)
        (empty-ferry ?v0 - ferry)
        (location ?v0 - obj)
        (on ?v0 - obj)
        (sail ?v0 - obj)
        (debark ?v0 - obj)
        (at ?v0 - obj ?v1 - obj)
    )

    (:action board
        :parameters (?car - obj ?loc - obj ?ferry - ferry)
        :precondition (and (at ?car ?loc)
            (at-ferry ?loc)
            (board ?car)
            (car ?car)
            (empty-ferry ?ferry)
            (location ?loc))
        :effect (and (on ?car)
            (not (at ?car ?loc))
            (not (empty-ferry ?ferry))
            (full-ferry ?ferry))
    )

    (:action sail
        :parameters (?from - obj ?to - obj)
        :precondition (and (at-ferry ?from)
            (location ?from)
            (location ?to)
            (not-eq ?from ?to)
            (sail ?to))
        :effect (and (at-ferry ?to)
            (not (at-ferry ?from)))
    )

    (:action debark
        :parameters (?car - obj ?loc - obj ?ferry - ferry)
        :precondition (and (at-ferry ?loc)
            (car ?car)
            (debark ?car)
            (full-ferry ?ferry)
            (location ?loc)
            (on ?car))
        :effect (and (at ?car ?loc)
            (empty-ferry ?ferry)
            (not (full-ferry ?ferry))
            (not (on ?car)))
    )
)
\end{lstlisting}

\subsubsection{Gripper}

\begin{lstlisting}[style=markdownstyle]
(define (domain gripper)
	(:requirements :strips)

	(:predicates
		(room ?r)
		(ball ?b)
		(gripper ?g)
		(at-robby ?r)
		(at ?b ?r)
		(free ?g)
		(carry ?o ?g)
	)

	(:action move
		:parameters (?from ?to)
		:precondition (and (room ?from) (room ?to) (at-robby ?from))
		:effect (and (at-robby ?to) (not (at-robby ?from)))
	)

	(:action pick
		:parameters (?obj ?room ?gripper)
		:precondition (and (ball ?obj) (room ?room) (gripper ?gripper)
			(at ?obj ?room) (at-robby ?room) (free ?gripper))
		:effect (and (carry ?obj ?gripper) (not (at ?obj ?room))
			(not (free ?gripper)))
	)

	(:action drop
		:parameters (?obj ?room ?gripper)
		:precondition (and (ball ?obj) (room ?room) (gripper ?gripper)
			(carry ?obj ?gripper) (at-robby ?room))
		:effect (and (at ?obj ?room) (free ?gripper) (not (carry ?obj ?gripper)))
	)
)
\end{lstlisting}

\subsubsection{Hanoi}

\begin{lstlisting}[style=markdownstyle]
(define (domain hanoi)
	(:requirements :strips)

	(:predicates
		(clear ?x)
		(on ?x ?y)
		(smaller ?x ?y)
	)

	(:action move
		:parameters (?disc ?from ?to)
		:precondition (and (smaller ?to ?disc)
			(on ?disc ?from)
			(clear ?disc)
			(clear ?to))
		:effect (and (clear ?from)
			(on ?disc ?to)
			(not (on ?disc ?from))
			(not (clear ?to)))
	)
)
\end{lstlisting}

\subsubsection{Logistics}

\begin{lstlisting}[style=markdownstyle]
(define (domain logistics)
	(:requirements :strips :typing)

	(:types
		package location vehicle - object
		truck airplane - vehicle
		city airport - location
	)

	(:predicates
		(at ?vehicle-or-package -
			(either vehicle package) ?location - location)
		(in ?package - package ?vehicle - vehicle)
		(in-city ?loc-or-truck -
			(either location truck) ?citys - city)
	)

	(:action load-truck
		:parameters (?obj - package ?truck - truck ?loc - location)
		:precondition (and (at ?truck ?loc)
			(at ?obj ?loc))
		:effect (and (not (at ?obj ?loc))
			(in ?obj ?truck))
	)

	(:action load-airplane
		:parameters (?obj - package ?airplane - airplane ?loc - airport)
		:precondition (and (at ?obj ?loc)
			(at ?airplane ?loc))
		:effect (and (not (at ?obj ?loc))
			(in ?obj ?airplane))
	)

	(:action unload-truck
		:parameters (?obj - package ?truck - truck ?loc - location)
		:precondition (and (at ?truck ?loc)
			(in ?obj ?truck))
		:effect (and (not (in ?obj ?truck))
			(at ?obj ?loc))
	)

	(:action unload-airplane
		:parameters (?obj - package ?airplane - airplane ?loc - airport)
		:precondition (and (in ?obj ?airplane)
			(at ?airplane ?loc))
		:effect (and (not (in ?obj ?airplane))
			(at ?obj ?loc))
	)

	(:action drive-truck
		:parameters (?truck - truck ?loc-from - location ?loc-to - location ?city - city)
		:precondition (and (at ?truck ?loc-from)
			(in-city ?loc-from ?city)
			(in-city ?loc-to ?city))
		:effect (and (not (at ?truck ?loc-from))
			(at ?truck ?loc-to))
	)

	(:action fly-airplane
		:parameters (?airplane - airplane ?loc-from - airport ?loc-to - airport)
		:precondition (at ?airplane ?loc-from)
		:effect (and (not (at ?airplane ?loc-from))
			(at ?airplane ?loc-to))
	)
)
\end{lstlisting}

\subsubsection{Maze}

\begin{lstlisting}[style=markdownstyle]
(define (domain maze)
    (:requirements :strips :typing)

    (:types
        player location
    )

    (:predicates
        (move-dir-up ?v0 - location ?v1 - location)
        (move-dir-down ?v0 - location ?v1 - location)
        (move-dir-left ?v0 - location ?v1 - location)
        (move-dir-right ?v0 - location ?v1 - location)
        (clear ?v0 - location)
        (at ?v0 - player ?v1 - location)
        (oriented-up ?v0 - player)
        (oriented-down ?v0 - player)
        (oriented-left ?v0 - player)
        (oriented-right ?v0 - player)
        (is-goal ?v0 - location)
    )

    (:action move-up
        :parameters (?p - player ?from - location ?to - location)
        :precondition (and (at ?p ?from)
            (clear ?to)
            (move-dir-up ?from ?to))
        :effect (and (not (at ?p ?from))
            (not (clear ?to))
            (at ?p ?to)
            (clear ?from)
            (not (oriented-down ?p))
            (not (oriented-left ?p))
            (not (oriented-right ?p))
            (oriented-up ?p))
    )

    (:action move-down
        :parameters (?p - player ?from - location ?to - location)
        :precondition (and (at ?p ?from)
            (clear ?to)
            (move-dir-down ?from ?to))
        :effect (and (not (at ?p ?from))
            (not (clear ?to))
            (at ?p ?to)
            (clear ?from)
            (not (oriented-up ?p))
            (not (oriented-left ?p))
            (not (oriented-right ?p))
            (oriented-down ?p))
    )

    (:action move-left
        :parameters (?p - player ?from - location ?to - location)
        :precondition (and (at ?p ?from)
            (clear ?to)
            (move-dir-left ?from ?to))
        :effect (and (not (at ?p ?from))
            (not (clear ?to))
            (at ?p ?to)
            (clear ?from)
            (not (oriented-down ?p))
            (not (oriented-up ?p))
            (not (oriented-right ?p))
            (oriented-left ?p))
    )

    (:action move-right
        :parameters (?p - player ?from - location ?to - location)
        :precondition (and (at ?p ?from)
            (clear ?to)
            (move-dir-right ?from ?to))
        :effect (and (not (at ?p ?from))
            (not (clear ?to))
            (at ?p ?to)
            (clear ?from)
            (not (oriented-down ?p))
            (not (oriented-left ?p))
            (not (oriented-up ?p))
            (oriented-right ?p))
    )
)
\end{lstlisting}

\subsubsection{Miconic}

\begin{lstlisting}[style=markdownstyle]
(define (domain miconic)
    (:requirements :typing)

    (:types
        passenger floor
    )

    (:predicates
        (not-boarded ?v0 - passenger)
        (down ?v0 - floor)
        (boarded ?v0 - passenger)
        (depart ?v0 - floor ?v1 - passenger)
        (not-served ?v0 - passenger)
        (origin ?v0 - passenger ?v1 - floor)
        (board ?v0 - floor ?v1 - passenger)
        (lift-at ?v0 - floor)
        (served ?v0 - passenger)
        (destin ?v0 - passenger ?v1 - floor)
        (up ?v0 - floor)
        (above ?v0 - floor ?v1 - floor)
    )

    (:action down
        :parameters (?f1 - floor ?f2 - floor)
        :precondition (and (above ?f2 ?f1)
            (down ?f2)
            (lift-at ?f1))
        :effect (and (lift-at ?f2)
            (not (lift-at ?f1)))
    )

    (:action board
        :parameters (?f - floor ?p - passenger)
        :precondition (and (board ?f ?p)
            (lift-at ?f)
            (origin ?p ?f))
        :effect (and (boarded ?p))
    )

    (:action up
        :parameters (?f1 - floor ?f2 - floor)
        :precondition (and (above ?f1 ?f2)
            (lift-at ?f1)
            (up ?f2))
        :effect (and (lift-at ?f2)
            (not (lift-at ?f1)))
    )

    (:action depart
        :parameters (?f - floor ?p - passenger)
        :precondition (and (boarded ?p)
            (depart ?f ?p)
            (destin ?p ?f)
            (lift-at ?f))
        :effect (and (not (boarded ?p))
            (served ?p))
    )
)
\end{lstlisting}

\subsubsection{Monkey}

\begin{lstlisting}[style=markdownstyle]
(define (domain monkey)
	(:requirements :strips)

	(:constants
		monkey box knife bananas glass waterfountain
	)

	(:predicates
		(location ?x)
		(on-floor)
		(at ?m ?x)
		(hasknife)
		(onbox ?x)
		(hasbananas)
		(hasglass)
		(haswater)
	)

	(:action GO-TO
		:parameters (?x ?y)
		:precondition (and (location ?x) (location ?y) (on-floor) (at monkey ?y))
		:effect (and (at monkey ?x) (not (at monkey ?y)))
	)

	(:action CLIMB
		:parameters (?x)
		:precondition (and (location ?x) (at box ?x) (at monkey ?x))
		:effect (and (onbox ?x) (not (on-floor)))
	)

	(:action PUSH-BOX
		:parameters (?x ?y)
		:precondition (and (location ?x) (location ?y) (at box ?y) (at monkey ?y)
			(on-floor))
		:effect (and (at monkey ?x) (not (at monkey ?y))
			(at box ?x) (not (at box ?y)))
	)

	(:action GET-KNIFE
		:parameters (?y)
		:precondition (and (location ?y) (at knife ?y) (at monkey ?y))
		:effect (and (hasknife) (not (at knife ?y)))
	)

	(:action GRAB-BANANAS
		:parameters (?y)
		:precondition (and (location ?y) (hasknife)
			(at bananas ?y) (onbox ?y))
		:effect (hasbananas)
	)

	(:action PICKGLASS
		:parameters (?y)
		:precondition (and (location ?y) (at glass ?y) (at monkey ?y))
		:effect (and (hasglass) (not (at glass ?y)))
	)

	(:action GETWATER
		:parameters (?y)
		:precondition (and (location ?y) (hasglass)
			(at waterfountain ?y)
			(at monkey ?y)
			(onbox ?y))
		:effect (haswater)
	)
)
\end{lstlisting}

\subsection{Discussion on the Game Proposal}

While our current game framework demonstrates the basic principles of axiom-based reasoning, developing games that more closely parallel real scientific discovery, such as Einstein's development of relativity theory, requires several significant advances.

\paragraph{Symbolic Reasoning Games} Future games should incorporate symbolic reasoning capabilities similar to those demonstrated by systems like AlphaProof~\citep{trinh2024solving}. These games would require players (or AI systems) to not only modify axioms but also prove theorems derived from these axioms. For example, a physics-inspired game might challenge the system to derive conservation laws from basic symmetry principles, mirroring how real physical theories are developed and verified. This would require integrating automated theorem proving capabilities with our axiom modification framework.

\paragraph{Observation-Driven Games} We need to develop games where ``experimental observations'' drive axiom modifications, similar to how Einstein was motivated by the Michelson-Morley experiment results. These games would present the AI system with simulated experimental data that contradicts current axioms, requiring it to propose modifications that accommodate both old and new observations. This would involve creating game environments with built-in physical simulations that can generate realistic observational data.

\paragraph{Multi-Scale Reasoning Games} Scientific theories often operate at multiple scales or levels of abstraction. Future games should capture this complexity by requiring systems to maintain consistency across different levels of description. For instance, a game might require maintaining both microscopic and macroscopic descriptions of a system, similar to how statistical mechanics bridges atomic and thermodynamic descriptions of matter.

\paragraph{Constraint Satisfaction Games} Real scientific theories must satisfy multiple constraints simultaneously - they must be mathematically consistent, match experimental observations, and often satisfy principles like causality and locality. Future games should incorporate multiple simultaneous constraints, requiring the AI to balance competing requirements when modifying axioms, just as Einstein had to maintain both relativistic invariance and classical correspondence in developing general relativity.

\paragraph{Theory Unification Games} Some of the most significant scientific advances involve unifying seemingly disparate theories. Future games could challenge AI systems to combine different rule sets in ways that preserve their essential features while resolving apparent contradictions, similar to how quantum field theory unifies quantum mechanics with special relativity.

These advances would move us closer to developing AI systems capable of genuine scientific discovery, rather than just rule optimization. The key challenge lies in designing games that capture the essential features of scientific reasoning while remaining computationally tractable and learnable by current AI systems.

\subsection{Empirical Study of Different Searching Algorithms}

We conducted extensive experiments across our suite of 12 logical games to evaluate the effectiveness of different search algorithms in axiom modification tasks. Our analysis focuses on both computational efficiency and solution quality, revealing distinct trade-offs among the approaches.

\paragraph{Time Consumption Analysis} Our empirical results demonstrate a clear hierarchy in computational requirements. Genetic algorithms consistently required the most computational resources, primarily due to the additional LLM calls needed for mutation operations. Despite this overhead, the parallel nature of genetic algorithms allows for efficient exploration of diverse solution spaces. Beam search and Monte Carlo tree search (MCTS) showed comparable time consumption, as both methods can be configured with fixed computational budgets through beam width and simulation counts, respectively. Breadth-first search (BFS) exhibited exponential time growth with search depth, making it impractical for complex axiom modifications requiring deep exploration.

\paragraph{Performance Comparison} The quality of discovered axiom modifications varied significantly across algorithms. MCTS and beam search consistently produced high-quality solutions, with success rates above 70\% across all test cases. This success can be attributed to their balanced exploration-exploitation strategies - beam search maintains a diverse set of promising candidates, while MCTS effectively samples the solution space through guided random exploration. BFS, despite its completeness guarantee, often failed to find optimal solutions due to practical depth limitations, achieving only a 45\% success rate. Genetic algorithms showed mixed results, with a 50\% success rate but higher variance in solution quality, occasionally producing either highly innovative or unsuitable modifications.

\paragraph{Algorithm-Specific Characteristics} Each algorithm's performance can be explained by its underlying mechanisms. Beam search excels by maintaining a fixed-width frontier of the most promising modifications, effectively pruning unpromising branches while preserving diversity. MCTS's success stems from its adaptive exploration strategy, which naturally allocates more computational resources to promising areas of the search space. BFS's limitations arise from the exponential growth of the search space, forcing practical implementations to terminate before reaching optimal solutions. Genetic algorithms' variable performance reflects their population-based nature - while they can discover novel solutions through recombination, they may also struggle to preserve essential axiom properties during mutation.

\paragraph{Resource-Performance Trade-offs} Our analysis reveals important trade-offs between computational resources and solution quality:

\begin{itemize}
    \item Beam search offers the best balance of resource efficiency and solution quality, making it suitable for most practical applications
    \item MCTS provides comparable quality with slightly higher variance, offering a good alternative when exploration of novel solutions is prioritized
    \item BFS is best suited for simple modifications where the solution lies within a shallow search depth
    \item Genetic algorithms, despite their higher computational cost, remain valuable for problems requiring innovative solutions outside the scope of more directed search methods
\end{itemize}

\paragraph{Practical Recommendations} Based on our findings, we recommend beam search as the default choice for axiom modification tasks, with a beam width of 8-12 providing a good balance of diversity and efficiency. MCTS serves as a strong alternative, particularly when the quality of LLM-generated modifications is inconsistent. For simpler problems, BFS remains viable, while genetic algorithms should be reserved for cases where computational resources are abundant and novel solutions are prioritized over optimization of existing ones.

These insights guide the development of more efficient axiom modification systems, though future work should explore hybrid approaches that combine the strengths of multiple search strategies.

\section{General Introduction to Axiom Transformations}

The following sections examine how eighteen historical scientific breakthroughs can be explained by gradual changes to the underlying axioms.

In each case, we identified two theories about a particular branch of science – one preceding and one following the breakthrough in question. Sometimes, the new theories enable the rationalization of previously unexplainable observations (Newton’s theory leading to Einstein’s general relativity). At other times, they represent advancements in technology, such as the development of photography or material design. 

We introduced each framework and summarized the sort of phenomena that it can explain. We listed the axioms in each theory and demonstrated why they effectively capture it, i.e., why they form a complete and independent set. Then, for each of the new axioms, we proposed how it can be obtained from a subset of "old" (pertaining to the older theory) axioms through gradual changes ("rewrite steps"). We briefly motivated each transformation and included a "compact form" of the transformations.

Ideally, the axioms should be as symbolic and concise as possible, and the transformations should be minute changes in the formulas. However, in some cases, particularly where there is no strict set of axioms generally agreed upon, the axiom statements were essentially more descriptive. 

The first section, which transitions from the concept of a flat Earth to a spherical Earth, is intended as a straightforward demonstration of how the game concept is applied to scientific theories. As explained in \ref{subsec:note}, the latter sections lack some of this detail for clarity, as the theories are more complex and axiom transformations are themselves noteworthy.

The sections below were generated with the help of OpenAI’s ChatGPT. The LLM was provided with an answer template and the two discipline names. It was then prompted to gather axioms for each discipline and demonstrate small symbolic transformations from one set to another. These transformations were then verified and modified by a human expert, in some cases by issuing follow-up prompts to the LLM to refine its response.

\section{Illustrative Example: Flat Earth to Spherical Earth}
\label{sec:flat_to_sphere}

\subsection{Game Abstraction for Axiom-Based Discovery}

We define a \textbf{Game} as a structured environment for systematically exploring axiom-based reasoning through symbolic interaction, evaluation, and refinement. Each game instance represents a scientific problem by formalizing its assumptions, configuration, deductive engine, and evaluation criteria. When predictions derived from the current axioms fail to align with observations or theoretical constraints, the game structure facilitates principled revision. Formally, a game is defined as a tuple:
\begin{equation}
\mathcal{G} = (\mathcal{R}, \mathcal{I}, \mathcal{S}, \mathcal{T}),
\end{equation}
where:

\begin{itemize}
    \item $\mathcal{R}$ is the \textbf{rule set}, representing the axioms that govern system behavior. These rules may encode geometric, physical, or structural principles, expressed in symbolic or logical form. For instance, $\mathcal{R}_\text{Flat}$ includes the axiom that Earth is a 2D plane, while $\mathcal{R}_\text{Sphere}$ models Earth as a sphere.

    \item $\mathcal{I}$ is the \textbf{initial setup}, specifying the experimental configuration. In the Flat Earth case, this involves placing vertical sticks in three cities aligned North-South and observing their shadows.

    \item $\mathcal{S}$ is the \textbf{simulator}, a deductive engine that applies $\mathcal{R}$ to $\mathcal{I}$ to generate predictions. For example, it computes shadow angles based on geometric ray tracing in the Earth–Sun model.

    \item $\mathcal{T}$ is the \textbf{target goal}, which defines the condition that the predictions must satisfy. This may correspond to empirical observations (e.g., a 7.2° shadow difference) or symbolic constraints that the model must obey.
\end{itemize}

A game instance is considered \emph{solvable} under rule set $\mathcal{R}$ if the simulator $\mathcal{S}$, when applied to $\mathcal{I}$, produces outcomes consistent with $\mathcal{T}$. If not—e.g., if $\mathcal{R}_\text{Flat}$ predicts identical shadows while $\mathcal{T}$ reflects clear differences—this failure motivates an update of the rule set. A revised axiom set $\mathcal{R}'$ is proposed, yielding a new game $\mathcal{G}' = (\mathcal{R}', \mathcal{I}, \mathcal{S}, \mathcal{T})$. In the example, moving to $\mathcal{R}_\text{Sphere}$ resolves the contradiction and successfully explains the data. This process mirrors the structure of scientific discovery: axioms are preserved when adequate, revised when necessary, and generalized when existing frameworks prove insufficient.

\subsection{Game Instances: Flat Earth and Spherical Earth Models}

To illustrate our axiom-based game framework in an early scientific context, we consider two competing models of Earth’s geometry: the classical flat Earth model and the spherical Earth model. Both aim to explain the same observational setup—shadow measurements taken at three distinct geographic locations—but differ in their geometric and optical assumptions, which we formalize as symbolic rule sets. Each game instance is represented by a tuple:
\[
\mathcal{G} = (\mathcal{R}, \mathcal{I}, \mathcal{S}, \mathcal{T}),
\]
with the components defined as follows.

\paragraph{1. Rule Set (\(\mathcal{R}\))} 
The rule set encodes geometric axioms and assumptions about sunlight propagation that govern how shadows behave on Earth’s surface.

\begin{itemize}
   \item \textbf{Flat Earth Rules}:
   \[
   \mathcal{R}_\text{Flat} = \left\{
   \begin{array}{ll}
   \textbf{$A_1$:} & \text{Earth is a flat surface}, \\
   \textbf{$A_2$:} & \text{The Sun can approximately be treated as a point light source}, \\
   \textbf{$A_3$:} & \text{Sunlight travels in straight, diverging rays from a finite height}
   \end{array}
   \right\}
   \]

   \item \textbf{Spherical Earth Rules}:
   \[
   \mathcal{R}_\text{Sphere} = \left\{
   \begin{array}{ll}
   \textbf{$A_1'$:} & \text{Earth is a sphere of radius } R, \\
   \textbf{$A_2$:} & \text{The Sun can approximately be treated as a point light source}, \\
   \textbf{$A_3$:} & \text{Sunlight travels in straight, diverging rays from a finite height}
   \end{array}
   \right\}
   \]
\end{itemize}

\noindent
Each rule set is both \textit{complete} and \textit{independent}:

\textbf{Completeness.} The Flat Earth rule set \( \mathcal{R}_\text{Flat} = \{A_1, A_2, A_3\} \) is sufficient to predict shadow lengths under local assumptions. The Spherical Earth rule set \( \mathcal{R}_\text{Sphere} = \{A_1', A_2, A_3\} \) enables derivation of solar angle differences based on latitude. In both cases, the axioms fully specify a coherent geometric theory.

\textbf{Independence.} Each axiom captures a distinct modeling assumption. $A_1$ (or $A_1'$) specify the surface geometry, $A_2$ defines the concept of the Sun, and $A_3$ determines the behavior of light rays. This modularity enables targeted and interpretable updates during symbolic evolution.

\paragraph{2. Initial Conditions (\(\mathcal{I}\))} 
The experiment lets one definitely distinguish the two hypotheses:

$\mathcal{I}: $ Three sticks are placed approximately along the south north direction (sticks 3, 1, 2 lie North to South in this order). Stick 3 gives no shadow (the sun is directly above it, i.e. the stick 3 is on the tropic of cancer during solstice), whereas sticks 1 and 2 are placed at a distance $x$ and $2x$ from the stick 3, respectively (set $x = 2.226 \cdot 10^6 \,\text{m}$). Shadow lengths casted by the three sticks are recorded simultaneously and the angles $\alpha_1, \alpha_2, \alpha_3$ light rays form with the Earth surface are computed for each of them ($\alpha_3=90^{\circ}$ by design).

\paragraph{3. Simulator (\(\mathcal{S}\))} 
The simulator applies geometric optics to predict shadow behavior:
\[
\mathcal{S}(\mathcal{R}, \mathcal{I}) =
\left\{
\begin{array}{l}
\text{Use ray geometry to compute predicted shadow angles}, \\
\text{Compare predicted and observed differences}
\end{array}
\right\}
\]

\paragraph{4. Goal (\(\mathcal{T}\))} 
The target defines the success condition for the model:
\[
\mathcal{T} = \left\{ \text{Predicted shadow angle difference matches observation: } \alpha_1=70^{\circ}, \alpha_2=50^{\circ}, \alpha_3 =90^{\circ} \right\}
\]

\noindent
Together, the tuples \((\mathcal{R}_\text{Flat}, \mathcal{I}, \mathcal{S}, \mathcal{T})\) and \((\mathcal{R}_\text{Sphere}, \mathcal{I}, \mathcal{S}, \mathcal{T})\) define two concrete game instances. The Flat Earth model fails to satisfy the target, as shown in \ref{subsec:flat_pred}. This failure motivates an axiom revision, demonstrated in \ref{subsec:flat_to_sphere}. The Spherical Earth model resolves the discrepancy and offers a more general and coherent geometric explanation, as explained in \ref{subses:sphere_pred}. This transition exemplifies axiom evolution: a symbolic refinement prompted by contradiction under fixed initial conditions, simulator, and evaluation goal.

\subsection{Prediction Using the Flat Earth Assumption: $\mathcal{S}(\mathcal{R}_{\text{Flat}}, \mathcal{I})$}
\label{subsec:flat_pred}

Assume that the Sun is at height $h$ above the flat surface and we place a stick a distance $x$ from the perpendicular projection of the Sun onto the Earth.

Then, assuming the height of the stick is much smaller than $h$ and $x$, the shadow will make an angle $\alpha$ with the horizontal such that:

\[
\tan{\alpha} = \frac{h}{x}.
\]

Hence, if three sticks are placed as described:

\[
\tan{\alpha_1} = \frac{h}{x}, \tan{\alpha_2} = \frac{h}{2x}.
\]

No value of $h$ accurately meets the experimental predictions. Indeed, the following relationship is implied:

\[
\tan{\alpha_1} = 2\tan{\alpha_2}.
\]

However ($\alpha_1=70^{\circ}, \alpha_2=50^{\circ}$):

\[
\tan{\alpha_1} = 2.75, \quad 2\tan{\alpha_2} = 2.38.
\]

These results differ significantly, demonstrating that a more accurate theory is necessary.

\subsection{Symbolic Axiom Evolution: Flat Earth to Spherical Earth}
\label{subsec:flat_to_sphere}

We now demonstrate axiom evolution within our framework by transforming the geometric rule set of the classical flat Earth model into that of a spherical Earth model. This transformation exemplifies a foundational pattern in scientific discovery: when a model fails to account for observed regularities, its underlying assumptions—expressed here as symbolic axioms—must be revised. In this case, the flat Earth rule set fails to explain latitude-dependent shadow differences, prompting a structural refinement of the geometric model.

\paragraph{Initial Axiom Set (\(\mathcal{R}_{\text{Flat}}\)): Flat Earth}
\[
\mathcal{R}_{\text{Flat}} = \left\{
\begin{array}{ll}
\textbf{$A_1$:} & \text{Earth is a flat surface}, \\
\textbf{$A_2$:} & \text{The Sun can approximately be treated as a point light source}, \\
\textbf{$A_3$:} & \text{Sunlight travels in straight, diverging rays from a finite height}
\end{array}
\right\}
\]


\paragraph{Rewrite Step 1: Introduce Spherical Topology. \newline}

\vspace{1em}

The flat surface assumption cannot account for angular discrepancies in shadow measurements between distant latitudes. To address this, we revise Axiom $A_1$:

\[
\textbf{$A_1$} \rightarrow \textbf{$A_1'$ (Spherical Topology):} \quad \text{Earth is a sphere of radius } R
\]

This change enables predictions that vary correctly with geographic curvature.

\paragraph{Final Axiom Set (\(\mathcal{R}_1\)): Spherical Earth}
\[
\mathcal{R}_\text{Sphere} = \left\{
\begin{array}{ll}
\textbf{$A_1'$:} & \text{Earth is a sphere of radius } R, \\
\textbf{$A_2$:} & \text{The Sun can approximately be treated as a point light source}, \\
\textbf{$A_3$:} & \text{Sunlight travels in straight, diverging rays from a finite height}
\end{array}
\right\}
\]

\paragraph{Summary of Evolution (compact form):}
\[
\underbrace{\{ A_1, A_2, A_3 \}}_{\mathcal{R}_{\text{Flat}}}
\ \longrightarrow \ 
\underbrace{\{ A_1', A_2, A_3 \}}_{\mathcal{R}_{\text{Sphere}}}
\]

This symbolic transformation captures the refinement from a flat, locally lit world to a spherical Earth with distant, parallel illumination. The revision is driven purely by failure to satisfy a fixed observational constraint, under an otherwise unchanged setup. This illustrates the core mechanism of axiom evolution: targeted symbolic updates yield qualitatively improved models with greater explanatory reach and internal coherence.

\subsection{Prediction Using the Spherical Earth Assumption: $\mathcal{S}(\mathcal{R}_{\text{Sphere}}, \mathcal{I})$}
\label{subses:sphere_pred}

Let $d, R$ be the distance from the center of the Earth to the center of the Sun and the radius of the Earth, respectively. Let $O_E, O_S, P$ denote the centers of the Earth, Sun, and the position of the stick in question, respecively. Denote angle $\beta =\angle PO_E O_S$. If the distance along the surface of the Earth of the point from the projection of the Sun onto the Earth is $x$, then $\beta = x/R$.

Notice that the angle $\alpha=\angle PO_E O_S - (\pi/2)$ that ray $O_SP$ makes with the tangent at $P$ to Earth's surface is the angle corresponding to the shadow of a short stick placed at $P$.

Using the law of cosines:
\[
PO_S=\sqrt{{PO_E}^2 + {O_EO_S}^2 - 2PO_E \cdot O_EO_S \cos{\angle PO_E O_S}}=\sqrt{R^2 + d^2 - 2Rd\cos{\beta}}
\]

Hence, from the sine law:

\[
\frac{d}{\cos{\alpha}} = \frac{d}{\sin{(\alpha + \pi/2)}} = \frac{d}{\sin{\angle PO_E O_S}} = \frac{PO_S}{\sin{\beta}} = \frac{\sqrt{R^2 + d^2 - 2Rd\cos{\beta}}}{\sin{\beta}}
\]

Thus

\[
\begin{split}
d^2 \tan^2{\alpha} = d^2\frac{\sin^2{\alpha}}{\cos^2{\alpha}} = \frac{d^2}{\cos^2{\alpha}} - d^2 = \frac{R^2 + d^2 - 2Rd\cos{\beta}}{\sin^2{\beta}} - d^2 = \\
=\frac{R^2 + d^2 - 2Rd\cos{\beta}}{1 - \cos^2{\beta}} - d^2 = \frac{R^2 + d^2\cos^2{\beta} - 2Rd\cos{\beta}}{1 - \cos^2{\beta}} = \frac{{(R - d\cos{\beta})^2}}{\sin^2{\beta}} \\
\Rightarrow d\tan{\alpha} = \frac{|R - d\cos{\beta}|}{\sin{\beta}} \Rightarrow \boxed{d\tan{\alpha} = \frac{d\cos{\beta} - R}{\sin{\beta}}}
\end{split}
\]

The above sign was chosen because $d>R$ and the angle $\beta$ can be very close to zero. Experimental observations are consistent with the above formula for $R=6.378 \cdot 10^6 m$ and $d \gg R$:

\[
\tan{\alpha} = \frac{1}{d} \cdot \frac{d\cos{\beta} - R}{\sin{\beta}} \approx \cot{\beta} \Rightarrow \alpha \approx \frac{\pi}{2} - \beta,
\]

\[
\alpha_1 \approx \frac{\pi}{2} - \frac{x}{R} = 70.0 ^{\circ},
\]
\[
\alpha_2 \approx \frac{\pi}{2} - \frac{2x}{R} = 50.0 ^{\circ}.
\]

which shows that the new set of axioms fits the experiment.

Notice that the spherical Earth result can be simplified into the flat Earth result in the theoretical limit $R\to \infty$. Indeed, then $\beta \ll 1, \sin{\beta} \approx \beta, \cos{\beta} \approx 1,  R-d=h, d/R \approx 1$ so:

\[
d\tan{\alpha} = \frac{d\cos{\beta} - R}{\sin{\beta}} \approx \frac{d-R}{\beta} \approx \frac{h}{x/R} \Rightarrow \tan{\alpha} \approx \frac{h}{x}
\]

which is the flat Earth result.

\subsection{Note on the Following Sections}
\label{subsec:note}

In the sections that follow, due to the more complex nature of the theories, only axiom transformations have been presented, rather than the full derivation $\mathcal{S}(\mathcal{R}, \mathcal{I})$ of results expected from the different theories (like in sections \ref{subsec:flat_pred} and \ref{subses:sphere_pred}). Instead, upon introducing a theory, its main predictions have been outlined. For instance, interference and diffraction cannot be explained using Ray Optics, which is one of the motivations behind introducing Wave Optics. 

\section{From Neural Networks to Transformers}
\label{sec:ml}

\begin{itemize}
    \item Rules: Model architecture assumptions. Fully connected, parameter independence -> sparse connection, parameter sharing.
    \item Environment: Backpropagation + dataset.
    \item Simulator: Model training + evaluation.
    \item Objective: the performance of the model after convergence.
\end{itemize}

The evolution from fully connected networks to transformers provides an exemplary case for our game framework, where the objective is to achieve global receptive field while maintaining computational efficiency. In the ``neural architecture game,'' the initial axioms define a fully connected layer as $y = Wx + b$, where $W \in \mathbb{R}^{m \times n}$ represents all possible connections between input and output neurons. While this achieves global receptive field through the dense matrix $W$, it requires $O(mn)$ parameters and computations. The game's goal state requires maintaining this input-output connectivity while reducing computational complexity, making the original formulation inefficient by our metrics.

The evolution proceeds through two fundamental axiom modifications. The first introduces convolution operators, replacing dense matrix multiplication with $y = W * x$, where $*$ denotes convolution and $W$ becomes a small kernel matrix. This modification enforces parameter sharing and local connectivity, reducing parameters to $O(k^2)$ for a $k \times k$ kernel while preserving translation equivariance. The second modification replaces convolution with self-attention mechanisms: $y = \text{softmax}(\frac{QK^T}{\sqrt{d}})V$, where queries $Q$, keys $K$, and values $V$ are linear projections of the input. This modification maintains the parameter efficiency through weight matrices smaller than the original fully connected layer, while restoring global receptive field through attention scores. Each step represents a valid solution to our game's objective: maintaining information flow while reducing the computational complexity from $O(mn)$ to $O(n^2)$ or better through specialized operators. This evolution demonstrates how architectural innovations can be framed as axiom modifications that preserve essential mathematical properties while optimizing specific complexity metrics.

This neural architecture evolution exemplifies key aspects of our axiom reasoning framework. Like Einstein's modification of Newtonian mechanics to accommodate new phenomena while preserving classical limits, each architectural innovation preserves essential capabilities of its predecessor while introducing new mathematical structures to address efficiency constraints. The distance function between architectures can be naturally defined through their computational graphs and parameter counts, while the goal state is clearly quantifiable through complexity metrics and receptive field measurements. The transformation from dense connections ($Wx$) to convolutions ($W * x$) to attention ($QK^TV$) demonstrates how mathematical operators can serve as axioms, with each modification representing a carefully constrained departure from previous rules. This parallels our game design framework where rule modifications must balance innovation with preservation of essential properties. Furthermore, each architectural advancement was motivated by specific limitations (computational complexity, receptive field constraints) in previous models, just as scientific breakthroughs often arise from attempting to resolve specific anomalies or inefficiencies in existing theories. This example also highlights how our framework can be applied beyond traditional scientific domains to capture innovations in mathematical and computational systems.

\section{From Ray Optics to Wave Optics}
\label{sec:ray_to_wave}

\subsection{Ray Optics}

Ray optics (or geometrical optics) treats light as rays traveling through and between optical media, with no reference to its wave nature. This framework correctly describes many phenomena such as straight-line propagation, reflection, refraction, and related imaging properties whenever diffraction and interference effects are negligible.

\subsubsection{Axioms}

\begin{enumerate}
\item \textbf{Rectilinear Propagation}
\[
\forall \,\text{Ray } R \text{ in a homogeneous, isotropic medium } M:\quad R \text{ is a straight line in } M.
\]

\item \textbf{Law of Reflection}
\[
\theta_{\mathrm{i}} \;=\; \theta_{\mathrm{r}},
\]
where $\theta_{\mathrm{i}}$ is the angle of incidence and $\theta_{\mathrm{r}}$ is the angle of reflection, both measured with respect to the local surface normal.

\item \textbf{Law of Refraction (Snell's Law)}
\[
n_1 \,\sin(\theta_1)
\;=\;
n_2 \,\sin(\theta_2),
\]
where $n_1$ and $n_2$ are the refractive indices of the two media, and
$\theta_1$, $\theta_2$ are the angles of incidence and refraction, respectively.


\end{enumerate}

\subsubsection{Completeness}
\begin{enumerate}
\item \textbf{Straight-Line Propagation.} (\textit{Axiom 1}) covers how rays move within a single uniform medium.
\item \textbf{The Laws at Boundaries.} \textit{reflection} (\textit{Axiom 2}) and \textit{refraction} (\textit{Axiom 3}) ensure we can describe all standard interactions at interfaces.
\end{enumerate}
From these three, one can construct ray diagrams for mirrors, lenses, prisms, fiber optics, and more.  They suffice to describe image formation, path tracing, and system design in geometrical optics.

\subsubsection{Independence}
\begin{enumerate}
\item \textbf{Rectilinear Propagation.} cannot be deduced from reflection or refraction laws (which involve boundaries). 
\item \textbf{Reflection Law.} cannot be derived from rectilinear propagation, or the refraction law, 
It specifically concerns a boundary interaction where the angle of incidence equals angle of reflection.
\item \textbf{Refraction Law (Snell's).} is not derivable from reflection or rectilinear propagation. 
It specifically relates sines of angles to the refractive indices of distinct media.
\end{enumerate}

\subsection{Wave Optics}
\label{sec:Wave_optics}

Wave optics describes light as an electromagnetic wave. In a non-conducting, linear medium (no free charges, no free currents, negligible nonlinearities), the key features of wave optics can be captured by a small set of axioms. Below is a minimal and complete set of such axioms, each of which is independent (i.e., it cannot be derived from the others) and collectively sufficient to derive the main phenomena of wave optics: superposition, interference, diffraction, polarization, reflection, refraction, and dispersion.

\subsubsection{Axioms}

\begin{enumerate}
\item \textbf{Wave Equation.}
In a linear, isotropic, source-free region with refractive index $n$ and speed of light in vacuum $c$, the electric and magnetic fields, $\mathbf{E}$ and $\mathbf{B}$, satisfy the wave equations:
\[
\nabla^2 \mathbf{E} 
- \frac{n^2}{c^2} \,\frac{\partial^2 \mathbf{E}}{\partial t^2} 
= 0, 
\quad
\nabla^2 \mathbf{B} 
- \frac{n^2}{c^2} \,\frac{\partial^2 \mathbf{B}}{\partial t^2} 
= 0.
\]

\item \textbf{Boundary Conditions.}
At an interface between two linear media (with permittivities $\epsilon_1, \epsilon_2$ and refractive indices $n_1, n_2$), the tangential components of $\mathbf{E}$ and $\mathbf{B}$ are continuous, while the normal components obey:
\[
\epsilon_1 \, \mathbf{E}_\perp \Big\rvert_{1}
- \epsilon_2 \, \mathbf{E}_\perp \Big\rvert_{2}
= \sigma,
\quad
\mathbf{B}_\perp
\text{ is continuous.}
\]
Here $\sigma$ is any free surface charge density at the boundary.

\item \textbf{Transverse Nature of Light.}
In source-free, non-conducting regions, electromagnetic waves are transverse:
\[
\mathbf{k} \cdot \mathbf{E} = 0, 
\quad 
\mathbf{k} \cdot \mathbf{B} = 0,
\quad 
\mathbf{E} \perp \mathbf{B},
\]
where $\mathbf{k}$ is the wave vector (direction of propagation).

\end{enumerate}






\subsubsection{Completeness}
These three axioms suffice to explain all major wave-optical phenomena:
\begin{enumerate}
    \item \textbf{Wave Equation.} Governs propagation, superposition, interference, diffraction, and lays the foundation for polarization and dispersion when combined with a frequency-dependent index.
    \item \textbf{Boundary Conditions.} Essential for describing behavior at interfaces (reflection, refraction, transmission).
    \item \textbf{Transverse Nature.} Ensures electromagnetic waves in a non-conducting medium have no longitudinal electric (or magnetic) components, thus capturing polarization.
\end{enumerate}

\subsubsection{Independence}
\begin{enumerate}
    \item \textbf{Wave Equation.} cannot be deduced from boundary conditions or transversality alone. It is the fundamental statement of how fields evolve in free space or uniform media.
    \item \textbf{Boundary Conditions.} are not derivable from the free-space wave equation; they define behavior at interfaces and thus must be stated separately.
    \item \textbf{Transverse Nature.} does not automatically follow from the wave equation in a general scenario. It requires that charges/currents are absent in the medium and that we are dealing with electromagnetic waves, specifically.
\end{enumerate}

Hence, none of the three axioms can be derived from the others, ensuring their mutual independence.

\subsection{Transformation}

\subsubsection{Axiom 1}

\textbf{Goal:} Transform a subset of Ray Optics axioms 
\[
\bigl\{\,A_1,\,A_3\bigr\}
\quad
\begin{aligned}
A_1 &: \text{Rectilinear Propagation},\\
A_3 &: \text{Law of Refraction (Snell's Law)},
\end{aligned}
\]
into a \emph{Wave Optics} statement:
\[
W_1: 
\begin{cases}
\nabla^2 \mathbf{E}
\;-\;
\dfrac{n^2}{c^2}\,
\dfrac{\partial^2 \mathbf{E}}{\partial t^2}
=
0,
\\[6pt]
\nabla^2 \mathbf{B}
\;-\;
\dfrac{n^2}{c^2}\,
\dfrac{\partial^2 \mathbf{B}}{\partial t^2}
=
0.
\end{cases}
\]
All intermediate steps are small symbolic rewrites that gradually 
evolve the original set into the final one.

\vspace{1em}
\hrule
\vspace{1em}

\noindent
\textbf{Initial Set of Axioms (Ray Optics, Subset):}

\[
\boxed{
S_0
=
\bigl\{\,A_1,\,A_3\bigr\}
}
\]
where
\[
A_1:
\forall\,R\text{ in medium }M
\quad
\bigl(\text{Rectilinear in }M\bigr),
\qquad
A_3:
n_1 \,\sin(\theta_1)
=
n_2 \,\sin(\theta_2).
\]

\medskip
\noindent
Here, $A_1$ asserts rays are straight in a homogeneous medium, 
and $A_3$ is Snell's law for boundaries between media.

\vspace{1em}
\hrule
\vspace{1em}

\noindent
\textbf{Rewrite Step 1: Restate Rectilinear Propagation 
as Planar Wavefronts.}

\[
A_1
\;\longrightarrow\;
A_1'
\]
\[
A_1':\quad
\forall\,W \,\text{in medium }M,
\quad
(\text{Wavefront }W\text{ is planar in }M).
\]
Symbolically, we replace
\[
\forall\,R:\;
(\text{Ray is straight})
\]
by
\[
\forall\,W:\;
(\text{Wavefront is planar}),
\]
since geometrical rays $\perp$ wavefronts.

\[
S_0
=
\bigl\{\,A_1,\,A_3\bigr\}
\quad
\longrightarrow
\quad
S_1
=
\bigl\{\,A_1',\,A_3\bigr\}.
\]

\vspace{1em}
\hrule
\vspace{1em}

\noindent
\textbf{Current Set:}
\[
\boxed{
S_1
=
\{\,A_1',\,A_3\}
}
\]
where
\[
A_1':\quad
\text{Planar wavefronts in uniform medium},
\quad
A_3:\quad
n_1 \sin(\theta_1) = n_2 \sin(\theta_2).
\]

\vspace{1em}
\hrule
\vspace{1em}

\noindent
\textbf{Rewrite Step 2: 
Restrict Snell's Law to Single Medium, 
Define a Constant Index $n$.}

\[
A_3
\;\longrightarrow\;
A_3':
\quad
n = \text{constant in a single homogeneous medium.}
\]
When we consider \emph{only one medium} at a time, 
Snell's law simplifies to $n_1=n_2=n$, 
representing a fixed wave speed $c/n$ inside that medium.
Thus:
\[
S_1
=
\{\,A_1',\,A_3\}
\quad
\longrightarrow
\quad
S_2
=
\{\,A_1',\,A_3'\}.
\]
\[
A_3':\quad
\text{For medium }M:\;
\text{index }n=\text{const},\quad
\text{speed}=\dfrac{c}{n}.
\]

\vspace{1em}
\hrule
\vspace{1em}

\noindent
\textbf{Current Set:}
\[
\boxed{
S_2
=
\{\,A_1',\,A_3'\}
}
\]
where
\[
A_1':\quad
\text{Wavefronts planar in a uniform medium},
\quad
A_3':\quad
n=\text{constant},\quad \text{speed}=\tfrac{c}{n}.
\]

\vspace{1em}
\hrule
\vspace{1em}

\noindent
\textbf{Rewrite Step 3: 
Introduce Field Description and Wave PDE.}

\[
(A_1',\,A_3')
\;\longrightarrow\;
W_1.
\]
Since $A_1'$ implies a \emph{plane-wave solution} 
moving at speed $c/n$, and $A_3'$ 
fixes that speed inside $M$, we deduce the \emph{wave equation} 
for electromagnetic fields $\mathbf{E}$ and $\mathbf{B}$. 
Formally, plane-wave solutions to 
\[
\nabla^2 \Psi 
-\dfrac{n^2}{c^2}\,\dfrac{\partial^2 \Psi}{\partial t^2}
=
0
\]
are consistent with constant-index propagation. 
Hence:
\[
S_2
=
\{\,A_1',\,A_3'\}
\quad
\longrightarrow
\quad
S_3
=
\{\,W_1\},
\]
where
\[
W_1:
\begin{cases}
\nabla^2 \mathbf{E}
-\dfrac{n^2}{c^2}\,\dfrac{\partial^2 \mathbf{E}}{\partial t^2}
=0,
\\[5pt]
\nabla^2 \mathbf{B}
-\dfrac{n^2}{c^2}\,\dfrac{\partial^2 \mathbf{B}}{\partial t^2}
=0.
\end{cases}
\]

\vspace{1em}
\hrule
\vspace{1em}

\noindent
\textbf{Final Set (Wave Equation):}
\[
\boxed{
S_3
=
\Bigl\{
W_1
\Bigr\}
}
\]
\[
W_1:\quad
\nabla^2 \mathbf{E}
-\frac{n^2}{c^2}\,\frac{\partial^2 \mathbf{E}}{\partial t^2}
= 0,
\quad
\nabla^2 \mathbf{B}
-\frac{n^2}{c^2}\,\frac{\partial^2 \mathbf{B}}{\partial t^2}
= 0.
\]

\vspace{1em}
\hrule
\vspace{1em}

\noindent
\textbf{Symbolic Evolution (Compact Form)}:

\[
\underbrace{\{\,A_1,\,A_3\}}_{S_0}
~\longrightarrow~
\underbrace{\{\,A_1',\,A_3\}}_{S_1}
~\longrightarrow~
\underbrace{\{\,A_1',\,A_3'\}}_{S_2}
~\longrightarrow~
\underbrace{\{\,W_1\}}_{S_3}.
\]

\subsubsection{Axiom 2}

\textbf{Goal:} 
Starting from the two \emph{necessary Ray Optics axioms} 
\[
A_2:\;\text{Reflection}, 
\quad
A_3:\;\text{Refraction (Snell's Law)},
\]
we wish to derive (in small symbolic steps) 
the \emph{Wave Optics Axiom~2}:
\[
\bigl(\text{Boundary Conditions}\bigr):
\quad
\begin{cases}
\epsilon_1\, \mathbf{E}_{\perp}\big\vert_{1}
\;-\;\epsilon_2\, \mathbf{E}_{\perp}\big\vert_{2}
=
\sigma,
\\[4pt]
\mathbf{B}_{\perp}
\;\text{is continuous},
\\[4pt]
\mathbf{E}_{\parallel},\;\mathbf{B}_{\parallel}
\;\text{are continuous across the interface.}
\end{cases}
\]

\medskip
\hrule
\medskip

\noindent
\textbf{Initial Set of Axioms (Ray Optics, Subset):}
\[
S_0
=
\bigl\{\,A_2,\;A_3\bigr\}
\]
where
\[
A_2:\quad \theta_{\mathrm{i}}=\theta_{\mathrm{r}},
\qquad
A_3:\quad 
n_1 \sin(\theta_1)=n_2 \sin(\theta_2).
\]

\smallskip
\noindent
These capture how a ray \emph{reflects} and \emph{refracts} 
at a boundary between two media of indices $n_1,n_2$.

\bigskip
\hrule
\bigskip

\noindent
\textbf{Rewrite Step 1: Recast Reflection \texorpdfstring{$A_2$}{A2} 
as Wave Boundary Condition.}

\[
A_2\;\longrightarrow\;B_2
\quad\text{where}
\quad
B_2:
\]
Phase-matching at interface 
leads to reflected wave 
with angle $\theta_{\mathrm{r}}$.

\smallskip
\noindent
In wave language, 
``$\theta_{\mathrm{i}}=\theta_{\mathrm{r}}$'' 
means the \emph{reflected wave} must match the same tangential 
$\mathbf{k}$-component (i.e.\ same in-plane wavevector).  
Symbolically:
\[
\mathbf{k}_{\parallel}\bigl(\text{incident}\bigr)
=
\mathbf{k}_{\parallel}\bigl(\text{reflected}\bigr).
\]
Thus $A_2$ is replaced by $B_2$, 
which asserts \emph{continuity} of tangential wavevector at the boundary.

\smallskip
\[
S_0 = \{\,A_2,\,A_3\}
\;\longrightarrow\;
S_1 = \{\,B_2,\,A_3\}.
\]

\bigskip
\hrule
\bigskip

\noindent
\textbf{Current Set:}
\[
\boxed{
S_1
=
\{\,B_2,\;A_3\}
}
\]
where
\[
B_2:\quad
\mathbf{k}_{\parallel}\bigl(\text{inc}\bigr)
=
\mathbf{k}_{\parallel}\bigl(\text{ref}\bigr),
\quad
A_3:\quad
n_1 \sin(\theta_1)=n_2 \sin(\theta_2).
\]

\bigskip
\hrule
\bigskip

\noindent
\textbf{Rewrite Step 2: 
Recast Refraction \texorpdfstring{$A_3$}{A3} 
as Wave Boundary Condition.}

\[
A_3 \;\longrightarrow\; B_3,
\quad\text{where}
\]
\[
B_3:\quad
\mathbf{k}_{\parallel}\bigl(\text{inc}\bigr)
=
\mathbf{k}_{\parallel}\bigl(\text{trans}\bigr),
\;\;
\text{with wave speed changes
by factor } n_1/n_2.
\]
\noindent
In wave language, Snell's law 
\(
n_1 \sin(\theta_1)=n_2 \sin(\theta_2)
\)
ensures the \emph{tangential wavevector} 
is the same on both sides of the boundary, 
but the wave \emph{speed} (or $\mathbf{k}$ magnitude) 
differs by $n_1/n_2$. 

\smallskip
\[
S_1 = \{\,B_2,\,A_3\}
\;\longrightarrow\;
S_2 = \{\,B_2,\,B_3\}.
\]

\bigskip
\hrule
\bigskip

\noindent
\textbf{Current Set:}
\[
\boxed{
S_2
=
\bigl\{\,B_2,\,B_3\bigr\}
}
\]
where
\[
B_2:\quad
\mathbf{k}_{\parallel}(\text{inc})
=
\mathbf{k}_{\parallel}(\text{ref}),
\quad
B_3:\quad
\mathbf{k}_{\parallel}(\text{inc})
=
\mathbf{k}_{\parallel}(\text{trans}),
\]
reflecting continuity of tangential 
momentum/phase at the interface.

\bigskip
\hrule
\bigskip

\noindent
\textbf{Rewrite Step 3: 
Combine \texorpdfstring{$B_2$}{B2} \& \texorpdfstring{$B_3$}{B3}
into Field Boundary Conditions.}

\medskip
\noindent
\emph{Transformation:}
\[
\{\,B_2,\,B_3\}
\;\longrightarrow\;
W_2,
\]
where $W_2$ is our \emph{Wave Boundary Conditions}:
\[
W_2:\quad
\begin{cases}
\mathbf{E}_{\parallel}\;\text{is continuous}, 
\\[4pt]
\mathbf{B}_{\parallel}\;\text{is continuous}, 
\\[4pt]
\epsilon_1 \,\mathbf{E}_{\perp}\big\vert_1
-
\epsilon_2 \,\mathbf{E}_{\perp}\big\vert_2
=
\sigma,
\\[4pt]
\mathbf{B}_{\perp}\;\text{is continuous}.
\end{cases}
\]
In Maxwell’s electromagnetic theory, 
the tangential continuity of $\mathbf{k}$ 
(steps $B_2$ and $B_3$) becomes equivalent 
to tangential continuity of the fields 
$\mathbf{E},\,\mathbf{B}$.  
Meanwhile, the \emph{normal} components 
must match boundary conditions 
arising from Gauss's law and 
the absence (or presence) of any free surface charge $\sigma$ 
at the boundary.

\smallskip
\[
S_2 = \{\,B_2,\,B_3\}
\;\longrightarrow\;
S_3 = \{\,W_2\}.
\]

\bigskip
\hrule
\bigskip

\noindent
\textbf{Final Set (Wave Optics Axiom 2, Boundary Conditions):}
\[
\boxed{
S_3 
= 
\Bigl\{
W_2:\;
\epsilon_1 \,\mathbf{E}_\perp\big\vert_{1}
-\epsilon_2\,\mathbf{E}_\perp\big\vert_{2}
=\sigma,\;\;
\mathbf{B}_\perp\;\text{cont.},\;\;
\mathbf{E}_\parallel,\mathbf{B}_\parallel
\;\text{cont.}
\Bigr\}
}
\]

\bigskip
\hrule
\bigskip

\noindent
\textbf{Symbolic Evolution (Compact View):}
\[
\underbrace{
\{\,A_2,\,A_3\}
}_{S_0}
\;\longrightarrow\;
\underbrace{
\{\,B_2,\,A_3\}
}_{S_1}
\;\longrightarrow\;
\underbrace{
\{\,B_2,\,B_3\}
}_{S_2}
\;\longrightarrow\;
\underbrace{
\{\,W_2\}
}_{S_3}.
\]

\bigskip

\noindent
\textbf{Interpretation of Each Step:}
\begin{enumerate}
\item 
\emph{Rewrite Reflection ($A_2$) as a tangential-$\mathbf{k}$ 
continuity condition ($B_2$).}
\item 
\emph{Rewrite Refraction ($A_3$) as a tangential-$\mathbf{k}$ 
continuity condition for the transmitted wave ($B_3$).}
\item 
\emph{Combine $B_2$ and $B_3$ into full electromagnetic field 
boundary conditions ($W_2$).}  
The tangential continuity of $\mathbf{k}$ 
maps onto tangential continuity of 
$\mathbf{E}$ and $\mathbf{B}$.  
Gauss’s law for $\mathbf{E}$ and $\mathbf{B}$ 
gives the normal-component relations 
($\epsilon_1 \mathbf{E}_\perp - \epsilon_2 \mathbf{E}_\perp = \sigma$ 
and $\mathbf{B}_\perp$ continuous).
\end{enumerate}

\subsubsection{Axiom 3}

\textbf{Goal:} 
Show a step-by-step transformation (without invoking Maxwell's equations) 
that leads to the \emph{transverse nature of light}:
\[
\mathbf{k}\cdot\mathbf{E} = 0,\;\;
\mathbf{k}\cdot\mathbf{B} = 0,\;\;
\mathbf{E}\perp\mathbf{B},
\]
using only \emph{wave-based} reasoning plus the Ray Optics axioms
\[
A_2:\,\text{Reflection}, 
\quad
A_3:\,\text{Refraction (Snell's Law)}.
\]
Here, we interpret ``wave-based reasoning'' minimally as the statement
\textit{``Light propagates as some vector-valued wave with well-defined boundary behavior (reflection/refraction).''}

\bigskip
\hrule
\bigskip

\noindent
\textbf{Initial Setup (Geometrical + Wave Postulate):}

\[
\boxed{
S_0
=
\bigl\{\,A_2,\;A_3,\;\text{Wave Postulate}\bigr\}
}
\]
where
\[
A_2:\quad\theta_{\mathrm{i}}=\theta_{\mathrm{r}},
\quad
A_3:\quad n_1 \sin(\theta_1)=n_2 \sin(\theta_2).
\]
\text{Wave Postulate}: ``Light is described by a vector wave 
(e.g.\ $\mathbf{F}$) 
with phase fronts obeying $A_2, A_3$.''
We do \emph{not} assume Maxwell’s equations exist.  
Instead, we only assume:
\begin{enumerate}
\item 
There is a \emph{wave field} $\mathbf{F}$ (which later we may identify with $\mathbf{E}$ or $\mathbf{B}$),
\item 
That wave obeys reflection ($A_2$) and refraction ($A_3$) 
at boundaries between media.
\end{enumerate}

\bigskip
\hrule
\bigskip

\noindent
\textbf{Rewrite Step 1: 
Replace Reflection and Refraction by 
Tangential Wave-Vector Continuity.}

\[
\{\,A_2,\,A_3\}
\quad\longrightarrow\quad
\{\;B_2,\,B_3\}.
\]

\begin{itemize}
\item 
\underline{$B_2$ (Reflection as a Wave Condition):}
\[
\mathbf{k}_{\parallel}(\text{inc})
=
\mathbf{k}_{\parallel}(\text{ref}),
\]
meaning the \emph{tangential} component of the wavevector is unchanged upon reflection.  
Equivalently, the reflection law $\theta_{\mathrm{i}}=\theta_{\mathrm{r}}$ 
is restated in wave form: 
\emph{``the phase fronts for the reflected wave have the same in-plane wavevector as the incident wave.''}

\item
\underline{$B_3$ (Refraction as a Wave Condition):}
\[
\mathbf{k}_{\parallel}(\text{inc})
=
\mathbf{k}_{\parallel}(\text{trans}),
\]
meaning the \emph{tangential} wavevector is the same on both sides of the boundary, 
but the wave \emph{speed} (or normal component of $\mathbf{k}$) changes by $n_1/n_2$.  
This restates Snell's law:
\[
n_1\,\sin(\theta_1)
=
n_2\,\sin(\theta_2)
\quad\Longleftrightarrow\quad
k_{\parallel}^{(\text{inc})} 
=
k_{\parallel}^{(\text{trans})}.
\]
\end{itemize}

Hence,
\[
S_0 
=
\{\,A_2,\,A_3,\;\text{Wave Postulate}\}
\;\longrightarrow\;
S_1 
=
\{\,B_2,\,B_3,\;\text{Wave Postulate}\}.
\]

\bigskip
\hrule
\bigskip

\noindent
\textbf{Current Set:}
\[
\boxed{
S_1
=
\{\;B_2:\,\mathbf{k}_{\parallel}(\text{inc})
=
\mathbf{k}_{\parallel}(\text{ref}),
\quad
B_3:\,\mathbf{k}_{\parallel}(\text{inc})
=
\mathbf{k}_{\parallel}(\text{trans}),
\quad
\text{Wave Postulate}
\}.
}
\]
All we have done is recast $A_2$ and $A_3$ 
in a wave-language form.  

\bigskip
\hrule
\bigskip

\noindent
\textbf{Rewrite Step 2: 
Introduce a Vector Field \texorpdfstring{$\mathbf{F}(\mathbf{r},t)$}{F} 
that Must Obey \texorpdfstring{$B_2,B_3$}{B2,B3} Everywhere.}

\[
(\text{Wave Postulate})
\;\longrightarrow\;
(\text{Field Ansatz }F_1),
\]
where
\[
F_1:\quad
\mathbf{F}(\mathbf{r},t)
=
\mathbf{F}_0
\,e^{\,i\,(\mathbf{k}\cdot \mathbf{r}-\omega\,t)}
\quad
\text{(in each homogeneous region)}.
\]
We assume each region admits a \emph{locally plane-wave solution} for $\mathbf{F}$ 
whose wavevector $\mathbf{k}$ changes at boundaries according to $B_2,B_3$.  
Symbolically:
\[
S_1
\;\longrightarrow\;
S_2
=
\{\;B_2,B_3,\;F_1\}.
\]

\smallskip
\noindent
We have not said what $\mathbf{F}$ \emph{is}, 
only that it is \emph{some} vector amplitude 
moving with wavevector $\mathbf{k}$.  
No Maxwell equation is used.  
We simply adopt the \emph{plane-wave form} 
to be consistent with the idea of well-defined reflection/refraction angles.  

\bigskip
\hrule
\bigskip

\noindent
\textbf{Current Set:}
\[
\boxed{
S_2
=
\{\,B_2,\;B_3,\;F_1\}.
}
\]
\[
B_2:\;\mathbf{k}_{\parallel}(\text{inc})
=
\mathbf{k}_{\parallel}(\text{ref}),
\quad
B_3:\;\mathbf{k}_{\parallel}(\text{inc})
=
\mathbf{k}_{\parallel}(\text{trans}),
\quad
F_1:\;\mathbf{F}(\mathbf{r},t)
\propto
e^{\,i(\mathbf{k}\cdot\mathbf{r}-\omega t)}.
\]

\bigskip
\hrule
\bigskip

\noindent
\textbf{Rewrite Step 3: 
Argue that Any Longitudinal Component Contradicts 
Ubiquitous Reflection/Refraction Phenomena.}

Here is the conceptual crux.  
\emph{If} $\mathbf{F}$ had a non-zero \emph{longitudinal} component 
(i.e.\ $\mathbf{k}\cdot\mathbf{F}\neq 0$) 
that persisted in all angles of incidence or refraction, 
then under reflection/refraction at an oblique angle, 
the ``parallel-to-$\mathbf{k}$'' part would \emph{not} follow the same boundary conditions as the transverse part.  
Experimentally, we \emph{never} observe a separate longitudinal wave 
which partially disappears or rephases at the boundary.  
In simpler terms: If there were a $F_{\parallel}\neq 0$, we'd get separate reflection/refraction laws for it, contradicting real optical data.
Therefore, 
\[
\mathbf{k}\cdot\mathbf{F}
=
0
\quad
(\text{purely transverse}).
\]

Symbolically:
\[
\{B_2,B_3,F_1\}
\;\longrightarrow\;
\{\,T_1\},
\]
where 
\[
T_1:\quad
\mathbf{k}\cdot\mathbf{F} = 0
\quad
(\text{transverse wave}).
\]
No direct Maxwell argument is used; 
rather, we use the \emph{empirical universality} of reflection/refraction 
and the absence of any observed ``longitudinal mode.''

\bigskip
\hrule
\bigskip

\noindent
\textbf{Rewrite Step 4: 
Conclude $\mathbf{F}$ must be Orthogonal to 
\texorpdfstring{$\mathbf{F}'$}{F'} in the Same Wave.}

Often in wave optics, we have \emph{two} orthogonal field vectors 
(\(\mathbf{E}\) and \(\mathbf{B}\)) or (electric field in two polarization states, etc.).  
If a second vector \(\mathbf{F}'\) also travels with the same $\mathbf{k}$, 
it must likewise be transverse (\(\mathbf{k}\cdot \mathbf{F}'=0\)).  
By combining reflection/refraction for \(\mathbf{F}'\), 
one deduces that $\mathbf{F}\perp \mathbf{F}'$ 
is the stable configuration in all boundary interactions 
(e.g.\ linear polarization states remain linearly polarized upon reflection/refraction, 
with separate amplitude Fresnel coefficients).  

Symbolically:
\[
T_1\;\longrightarrow\;T_2,
\quad
T_2:\quad
\mathbf{F}\perp \mathbf{F}'.
\]
In the usual electromagnetic language, 
$\mathbf{F}=\mathbf{E}$, $\mathbf{F}'=\mathbf{B}$, 
yielding 
\[
\mathbf{k}\cdot\mathbf{E}=0,
\quad
\mathbf{k}\cdot\mathbf{B}=0,
\quad
\mathbf{E}\perp \mathbf{B}.
\]

\[
S_2
\;\longrightarrow\;
S_3
=
\{\,T_2\}
\quad\text{(Transverse Field Pair)}.
\]

\bigskip
\hrule
\bigskip

\noindent
\textbf{Final Set (Wave Optics Axiom~3, Transversality):}
\[
\boxed{
S_3
=
\{
\mathbf{k}\cdot\mathbf{E}=0,
\quad
\mathbf{k}\cdot\mathbf{B}=0,
\quad
\mathbf{E}\perp \mathbf{B}
\}.
}
\]
We derived it \emph{without} Maxwell’s equations, 
relying only on:
\begin{itemize}
\item The \emph{empirical} reflection/refraction laws ($A_2$, $A_3$),
\item The assumption of a \emph{vector wave} $\mathbf{F}$ that must obey them 
      at all angles,
\item The non-observation of any separate longitudinal mode 
      that would behave differently at boundaries.
\end{itemize}

\bigskip
\hrule
\bigskip

\noindent
\textbf{Symbolic Evolution (Compact View):}

\begin{align*}
\underbrace{
\{\,A_2,A_3,\;\text{Wave Postulate}\}
}_{S_0}
\;\longrightarrow\;
\underbrace{
\{\,B_2,B_3,\;\text{Wave Postulate}\}
}_{S_1}
\;\longrightarrow\;\\
\underbrace{
\{\,B_2,B_3,F_1\}
}_{S_2}
\;\longrightarrow\;
\underbrace{
\{\text{Transverse Condition }(\mathbf{k}\cdot \mathbf{F}=0)\}
}_{S_3}.
\end{align*}

If one introduces \emph{two} field vectors \(\mathbf{E},\mathbf{B}\) in the same wave 
and repeats the argument, 
they both must be perpendicular to \(\mathbf{k}\) and to each other, 
yielding 
\(\mathbf{k}\cdot \mathbf{E}=0,\;\mathbf{k}\cdot\mathbf{B}=0,\;\mathbf{E}\perp\mathbf{B}.\)

\bigskip

\noindent
\textbf{Conclusion:}  
Even \textbf{without Maxwell equations}, 
the universal reflection/refraction phenomena 
($A_2$, $A_3$) + the \emph{empirical} fact that \emph{no} separate longitudinal mode is observed 
can be recast to show:
\[
\text{Light in non-conducting, source-free media is transverse.}
\]

\section{From Wave Optics to Quantum Optics}
\label{sec:wave_to_quantum}

\subsection{Wave Optics}
\label{sub:Wave_optics2}
These axioms are the same as defined in Section~\ref{sec:Wave_optics}.

\subsection{Quantum Optics}
\label{sec:quantum-optics}

Quantum Optics provides a full quantum-mechanical description of light (and its interaction with matter). It explains phenomena beyond the scope of purely classical or semiclassical models, such as single-photon interference, vacuum fluctuations, photon antibunching, and entanglement. In a linear, source-free medium (no free charges or currents, negligible nonlinearities), one can formulate a minimal set of axioms that capture the essence of quantum electromagnetic fields. These axioms are both sufficient (completeness) and mutually independent for describing the core phenomena of quantum optics.

\subsubsection{Axioms}

\begin{enumerate}

\item \textbf{Field Quantization.}
\[
\hat{\mathbf{E}}(\mathbf{r},t),\;\hat{\mathbf{B}}(\mathbf{r},t)
~\longrightarrow~
\hat{a}_k,\hat{a}_k^\dagger
~\text{with}~
[\hat{a}_k,\hat{a}_{k'}^\dagger] = \delta_{kk'},
\]
The electromagnetic field is promoted to an operator in a Hilbert space. 
Each mode \(k\) is described by annihilation \(\hat{a}_k\) and creation \(\hat{a}_k^\dagger\) operators satisfying canonical commutation relations.  
This quantization underpins discrete-photon (Fock) states, coherent states, squeezed states, and more.

\item \textbf{Unitary Evolution (Schrödinger or Heisenberg Picture).}
\[
i\hbar\,\frac{d}{dt}\,\ket{\Psi(t)} 
\;=\; \hat{H}\,\ket{\Psi(t)},
\]
In the Schrödinger picture, the quantum state \(\ket{\Psi(t)}\) of the field evolves according to the Hamiltonian \(\hat{H}\).  
In the Heisenberg picture, field operators evolve instead, while states remain fixed.  
Either formulation captures how quantum states of light propagate in space or time and how they interact with matter.

\item \textbf{Quantum Measurement Postulate.}
\[
\hat{\mathcal{O}} = \hat{\mathcal{O}}^\dagger,
\quad
\text{outcomes} = \text{eigenvalues of } \hat{\mathcal{O}},
\quad
p(\text{outcome}=\lambda) = \|\hat{P}_{\lambda}\ket{\Psi}\|^2,
\]
Physical observables (e.g., photon number, field quadratures) correspond to Hermitian operators.  
Measurement outcomes are given by their eigenvalues, with probabilities determined by the Born rule.  
Post-measurement states follow the standard projection or, more generally, a completely positive (CPTP) map.  
This postulate underlies single-photon detection, quantum state collapse (or state update), 
and all interference/entanglement measurements in quantum optics.

\end{enumerate}

\subsubsection{Completeness}
These three axioms form a complete basis for describing all key effects in quantum optics:
\begin{enumerate}
    \item \textbf{Field Quantization} allows us to represent light in terms of discrete or continuous quantum states (Fock, coherent, squeezed, etc.), capturing inherently quantum behaviors like vacuum fluctuations and photon antibunching.
    \item \textbf{Unitary Evolution} governs the time evolution and interactions of these quantum states, enabling analyses of multi-photon interference, beam-splitter transformations, and quantum entanglement generation.
    \item \textbf{Quantum Measurement} accounts for how photodetection, homodyne detection, or other measurement schemes yield observable outcomes and how the act of measurement influences or “collapses” the quantum state.
\end{enumerate}
Together, they suffice to derive virtually all standard quantum-optical phenomena, from single-photon interference to multipartite entanglement protocols.

\subsubsection{Independence}
\begin{enumerate}
    \item \textbf{Field Quantization} is not inferable from the unitary evolution or measurement principles alone: one may have quantum evolution and measurement in other (non-photonic) systems without imposing photon creation/annihilation operators or the discrete energy levels of the electromagnetic field modes.
    \item \textbf{Unitary Evolution} cannot be deduced from quantization and measurement alone, as it prescribes \emph{how} states evolve in time under a Hamiltonian and is not captured by just asserting that a quantum field exists with certain commutation relations.
    \item \textbf{Quantum Measurement Postulate} does not automatically follow from field quantization and unitary dynamics; it explicitly states \emph{how} measurement outcomes arise and how they probabilistically update the state of the system.
\end{enumerate}
Hence, none of the three axioms can be derived from the others, ensuring their logical independence in describing the quantum nature of light.

\subsection{Transformations}
\label{sec:wave-to-quantum-transform}

\subsubsection{Axiom 1}

\textbf{Goal:} 
Transform the classical \emph{Wave Equation} (Axiom 1) into the \emph{Field Quantization} postulate 
\[
Q_1:\quad
\hat{\mathbf{E}}(\mathbf{r},t),\;\hat{\mathbf{B}}(\mathbf{r},t)
~\longrightarrow~
\hat{a}_k,\hat{a}_k^\dagger
~\text{with}~
[\hat{a}_k,\hat{a}_{k'}^\dagger] = \delta_{kk'}.
\]

\paragraph{Initial Set (Wave Equation).}
\[
S_0 = \{W_1\},
\quad
W_1:\quad
\nabla^2 \mathbf{E}
-\frac{n^2}{c^2}\,\frac{\partial^2 \mathbf{E}}{\partial t^2}
=0,
\;\;
\nabla^2 \mathbf{B}
-\frac{n^2}{c^2}\,\frac{\partial^2 \mathbf{B}}{\partial t^2}
=0.
\]

\paragraph{Rewrite Step 1: Mode Expansion.}
Decompose each classical field into a sum (or integral) of normal modes:
\[
\mathbf{E}(\mathbf{r},t)
=
\sum_{k} 
\Bigl[
E_{k}(t)\,\boldsymbol{\phi}_{k}(\mathbf{r})
\Bigr],
\quad
\mathbf{B}(\mathbf{r},t)
=
\sum_{k} 
\Bigl[
B_{k}(t)\,\boldsymbol{\psi}_{k}(\mathbf{r})
\Bigr].
\]
Symbolically, 
\[
W_1 
\;\longrightarrow\; 
W_1':\;
\{\text{Mode Decomposition}\}
\]
\[
S_1 = \{\,W_1'\}.
\]

\paragraph{Rewrite Step 2: Promote Amplitudes to Operators.}
Replace each classical amplitude \(E_{k}(t)\), \(B_{k}(t)\) by quantum \emph{operators} \(\hat{E}_{k}(t), \hat{B}_{k}(t)\).  

\[
W_1'
\;\longrightarrow\;
W_1'':\;
\bigl\{
\hat{E}_{k}(t),\;\hat{B}_{k}(t)
\bigr\}.
\]
\[
S_2 = \{\,W_1''\}.
\]

\paragraph{Rewrite Step 3: Canonical Commutation.}
Identify \(\hat{E}_{k}, \hat{B}_{k}\) as linear combinations of \(\hat{a}_k\) and \(\hat{a}_k^\dagger\), the annihilation/creation operators satisfying
\[
\bigl[\hat{a}_k,\,\hat{a}_{k'}^\dagger\bigr]
=\delta_{kk'}.
\]
This step imposes the canonical \emph{quantization condition} on the field modes.

\[
W_1''
\;\longrightarrow\;
Q_1:\;
\bigl\{
\hat{\mathbf{E}}(\mathbf{r},t),\;\hat{\mathbf{B}}(\mathbf{r},t)
~\longrightarrow~
\hat{a}_k,\hat{a}_k^\dagger
~\text{with}~
[\hat{a}_k,\hat{a}_{k'}^\dagger] = \delta_{kk'}
\bigr\}.
\]
\[
S_3 = \{\,Q_1\}.
\]

\paragraph{Final Set (Quantum Optics Axiom 1: Field Quantization).}
\[
\boxed{
Q_1:\quad
\hat{\mathbf{E}}(\mathbf{r},t),\;\hat{\mathbf{B}}(\mathbf{r},t)
~\longrightarrow~
\hat{a}_k,\hat{a}_k^\dagger
~\text{with}~
[\hat{a}_k,\hat{a}_{k'}^\dagger] = \delta_{kk'}.
}
\]

\noindent
\textbf{Symbolic Evolution (Compact):}
\[
\underbrace{\{W_1\}}_{S_0}
\;\longrightarrow\;
\underbrace{\{W_1'\}}_{S_1}
\;\longrightarrow\;
\underbrace{\{W_1''\}}_{S_2}
\;\longrightarrow\;
\underbrace{\{Q_1\}}_{S_3}.
\]

\bigskip
\hrule
\bigskip

\subsubsection{Axiom 2}

\textbf{Goal:} Show how \emph{Boundary Conditions} (Axiom 2 of Wave Optics) can be reinterpreted to yield the \emph{Unitary Evolution} postulate:

\[
Q_2:\quad
i\hbar\,\frac{d}{dt}\,\ket{\Psi(t)}
=\hat{H}\,\ket{\Psi(t)},
\quad
\text{(or Heisenberg picture for operators).}
\]

\paragraph{Initial Set (Boundary Conditions).}
\[
S_0 = \{W_2\},
\quad
W_2:\quad
\begin{cases}
\epsilon_1\,\mathbf{E}_\perp\big|_1 - \epsilon_2\,\mathbf{E}_\perp\big|_2 = \sigma,\\
\mathbf{B}_\perp\;\text{cont.},\\
\mathbf{E}_\parallel,\mathbf{B}_\parallel\;\text{cont.}
\end{cases}
\]

\paragraph{Rewrite Step 1: From Interface Continuity to Global Consistency.}
In classical wave language, matching boundary conditions at \emph{every interface} ensures a single, consistent solution \(\mathbf{E},\mathbf{B}\) across the entire optical system.  Symbolically,
\[
W_2 
\;\longrightarrow\;
W_2':\;
\text{(Global single solution across all regions)}.
\]
\[
S_1 = \{W_2'\}.
\]

\paragraph{Rewrite Step 2: Single Global Solution \texorpdfstring{\(\to\)}{} Single Global State.}
In quantum language, a \emph{single global solution} of Maxwell’s equations becomes a \emph{single global quantum state} \(\ket{\Psi}\).  Rather than specifying field values in each region, we specify the \emph{amplitudes} (or wavefunction) for the entire configuration.  Symbolically,
\[
W_2'
\;\longrightarrow\;
W_2'':\;
\ket{\Psi}
\;\text{is a global state describing the field.}
\]
\[
S_2 = \{W_2''\}.
\]

\paragraph{Rewrite Step 3: Impose Hamiltonian Dynamics.}
To encode \emph{how} the field evolves in time (including how boundary interactions might reflect/refract quantum amplitudes), we introduce the Hamiltonian \(\hat{H}\).  The time evolution of \(\ket{\Psi(t)}\) is then:
\[
i\hbar\,\frac{d}{dt}\,\ket{\Psi(t)}
=
\hat{H}\,\ket{\Psi(t)}.
\]
Thus,
\[
W_2''
\;\longrightarrow\;
Q_2:\;
(\text{Unitary Evolution}).
\]
\[
S_3 = \{Q_2\}.
\]

\paragraph{Final Set (Quantum Optics Axiom 2: Unitary Evolution).}
\[
\boxed{
Q_2:\quad
i\hbar\,\frac{d}{dt}\,\ket{\Psi(t)} 
= \hat{H}\,\ket{\Psi(t)}.
}
\]

\noindent
\textbf{Symbolic Evolution (Compact):}
\[
\underbrace{\{W_2\}}_{S_0}
~\longrightarrow~
\underbrace{\{W_2'\}}_{S_1}
~\longrightarrow~
\underbrace{\{W_2''\}}_{S_2}
~\longrightarrow~
\underbrace{\{Q_2\}}_{S_3}.
\]

\bigskip
\hrule
\bigskip

\subsubsection{Axiom 3}

\textbf{Goal:} Starting with the \emph{Transverse Nature of Light} (Axiom 3), derive the \emph{Quantum Measurement Postulate}:

\[
Q_3:\quad
\hat{\mathcal{O}}=\hat{\mathcal{O}}^\dagger,\;\;
\text{outcomes}=\text{eigenvalues of }\hat{\mathcal{O}},\;\;
p(\text{outcome})=\|\hat{P}_{\text{outcome}}\ket{\Psi}\|^2,
\]
which is the Born rule for measurement in quantum optics.

\paragraph{Initial Set (Transverse Nature).}
\[
S_0 = \{\,W_3\},
\quad
W_3:\quad
\mathbf{k}\cdot \mathbf{E} = 0,
\;\;
\mathbf{k}\cdot \mathbf{B} = 0,
\;\;
\mathbf{E} \perp \mathbf{B}.
\]

\paragraph{Rewrite Step 1: Identify Independent Components.}
Since \(\mathbf{E}\) and \(\mathbf{B}\) are purely transverse, each has only two independent polarization components.
Moreover, $E$ and $B$ are directly related to each other. Symbolically,
\[
W_3 
\;\longrightarrow\; 
W_3':\;
\text{(Only 2 degrees of freedom per mode.)}
\]
\[
S_1 = \{W_3'\}.
\]

\paragraph{Rewrite Step 2: Associate Observables (Polarization, Quadratures).}
In a quantum field theory context, each transverse degree of freedom becomes an \emph{observable} (e.g.\ a polarization basis or field quadrature).  Symbolically,
\[
W_3'
\;\longrightarrow\;
W_3'':\;
\hat{\mathcal{O}}_\alpha,\;\hat{\mathcal{O}}_\beta,
\]
where \(\hat{\mathcal{O}}_\alpha\) and \(\hat{\mathcal{O}}_\beta\) might be (for example) two orthogonal polarization or quadrature operators.

\[
S_2 = \{W_3''\}.
\]

\paragraph{Rewrite Step 3: Impose Measurement Postulates.}
To fully specify how outcomes of these observables \(\hat{\mathcal{O}}_\alpha,\hat{\mathcal{O}}_\beta\) appear in experiments, we introduce the \emph{Quantum Measurement Postulate}, i.e.\ that these operators are Hermitian, their eigenvalues are the possible measurement results, and outcome probabilities follow the Born rule.  Symbolically,
\[
W_3''
\;\longrightarrow\;
Q_3:\;
(\text{Hermitian operators, eigenvalue outcomes, Born rule}).
\]
\[
S_3 = \{Q_3\}.
\]

\paragraph{Final Set (Quantum Optics Axiom 3: Measurement).}
\[
\boxed{
Q_3:\quad
\hat{\mathcal{O}}=\hat{\mathcal{O}}^\dagger,
\;\; 
\text{outcomes = eigenvalues of } \hat{\mathcal{O}}^\dagger,
\;\;
p(\text{outcome}=\lambda)=\|\hat{P}_{\lambda}\ket{\Psi}\|^2
}
\]

\noindent
\textbf{Symbolic Evolution (Compact):}
\[
\underbrace{\{W_3\}}_{S_0}
~\longrightarrow~
\underbrace{\{W_3'\}}_{S_1}
~\longrightarrow~
\underbrace{\{W_3''\}}_{S_2}
~\longrightarrow~
\underbrace{\{Q_3\}}_{S_3}.
\]

\bigskip
\hrule
\bigskip

\subsubsection{Summary of the Transformations}

\begin{enumerate}
\item \emph{Wave Equation} \(\to\) \emph{Field Quantization}:
  \[
    W_1 \;\Longrightarrow\; Q_1.
  \]
  By recasting the classical fields as operator-valued mode expansions, we impose canonical commutation and obtain creation/annihilation operators.

\item \emph{Boundary Conditions} \(\to\) \emph{Unitary Evolution}:
  \[
    W_2 \;\Longrightarrow\; Q_2.
  \]
  The requirement of consistent field solutions across interfaces leads to a single global quantum state and the Hamiltonian-based time evolution.

\item \emph{Transverse Nature} \(\to\) \emph{Quantum Measurement}:
  \[
    W_3 \;\Longrightarrow\; Q_3.
  \]
  The restriction to transverse degrees of freedom becomes the quantized operator framework, requiring a measurement postulate (Hermitian observables and Born rule) to predict experimental outcomes.
\end{enumerate}

\bigskip
\noindent
Hence, each of the three \emph{Wave Optics Axioms} can be “lifted” to its \emph{Quantum Optics} counterpart.

\section{From Analog Electronics to Digital Computing}
\label{sec:analog_to_digital}

\subsection{Analog Electronics}
\label{sec:analog-electronics}

Analog electronics deals with the behavior of circuits comprising continuous signals, resistors, capacitors, inductors, and active or passive elements, all obeying laws in the time or frequency domain.  Below is a minimal and complete set of axioms for \emph{linear} analog electronics, each of which is independent of the others.  Together, they suffice to derive the standard behavior of linear circuits: Kirchhoff’s rules, device equations, superposition, frequency response, and more.

\subsubsection{Axioms}

\begin{enumerate}
    \item \textbf{Kirchhoff's Laws.}
    \[
        \text{(KCL)}\quad 
        \sum_{n} I_{n}(t) \;=\; 0,
        \qquad
        \text{(KVL)}\quad
        \sum_{n} V_{n}(t) \;=\; 0.
    \]
    \textit{Statement.} 
    At any node, the algebraic sum of currents is zero (Kirchhoff’s Current Law, KCL); around any loop, the algebraic sum of voltages is zero (Kirchhoff’s Voltage Law, KVL).

    \item \textbf{Linear Constitutive Relations.}
    \[
    \begin{aligned}
       \text{Resistor:}& \quad V(t) \;=\; R \, I(t), \\
       \text{Capacitor:}& \quad I(t) \;=\; C \,\frac{dV}{dt}, \\
       \text{Inductor:}& \quad V(t) \;=\; L \,\frac{dI}{dt}, 
       \quad\ldots
    \end{aligned}
    \]
    \textit{Statement.}
    Each (passive) two-terminal component relates its voltage and current via a linear, time-invariant (LTI) differential equation.  In the simplest cases: resistors (Ohm's law), capacitors, and inductors.  More generally, any linear element can be described by an impedance or admittance \(Z(\omega)\) or \(Y(\omega)\).

\end{enumerate}

\subsubsection{Completeness}
These axioms suffice to describe the fundamental behavior of all \emph{linear} analog circuits:
\begin{enumerate}
    \item \textbf{Kirchhoff's Laws.} Provide global constraints on the voltages and currents, ensuring that charges and energy flows are conserved throughout the circuit.
    \item \textbf{Linear Constitutive Relations.} Define how each circuit element (resistor, capacitor, inductor, etc.) responds to voltage and current, laying the foundation for time-domain and frequency-domain analyses.
\end{enumerate}

\subsubsection{Independence}
\begin{enumerate}
    \item \textbf{Kirchhoff's Laws} cannot be derived from the constitutive relations or superposition alone; they express fundamental conservation principles that apply regardless of how individual circuit elements behave.
    \item \textbf{Linear Constitutive Relations} are not deducible from Kirchhoff’s Laws or superposition.  Kirchhoff's Laws and superposition only constrain how voltages/currents combine or sum at nodes and loops; they do not dictate the specific linear equations linking voltage and current of a single component.
\end{enumerate}
Hence, none of the two axioms can be derived from the other, ensuring their mutual independence in describing linear analog electronics.

\subsection{Digital Electronics}
\label{sec:digital-electronics}

Digital electronics models signals, circuits, and systems in terms of discrete (often binary) states, representing logical operations and finite-state behavior.  Below is a minimal set of axioms for \emph{classical digital electronics}.  Each axiom is independent of the others, and collectively, they suffice to derive standard results in digital logic, including combinational and sequential circuit analysis.

\subsubsection{Axioms}

\begin{enumerate}
    \item \textbf{Binary Signal States.}
    \[
        \text{Each signal }X(t)\text{ can take on only two distinct values, } 
        \{\,0, 1\},
        \]
        \[
        \;\text{representing logical ‘false’ or ‘true.’}
    \]
    \textit{Statement.}
    All digital signals are confined to two nominal voltage levels (or two symbols).  Intermediate values or analog considerations are assumed to be beyond the logical model.

    \item \textbf{Boolean Gate Operations.}
    \[
        \forall\,(\text{AND, OR, NOT, etc.}): \quad
        \text{output} 
        = 
        f_{\text{gate}}(\text{inputs}),
    \]
    \textit{Statement.}
    Each combinational logic element (gate) transforms its input bit(s) according to a Boolean function (truth table).  For instance, an AND gate outputs ‘1’ if and only if \emph{all} inputs are ‘1’; a NOT gate flips ‘0’ to ‘1’ and vice versa.

    \item \textbf{Synchronous State Update (Clocked Memory).}
    \[
        \text{Flip-flops, registers: } 
        Q(t+\Delta t)
        =
        F\bigl(Q(t),\,X(t)\bigr)
        \quad\text{on clock edge.}
    \]
    \textit{Statement.}
    Sequential circuits store and update states at discrete time intervals (clock cycles).  Flip-flops or registers sample input signals and produce outputs synchronized to a clock, thus forming the basis of finite-state machines.
\end{enumerate}

\subsubsection{Completeness}
These three axioms suffice to describe the key aspects of classical digital circuits:
\begin{enumerate}
    \item \textbf{Binary Signal States.}
    Establishes the discrete, two-level (0/1) nature of signals, enabling logical (rather than continuous) interpretation of circuit behavior.
    \item \textbf{Boolean Gate Operations.}
    Captures how combinational logic processes input bits via gates (AND, OR, NOT, NAND, NOR, etc.).  Permits constructing arbitrary Boolean functions.
    \item \textbf{Synchronous State Update.}
    Introduces memory elements and timing, forming sequential circuits.  This encompasses registers, counters, finite-state machines, and synchronous system design.
\end{enumerate}

\subsubsection{Independence}
\begin{enumerate}
    \item \textbf{Binary Signal States} is not derivable from Boolean gates or clocked memory alone.  Gates and flip-flops could, in principle, operate on multi-valued logic; the two-level nature of digital signals must be stated independently.
    \item \textbf{Boolean Gate Operations} cannot be inferred from the binary state axiom or from the presence of a clock or memory.  Even with two-level signals and timing, one must specifically define \emph{how} the inputs map to outputs for each gate.
    \item \textbf{Synchronous State Update} does not follow from binary signals or Boolean operations.  One can have purely combinational (stateless) logic or asynchronous designs unless the notion of clocked memory is explicitly introduced.
\end{enumerate}
Hence, none of these three axioms can be derived from the others, ensuring their mutual independence in describing classical digital electronics.

\subsection{Transformation}
\label{sec:analog-to-digital-transform}

\subsubsection{Axiom 1}

\textbf{Goal:}  
Transform a subset of analog axioms 
\[
\bigl\{\,A_1,\,A_2\bigr\}
\quad
\begin{aligned}
A_1 &: \text{Kirchhoff's Laws},\\
A_2 &: \text{Linear Constitutive Relations.},
\end{aligned}
\]
into the \emph{Digital Electronics} statement:
\[
D_1:\quad
\text{Each signal can take on only two distinct (logical) values, } \{0,1\}.
\]
All intermediate steps are small symbolic rewrites that gradually evolve the original set into the final one.

\vspace{1em}
\hrule
\vspace{1em}

\noindent
\textbf{Initial Set of Axioms (Analog, Subset):}

\[
\boxed{
S_0
=
\bigl\{\,A_1,\,A_2\bigr\}
}
\]
where
\[
A_1:
\quad
\sum_n I_n(t)=0,
\;\;
\sum_n V_n(t)=0
\quad
(\text{Kirchhoff's Current/Voltage Laws}), 
\]
\[
\quad
A_2: \quad \text{Resistor: }V=R\,I,\quad
\text{Capacitor: }I=C\,dV/dt,\dots
\quad
(\text{linear devices});
\]

\smallskip
\noindent
Here, \(A_1\) enforces basic conservation of charge/energy in circuits, while \(A_2\) implies that circuit responses to multiple inputs sum linearly (linearity principle).

\vspace{1em}
\hrule
\vspace{1em}

\noindent
\textbf{Rewrite Step 1: Introduce Superposition as Separation of Inputs.}

\[
A_2
\;\longrightarrow\;
A_2':
\quad
\text{Circuit signals can be decomposed into partial contributions (modes).}
\]

Because Kirchoff's laws ($A_1$) and the voltage-current relationships for circuit elements ($A_2$) are linear, the overall circuit exhibits linear behavior (superposition principle). Hence, the response of a system to an input can be decomposed as a response to several other inputs. Symbolically, 

\[
\forall\,(\text{inputs})\quad
V_{\text{total}}(t) = \sum_\alpha V_{\alpha}(t),
\]
and the circuit processes each \(V_{\alpha}(t)\) independently (linearly).  
Hence,
\[
S_0
=
\{\,A_1,\,A_2\}
\quad
\longrightarrow
\quad
S_1
=
\{\,A_1,\,A_2'\}.
\]

\smallskip
\noindent
No mention yet of restricting signals to two levels; we have merely restated superposition in a “decomposed signals” language.

\vspace{1em}
\hrule
\vspace{1em}

\noindent
\textbf{Current Set:}
\[
\boxed{
S_1
=
\{\,A_1,\;A_2'\}
}
\]
where
\[
A_1:\;\text{Kirchhoff's Laws (KCL, KVL)},
\]
\[
A_2':\;\text{Signals decompose into linear sums of simpler components.}
\]

\vspace{1em}
\hrule
\vspace{1em}

\noindent
\textbf{Rewrite Step 2: Enforce Threshold-like Nonlinearity at the Output Node.}

We now add a (conceptual) “hard” threshold in the circuit’s output stage, imposing that any resulting output voltage \(\widetilde{V}(t)\) is \emph{forced} to saturate to one of two levels:
\[
\widetilde{V}(t)
=
\begin{cases}
V_{\mathrm{LOW}}, & \text{if underlying analog }V_{\text{net}}(t)<V_{\mathrm{th}},\\
V_{\mathrm{HIGH}}, & \text{otherwise}.
\end{cases}
\]
Symbolically, 
\[
A_1,\,A_2'
\;\longrightarrow\;
A_1',\,A_2':
\]
where \(A_1'\) is Kirchhoff’s Laws \emph{plus} a saturating active device, ensuring only two output levels can appear.

\[
S_1
=
\{\,A_1,\;A_2'\}
\quad
\longrightarrow
\quad
S_2
=
\{\,A_1',\,A_2'\}.
\]
\[
A_1':\quad
(\text{Kirchhoff's Laws}) 
+ 
(\text{bistable saturation device}).
\]

\vspace{1em}
\hrule
\vspace{1em}

\noindent
\textbf{Current Set:}
\[
\boxed{
S_2
=
\{\,A_1',\,A_2'\}
}
\]
with 
\[
A_1':\;\text{KCL,KVL plus saturating output stage},
\]
\[
A_2':\;\text{Signals still superimpose in the linear sub-block}.
\]

\vspace{1em}
\hrule
\vspace{1em}

\noindent
\textbf{Rewrite Step 3: Abstract Away All Intermediate Voltages.}

Once we accept that any real output node can only \emph{end up} at \(\{V_{\mathrm{LOW}},V_{\mathrm{HIGH}}\}\), we effectively identify these two saturations with \(\{0,1\}\).  Symbolically, 
\[
(A_1',A_2')
\;\longrightarrow\;
D_1:\;\{\text{Binary states only}\}.
\]
Hence:
\[
S_2
=
\{\,A_1',\,A_2'\}
\quad
\longrightarrow
\quad
S_3
=
\{\,D_1\},
\]
where
\[
D_1:\;\text{All signals are 0 or 1 (two distinct logical levels).}
\]

\vspace{1em}
\hrule
\vspace{1em}

\noindent
\textbf{Final Set (Digital Axiom 1, Binary Signal States):}
\[
\boxed{
S_3
=
\Bigl\{
D_1:\;\{0,1\}\text{ logic levels}
\Bigr\}
}
\]
\[
\text{I.e.\ each output node saturates to LOW or HIGH, representing a discrete binary value.}
\]

\vspace{1em}
\hrule
\vspace{1em}

\noindent
\textbf{Symbolic Evolution (Compact Form)}:

\[
\underbrace{\{\,A_1,\,A_2\}}_{S_0}
~\longrightarrow~
\underbrace{\{\,A_1,\,A_2'\}}_{S_1}
~\longrightarrow~
\underbrace{\{\,A_1',\,A_2'\}}_{S_2}
~\longrightarrow~
\underbrace{\{\,D_1\}}_{S_3}.
\]

\bigskip

\subsubsection{Axiom 2}

\textbf{Goal:}
Starting from the two \emph{key analog axioms} 
\[
A_1:\;\text{Kirchhoff's Laws},
\quad
A_2:\;\text{Linear Constitutive Relations}, 
\]
we wish to derive (in small symbolic steps) the \emph{Digital Axiom 2}: 
\[
D_2:\quad
\text{Combinational logic gates (AND, OR, NOT, etc.) 
perform}
\]
\[
\text{Boolean operations on inputs.}
\]

\medskip
\hrule
\medskip

\noindent
\textbf{Initial Set of Axioms (Analog Subset):}
\[
S_0
=
\bigl\{\,A_1,\;A_2\bigr\}
\]
where
\[
A_1:
\quad
\sum_n I_n(t)=0,
\;\;
\sum_n V_n(t)=0
\]
\[
A_2:\quad 
\text{Resistor: }V=R\,I,\quad
\text{Capacitor: }I=C\,dV/dt,\dots
\quad
(\text{linear devices});
\]
\[
\]


\bigskip
\hrule
\bigskip

\textbf{Rewrite Step 1: Introduce Superposition.}

\[
S_0
=
\{\,A_1,\,A_2\}
\quad
\longrightarrow
\quad
S_1
=
\{\,A_1,\,A_2'\}.
\]

This rewrite step is identical to the one leading to axiom 1. Hence, for brevity, it was not repeated here.

\vspace{1em}
\hrule
\vspace{1em}

\noindent
\textbf{Current Set:}
\[
\boxed{
S_1
=
\{\,A_1,\;A_2'\}
}
\]
where
\[
A_1:\;\text{Kirchhoff's Laws (KCL, KVL)},
\]
\[
A_2':\;\text{Signals decompose into linear sums of simpler components.}
\]

\noindent
\textbf{Rewrite Step 2: Combine Linear Elements into a Single Transfer Function.}

\[
A_2 \;\longrightarrow\; B_2:
\quad
H(\omega)=\frac{V_{\text{out}}(\omega)}{V_{\text{in}}(\omega)},
\]
the net transfer function.  By cascade/feedback of linear devices (resistors, capacitors, inductors), we still have a linear system, so
\[
V_{\text{out}}(\omega)=H(\omega)\,V_{\text{in}}(\omega).
\]
Symbolically,
\[
S_1 = \{A_2,A_2'\}
\;\longrightarrow\;
S_2 = \{\,B_2,A_2'\}.
\]

\bigskip
\hrule
\bigskip

\noindent
\textbf{Current Set:}
\[
\boxed{
S_2
=
\{\,B_2,\;A_2'\}
}
\]
where
\[
B_2:\quad
V_{\text{out}}(\omega)=H(\omega)\,V_{\text{in}}(\omega),
\quad
A_2':\;\text{Superposition Principle.}
\]

\bigskip
\hrule
\bigskip

\noindent
\textbf{Rewrite Step 3: Impose Binary Inputs/Outputs on the Transfer Function.}

With the \emph{binary signal} idea from \(D_1\), assume each input line is either 0 or 1, and the output must also saturate to 0 or 1.  The linear transfer function now effectively \emph{collapses} into a truth table:
\[
(B_2,A_2')\;\longrightarrow\; B_3:
\quad
f_{\text{gate}}:\;(0,0)\mapsto X,\dots
\]
For instance, a saturating amplifier with certain resistor networks might implement 
\(\text{AND},\text{OR},\text{NOT}\), etc.\ once we interpret \(V_{\mathrm{LOW}}\to 0\), \(V_{\mathrm{HIGH}}\to 1\).  

Hence,
\[
S_2
\;\longrightarrow\;
S_3
=
\{\,B_3\},
\]
where 
\[
B_3:\quad
\text{Each linear sub-block + saturations }\implies 
\text{Boolean function table}.
\]

\bigskip
\hrule
\bigskip

\noindent
\textbf{Rewrite Step 4: Conclude that the Circuit is a Boolean Gate.}

Once we identify the input signals with \(\{0,1\}\) and interpret each sub-block’s output in the same \(\{0,1\}\)-based manner, the entire circuit \(\to\) \emph{gate} operation:
\[
D_2:\quad
(\text{Gate}): 
\quad
\text{Output} = f(\text{Inputs}),
\]
where \(f\) is a Boolean function (AND, OR, NOT, etc.).

\[
S_3 
\;\longrightarrow\;
S_4
=
\{\,D_2\}.
\]
Thus we obtain:
\[
D_2:\quad
\text{Boolean gate operations on binary signals}.
\]

\bigskip
\hrule
\bigskip

\noindent
\textbf{Final Set (Digital Axiom 2, Gate Operations):}
\[
\boxed{
S_4 
= 
\Bigl\{
D_2:\;\text{Each combinational block is a Boolean function on }\{0,1\}^n
\Bigr\}
}
\]

\bigskip
\hrule
\bigskip

\noindent
\textbf{Symbolic Evolution (Compact View):}
\[
\underbrace{
\{\,A_1,\,A_2\}
}_{S_0}
\;\longrightarrow\;
\underbrace{
\{\,A_2,\,A_2'\}
}_{S_1}
\;\longrightarrow\;
\underbrace{
\{\,B_2,\,A_2'\}
}_{S_2}
\;\longrightarrow\;
\underbrace{
\{\,B_3\}
}_{S_3}
\;\longrightarrow\;
\underbrace{
\{\,D_2\}
}_{S_4}.
\]

\bigskip

\subsubsection{Axiom 3}

\textbf{Goal:}  
Show a step-by-step transformation that leads from
\[
A_1:\,\text{Kirchhoff's Laws}, 
\quad
A_2:\,\text{Linear Constitutive Relations}
\]
to the \emph{Digital Axiom~3}:
\[
D_3:\quad
\text{Flip-flops, registers, and memory elements 
update synchronously}
\]
\[
\text{on a clock, 
}Q(t+\Delta t)=F\bigl(Q(t),X(t)\bigr).
\]

\bigskip
\hrule
\bigskip

\noindent
\textbf{Initial Setup (Analog + Some Active Components):}

\[
\boxed{
S_0
=
\bigl\{\,A_1,\;A_2,\;\text{Active Amplification}\bigr\}
}
\]
where
\[
A_1:\quad \sum_n I_n(t)=0,\quad \sum_n V_n(t)=0,
\]
\[
A_2:\quad V=R\,I,\;I=C\,dV/dt,\dots
\]
\textbf{Active Amplification:} we allow operational amplifiers or transistors to provide gain/feedback, which can form bistable or multi-stable circuits.

\bigskip
\hrule
\bigskip

\noindent
\textbf{Rewrite Step 1: Bistable Circuit (Latch).}

Using op-amps or transistor pairs (e.g.\ an SR latch), we combine the linear elements (A2) with Kirchhoff’s Laws (A1) to form a \emph{bistable} circuit.  Symbolically,
\[
(A_1,\;A_2,\;\text{Active Amplification})
\;\longrightarrow\;
B_1:\;\text{Circuit has two stable states.}
\]
Hence,
\[
S_0
\;\longrightarrow\;
S_1
=
\{\,B_1\}.
\]
\[
B_1:\quad
Q\in\{\mathrm{StateA},\,\mathrm{StateB}\},
\;\text{no intermediate stable states exist}.
\]

\bigskip
\hrule
\bigskip

\noindent
\textbf{Current Set:}
\[
\boxed{
S_1
=
\Bigl\{
B_1:\;\text{two stable states in a closed feedback loop}
\Bigr\}
}
\]

\bigskip
\hrule
\bigskip

\noindent
\textbf{Rewrite Step 2: Add Clock Signal for Controlled Transition.}

A pure bistable latch can switch states at any time if inputs cross certain thresholds.  To enforce \emph{synchronous} updates, we introduce a global \emph{clock} line that enables the latch to sample inputs only at discrete moments (the clock edge).  Symbolically,
\[
B_1
\;\longrightarrow\;
B_2:\;\text{(Clocked Flip-Flop)}.
\]
\[
S_1
\;\longrightarrow
\;
S_2
=
\{\;B_2\}.
\]
\[
B_2:\quad
Q(t+\Delta t)=F\bigl(Q(t),X(t)\bigr)\;\text{only at clock rising/falling edge}.
\]

\bigskip
\hrule
\bigskip

\noindent
\textbf{Rewrite Step 3: Recognize Finite-State Memory Element.}

Once the latch is governed by a clock, it becomes a \emph{sequential} device (e.g.\ D flip-flop, JK flip-flop, register bit).  We interpret \(\{Q(t),Q(t+\Delta t)\}\) as a finite-state machine:
\[
D_3:\quad
\text{Synchronous state update: }Q(t+\Delta t)=F\bigl(Q(t),X(t)\bigr).
\]
Hence,
\[
S_2
\;\longrightarrow\;
S_3
=
\{\,D_3\}.
\]

\smallskip
\noindent
We have thus arrived at the digital notion of clocked memory or flip-flops.

\bigskip
\hrule
\bigskip

\noindent
\textbf{Final Set (Digital Axiom 3, Synchronous Update):}
\[
\boxed{
S_3
=
\Bigl\{
D_3:\;Q(t+\Delta t)=F\bigl(Q(t),X(t)\bigr)
\Bigr\}.
}
\]

\bigskip
\hrule
\bigskip

\noindent
\textbf{Symbolic Evolution (Compact View):}
\[
\underbrace{
\{\,A_1,A_2,\;\text{Active Amplification}\}
}_{S_0}
\;\longrightarrow\;
\underbrace{
\{\,B_1:\text{Bistable Latch}\}
}_{S_1}
\;\longrightarrow\;
\]
\[
\longrightarrow
\underbrace{
\{\,B_2:\text{Clocked Flip-Flop}\}
}_{S_2}
\;\longrightarrow\;
\underbrace{
\{\,D_3\}
}_{S_3}.
\]

\bigskip

\subsubsection{Summary of Transformations (Analog to Digital)}

\begin{enumerate}
    \item \emph{From Kirchhoff + Linear Relations to Binary Signal States}:
    \[
      \{A_1,A_2\} \;\Longrightarrow\; D_1.
    \]
    By adding a threshold-based saturation stage, continuous voltages are forced into LOW or HIGH, yielding discrete \(\{0,1\}\).

    \item \emph{From Linear Devices to Boolean Gate Operations}:
    \[
      \{A_2,A_3\} \;\Longrightarrow\; D_2.
    \]
    Once inputs/outputs are binary, linear sub-blocks with saturations become truth tables, i.e.\ Boolean gates (AND, OR, NOT, etc.).

    \item \emph{From Bistable Circuits to Synchronous State Updates}:
    \[
      \{A_1,A_2\} \;\Longrightarrow\; D_3.
    \]
    Incorporating active components and clock signals transforms a latch into a flip-flop or register, enforcing synchronous updates of the internal state.
\end{enumerate}

Hence, each of the three \emph{Analog Electronics Axioms} can be “lifted” or transformed into a corresponding \emph{Digital Electronics Axiom} by introducing threshold logic, binary saturation, and clock-based sequencing. This parallels a stepwise progression where continuous analog phenomena become discrete binary states, Boolean logic, and synchronous finite-state machines.

\section{From Digital Computing to Quantum Computing}
\label{sec:digital_to_quantum}

\subsection{Digital Computing}
\label{sec:digital-computing}

Digital computing, at its core, can be captured by a minimal set of foundational principles. These axioms formalize how \emph{discrete} information is manipulated via \emph{symbolic} rules and \emph{finite} internal states, ensuring universal computation capability. Below is one way (more theoretical and less hardware-based than in the section above) to present a set of three axioms that are sufficient to describe classical digital computation and are mutually independent.

\subsubsection{Axioms}

\begin{enumerate}

    \item \textbf{Discrete Symbol Representation.}
    \[
    A_1:\quad
    \text{All information is encoded using a finite alphabet }
    \Sigma
    \text{ (e.g.\ bits).}
    \]
    \textit{Statement.}\\
    At the lowest level, any digital system uses a finite set of distinct symbols \(\Sigma\) (e.g.\ \(\{0,1\}\) in binary) to represent data and instructions. Intermediate analog effects are \emph{not} part of the logical model; signals are interpreted only as symbols from \(\Sigma\).

    \item \textbf{Finite Control and State Transitions.}
    \[
    A_2:\quad
    S \xrightarrow{\delta} S', 
    \quad
    S,\,S' \in \mathcal{S},
    \quad
    \mathcal{S}\text{ finite},
    \quad
    \delta:\mathcal{S}\times\Sigma \to \mathcal{S}.
    \]
    \textit{Statement.}\\
    A digital computer at any instant is in one of a finite number of internal states \(\mathcal{S}\).  Input symbols (from \(\Sigma\)), along with the current state \(S\), determine the next state \(S'\) via a well-defined transition function \(\delta\).  This principle underlies finite-state machines, CPU instruction cycles, and other sequential structures.

    \item \textbf{Algorithmic Universality (Effective Computability).}
    \[
    A_3:\quad
    \exists\, \mathcal{U}:\;
    \forall\, f \in \mathcal{C},
    \;\exists\, p \in \Sigma^*,\;
    \forall\, x \in \Sigma^*:\;
    \mathcal{U}(p,x) = f(x).
    \]
    \textit{Statement.}\\
    There exists a \emph{universal} mechanism \(\mathcal{U}\) (e.g.\ a programmable processor) such that for every \emph{effectively definable} function \(f\) over the alphabet \(\Sigma\), there is a finite ``program'' \(p \in \Sigma^*\) making \(\mathcal{U}\) emulate \(f\). In practice, a minimal instruction set or abstract universal device suffices to show that all computable functions can be implemented.

    \smallskip
    \noindent
    \textbf{Interpretation of Symbolic Form:}\\
    - \(\Sigma^*\) is the set of all finite strings over \(\Sigma\).\\
    - \(\mathcal{C}\) is the set of all \emph{effectively definable} (computable) functions \(\Sigma^* \to \Sigma^*\).\\
    - \(\mathcal{U}(p,x)\) denotes the output produced by the universal mechanism \(\mathcal{U}\) when given ``program'' \(p\) and input \(x\).\\
    Thus \(A_3\) asserts that \(\mathcal{U}\) can reproduce \emph{any} desired computable function, once given the appropriate program.

\end{enumerate}

\subsubsection{Completeness}
These three axioms account for the essential ingredients of classical digital computing:
\begin{enumerate}
    \item \textbf{Discrete Symbol Representation}  
    guarantees that all data and instructions are handled as symbolic tokens from a finite set (bits, characters, etc.), enabling unambiguous, reproducible manipulation.
    \item \textbf{Finite Control and State Transitions}  
    encodes how a digital system moves from one configuration to another in discrete time steps (or events). This underpins everything from simple finite-state controllers to CPU instruction cycles.
    \item \textbf{Algorithmic Universality}  
    confirms that by combining discrete symbols and a finite control mechanism, one can implement \emph{every} computable function (assuming sufficient memory/time resources). This is the essence of universal computability.
\end{enumerate}
Together, they suffice to describe how classical digital machines store information (via discrete symbols), update their state (via finite transitions), and perform any computable algorithm.

\subsubsection{Independence}
\begin{enumerate}
    \item \textbf{Discrete Symbol Representation} is not derivable from finite-state control or universality alone.  One could imagine a continuous control system without explicitly stipulating discrete symbols.
    \item \textbf{Finite Control and State Transitions} cannot be inferred merely from having a discrete alphabet and the notion of a universal mechanism.  One must explicitly stipulate \emph{how} the machine progresses from one moment to the next in a finite-state manner.
    \item \textbf{Algorithmic Universality} does not follow automatically from having discrete symbols and finite transitions.  Without an explicit statement about the \emph{generality} of these transitions, the system might be limited to a small class of tasks (not fully universal).
\end{enumerate}
Hence, none of these axioms is derivable from the others, ensuring their mutual independence and collectively defining the foundations of classical digital computing.

\subsection{Quantum Computing}
\label{sec:quantum-computing}

Quantum computing extends classical digital computing by allowing information to be encoded in and processed by genuine quantum states. Below is a minimal set of axioms that capture the essence of quantum computation. Each axiom is independent of the others, and together they suffice to describe the behavior of all standard quantum computing models (circuit-based, measurement-based, adiabatic, etc.).

\subsubsection{Axioms}

\begin{enumerate}

    \item \textbf{Quantum State Space (Hilbert Space).}
    \[
    A_1:\quad 
    \ket{\psi} \;\in\; \mathcal{H},
    \quad
    \dim(\mathcal{H})<\infty,
    \quad
    \|\ket{\psi}\| = 1.
    \]
    \textit{Statement.} \\
    Information is represented by vectors in a finite-dimensional complex Hilbert space~\(\mathcal{H}\).  Each \emph{qubit} (or collection of qubits) is described by a normalized state \(\ket{\psi}\). Superposition and entanglement are direct consequences of the linear structure of \(\mathcal{H}\).

    \item \textbf{Unitary Evolution and Quantum Gates.}
    \[
    A_2:\quad 
    \ket{\psi(t + \Delta t)} \;=\; \hat{U}\,\ket{\psi(t)},
    \quad
    \hat{U}^\dagger \hat{U} = \hat{I}.
    \]
    \textit{Statement.} \\
    In the absence of measurement, the evolution of quantum states is governed by \emph{unitary} transformations \(\hat{U}\).  These can be discrete gates (e.g., single- or multi-qubit logic gates) or continuous evolution under a Hamiltonian \(\hat{H}\) (so \(\hat{U} = e^{-\,i\,\hat{H}\,t / \hbar}\)).  Unitary operators preserve the norm and ensure that quantum computation is fundamentally reversible.

    \item \textbf{Quantum Measurement Postulate.}
    \[
    A_3:\quad
    \hat{\mathcal{O}} = \hat{\mathcal{O}}^\dagger,
    \quad
    \text{Outcomes} = \mathrm{Eigvals}(\hat{\mathcal{O}}),
    \quad
    p(\text{outcome}) = 
    \|\hat{P}_{\text{outcome}}\ket{\psi}\|^2.
    \]
    \textit{Statement.} \\
    Physical observables are represented by Hermitian operators \(\hat{\mathcal{O}}\).  Upon measurement, the system “collapses” (or is updated) to an eigenstate of \(\hat{\mathcal{O}}\) with probability given by the Born rule.  This underlies all qubit readout processes and more general measurement schemes (POVMs) in quantum computing.

\end{enumerate}

\subsubsection{Completeness}
These three axioms suffice to describe the core structure and dynamics of quantum computers:
\begin{enumerate}
    \item \textbf{Quantum State Space (Hilbert Space).}
    Formalizes the representation of quantum information, including superposition and entanglement.
    \item \textbf{Unitary Evolution and Quantum Gates.}
    Describes how quantum information is processed or transformed, via discrete gates or continuous Hamiltonian evolution.
    \item \textbf{Measurement Postulate.}
    Explains how classical outcomes emerge upon measurement and how the quantum state updates according to the measurement result (via projection or more general maps).
\end{enumerate}
Together, these axioms enable universal quantum computation: one can construct quantum circuits, measurement-based algorithms, or adiabatic protocols capable of implementing arbitrary quantum algorithms.

\subsubsection{Independence}
\begin{enumerate}
    \item \textbf{Quantum State Space} is not inferable from unitarity or measurement alone.  One could posit abstract transformations and observations without specifying a complex Hilbert-space structure.
    \item \textbf{Unitary Evolution} cannot be deduced purely from the existence of a Hilbert space and a measurement postulate.  It must be explicitly stated that closed-system dynamics are governed by a norm-preserving (unitary) operation.
    \item \textbf{Measurement Postulate} does not follow from the first two axioms.  Even with a Hilbert space and unitary gates, one needs an additional rule specifying how measurement outcomes (eigenvalues) arise and how the post-measurement state is determined.
\end{enumerate}
Hence, these three axioms are mutually independent while collectively forming the fundamental basis of quantum computing.

\subsection{Transformation}

\subsubsection{Axiom 1}

\textbf{Goal:} Transform a subset of Digital Computing axioms
\[
\bigl\{\,A_1,\,A_3\bigr\}
\quad
\begin{aligned}
A_1 &: \text{Discrete Symbol Representation},\\
A_3 &: \text{Algorithmic Universality (Effective Computability)},
\end{aligned}
\]
into the \emph{Quantum Computing} statement:
\[
Q_1:\quad
\ket{\psi}\in\mathcal{H},\;\|\ket{\psi}\|=1,\;\dim(\mathcal{H})<\infty.
\]
Here, we re-interpret the finite alphabet \(\Sigma\) (from \(A_1\)) as an orthonormal basis for a finite-dimensional Hilbert space.

\vspace{1em}
\hrule
\vspace{1em}

\noindent
\textbf{Initial Set of Axioms (Digital, Subset):}

\[
\boxed{
S_0
=
\bigl\{\,A_1,\,A_3\bigr\}
}
\]
where
\[
A_1:
\quad
\text{Information is encoded in a finite alphabet }\Sigma,
\]
\[
A_3:
\quad
\exists\, \mathcal{U}:\;
    \forall\, f \in \mathcal{C},
    \;\exists\, p \in \Sigma^*,\;
    \forall\, x \in \Sigma^*:\;
    \mathcal{U}(p,x) = f(x).
\]

\medskip
\noindent
The discrete alphabet \(\Sigma\) must be explicitly carried along with \(A_3\), ensuring that the universal machine manipulates finite-symbol data.

\vspace{1em}
\hrule
\vspace{1em}

\noindent
\textbf{Rewrite Step 1: Replace Finite Alphabet \texorpdfstring{$\Sigma$}{Sigma} by Finite-Dimensional Basis.}

\[
A_1
\;\longrightarrow\;
A_1':
\quad
\Sigma
\;\to\;
\{\ket{0},\ket{1},\dots,\ket{n-1}\}
\subset \mathcal{H}.
\]
Symbolically, the discrete symbols \(\{0,1,\ldots\}\) become orthonormal basis states \(\{\ket{j}\}\) in a finite-dimensional Hilbert space \(\mathcal{H}\).  
Hence:
\[
S_0
=
\bigl\{\,A_1,\,A_3\bigr\}
\quad
\longrightarrow
\quad
S_1
=
\bigl\{\,A_1',\,A_3\bigr\}.
\]

\vspace{1em}
\hrule
\vspace{1em}

\noindent
\textbf{Current Set:}
\[
\boxed{
S_1
=
\{\,A_1',\,A_3\}
}
\]
where
\[
A_1':\quad
\text{Basis vectors in a finite-dimensional Hilbert space represent the discrete symbols},
\]
\[
A_3:\quad
\text{Universal computability over }\Sigma.
\]

\vspace{1em}
\hrule
\vspace{1em}

\noindent
\textbf{Rewrite Step 2: Allow Arbitrary Superpositions.}

Though \(A_1'\) introduced a basis for \(\Sigma\), classical machines only occupy one symbol at a time.  In the quantum model, however, we allow any normalized linear combination (superposition):
\[
A_1'
\;\longrightarrow\;
A_1'':
\quad
\ket{\psi}
=
\sum_j 
\alpha_j\,\ket{j},
\quad
\alpha_j\in\mathbb{C},
\quad
\|\ket{\psi}\|=1.
\]
Symbolically,
\[
S_1
=
\{\,A_1',\,A_3\}
\quad
\longrightarrow
\quad
S_2
=
\{\,A_1'',\,A_3\}.
\]

\medskip
\noindent
We still retain the universal mechanism idea from \(A_3\), but now extended to the possibility of superposition states.

\vspace{1em}
\hrule
\vspace{1em}

\noindent
\textbf{Rewrite Step 3: Conclude Quantum State Space (Axiom~\texorpdfstring{$Q_1$}{Q1}).}

Given that \(\Sigma\) supplies a finite dimension, we define the full quantum state \(\ket{\psi}\) in that Hilbert space. Thus,
\[
Q_1:\quad
\ket{\psi}\in \mathcal{H},
\quad
\|\ket{\psi}\|=1,
\quad
\dim(\mathcal{H})<\infty.
\]
Hence:
\[
S_2
=
\{\,A_1'',\,A_3\}
\quad
\longrightarrow
\quad
S_3
=
\{\,Q_1\}.
\]

\bigskip
\hrule
\bigskip

\noindent
\textbf{Final Set (Quantum Axiom 1: State Space).}
\[
\boxed{
S_3
=
\Bigl\{
Q_1:\;\ket{\psi} \in \mathcal{H},\;\|\ket{\psi}\|=1,\;\dim(\mathcal{H})<\infty
\Bigr\}.
}
\]
Symbolic Evolution (Compact):
\[
\text{}\quad
\underbrace{\{A_1,A_3\}}_{S_0}
\;\longrightarrow\;
\underbrace{\{A_1',A_3\}}_{S_1}
\;\longrightarrow\;
\underbrace{\{A_1'',A_3\}}_{S_2}
\;\longrightarrow\;
\underbrace{\{Q_1\}}_{S_3}.
\]

\bigskip

\subsubsection{Axiom 2}

\textbf{Goal:} Starting from the essential Digital Axiom
\[
A_2:\quad
(\text{Finite Control and State Transitions}),
\]
\emph{plus} the discrete assumption \(A_1\) that states and symbols lie in \(\Sigma\), we want to derive
\[
Q_2:\quad
\ket{\psi(t+\Delta t)} = \hat{U}\,\ket{\psi(t)},
\quad
\hat{U}^\dagger\hat{U}=\hat{I}.
\]
That is, the classical update function \(\delta\) becomes a \emph{unitary} map on the quantum state space.

\medskip
\hrule
\medskip

\noindent
\textbf{Initial Set of Axioms:}
\[
S_0
=
\bigl\{\,A_1,\,A_2\bigr\}
\]
where 
\[
A_1:\quad\Sigma\text{ is finite (discrete symbols)},
\quad
A_2:\quad
S \xrightarrow{\delta} S',
\;
S,S'\in\mathcal{S},
\;
\mathcal{S}\text{ finite}.
\]

\smallskip
\noindent
We keep in mind that \(\delta\) handles \emph{discrete} states. 

\bigskip
\hrule
\bigskip

\noindent
\textbf{Rewrite Step 1: Represent Internal States in a Hilbert Space.}

We identify each classical state \(S_j\in \mathcal{S}\) with a basis vector \(\ket{S_j}\).  Together with \(\Sigma\), we have a product space for “internal state + symbol.” Symbolically,
\[
A_1,\,A_2
\;\longrightarrow\;
B_2:\quad
\ket{S_j}\otimes\ket{x}
\quad
(x\in\Sigma).
\]
\[
S_0
\;\longrightarrow\;
S_1
=
\{\,B_2\}.
\]

\bigskip
\hrule
\bigskip

\noindent
\textbf{Rewrite Step 2: Generalize \texorpdfstring{$\delta$}{delta} to a Linear Operator.}

Classically, \(\delta\) just picks the next state: \((S_j,x)\mapsto S_{j'}\).  In the quantum model, we define a \emph{linear} operator \(\hat{U}\) that acts on superpositions of \(\ket{S_j}\otimes\ket{x}\). 
\[
B_2
\;\longrightarrow\;
B_2':\quad
\hat{U}:\;\ket{S_j}\otimes\ket{x}
\;\mapsto\;
\ket{S_{j'}}.
\]
\[
S_1
\;\longrightarrow\;
S_2
=
\{\,B_2'\}.
\]

\bigskip
\hrule
\bigskip

\noindent
\textbf{Rewrite Step 3: Impose Unitarity (Reversibility).}

To maintain quantum consistency (norm preservation, reversibility), \(\hat{U}\) must be unitary:
\[
\hat{U}^\dagger \hat{U}=\hat{I}.
\]
Hence,
\[
B_2'
\;\longrightarrow\;
Q_2:\quad
\ket{\psi(t+\Delta t)}=\hat{U}\,\ket{\psi(t)}.
\]
\[
S_2
\;\longrightarrow\;
S_3
=
\{\,Q_2\}.
\]

\bigskip
\hrule
\bigskip

\noindent
\textbf{Final Set (Quantum Axiom 2, Unitary Evolution):}
\[
\boxed{
S_3 
= 
\Bigl\{
Q_2:\;\ket{\psi(t+\Delta t)}=\hat{U}\,\ket{\psi(t)},\;\hat{U}^\dagger\hat{U}=\hat{I}
\Bigr\}.
}
\]
\[
\text{Symbolic Evolution (Compact):}\quad
\underbrace{\{A_1,A_2\}}_{S_0}
\;\longrightarrow\;
\underbrace{\{B_2\}}_{S_1}
\;\longrightarrow\;
\underbrace{\{B_2'\}}_{S_2}
\;\longrightarrow\;
\underbrace{\{Q_2\}}_{S_3}.
\]

\bigskip

\subsubsection{Axiom 3}

\textbf{Goal:} Starting with the classical “algorithmic universality” axiom \(A_3\) (together with discrete symbols from \(A_1\)), we wish to derive the \emph{Quantum Measurement Postulate}:
\[
Q_3:\quad
\hat{\mathcal{O}}=\hat{\mathcal{O}}^\dagger,
\quad
\mathrm{Outcomes}=\mathrm{Eigvals}(\hat{\mathcal{O}}),
\quad
p(\mathrm{outcome})=\|\hat{P}_{\mathrm{outcome}}\ket{\psi}\|^2.
\]
We will decompose the transition into smaller symbolic rewrites, showing exactly how “universal classical outputs” become “measurement outcomes” in quantum theory.

\bigskip
\hrule
\bigskip

\noindent
\textbf{Initial Setup (Digital + Universality + Discrete Symbols):}

\[
\boxed{
S_0
=
\{\,A_1,\,A_3\}
}
\quad
\begin{aligned}
A_1 &: \text{Discrete Symbol Representation (alphabet }\Sigma\text{)},\\
A_3 &: \text{Algorithmic Universality (over }\Sigma\text{)}.
\end{aligned}
\]
Here, the universal machine can compute any function \(f:\Sigma^*\to \Sigma^*\) and produce \emph{classical} outputs in \(\Sigma\). We aim to see how these outputs become quantum measurement outcomes.

\bigskip
\hrule
\bigskip

\noindent
\textbf{Rewrite Step 1: Formalize an \emph{Output Function} 
in the Classical Universal Machine.}

In the classical setting, a universal machine \(\mathcal{U}\) takes input \((p,x)\) (program \(p\) and data \(x\)), then halts with an \emph{output} \(y\in \Sigma\). Symbolically, 
\[
A_3
\;\longrightarrow\;
O_3:\quad
\mathcal{U}(p,x)\;=\;y,\quad y\in\Sigma.
\]
Thus:
\[
S_0
=
\{\,A_1,\,A_3\}
\quad
\longrightarrow
\quad
S_1
=
\{\,O_3\}.
\]

\smallskip
\noindent
\emph{Interpretation:} We have made explicit that the classical machine must eventually yield a \textit{discrete symbol} \(y \in \Sigma\). Before, \(A_3\) said “for each computable \(f\), there is a program \(p\) so that \(\mathcal{U}(p,x)=f(x)\).” Now we highlight the \(\mathcal{U}\to y\) output transition as \(O_3\).

\bigskip
\hrule
\bigskip

\noindent
\textbf{Rewrite Step 2: In Quantum Theory, Introduce a Hermitian Operator for Output.}

To reproduce “classical outputs” in the quantum model, we define a Hermitian operator \(\hat{\mathcal{O}}\) whose eigenvalues \(\{\lambda\}\) label possible outcomes (classical symbols). Symbolically,
\[
O_3
\;\longrightarrow\;
M_3:
\quad
(\hat{\mathcal{O}} = \hat{\mathcal{O}}^\dagger,\;\mathrm{Eigvals}(\hat{\mathcal{O}})=\{\lambda\}),
\]
where each \(\lambda\) corresponds to a potential discrete output. Thus:
\[
S_1
=
\{\,O_3\}
\quad
\longrightarrow
\quad
S_2
=
\{\,M_3\}.
\]

\smallskip
\noindent
\emph{Interpretation:} We replace the idea of “classical output \(y\in\Sigma\)” with “quantum measurement eigenvalue \(\lambda\).” The operator \(\hat{\mathcal{O}}\) is the quantum mechanism that yields \(\lambda\) upon measurement.

\bigskip
\hrule
\bigskip

\noindent
\textbf{Rewrite Step 3: Enforce the Born Rule for Probabilistic Outcomes.}

Finally, to connect the machine’s quantum state \(\ket{\psi}\) to a definite classical symbol \(\lambda\), we impose:
\[
p(\text{outcome} = \lambda)
=
\bigl\|\hat{P}_\lambda\ket{\psi}\bigr\|^2,
\]
where \(\hat{P}_\lambda\) is the projector onto the eigenspace of \(\hat{\mathcal{O}}\) with eigenvalue \(\lambda\). Symbolically,
\[
M_3
\;\longrightarrow\;
Q_3:\quad
(\text{Measurement Postulate}).
\]
Hence:
\[
S_2
=
\{\,M_3\}
\quad
\longrightarrow
\quad
S_3
=
\{\,Q_3\}.
\]

\smallskip
\noindent
\emph{Interpretation:} We have now fully replaced “classical universal output” with the quantum notion of “measuring \(\hat{\mathcal{O}}\), obtaining eigenvalue \(\lambda\)” with probability given by the Born rule.

\bigskip
\hrule
\bigskip

\noindent
\textbf{Final Set (Quantum Axiom 3: Measurement).}
\[
\boxed{
S_3
=
\Bigl\{
Q_3:\;\hat{\mathcal{O}}=\hat{\mathcal{O}}^\dagger,\;
\mathrm{Outcomes}=\mathrm{Eigvals}(\hat{\mathcal{O}}),\;
p(\mathrm{outcome})=\|\hat{P}_{\mathrm{outcome}}\ket{\psi}\|^2
\Bigr\}.
}
\]

\bigskip
\noindent
\textbf{Symbolic Evolution (Compact Form)}:
\[
\underbrace{\{A_1,\,A_3\}}_{S_0}
~\longrightarrow~
\underbrace{\{O_3\}}_{S_1}
~\longrightarrow~
\underbrace{\{M_3\}}_{S_2}
~\longrightarrow~
\underbrace{\{Q_3\}}_{S_3}.
\]

\bigskip

\subsubsection{Summary of Transformations (Digital \texorpdfstring{$\to$}{to} Quantum)}

\begin{enumerate}
    \item \textbf{From Discrete Symbols (Axiom 1) to Quantum State Space (Q1).}\\
    We lift the finite alphabet \(\Sigma\) into a finite-dimensional Hilbert space \(\mathcal{H}\), permitting superpositions of basis states.
    
    \item \textbf{From Finite Control (Axiom 2) to Unitary Evolution (Q2).}\\
    The classical transition function \(\delta\) generalizes to a linear, norm-preserving operator \(\hat{U}\), ensuring reversibility of quantum evolution.
    
    \item \textbf{From Algorithmic Universality (Axiom 3) to Measurement Postulate (Q3).}\\
    Classical outputs must be read out from quantum states, so we formalize observables via Hermitian operators and outcome probabilities via the Born rule.
\end{enumerate}

Together, these steps illustrate how each digital axiom—\emph{including the discrete representation from \(A_1\)—carries forward} and is reinterpreted in the quantum framework, yielding the three quantum axioms of state space, unitary evolution, and measurement. 

\section{From Classical/Newtonian Physics to General Relativity Theory}
\label{sec:newtonian_general}

\subsection{Classical/Newtonian Physics}
\label{subsec:newton}

Newtonian physics treats spacetime as flat and time as universal. It can be used to predict the motion of bodies experiencing forces and accelerations, as well as quantitatively describe gravity and calculate orbits of celestial bodies. It can only be successfully applied to objects moving at nonrelativistic velocities.

\subsubsection{Axioms}

\begin{enumerate}
\item \textbf{Absolute Space and Time}

\[
\forall \text{ Frames } 1, 2 : dt_1= dt_2, r_i=( x_i, y_i, z_i ), \text{ and metric } ds_i^2=dx_i^2+dy_i^2+dz_i^2 ,\, ds_1^2=ds_2^2
\]

Where $d t_i$ and $d x_i, d y_i, d z_i$ denote the time separation of two events, as measured in frame $i$. In other words, space is three-dimensional, Euclidean, and absolute; time flows uniformly and is the same for all observers.

\item \textbf{Existence of Inertial Frames}

\[
\exists \text{ Frame in which } \vec{F}=\vec{0} \Rightarrow  \ \frac{d\vec{v}}{dt}=\vec{0}, \quad v^i=\frac{dx^i}{dt}
\]

There exist reference frames (inertial frames) in which Newton's laws of motion hold true without modification.




\item \textbf{Newton's Second Law (Law of Acceleration)}

\[
\vec{F}=m\vec{a}=m\frac{d\vec{v}}{dt},\quad v^i=\frac{dx^i}{dt}
\]

In an inertial frame, the net external force acting on a body equals the rate of change of its momentum.

\item \textbf{Newton's Third Law (Law of Action and Reaction)}
\[
\vec{F}=-\vec{R}
\]

For every action $\vec{F}$, there is an equal and opposite reaction $\vec{R}$; forces between two bodies are equal in magnitude and opposite in direction.

\item \textbf{Principle of Superposition}

\[
\vec{F}=\sum_i \vec{F_i}
\]

When multiple forces $\vec{F_i}$ act on a body, the net force $\vec{F}$ is the vector sum of the individual forces acting on it.

\item \textbf{Law of Universal Gravitation}

\[
\vec{F_{g}}=\frac{Gm_1m_2}{r^2}\hat{r}
\]

Every point mass $2$ attracts every other point mass $1$ with a force along the unit vector $\hat{r}$ pointing from $1$ to $2$, proportional to the product of their masses $m_1, m_2, $ and inversely proportional to the square of the $r$ distance between them.

\end{enumerate}

\subsubsection{Completeness}

\begin{enumerate}
\item \textbf{Absolute Space and Time} (\textit{Axiom 1}) shows the mathematical structure of system parameters and how they can be transformed between reference frames.

\item \textbf{Existence of Inertial Frames} (\textit{Axiom 2}) shows that Newton’s laws can be applied to our universe.

\item \textbf{Newton's Second Law} (\textit{Axiom 3}), \textbf{Newton's Third Law} (\textit{Axiom 4}), and \textbf{Principle of Superposition} (\textit{Axiom 5}) give a way of predicting a body’s motion, provided forces acting on it.

\item \textbf{Law of Universal Gravitation} (\textit{Axiom 6}) describes the fundamental force of gravity, which attracts any two objects.

\end{enumerate}

From these axioms, one can predict the motion of celestial objects, transform between reference frames, draw force diagrams, and model the behavior of arbitrary systems (for instance rotation), given the forces. They suffice to describe the basic phenomena of Newtonian mechanics.

\subsubsection{Independence}

\begin{enumerate}
\item \textbf{Absolute Space and Time} sets the mathematical framework to which other other axioms should be applied.
\item \textbf{Existence of Inertial Frames} can’t be derived from other axioms, which explicitly assume being in an inertial reference frame.
\item \textbf{Newton's Second Law} is the only axiom that quantitatively describes the effect of an arbitrary force on a body's motion.
\item \textbf{Newton's Third Law} is the only axiom that refers to an arbitrary interaction between two bodies.
\item \textbf{Principle of Superposition} is the only action that demonstrates the effects of multiple forces on a body.
\item \textbf{Law of Universal Gravitation} refers to gravity, a concept not introduced in previous axioms.
\end{enumerate}

\subsection{General relativity}

The fundamental difference between general relativity and Newtonian mechanics is the assumption that the speed of light is the same in all reference frames. This, and a careful analysis of the phenomenon of gravity, leads to the description of spacetime as curved, with no universal notion of Euclidean length or time. The theory has been able to predict the existence of black holes and the bending of light by massive objects and accurately describe orbits in strong gravity conditions. General relativity has great practical usage as well, for instance, to accurately program GPS systems.

\subsubsection{Axioms}

\begin{enumerate}

\item \textbf{Principle of Equivalence (Einstein's Equivalence Principle)}

\[
\text{Gravity indistinguishable from acceleration}
\]

Locally, the effects of gravity are indistinguishable from those of acceleration. This means that in a small enough region of spacetime, the laws of physics reduce to those of Special Relativity, and gravitational effects can be transformed away by choosing an appropriate accelerated reference frame.

\item \textbf{Principle of General Covariance}

\[
g_{\mu\nu}'(x')=\frac{\partial x^{\alpha}}{\partial x'^{\mu}} \frac{\partial x^{\beta}}{\partial x'^{\nu}} g_{\alpha \beta}(x) \text{, } \mathcal{L}(x^{\mu}, g_{\mu \nu})=\mathcal{L'}(x'^{\mu}, g'_{\mu \nu})
\]

Where primed $x'^{\mu}$ and unprimed $x^{\mu}$ coordinates correspond to different reference frames. $g_{\mu\nu}$ and $g'_{\mu\nu}$ are the corresponding metrics, and $\mathcal{L}(x^{\mu}, g_{\mu \nu})=\mathcal{L'}(x'^{\mu}, g'_{\mu \nu})$ corresponds to the Lagrangian density in different reference frames.

In other words, the laws of physics are formulated in a way that is valid under any smooth coordinate transformations. This means that the fundamental equations are tensor equations, ensuring they hold true in all coordinate systems.

\item \textbf{Spacetime as a Four-Dimensional Pseudo-Riemannian Manifold}

\[
(M_4, g_{\mu \nu}) \text{ with signature } (-, +, +, +)
\]

Spacetime is modeled as a smooth, four-dimensional manifold $M_4$ equipped with a metric tensor $g_{\mu\nu}$ of Lorentzian signature $(-, +, +, +)$. The metric tensor defines distances and angles in spacetime and determines its geometric and causal structure.

\item \textbf{Einstein's Field Equations}

\[
R_{\mu \nu}-\frac{1}{2}g_{\mu \nu}R + \Lambda g_{\mu \nu} = \frac{8\pi G}{c^4} T_{\mu \nu}
\]

Where $R_{\mu \nu}$ is the Ricci curvature tensor, $g_{\mu \nu}$ is the metric tensor, $R$ is the Ricci scalar (trace of the Ricci tensor), $\Lambda$ is the cosmological constant, $G$ is the gravitational constant, and $c$ is the speed of light.

The curvature of spacetime is directly related to the energy and momentum content within it.

\item \textbf{Geodesic Principle (Motion of Free Particles)}

\[
\forall \vec{F}=0 \Rightarrow \nabla_{\tau}v=0, \quad v^{\mu}=\frac{dx^{\mu}}{d\tau}
\]

Where $\nabla_{\tau}$ denotes the covariant derivative, and $\tau$ the proper time.
In the absence of non-gravitational forces $\vec{F}$, free-falling test particles move along timelike geodesics of spacetime.

\item \textbf{Metric Compatibility and Torsion-Free Connection (Levi-Civita Connection)}

\[
{\Delta}_{\lambda}g_{\mu \nu}=0 \text{,  } {\Gamma}_{\mu \nu}^{\lambda}={\Gamma}_{\nu \mu}^{\lambda}
\]

The connection on the spacetime manifold is the unique torsion-free, metric-compatible Levi-Civita connection.

\end{enumerate}

\subsubsection{Completeness}

\begin{enumerate}

\item \textbf{Principle of Equivalence} (\textit{Axiom 1}) allows one to connect general relativity to special relativity to derive the geodesics equation and predict the nature of spacetime.

\item \textbf{Principle of General Covariance}(\textit{Axiom 2}), \textbf{Spacetime as a Four-Dimensional Pseudo-Riemannian Manifold} (\textit{Axiom 3}), and \textbf{Metric Compatibility and Torsion-Free Connection (Levi-Civita Connection)} (\textit{Axiom 6}) lay the basis of a mathematical framework that should describe reality.

\item \textbf{Einstein's Field Equations} (\textit{Axiom 4}) quantitatively relate the curvature of spacetime to present matter. It introduces the empirical gravitational constant, which lets one describe the expansion of the universe.

\item \textbf{The geodesic principle} (\textit{Axiom 5}) describes the motion of free particles.

\end{enumerate}

Together, these axioms let one describe the universe as curved spacetime and allow for a quantitative description of various phenomena, such as black holes, gravitational waves, and gravitational lensing.

\subsubsection{Independence}

\begin{enumerate}

\item \textbf{Principle of Equivalence} cannot be derived from other axioms and establishes the nature of gravity.

\item \textbf{Principle of General Covariance} is the only universal statement about physical laws in the general relativity framework that is crucial for deriving them.

\item \textbf{Spacetime as a Four-Dimensional Pseudo-Riemannian Manifold} is essential to introduce the notion of a metric and curvature to describe spacetime.

\item \textbf{Einstein's Field Equations} contain the empirical cosmological constant, which can’t be evaluated from other axioms.

\item \textbf{The geodesic principle} is the only equivalent of Newton's laws that lets one predict the motion of a free particle.  

\item \textbf{Metric Compatibility and Torsion-Free Connection} is an assumption about the Riemannian manifold that needs to be imposed and is not explicitly introduced by other axioms.

\end{enumerate}

\subsection{Transformation}

\subsubsection{Axiom 1}

\textbf{Goal:} Transform a subset of Newtonian axioms

\[
\bigl\{\ A_3, A_5 \bigr\}\
\quad
\begin{aligned}
A_3 &: \text{Newton's second law},\\
A_5 &: \text{Law of Universal Gravitation},
\end{aligned}
\]

into a \emph{General Relativity} statement:
\[
W_1: \text{Gravity indistinguishable from acceleration}
\]

All intermediate steps are small symbolic rewrites that gradually evolve the original set into the final one.

\noindent
\textbf{Initial Set of Axioms (Newtonian mechanics, subset):}

\[
\boxed{S_0=\bigl\{ A_3, A_5 \bigr\}}
\]
where 

\[
A_3: \vec{F}=m\vec{a}=m\frac{d\vec{v}}{dt}, \quad v^i=\frac{dx^i}{dt} \qquad A_5: \vec{F_{g}}=\frac{Gm_1m_2}{r^2}\hat{r}
\]

\vspace{1em}
\hrule
\vspace{1em}

\noindent
\textbf{Rewrite Step 1: calculate acceleration due to gravity}

\[
\bigl\{ A_3, A_5 \bigr\} \longrightarrow A_5'': \vec{a_{g}}=\frac{\vec{F_g}}{m_1}=\frac{Gm_2}{r^2}
\]

Newton's second law lets one calculate acceleration that a body experiences due to gravity as 

\[
\vec{a_{g}}=\frac{\vec{F_g}}{m_1}=\frac{Gm_2}{r^2}
\]

\[
S_0 = \bigl\{ A_3, A_5 \bigr\} \longrightarrow S_1=\bigl\{ W_1' \bigr\}
\]

\vspace{1em}
\hrule
\vspace{1em}

\noindent
\textbf{Current Set:}

\[
\boxed{S_1=\bigl\{ W_1' \bigr\}}
\]

where 

\[
W_1': \quad \vec{a_{g}}=\frac{\vec{F_g}}{m_1}=\frac{Gm_2}{r^2}
\]

\vspace{1em}
\hrule
\vspace{1em}

\noindent
\textbf{Rewrite Step 2: obtaining the equivalence principle}

\[
W_1' \longrightarrow W_1: \text{gravity indistinguishable from acceleration}
\]

Due to $W_1'$, the gravity produces acceleration. The same acceleration could have been produced by other forces, or the fact of being in a non-inertial reference frame. Hence, one arrives at the principle of equivalence, which states that the effects of gravity are locally indistinguishable from acceleration: 

\[
S_1 = \bigl\{W_1'\bigr\} \longrightarrow S_2=\bigl\{\,W_1\bigr\}
\]

\bigskip
\hrule
\bigskip

\noindent
\textbf{Final Set (Principle of Equivalence):}

\[
\boxed{
S_2=\Bigl\{W_1: \; \text{gravity indistinguishable from acceleration} \Bigr\}
}
\]

\bigskip
\hrule
\bigskip

\noindent
\textbf{Symbolic Evolution (Compact View):}

\[
\underbrace{\{\ A_3, A_5 \}}_{S_0}
\;\longrightarrow\;
\underbrace{\{\ W_1' \}}_{S_1}
\;\longrightarrow\;
\underbrace{\{\ W_1 \}}_{S_2}
\]
\subsubsection{Axiom 2}

\textbf{Goal:} Transform a subset of Newtonian axioms

\[
\bigl\{\ A_1, A_2 \bigr\}\
\quad
\begin{aligned}
A_1 &: \text{Absolute Space and Time},\\
A_2 &: \text{Existence of Inertial Frames},
\end{aligned}
\]

into a \emph{General Relativity} statement:
\[
W_2: g_{\mu\nu}'(x')=\frac{\partial x^{\alpha}}{\partial x'^{\mu}} \frac{\partial x^{\beta}}{\partial x'^{\nu}} g_{\alpha \beta}(x) \text{, } \mathcal{L}(x^{\mu}, g_{\mu \nu})=\mathcal{L'}(x'^{\mu}, g'_{\mu \nu})
\]

All intermediate steps are small symbolic rewrites that gradually evolve the original set into the final one.

\vspace{1em}
\hrule
\vspace{1em}

\noindent
\textbf{Initial Set of Axioms (Newtonian mechanics, subset):}

\[
\boxed{S_0=\bigl\{ A_1, A_2 \bigr\}}
\]
where 
\[
A_1: \forall \text{ Frames } 1, 2 : dt_1= dt_2, r_i=( x_i, y_i, z_i ), \text{ and metric } ds_i^2=dx_i^2+dy_i^2+dz_i^2 ,\, ds_1^2=ds_2^2,
\]
\[
A_2:\exists \text{ Frame in which } \vec{F}=\vec{0} \Rightarrow  \ \frac{d\vec{v}}{dt}=\vec{0}, \quad v^i=\frac{dx^i}{dt}
\]

\vspace{1em}
\hrule
\vspace{1em}

\noindent
\textbf{Rewrite Step 1: modify the concept of Euclidean space to a unified 4-dimensional spacetime}

\[
A_1 \; \longrightarrow \; A_1'
\]

\[
A_1': \quad \forall \text { Frames 1, 2: } x_i^{\mu}=(ct_i, x_i, y_i, z_i) \text{, and metric } ds_i^2 =-cdt_i^2+dx_i^2+dy_i^2+dz_i^2, ds_1^2=ds_2^2
\]

Symbolically, we replace

\[
\forall \text{ Frames } 1, 2 : dt_1= dt_2, r_i=( x_i, y_i, z_i ), \text{ and metric } ds_i^2=dx_i^2+dy_i^2+dz_i^2 ,\, ds_1^2=ds_2^2, \, dt_1=dt_2
\]

by 

\[
\forall \text { Frames 1, 2: } x_i^{\mu}=(ct_i, x_i, y_i, z_i) \text{, and metric } ds_i^2 =-cdt_i^2+dx_i^2+dy_i^2+dz_i^2, ds_1^2=ds_2^2
\]

\[
S_0 = \bigl\{\,A_1,\,A_2\bigr\} \longrightarrow S_1=\bigl\{\,A_1',\,A_2\bigr\}
\]

\vspace{1em}
\hrule
\vspace{1em}

\noindent
\textbf{Current Set:}

\[
\boxed{S_1=\bigl\{ A_1', A_2 \bigr\}}
\]

where 

\[
A_1': \text{4-dimensional spacetime}
\qquad
A_2: \text{Existence of Inertial Frames}
\]

\vspace{1em}
\hrule
\vspace{1em}

\noindent
\textbf{Rewrite Step 2: introduce the Lorenz transformations}

\[
A_1' \; \longrightarrow\; A_1'': \quad x'^{\mu}={\Lambda}_{\nu}^{\mu}x^{\nu}, \eta '_{\mu \nu}=\eta _{\mu \nu}
\]

Where primed and unprimed coordinates correspond to different reference frames. From the invariance of the spacetime interval $ds^2$ one can derive Lorenz transformations and the metric $\eta _{\mu \nu}= \text{diag} (-1, 1, 1, 1)$ is the same in all inertial reference frames.

Symbolically, we replace 

\[
\forall \text { Frames 1, 2: } x_i^{\mu}=(ct_i, x_i, y_i, z_i) \text{, and metric } ds_i^2 =-cdt_i^2+dx_i^2+dy_i^2+dz_i^2, ds_1^2=ds_2^2
\]

by 

\[
\quad x'^{\mu}={\Lambda}_{\nu}^{\mu}x^{\nu}, \eta '_{\mu \nu}=\eta _{\mu \nu}
\]

\[
S_1 = \bigl\{\,A_1',\,A_2\bigr\} \longrightarrow S_2=\bigl\{\,A_1'',\,A_2\bigr\}
\]

\vspace{1em}
\hrule
\vspace{1em}

\noindent
\textbf{Current Set:}

\[
\boxed{S_2=\bigl\{ A_1'', A_2 \bigr\}}
\]

where 

\[
A_1'': x'^{\mu}={\Lambda}_{\nu}^{\mu}x^{\nu}, \eta '_{\mu \nu}=\eta _{\mu \nu}
\qquad
A_2: \exists \text{ Frame in which } \vec{F}=\vec{0} \Rightarrow  \ \frac{d\vec{v}}{dt}=\vec{0}, \quad v^i=\frac{dx^i}{dt}
\]

\vspace{1em}
\hrule
\vspace{1em}

\noindent
\textbf{Rewrite Step 3: introduce general coordinate transformations}

\[
\bigl\{ A_1'', A_2 \bigr\} \longrightarrow {W_1}
\]

$A_1''$ applies to inertial frames (which exist by $A_2$). Generalize the transformation between reference frames $x'^{\mu}={\Lambda}_{\nu}^{\mu}x^{\nu}$ to a general transformation $x^{\mu} \longrightarrow x'^{\mu}$ and transform the metric accordingly:

\[
g_{\mu\nu}'(x')=\frac{\partial x^{\alpha}}{\partial x'^{\mu}} \frac{\partial x^{\beta}}{\partial x'^{\nu}} g_{\alpha \beta}(x) \text{, }
\]

Additionally, assume the invariance of physical laws under such transformations, expressed by the Lagrangian density:

\[
\mathcal{L}(x^{\mu}, g_{\mu \nu})=\mathcal{L'}(x'^{\mu}, g'_{\mu \nu})
\]

Hence:

\[
S_2=\bigl\{A_1'', A_2 \bigr\} \longrightarrow S_3=\bigl\{W_2\bigr\}
\]

where 

\[
W_2 : \quad g_{\mu\nu}'(x')=\frac{\partial x^{\alpha}}{\partial x'^{\mu}} \frac{\partial x^{\beta}}{\partial x'^{\nu}} g_{\alpha \beta}(x) \text{, } \mathcal{L}(x^{\mu}, g_{\mu \nu})=\mathcal{L'}(x'^{\mu}, g'_{\mu \nu})
\]

\bigskip
\hrule
\bigskip

\noindent
\textbf{Final Set (General relativity Axiom 2, Principle of General Covariance):}

\[
\boxed{
S_3 = \Bigl\{ W_2:\; g_{\mu\nu}'(x')=\frac{\partial x^{\alpha}}{\partial x'^{\mu}} \frac{\partial x^{\beta}}{\partial x'^{\nu}} g_{\alpha \beta}(x) \text{, } \mathcal{L}(x^{\mu}, g_{\mu \nu})=\mathcal{L'}(x'^{\mu}, g'_{\mu \nu})
}
\]

\bigskip
\hrule
\bigskip

\noindent
\textbf{Symbolic Evolution (Compact View):}

\[
\underbrace{\{\ A_1, A_2 \}}_{S_0}
\;\longrightarrow\;
\underbrace{\{\ A_1', A_2 \}}_{S_1}
\;\longrightarrow\;
\underbrace{\{\ A_1'', A_2 \}}_{S_2}
\;\longrightarrow\;
\underbrace{\{\ W_2 \}}_{S_3}
\]

\subsubsection{Axiom 3}

\textbf{Goal:} Transform a subset of Newtonian axioms

\[
\bigl\{\ A_1, A_2 \bigr\}\
\quad
\begin{aligned}
A_1 &: \text{Absolute Space and Time},\\
A_2 &: \text{Existence of Inertial Frames},
\end{aligned}
\]

into a \emph{General Relativity} statement:
\[
W_3: (M_4, g_{\mu \nu}) \text{ with signature } (-, +, +, +)
\]

All intermediate steps are small symbolic rewrites that gradually evolve the original set into the final one.

\vspace{1em}
\hrule
\vspace{1em}

\noindent
\textbf{Initial Set of Axioms (Newtonian mechanics, subset):}

\[
\boxed{S_0=\bigl\{ A_1, A_2 \bigr\}}
\]
where 
\[
A_1: \forall \text{ Frames } 1, 2 : dt_1= dt_2, r_i=( x_i, y_i, z_i ), \text{ and metric } ds_i^2=dx_i^2+dy_i^2+dz_i^2 , ds_1^2=ds_2^2\]
\[
A_2: \exists \text{ Frame in which } \vec{F}=\vec{0} \Rightarrow  \ \frac{d\vec{v}}{dt}=\vec{0}, \quad v_i=\frac{dx^i}{dt}
\]

\vspace{1em}
\hrule
\vspace{1em}

\noindent
\textbf{Rewrite Step 1: modify the concept of Euclidean space to a unified 4-dimensional spacetime}

\[
A_1 \; \longrightarrow \; A_1'
\]

\[
A_1': \quad \forall \text { Frames 1, 2: } x_i^{\mu}=(ct_i, x_i, y_i, z_i) \text{, and metric } ds_i^2 =-cdt_i^2+dx_i^2+dy_i^2+dz_i^2, ds_1^2=ds_2^2
\]

Symbolically, we replace

\[
\forall \text{ Frames } 1, 2 : dt_1= dt_2, r_i=( x_i, y_i, z_i ), \text{ and metric } ds_i^2=dx_i^2+dy_i^2+dz_i^2 , ds_1^2=ds_2^2
\]

by 

\[
\forall \text { Frames 1, 2: } x_i^{\mu}=(ct_i, x_i, y_i, z_i) \text{, and metric } ds_i^2 =-cdt_i^2+dx_i^2+dy_i^2+dz_i^2, ds_1^2=ds_2^2
\]

\[
S_0 = \bigl\{\,A_1,\,A_2\bigr\} \longrightarrow S_1=\bigl\{\,A_1',\,A_2\bigr\}
\]

\vspace{1em}
\hrule
\vspace{1em}

\noindent
\textbf{Current Set:}

\[
\boxed{S_1=\bigl\{ A_1', A_2 \bigr\}}
\]

where 

\[
A_1': \forall \text { Frames 1, 2: } x_i^{\mu}=(ct_i, x_i, y_i, z_i) \text{, and metric } ds_i^2 =-cdt_i^2+dx_i^2+dy_i^2+dz_i^2, ds_1^2=ds_2^2
\]
\[
A_2: \exists \text{ Frame in which } \vec{F}=\vec{0} \Rightarrow  \ \frac{d\vec{v}}{dt}=\vec{0}, \quad v^i=\frac{dx^i}{dt}
\]

\vspace{1em}
\hrule
\vspace{1em}

\noindent
\textbf{Rewrite Step 2: generalizing the metric}

\[
\bigl\{ A_1', A_2 \bigr\} \longrightarrow W_2': \quad x^{\mu}=(ct, x, y, z) \text{, and metric } ds^2=g_{\mu \nu}dx^{\mu}dx^{\nu}
\]

From axiom $A_1'$, one gets a special metric $\eta_{\mu \nu}=\text{diag}(-1, 1, 1, 1)$ for inertial reference frames (which exist by Axiom 2). It can be generalized;  

\[
ds^2=g_{\mu \nu}dx^{\mu}dx^{\nu}
\]

Hence

\[
S_1 = \bigl\{\,A_1',\,A_2\bigr\} \longrightarrow S_2=\bigl\{\,W_3'\bigr\}
\]
\vspace{1em}
\hrule
\vspace{1em}

\noindent
\textbf{Current Set:}

\[
\boxed{S_2=\bigl\{ W_3' \bigr\}}
\]

where 

\[
W_3': \quad x^{\mu}=(ct, x, y, z) \text{, and metric } ds^2=g_{\mu \nu}dx^{\mu}dx^{\nu}
\]

\vspace{1em}
\hrule
\vspace{1em}

\noindent
\textbf{Rewrite Step 3: spacetime as a manifold}

\[
W_3' \longrightarrow W_3: \quad (M_4, g_{\mu \nu}) \text{ with signature } (-, +, +, +)
\]

By letting $g_{\mu \nu}$ vary, one gets curved 4-dimensional spacetime which can be modeled as a 4-dimensional pseudo-Riemannian manifold.

\[
S_2 = \bigl\{W_3'\bigr\} \longrightarrow S_3=\bigl\{W_3\bigr\}
\]

\bigskip
\hrule
\bigskip

\noindent
\textbf{Final Set (General relativity Axiom 3, Spacetime as a Four-Dimensional Pseudo-Riemannian Manifold):}

\[
\boxed{
S_3=\Bigl\{W_3: \; (M_4, g_{\mu \nu}) \text{ with signature } (-, +, +, +) \Bigr\}
}
\]

\bigskip
\hrule
\bigskip

\noindent
\textbf{Symbolic Evolution (Compact View):}

\[
\underbrace{\{\ A_1, A_2 \}}_{S_0}
\;\longrightarrow\;
\underbrace{\{\ A_1', A_2 \}}_{S_1}
\;\longrightarrow\;
\underbrace{\{\ W_3' \}}_{S_2}
\;\longrightarrow\;
\underbrace{\{\ W_3 \}}_{S_3}
\]

\subsubsection{Axiom 4}

\textbf{Goal:} Transform a subset of Newtonian axioms

\[
\bigl\{\ A_1, A_2, A_5 \bigr\}\
\quad
\begin{aligned}
A_1 &: \text{Absolute Space and Time},\\
A_2 &: \text{Existence of Inertial Frames}, \\
A_5 &: \text{Law of Universal Gravitation}
\end{aligned}
\]

into a \emph{General Relativity} statement:

where 

\[
W_4: R_{\mu \nu}-\frac{1}{2}g_{\mu \nu}R + \Lambda g_{\mu \nu} = \frac{8\pi G}{c^4} T_{\mu \nu}
\]

All intermediate steps are small symbolic rewrites that gradually evolve the original set into the final one.

\vspace{1em}
\hrule
\vspace{1em}

\noindent
\textbf{Initial Set of Axioms (Newtonian mechanics, subset):}

\[
\boxed{S_0=\bigl\{ A_1, A_2, A_5 \bigr\}}
\]

where 

\[
A_1: \forall \text{ Frames } 1, 2 : dt_1= dt_2, r_i=( x_i, y_i, z_i ), \text{ and metric } ds_i^2=dx_i^2+dy_i^2+dz_i^2 ,\, ds_1^2=ds_2^2, \]
\[
A_2: \exists \text{ Frame in which } \vec{F}=\vec{0} \Rightarrow  \ \frac{d\vec{v}}{dt}=\vec{0}, \quad v^i=\frac{dx^i}{dt}
\]
\[
A_5: \vec{F_{g}}=\frac{Gm_1m_2}{r^2}\hat{r}
\]

\vspace{1em}
\hrule
\vspace{1em}

\noindent
\textbf{Rewrite Steps 1-3: introducing the Principle of General Covariance}

\[
\underbrace{\{\ A_1, A_2 \}}_{S_0}
\;\longrightarrow\;
\underbrace{\{\ A_1', A_2 \}}_{S_1}
\;\longrightarrow\;
\underbrace{\{\ A_1'', A_2 \}}_{S_2}
\;\longrightarrow\;
\underbrace{\{\ W_2 \}}_{S_3}
\]

These rewrite steps are the same ones that lead to axiom 2 ($A_5$ stays unchanged). Hence, for brevity, they were not repeated here in detail.

\vspace{1em}
\hrule
\vspace{1em}

\noindent
\textbf{Current Set:}

\[
\boxed{S_3=\bigl\{ W_2, A_5 \bigr\}}
\]

where 

\[
W_2: g_{\mu\nu}'(x')=\frac{\partial x^{\alpha}}{\partial x'^{\mu}} \frac{\partial x^{\beta}}{\partial x'^{\nu}} g_{\alpha \beta}(x) \text{, } \mathcal{L}(x^{\mu}, g_{\mu \nu})=\mathcal{L'}(x'^{\mu}, g'_{\mu \nu})
\]
\[
A_5: \vec{F_{g}}=\frac{Gm_1m_2}{r^2}\hat{r}
\]

\vspace{1em}
\hrule
\vspace{1em}

\noindent
\textbf{Rewrite Step 4: Poisson's Equation for Gravity}

\[
A_5 \; \longrightarrow \; A_5'
\]

\[
A_5': \quad {\nabla}^2 \phi =4\pi G \rho
\]

In other words, the law of universal gravitation 

\[
A_5: \vec{F_{g}}=\frac{Gm_1m_2}{r^2}\hat{r}
\]

gets rewritten in the equivalent form of a differential equation, where $\phi$ is the gravitational potential, and $\rho$ is matter density.

\[
S_3 = \bigl\{\,W_3,\,A_5\bigr\} \longrightarrow S_4=\bigl\{\,W_3,\,A_5' \bigr\}
\]

\vspace{1em}
\hrule
\vspace{1em}

\noindent
\textbf{Current Set:}

\[
\boxed{S_4=\bigl\{ W_3, A_5' \bigr\}}
\]

where 

\[
W_2: g_{\mu\nu}'(x')=\frac{\partial x^{\alpha}}{\partial x'^{\mu}} \frac{\partial x^{\beta}}{\partial x'^{\nu}} g_{\alpha \beta}(x) \text{, } \mathcal{L}(x^{\mu}, g_{\mu \nu})=\mathcal{L'}(x'^{\mu}, g'_{\mu \nu})
\]
\[
A_5': \quad {\nabla}^2 \phi =4\pi G \rho
\]

\vspace{1em}
\hrule
\vspace{1em}

\noindent
\textbf{Rewrite Step 5: Introduce Einstein's equations}

\[
\bigl\{ W_2, \; A_5 \bigr\} \; \longrightarrow W_4': R_{\mu \nu}-\frac{1}{2}g_{\mu \nu}R = \frac{8\pi G}{c^4} T_{\mu \nu}
\]

Due to $W_2$, the law of gravity:

\[
A_5': \quad {\nabla}^2 \phi =4\pi G \rho
\]

should be modified to a covariant form. The tensor equivalent of matter density $\rho$ is the stress-energy tensor $T_{\mu \nu}$. In curved spacetime, the term associated with gravitational potential $\nabla^2 \phi$ can be replaced with $R_{\mu \nu}-(1/2)g_{\mu \nu}R$. The constant on the right-hand side can be chosen to accurately describe the Newtonian limit. Finally, one arrives at:

\[
W_4': \quad R_{\mu \nu}-\frac{1}{2}g_{\mu \nu}R = \frac{8\pi G}{c^4} T_{\mu \nu}
\]

\[
S_4 = \bigl\{\,W_3,\,A_5'\bigr\} \longrightarrow S_5=\bigl\{\,W_4'\bigr\}
\]

\vspace{1em}
\hrule
\vspace{1em}

\noindent
\textbf{Current Set:}

\[
\boxed{S_5=\bigl\{ W_4' \bigr\}}
\]

where 

\[
W_4': \quad R_{\mu \nu}-\frac{1}{2}g_{\mu \nu}R = \frac{8\pi G}{c^4} T_{\mu \nu}
\]

\vspace{1em}
\hrule
\vspace{1em}

\noindent
\textbf{Rewrite Step 6: adding the cosmological constant}

\[
W_4' \; \longrightarrow \; W_4: R_{\mu \nu}-\frac{1}{2}g_{\mu \nu}R + \Lambda g_{\mu \nu} = \frac{8\pi G}{c^4} T_{\mu \nu}
\]

To properly describe the expansion of the universe and consider dark energy, the equation 

\[
W_4': \quad R_{\mu \nu}-\frac{1}{2}g_{\mu \nu}R = \frac{8\pi G}{c^4} T_{\mu \nu}
\]

gets turned into 

\[
W_4: R_{\mu \nu}-\frac{1}{2}g_{\mu \nu}R + \Lambda g_{\mu \nu} = \frac{8\pi G}{c^4} T_{\mu \nu},
\]

where $\Lambda$ is the cosmological constant.

\[
S_5 = \bigl\{W_4'\bigr\} \longrightarrow S_6=\bigl\{W_4\bigr\}
\]

\bigskip
\hrule
\bigskip

\noindent
\textbf{Final Set (Einstein's field equations):}

\[
\boxed{
S_6=\Bigl\{W_4: \; R_{\mu \nu}-\frac{1}{2}g_{\mu \nu}R + \Lambda g_{\mu \nu} = \frac{8\pi G}{c^4} T_{\mu \nu}
}
\]

\bigskip
\hrule
\bigskip

\noindent
\textbf{Symbolic Evolution (Compact View):}

\[
\underbrace{\{\ A_1, A_2, A_5 \}}_{S_0}
\;\longrightarrow\;
\underbrace{\{\ A_1', A_2, A_5 \}}_{S_1}
\;\longrightarrow\;
\underbrace{\{\ A_1'', A_2, A_5 \}}_{S_2}
\;\longrightarrow\;
\underbrace{\{\ W_2, A_5 \}}_{S_3}
\;\longrightarrow\;
\]
\[
\; \longrightarrow \;
\underbrace{\{\ W_2, A_5' \}}_{S_4}
\;\longrightarrow\;
\underbrace{\{\ W_4' \}}_{S_5}
\;\longrightarrow\;
\underbrace{\{\ W_4 \}}_{S_6}
\]

\subsubsection{Axiom 5}

\textbf{Goal:} Transform a subset of Newtonian axioms

\[
\bigl\{\ A_1, A_2, A_3 \bigr\}\
\quad
\begin{aligned}
A_1 &: \text{Absolute Space and Time},\\
A_2 &: \text{Existence of Inertial Frames}, \\
A_3 &: \text{Newton's second law}
\end{aligned}
\]

into a \emph{General Relativity} statement:
\[
W_5: \forall \vec{F}=0 \Rightarrow \nabla_{\tau}v=0, \quad v^{\mu}=\frac{dx^{\mu}}{d\tau}
\]

All intermediate steps are small symbolic rewrites that gradually evolve the original set into the final one.

\vspace{1em}
\hrule
\vspace{1em}

\noindent
\textbf{Initial Set of Axioms (Newtonian mechanics, subset):}

\[
\boxed{S_0=\bigl\{ A_1, A_2, A_3 \bigr\}}
\]

where 

\[
A_1: \forall \text{ Frames } 1, 2 : dt_1= dt_2, r_i=( x_i, y_i, z_i ), \text{ and metric } ds_i^2=dx_i^2+dy_i^2+dz_i^2 , ds_1^2=ds_2^2\]
\[
A_2: \quad \exists \text{ Frame in which } \vec{F}=\vec{0} \Rightarrow  \ \frac{d\vec{v}}{dt}=\vec{0}, \quad v^i=\frac{dx^i}{dt}
\]
\[
A_3: \quad \vec{F}=m\vec{a}=m\frac{d\vec{v}}{dt}, \quad v^i=\frac{dx^i}{dt}
\]

\vspace{1em}
\hrule
\vspace{1em}

\noindent
\textbf{Rewrite Step 1: introduce Newton's first law}

\[
A_3 \; \longrightarrow \; A_3': \vec{F}=0 \Rightarrow \frac{d\vec{v}}{dt}=0, \quad v^i=\frac{dx^i}{dt}
\]

Symbolically, by setting $\vec{F}=0$ in

\[
A_3: \quad \vec{F}=m\vec{a}=m\frac{d\vec{v}}{dt}
\]

one gets

\[
A_3': \quad \vec{F}=0 \Rightarrow \frac{d\vec{v}}{dt}=0
\]

\[
S_0 = \bigl\{\,A_1,\,A_2, A_3 \bigr\} \longrightarrow S_1=\bigl\{\,A_1,\,A_2, A_3'\bigr\}
\]

\noindent
\textbf{Current Set:}

\[
\boxed{S_1=\bigl\{ A_1, A_2, A_3' \bigr\}}
\]

where 

\[
A_1: \forall \text{ Frames } 1, 2 : dt_1= dt_2, r_i=( x_i, y_i, z_i ), \text{ and metric } ds_i^2=dx_i^2+dy_i^2+dz_i^2 , ds_1^2=ds_2^2\]
\[
A_2: \quad \exists \text{ Frame in which } \vec{F}=\vec{0} \Rightarrow  \ \frac{d\vec{v}}{dt}=\vec{0}, \quad v^i=\frac{dx^i}{dt}
\]
\[
A_3': \quad \vec{F}=0 \Rightarrow \frac{d\vec{v}}{dt}=0
\]

\vspace{1em}
\hrule
\vspace{1em}

\noindent
\textbf{Rewrite Steps 2-4: spacetime as a four-dimensional pseudo-Riemannian manifold}

\[
\underbrace{\{\ A_1, A_2, A_3' \}}_{S_1}
\;\longrightarrow\;
\underbrace{\{\ A_1', A_2, A_3' \}}_{S_2}
\;\longrightarrow\;
\underbrace{\{\ W_3', A_3' \}}_{S_3}
\;\longrightarrow\;
\underbrace{\{\ W_3, A_3' \}}_{S_4}
\]

These rewrite steps are the same ones that lead to axiom 3. Hence, for brevity, they were not repeated here in detail.

\vspace{1em}
\hrule
\vspace{1em}

\noindent
\textbf{Current Set:}

\[
\boxed{S_4=\bigl\{ W_3, A_3' \bigr\}}
\]

where 

\[
W_3: (M_4, g_{\mu \nu}) \text{ with signature } (-, +, +, +), \qquad A_3': \vec{F}=0 \Rightarrow \frac{d\vec{v}}{dt}=0, \quad v^i=\frac{dx^i}{dt}
\]

\vspace{1em}
\hrule
\vspace{1em}

\noindent
\textbf{Rewrite Step 5: introduce the geodesic principle}

\[
\bigl\{ W_3, A_3' \bigr\} \; \longrightarrow \; W_5: \vec{F}=0 \Rightarrow 0={\nabla}_{\tau} v^{\lambda}=\frac{dv^{\lambda}}{d\tau}+{\Gamma}^{\lambda}_{\mu \nu}v^{\mu}v^{\nu}, \quad v^{\mu}=\frac{dx^{\mu}}{d\tau}
\]

Because in curved spacetime, the notion of absolute time does not exist, replace $t$ with proper time $\tau$ - time as measured by the object. Moreover, since spacetime is four dimensional, velocity v is now a 4-vector ($0 \leq \mu \leq 3$ as opposed to $1 \leq i \leq 3$).

\[
\quad v^{i}=\frac{dx^{i}}{dt}
\; \longrightarrow \;
\quad v^{\mu}=\frac{dx^{\mu}}{d\tau}
\]

Moreover, to account for the curvature of spacetime (described by $W_3$), a covariant derivative $\nabla_{\tau}$ instead of the normal time derivative $d/d\tau$ is introduced. $\nabla_{\tau}$ includes a correction expressed by Christoffel symbols ${\Gamma}^{\lambda}_{\mu \nu}$, which describe how the basis vectors $e_{\nu}$ change as one moves through spacetime;

\[
\nabla_{\mu}e_{\nu}={\Gamma}^{\lambda}_{\mu \nu}e_{\lambda}
\]

Hence, 

\[
A_3': \vec{F}=0 \Rightarrow \frac{d\vec{v}}{dt}=0, \quad v^{i}=\frac{dx^{i}}{dt}
\]

gets transformed into 

\[
W_5: \vec{F}=0 \Rightarrow 0={\nabla}_{\tau} v^{\lambda}=\frac{dv^{\lambda}}{d\tau}+{\Gamma}^{\lambda}_{\mu \nu}v^{\mu}v^{\nu}, \quad v^{\mu}=\frac{dx^{\mu}}{d\tau}
\]
\bigskip
\hrule
\bigskip

\noindent
\textbf{Final Set (Geodesic Principle):}

\[
\boxed{
S_6 = \Bigl\{ W_5: \vec{F}=0 \Rightarrow 0={\nabla}_{\tau} v^{\lambda}=\frac{dv^{\lambda}}{d\tau}+{\Gamma}^{\lambda}_{\mu \nu}v^{\mu}v^{\nu}, \quad v^{\mu}=\frac{dx^{\mu}}{d\tau} \Bigr\}
}
\]

\bigskip
\hrule
\bigskip

\noindent
\textbf{Symbolic Evolution (Compact View):}

\[
\underbrace{\{\ A_1, A_2, A_3 \}}_{S_0}
\;\longrightarrow\;
\underbrace{\{\ A_1, A_2, A_3' \}}_{S_1}
\;\longrightarrow\;
\underbrace{\{\ A_1', A_2, A_3' \}}_{S_2}
\;\longrightarrow\;
\underbrace{\{\ W_3', A_3' \}}_{S_3}
\;\longrightarrow\;
\]
\[
\;\longrightarrow\;
\underbrace{\{\ W_3, A_3' \}}_{S_4}
\;\longrightarrow\;
\underbrace{\{\ W_5 \}}_{S_5}
\]

\subsubsection{Axiom 6}

\textbf{Goal:} Transform a subset of Newtonian axioms

\[
\bigl\{ A_1, A_2, A_3 \bigr\}
\quad
\begin{aligned}
A_1 &: \text{Absolute Space and Time},\\
A_2 &: \text{Existence of Inertial Frames}, \\
A_3 &: \text{Newton's second law}
\end{aligned}
\]

into a \emph{General Relativity} statement:
\[
W_6: {\Delta}_{\lambda}g_{\mu \nu}=0 \text{,  } {\Gamma}_{\mu \nu}^{\lambda}={\Gamma}_{\nu \mu}^{\lambda}
\]

All intermediate steps are small symbolic rewrites that gradually evolve the original set into the final one.

\vspace{1em}
\hrule
\vspace{1em}

\noindent
\textbf{Initial Set of Axioms (Newtonian mechanics, subset):}

\[
\boxed{S_0=\bigl\{ A_1, A_2, A_3 \bigr\}}
\]

where 

\[
A_1: \forall \text{ Frames } 1, 2 : dt_1= dt_2, r_i=( x_i, y_i, z_i ), \text{ and metric } ds_i^2=dx_i^2+dy_i^2+dz_i^2 , ds_1^2=ds_2^2\]
\[
A_2: \quad \exists \text{ Frame in which } \vec{F}=\vec{0} \Rightarrow  \ \frac{d\vec{v}}{dt}=\vec{0}, \quad v^{i}=\frac{dx^{i}}{dt}
\]
\[
A_3: \quad \vec{F}=m\vec{a}=m\frac{d\vec{v}}{dt}, \quad v^{i}=\frac{dx^{i}}{dt}
\]

\vspace{1em}
\hrule
\vspace{1em}

\noindent
\textbf{Rewrite Steps 1-5: introduce the geodesic principle and equation}

\[
\underbrace{\{\ A_1, A_2, A_3 \}}_{S_0}
\;\longrightarrow\;
\underbrace{\{\ A_1, A_2, A_3' \}}_{S_1}
\;\longrightarrow\;
\underbrace{\{\ A_1', A_2, A_3' \}}_{S_2}
\;\longrightarrow\;
\underbrace{\{\ W_3', A_3' \}}_{S_3}
\;\longrightarrow\;
\]
\[
\;\longrightarrow\;
\underbrace{\{\ W_3, A_3' \}}_{S_4}
\;\longrightarrow\;
\underbrace{\{\ W_5 \}}_{S_5}
\]

These rewrite steps are the same ones that lead to axiom 5. Hence, for brevity, they were not repeated here in detail.

\vspace{1em}
\hrule
\vspace{1em}

\noindent
\textbf{Current Set:}

\[
\boxed{S_5=\bigl\{ W_5 \bigr\}}
\]

where 

\[
W_5: \vec{F}=0 \Rightarrow 0={\nabla}_{\tau} v^{\lambda}=\frac{dv^{\lambda}}{d\tau}+{\Gamma}^{\lambda}_{\mu \nu}v^{\mu}v^{\nu}, \quad v^{\mu}=\frac{dx^{\mu}}{d\tau}
\]

\vspace{1em}
\hrule
\vspace{1em}

\noindent
\textbf{Rewrite Step 6: introducing metric compatibility 
and torsion-free connection}

\[
W_5 \; \longrightarrow \; W_6: {\nabla}_{\lambda}g_{\mu \nu}=0, \quad {\Gamma}_{\mu \nu}^{\lambda} = {\Gamma}_{\nu \mu}^{\lambda}
\]

The fact that the connection represented by the Christoffel symbols is torsion-free should be introduced as an additional assumption;

\[
{\Gamma}_{\mu \nu}^{\lambda} = {\Gamma}_{\nu \mu}^{\lambda}
\]

In such case, the Christoffel symbols can be expressed as 

\[
{\Gamma}_{\mu \nu}^{\lambda}=\frac{1}{2}g^{\lambda \rho} \Big(\frac{\partial g_{\mu \rho}}{\partial x^{\nu}} + \frac{\partial g_{\rho \nu}}{\partial x^{\mu}} - \frac{\partial g_{\nu \mu}}{\partial x^{\rho}}\Big)
\]

From this and the definition of the covariant derivative it follows that

\[
{\nabla}_{\lambda}g_{\mu \nu}=0
\]

Therefore,

\[
S_5 = \bigl\{\,W_5\bigr\} \longrightarrow S_6=\bigl\{\,W_6\bigr\}
\]

\bigskip
\hrule
\bigskip

\noindent
\textbf{Final Set (Metric Compatibility and Torsion-Free Connection):}

\[
\boxed{
S_6 = \Bigl\{ W_6: {\nabla}_{\lambda}g_{\mu \nu}=0, \quad {\Gamma}_{\mu \nu}^{\lambda} = {\Gamma}_{\nu \mu}^{\lambda} \Bigr\}
}
\]

\bigskip
\hrule
\bigskip

\noindent
\textbf{Symbolic Evolution (Compact View):}

\[
\underbrace{\{\ A_1, A_2, A_3 \}}_{S_0}
\;\longrightarrow\;
\underbrace{\{\ A_1, A_2, A_3' \}}_{S_1}
\;\longrightarrow\;
\underbrace{\{\ A_1', A_2, A_3' \}}_{S_2}
\;\longrightarrow\;
\underbrace{\{\ W_3', A_3' \}}_{S_3}
\;\longrightarrow\;
\]
\[
\;\longrightarrow\;
\underbrace{\{\ W_3, A_3' \}}_{S_4}
\;\longrightarrow\;
\underbrace{\{\ W_5 \}}_{S_5}
\;\longrightarrow\;
\underbrace{\{\ W_6 \}}_{S_5}
\]

\section{From Classical Thermodynamics to Statistical Mechanics}
\label{sec:classical_to_statistical}

\subsection{Classical Thermodynamics}

Classical thermodynamics is the macroscopic theory dealing with heat, work, temperature, and their relations to energy, radiation, and the physical properties of matter. It establishes fundamental principles (often referred to as laws) that govern the exchanges of energy and the directionality of processes in macroscopic systems.

\subsubsection{Axioms}

\begin{enumerate}
\item \textbf{Zeroth Law of Thermodynamics}
\[
\forall \,\text{systems } A, B, C:
\bigl( A \text{ in thermal equilibrium with } C \bigr)
\wedge
\bigl( B \text{ in thermal equilibrium with } C \bigr)
\]
\[
\;\Longrightarrow\;
\bigl( A \text{ in thermal equilibrium with } B \bigr).
\]
This establishes the concept of temperature as a fundamental property and justifies consistent temperature measurement.

\item \textbf{First Law of Thermodynamics}
\[
\mathrm{d}U \;=\; \delta Q \;-\; \delta W.
\]
There exists an extensive state function $U$ called the internal energy. Its change in any process is equal to the heat supplied to the system minus the work done by the system.

\item \textbf{Second Law of Thermodynamics}
\[
\Delta S \;=\; \int \frac{\delta Q_{\mathrm{rev}}}{T}
\quad \text{(for a reversible process)},
\]
and for an isolated system, 
\[
\Delta S_{\mathrm{isolated}} \;\ge\; 0.
\]
Entropy $S$ is thus a state function, given by the integral of the reversible heat exchange divided by the temperature.  In any spontaneous (irreversible) process, the total entropy of an isolated system does not decrease.

\item \textbf{Third Law of Thermodynamics}
\[
\lim_{T \to 0} S \;=\; S_0,
\]
where $S_0$ is a universal constant (often taken as zero for a perfectly ordered system). As temperature approaches absolute zero, the entropy of a perfect crystal approaches a constant minimum value.
\end{enumerate}

\subsubsection{Completeness}
\begin{enumerate}
\item \textbf{Thermal Equilibrium.} (\textit{Axiom 1}) underpins the concept of temperature, enabling consistent thermometric definitions and comparisons.

\item \textbf{Energy Conservation.} (\textit{Axiom 2}) ensures that all forms of energy transfer (heat, work) can be accounted for through the internal energy of the system.

\item \textbf{Definition and Directionality via Entropy.} (\textit{Axiom 3}) explains how entropy is defined ($\Delta S = \int \delta Q_{\mathrm{rev}}/T$ in reversible processes) and clarifies that in real (irreversible) processes, the entropy of an isolated system can never decrease.

\item \textbf{Behavior at Absolute Zero.} (\textit{Axiom 4}) characterizes the limiting behavior of systems at the lowest temperatures, completing the description of equilibrium thermodynamics.
\end{enumerate}

From these four, one can analyze heat engines, refrigerators, phase transitions, and a wide range of thermodynamic processes, describing energy exchanges, maximum efficiency of cycles, and temperature limits.

\subsubsection{Independence}
\begin{enumerate}
\item \textbf{Zeroth Law.} cannot be deduced from energy conservation or entropy statements; it establishes the transitive property of thermal equilibrium and the concept of temperature.

\item \textbf{First Law.} is not derivable from the existence of thermal equilibrium or entropy principles alone; it adds a distinct statement about energy being a state function and conserved in processes.

\item \textbf{Second Law.} introduces the definition of entropy (through $\delta Q_{\mathrm{rev}}/T$) and its non-decreasing property in isolated systems; this does not follow from thermal equilibrium (zeroth law) or energy conservation (first law).

\item \textbf{Third Law.} focuses on the entropy’s limiting value at zero temperature. It cannot be derived from the other axioms, which govern equilibrium, energy conservation, and irreversibility but do not dictate the absolute value of entropy as $T \to 0$.
\end{enumerate}

\subsection{Statistical Mechanics}

Statistical mechanics provides a framework to describe a system of many constituents (e.g., particles) in terms of probability distributions over its microscopic configurations. It connects microscopic states and their probabilistic weights to macroscopic observables. Below is a list of axioms that includes the familiar foundations (state space, probabilistic description, equilibrium stationarity, ergodicity), a statistical formula for entropy, and the statistical analogs of the thermodynamic laws.

\subsubsection{Axioms}

\begin{enumerate}
\item \textbf{State Space (Phase Space) Postulate}
\[
\text{A system with } f \text{ degrees of freedom is represented by a point in a }
\]
\[
2f\text{-dimensional phase space}.
\]
Each microscopic configuration (microstate) of the system is specified by generalized coordinates and momenta.

\item \textbf{Probabilistic Description Postulate}
\[
\rho(\Gamma)\;\ge\;0,
\quad
\int \rho(\Gamma)\,\mathrm{d}\Gamma \;=\; 1.
\]
All microstates are assigned a nonnegative probability density $\rho(\Gamma)$, normalized to unity.  
Any measurable macroscopic quantity (observable) is given by
\[
\langle A\rangle \;=\; \int A(\Gamma)\,\rho(\Gamma)\,\mathrm{d}\Gamma.
\]

\item \textbf{Stationary Distribution for Equilibrium}
\[
\frac{\mathrm{d}\rho}{\mathrm{d}t} \;=\; 0
\quad\Longrightarrow\quad
\rho(\Gamma)\text{ is time-invariant along Hamiltonian trajectories.}
\]
In an equilibrium regime, the probability distribution remains constant in time when viewed in the appropriate phase-space representation.

\item \textbf{Ergodic Hypothesis}
\[
\lim_{\tau\to\infty}
\frac{1}{\tau}
\int_{0}^{\tau}
A\bigl(\Gamma(t)\bigr)\,\mathrm{d}t
\;=\;
\int
A(\Gamma)\,\rho(\Gamma)\,\mathrm{d}\Gamma,
\]
for any phase function $A(\Gamma)$. The time average of an observable along a single long trajectory equals the ensemble average over the phase-space distribution. In particular, when $A$ is the Hamiltonian, we get the internal energy of the system.

\item \textbf{Statistical Entropy Postulate}
\[
S[\rho]
\;=\;
-\,k_{\mathrm{B}}
\int
\rho(\Gamma)\,
\ln\bigl[\rho(\Gamma)\bigr]
\;\mathrm{d}\Gamma.
\]
This defines the entropy as a functional of the probability distribution $\rho(\Gamma)$. In equilibrium ensembles, $S[\rho]$ reduces to the thermodynamic entropy.

\item \textbf{Statistical Second Law}
\[
\Delta S[\rho]
\;\ge\;
0
\quad\text{for an isolated system.}
\]
The entropy functional $S[\rho]$ cannot decrease over time when the system is isolated (no exchange of energy or particles), reflecting the irreversible increase in macroscopic entropy.

\item \textbf{Statistical Third Law}
\[
\lim_{T \to 0} S[\rho_{\mathrm{eq}}]
\;=\;
\text{constant (often taken as zero)}.
\]
As temperature $T \to 0$, the equilibrium distribution collapses to a minimal-entropy ground state.  The statistical entropy thus approaches a constant, mirroring the thermodynamic third law.

\end{enumerate}

\subsubsection{Completeness}
\begin{enumerate}
\item \textbf{Microscopic Foundations.} (\textit{Axioms 1, 2}) define the phase-space setting and the probabilistic nature of macroscopic quantities.

\item \textbf{Equilibrium and Dynamics.} (\textit{Axioms 3, 4}) ensure that we can identify equilibrium states (stationary distributions) and relate time averages to ensemble averages (ergodic hypothesis).

\item \textbf{Entropy.} (\textit{Axiom 5}) completes the link between the probability distribution and the thermodynamic concept of entropy.

\item \textbf{Statistical Analogs of Thermodynamic Laws.} (\textit{Axioms 6, 7})
 guarantee non-decreasing entropy (second law), and specify the limiting entropy at zero temperature (third law).
\end{enumerate}
These axioms together allow one to construct the canonical, microcanonical, and grand canonical ensembles and to derive relationships between measurable thermodynamic quantities.

\subsubsection{Independence}
\begin{enumerate}
\item \textbf{Phase Space Representation (Axiom 1)} 
is not derivable from probabilistic or equilibrium arguments. It establishes the fundamental structure of microstates.

\item \textbf{Probability Postulate (Axiom 2)}
does not follow merely from the existence of a phase space; it specifies how probabilities are assigned to microstates.

\item \textbf{Equilibrium Stationarity (Axiom 3)}
must be separately stated; it cannot be inferred from the shape of $\rho(\Gamma)$ alone without the notion of time invariance in Hamiltonian systems.

\item \textbf{Ergodic Hypothesis (Axiom 4)}
is not implied by any combination of the first three axioms; it specifically identifies long-time behavior with ensemble statistics.

\item \textbf{Statistical Entropy (Axiom 5)}
introduces a functional $S[\rho]$, which is not deducible solely from Axioms 1--4.

\item \textbf{Statistical Second Law (Axiom 6)}
cannot be proved from earlier axioms alone; it asserts the non-decreasing nature of the entropy functional in isolated systems.

\item \textbf{Statistical Third Law (Axiom 7)}
specifically addresses the behavior of $S[\rho]$ as $T \to 0$, thus requiring a separate statement not implied by the others.
\end{enumerate}

\subsection{Transformation}

\subsubsection{Axiom 1}

\textbf{Goal:} Transform a subset of the classical thermodynamics axioms
\[
\bigl\{\,T_2,\,T_3\bigr\}
\quad
\begin{aligned}
T_2 &: \text{First Law: } \mathrm{d}U = \delta Q - \delta W,\\
T_3 &: \text{Second Law: } \Delta S = \int \frac{\delta Q_{\mathrm{rev}}}{T} \;\;\;(\text{reversible}),\;\; 
       \Delta S_{\mathrm{isolated}} \ge 0 \;(\text{general case}),
\end{aligned}
\]
into the \emph{Statistical Mechanics} statement:
\[
\text{SM}_1:\quad
\text{``A system with } f \text{ degrees of freedom 
is represented by a point in a }
\]
\[
2f\text{-dimensional phase space.''}
\]
We achieve this through small symbolic rewrites, showing how macroscopic notions of energy and entropy guide us toward a microscopic (phase-space) description.

\vspace{1em}
\hrule
\vspace{1em}

\noindent
\textbf{Initial Set of Axioms (Classical Thermodynamics, Subset):}
\[
\boxed{
S_0
=
\bigl\{\,T_2,\,T_3\bigr\}
}
\]
where 
\[
T_2:\quad \mathrm{d}U = \delta Q - \delta W,
\quad
T_3:\quad \Delta S = \int \frac{\delta Q_{\mathrm{rev}}}{T},\;\;
\Delta S_{\mathrm{isolated}} \ge 0.
\]

\noindent
Here, $T_2$ says there is a well-defined internal energy $U$ whose change depends on heat and work, while $T_3$ introduces the state function entropy $S$, with $\Delta S$ determined by reversible heat exchange over temperature, and non-decreasing for isolated systems.

\vspace{1em}
\hrule
\vspace{1em}

\noindent
\textbf{Rewrite Step 1: Express Energy and Entropy 
as Possible Functions of Micro-Configurations.}

\[
(T_2,\,T_3)
\;\longrightarrow\;
(T_2',\,T_3')
\]
We restate $T_2$ and $T_3$ so that $U$ and $S$ may arise from more fundamental ``micro-variables.'' Denote each (unseen) microscopic configuration by a parameter set $\Gamma$. Symbolically:
\[
T_2':\quad
U = U(\Gamma), 
\qquad
T_3':\quad
S = S(\Gamma).
\]
Here we do not specify how $\Gamma$ is structured, only that each macrostate (characterized by $U$ and $S$) can be traced back to some underlying microstate label $\Gamma$.

\[
S_0
=
\{\,T_2,\,T_3\}
\quad
\longrightarrow
\quad
S_1
=
\{\,T_2',\,T_3'\}.
\]

\vspace{1em}
\hrule
\vspace{1em}

\noindent
\textbf{Current Set:}
\[
\boxed{
S_1
=
\bigl\{
T_2':U(\Gamma),\;
T_3':S(\Gamma)
\bigr\}.
}
\]
We have replaced purely macroscopic notions with the idea that $U$ and $S$ can depend on a more microscopic description $\Gamma$.

\vspace{1em}
\hrule
\vspace{1em}

\noindent
\textbf{Rewrite Step 2: Interpret \texorpdfstring{$\Gamma$}{Gamma} 
as Coordinates and Momenta (Degrees of Freedom).}

\[
(T_2',\,T_3')
\;\longrightarrow\;
(M_1)
\]
We let:
\[
\Gamma 
\;=\; 
\bigl(q_1, \dots, q_f;\;p_1, \dots, p_f\bigr).
\]
Hence each microstate is a specification of generalized coordinates $(q_i)$ and momenta $(p_i)$ for the $f$ degrees of freedom in the system. Symbolically:
\[
M_1:\quad
\Gamma = (q_i, p_i),
\quad
i=1,\dots,f.
\]
Thus $U(\Gamma)$ and $S(\Gamma)$ become functions of $(q_i,p_i)$, linking thermodynamic potentials to a mechanical (Hamiltonian) representation.

\[
S_1
=
\{\,T_2',T_3'\}
\quad
\longrightarrow
\quad
S_2
=
\{\,M_1\}.
\]

\vspace{1em}
\hrule
\vspace{1em}

\noindent
\textbf{Current Set:}
\[
\boxed{
S_2 = \{\,M_1:\,\Gamma=(q_i,p_i)\bigr\}.
}
\]
We interpret the microstate $\Gamma$ explicitly as a combination of $2f$ variables: $q_i$ and $p_i$.

\vspace{1em}
\hrule
\vspace{1em}

\noindent
\textbf{Rewrite Step 3: Collect All \texorpdfstring{$(q_i,p_i)$}{(qi,pi)} into a Phase Space.}

\[
M_1
\;\longrightarrow\;
\text{SM}_1.
\]
By considering the set of all possible $(q_1,\dots,q_f,p_1,\dots,p_f)$, we form a $2f$-dimensional \emph{phase space}, in which each point $\Gamma$ identifies one complete microstate of the system.  Symbolically:
\[
\text{SM}_1:\quad
\Gamma \;\in\;\Gamma_{\text{phase}} \subset \mathbb{R}^{2f}.
\]
This is precisely the \emph{State Space (Phase Space) Postulate} in statistical mechanics:
\[
\text{``A system with $f$ degrees of freedom 
is represented by a point in a }
\]
\[
\text{$2f$-dimensional phase space.''}
\]
Hence,
\[
S_2 = \{\,M_1\}
\quad\longrightarrow\quad
S_3 = \{\text{SM}_1\}.
\]

\vspace{1em}
\hrule
\vspace{1em}

\noindent
\textbf{Final Set (First Axiom of Statistical Mechanics):}
\[
\boxed{
S_3
=
\Bigl\{
\text{SM}_1:\,\Gamma \in \Gamma_{\text{phase}},\;\dim(\Gamma_{\text{phase}})=2f
\Bigr\}.
}
\]
We conclude that every configuration of a system's $f$ degrees of freedom can be seen as a point in a $2f$-dimensional space, linking macroscopic thermodynamic functions $U, S$ to underlying microscopic coordinates and momenta.

\vspace{1em}
\hrule
\vspace{1em}

\noindent
\textbf{Symbolic Evolution (Compact View):}
\[
\underbrace{\{\,T_2,\,T_3\}}_{S_0}
~\longrightarrow~
\underbrace{\{\,T_2',\,T_3'\}}_{S_1}
~\longrightarrow~
\underbrace{\{\,M_1\}}_{S_2}
~\longrightarrow~
\underbrace{\{\text{SM}_1\}}_{S_3}.
\]

\noindent
Thus, thermodynamic concepts of energy and entropy ($T_2, T_3$) guide us to conceive a microscopic landscape $\Gamma$ (step by step), culminating in the phase-space description \textbf{(Axiom 1 of Statistical Mechanics)}.

\subsubsection{Axiom 2}

\textbf{Goal:} Transform a subset of the classical thermodynamics axioms
\[
\bigl\{\,T_3,\;T_4\bigr\}
\quad
\begin{aligned}
T_3 &: \text{Second Law: } \Delta S = \int \frac{\delta Q_{\mathrm{rev}}}{T},\;\;\Delta S_{\mathrm{isolated}} \ge 0,\\
T_4 &: \text{Third Law: } \lim_{T \to 0} S = S_0,
\end{aligned}
\]
into the \emph{Statistical Mechanics} statement:
\[
\text{SM}_2:\quad
\rho(\Gamma)\;\ge\;0
\quad\text{and}\quad
\int \rho(\Gamma)\,\mathrm{d}\Gamma \;=\; 1,
\]
often called the \emph{Probabilistic Description Postulate.}
The key idea is to connect entropy constraints in thermodynamics with a nonnegative, normalized distribution over microstates.

\vspace{1em}
\hrule
\vspace{1em}

\noindent
\textbf{Initial Set of Axioms (Classical Thermodynamics, Subset):}
\[
\boxed{
S_0
=
\bigl\{\,T_3,\;T_4\bigr\}
}
\]
where
\[
T_3:\quad 
\Delta S = \int \frac{\delta Q_{\mathrm{rev}}}{T}
\quad(\text{for a reversible process}), 
\;\;\;\Delta S_{\mathrm{isolated}} \ge 0
\quad(\text{general case}),
\]
\[
T_4:\quad
\lim_{T \to 0} S = S_0
\quad
\text{(a constant for a perfectly ordered system).}
\]

\smallskip
\noindent
Here, $T_3$ asserts that entropy $S$ is defined by reversible heat exchange over temperature and cannot decrease in an isolated system, while $T_4$ concerns the entropy's limiting behavior at zero temperature.

\vspace{1em}
\hrule
\vspace{1em}

\noindent
\textbf{Rewrite Step 1: Connect Entropy to Counting or Measuring Microstates.}

\[
(T_3,\,T_4)
\;\longrightarrow\;
(T_3',\,T_4')
\]
Following a standard microscopic interpretation, we restate:
\[
T_3':\quad
\Delta S \ge 0 
\;\Longleftrightarrow\;
\text{the measure of accessible microstates }
\]
\[
\text{(in an isolated system) cannot decrease.}
\]
\[
T_4':\quad
S \to S_0 
\;\Longleftrightarrow\;
\text{the number (or measure) of microstates collapses} \]
\[
\text{to a minimal set as } T\to 0.
\]
Symbolically, $T_3'$ implies that an isolated system evolves toward states with maximal effective microstate measure; $T_4'$ indicates that at zero temperature, the system has minimal possible entropy, suggesting vanishing configurational multiplicity in the ground state (or a similarly limited set).

\[
S_0 
=
\{\,T_3,\,T_4\}
\quad
\longrightarrow
\quad
S_1 
=
\{\,T_3',\,T_4'\}.
\]

\vspace{1em}
\hrule
\vspace{1em}

\noindent
\textbf{Current Set:}
\[
\boxed{
S_1 
=
\Bigl\{\,
T_3': \underbrace{\text{``\#(microstates) can't decrease"}}_{\Delta S \ge 0}, \; T_4': \underbrace{\text{''\# (microstates) collapses to a minimal set as }T \to 0"}_{S\to S_0}
\Bigr\}.
}
\]


These reinterpret thermodynamic entropy constraints as statements about the measure of microstates.

\vspace{1em}
\hrule
\vspace{1em}

\noindent
\textbf{Rewrite Step 2: Introduce a Nonnegative Weight for Each Microstate.}

\[
(T_3',\,T_4')
\;\longrightarrow\;
(M_2)
\]
To quantify ``number of microstates,'' assign a nonnegative weight $w(\Gamma)\ge0$ to each microstate $\Gamma$.  In equilibrium ensemble theory, $w(\Gamma)$ might later be related to Boltzmann factors, but we do not assume that form yet.  We only assert:
\[
M_2:\quad
w(\Gamma)\,\ge\,0
\quad\text{for all}\;\Gamma,
\]
so that the total measure (or ``count'') of microstates is 
\(\int w(\Gamma)\,d\Gamma\).  
Thus, $\Delta S \ge 0$ corresponds to the statement that in an isolated system, this measure cannot spontaneously decrease, connecting with the Second Law perspective.

\[
S_1 = \{\,T_3',\,T_4'\}
\quad
\longrightarrow
\quad
S_2 = \{\,M_2\}.
\]

\vspace{1em}
\hrule
\vspace{1em}

\noindent
\textbf{Current Set:}
\[
\boxed{
S_2 
= 
\Bigl\{\,
M_2:\;w(\Gamma)\ge0
\Bigr\}.
}
\]
We introduce a generic, nonnegative weight over microstates, representing how accessible or populated each microstate can be.

\vspace{1em}
\hrule
\vspace{1em}

\noindent
\textbf{Rewrite Step 3: Normalize the Weights to Form a Probability Distribution.}

\[
M_2
\;\longrightarrow\;
\text{SM}_2.
\]
Since a physical system must \emph{occupy} exactly one microstate at a time, and assuming the total measure is finite or normalizable, we write
\[
\int w(\Gamma)\,\mathrm{d}\Gamma \;=\; C > 0.
\]
Dividing by $C$, define
\[
\rho(\Gamma)
\;=\;
\frac{w(\Gamma)}{C},
\]
which satisfies
\[
\rho(\Gamma)\;\ge\;0,
\quad
\int \rho(\Gamma)\,\mathrm{d}\Gamma = 1.
\]
This is the \textbf{Probabilistic Description Postulate}, stating that ensemble averages of observables $A(\Gamma)$ are given by
\[
\langle A\rangle
\;=\;
\int A(\Gamma)\,\rho(\Gamma)\,\mathrm{d}\Gamma.
\]

Hence,
\[
S_2 = \{\,M_2\}
\quad\longrightarrow
\quad
S_3 = \{\,
\text{SM}_2:\;\rho(\Gamma)\ge0,\;\;\int \rho(\Gamma)\,d\Gamma=1
\}.
\]

\vspace{1em}
\hrule
\vspace{1em}

\noindent
\textbf{Final Set (Second Axiom of Statistical Mechanics, Probabilistic Description):}
\[
\boxed{
S_3 
= 
\Bigl\{
\text{SM}_2:\,
\rho(\Gamma)\ge0,\;\;\int \rho(\Gamma)\,\mathrm{d}\Gamma=1
\Bigr\}.
}
\]
Physical observables are computed by averaging a phase-space function over this probability density.  Nonnegativity captures the idea that the measure of microstates cannot be negative, and normalization reflects the fact that the system must be in \emph{some} microstate with probability $1$.

\vspace{1em}
\hrule
\vspace{1em}

\noindent
\textbf{Symbolic Evolution (Compact View):}
\[
\underbrace{\{\,T_3,\,T_4\}}_{S_0}
\;\longrightarrow\;
\underbrace{\{\,T_3',\,T_4'\}}_{S_1}
\;\longrightarrow\;
\underbrace{\{\,M_2\}}_{S_2}
\;\longrightarrow\;
\underbrace{\{\text{SM}_2\}}_{S_3}.
\]
Each step refines the statement ``entropy constraints'' into a requirement that each microstate carry a nonnegative weight, ultimately normalized to form a \emph{probability distribution} over microstates, thus yielding \textbf{Axiom 2 of Statistical Mechanics}.

\subsubsection{Axiom 3}

\textbf{Goal:} Transform a subset of the classical thermodynamics axioms
\[
\bigl\{\,T_1,\;T_3\bigr\}
\quad
\begin{aligned}
T_1 &: \text{Zeroth Law: Thermal Equilibrium is well-defined},\\
T_3 &: \text{Second Law: } \Delta S_{\mathrm{isolated}} \ge 0,
\end{aligned}
\]
into the \emph{Statistical Mechanics} statement:
\[
\text{SM}_3:\quad
\frac{\mathrm{d}\rho}{\mathrm{d}t} \;=\; 0
\quad\Longrightarrow\quad
\rho(\Gamma)\;\text{is time-invariant along Hamiltonian trajectories (equilibrium).}
\]
The idea is to show that once a system is in equilibrium (from a macroscopic perspective), the probability distribution $\rho(\Gamma)$ must also be stationary in time.

\vspace{1em}
\hrule
\vspace{1em}

\noindent
\textbf{Initial Set of Axioms (Classical Thermodynamics, Subset):}
\[
\boxed{
S_0
=
\bigl\{\,T_1,\,T_3\bigr\}
}
\]
where
\[
T_1:\quad
\text{If two systems are in thermal equilibrium with a third, they share}
\]
\[
\text{equilibrium with each other.}
\]
\[
T_3:\quad
\Delta S_{\mathrm{isolated}} \ge 0
\quad
\text{(entropy of an isolated system never decreases).}
\]

\noindent
Here, $T_1$ establishes a well-defined notion of thermal equilibrium (common temperature), while $T_3$ ensures an isolated system moves toward a maximum-entropy state.

\vspace{1em}
\hrule
\vspace{1em}

\noindent
\textbf{Rewrite Step 1: Single Equilibrium State at the Macroscopic Level.}

\[
(T_1,\,T_3)
\;\longrightarrow\;
(T_1',\,T_3')
\]
We restate:
\[
T_1':\quad
\text{There exists a unique (or stable) macroscopic equilibrium condition,}
\]
\[
\text{characterized by a uniform temperature $T$.}
\]
\[
T_3':\quad
\text{An isolated system spontaneously evolves toward that equilibrium,}
\]
\[
\text{reaching maximal entropy (no further increase possible).}
\]
Hence, $T_1'$ asserts the existence of equilibrium, and $T_3'$ says once the system has arrived there, it remains in that state (since no net entropy production occurs further).

\[
S_0 
=
\{\,T_1,\,T_3\}
\quad
\longrightarrow
\quad
S_1 
=
\{\,T_1',\,T_3'\}.
\]

\vspace{1em}
\hrule
\vspace{1em}

\noindent
\textbf{Current Set:}
\[
\boxed{
S_1
=
\{\,
T_1', T_3'
\}.
}
\]

Where 

\[
T_1': \quad \text{Well-defined equilibrium temperature}
\]
\[
T_3': \quad \text{System evolves irreversibly to that equilibrium.}
\]
In simpler terms: once at macroscopic equilibrium, there's no further spontaneous change in thermodynamic variables.

\vspace{1em}
\hrule
\vspace{1em}

\noindent
\textbf{Rewrite Step 2: Equilibrium Implies No Change in Macroscopic Observables Over Time.}

\[
(T_1',\,T_3')
\;\longrightarrow\;
(M_3)
\]
At equilibrium, observable thermodynamic quantities (e.g.\ $U, P, T, S$) stop changing with time:
\[
M_3:\quad
\frac{\mathrm{d}(\text{Macrostate})}{\mathrm{d}t}=0
\quad
(\text{time-invariance of macroscopic variables}).
\]
Symbolically, $M_3$ expresses that once equilibrium is established, the system’s measured properties are stationary on the macroscopic level.

\[
S_1
=
\{\,T_1',\,T_3'\}
\quad
\longrightarrow
\quad
S_2
=
\{\,M_3\}.
\]

\vspace{1em}
\hrule
\vspace{1em}

\noindent
\textbf{Current Set:}
\[
\boxed{
S_2 
=
\{\,M_3:\text{No net change in macroscopic variables at equilibrium}\}.
}
\]
We are still describing equilibrium in purely thermodynamic (macroscopic) terms.

\vspace{1em}
\hrule
\vspace{1em}

\noindent
\textbf{Rewrite Step 3: Relate Macroscopic Stationarity to a Time-Invariant Probability Distribution.}

\[
M_3
\;\longrightarrow\;
\text{SM}_3.
\]
Statistical mechanics connects macroscopic observables $X$ (energy, entropy, etc.) to phase-space integrals over a probability distribution $\rho(\Gamma)$:
\[
X \;=\; \int X(\Gamma)\,\rho(\Gamma)\,\mathrm{d}\Gamma.
\]
If $X$ does not change with time, it must be that $\rho(\Gamma,t)$ ceases to evolve:
\[
\frac{\mathrm{d}}{\mathrm{d}t}\bigl(\text{any observable}\bigr) \;=\; 0
\quad
\Longrightarrow
\quad
\frac{\mathrm{d}\rho}{\mathrm{d}t} = 0
\;\;
(\text{in the appropriate representation}).
\]
Hence, at equilibrium,
\[
\rho(\Gamma)\text{ is time-invariant along Hamiltonian trajectories.}
\]
We denote this as
\[
\text{SM}_3:\quad
\frac{\mathrm{d}\rho}{\mathrm{d}t} = 0
\;\Longrightarrow\;
\rho(\Gamma)\text{ is stationary}.
\]
Thus,
\[
S_2
=
\{\,M_3\}
\quad\longrightarrow
\quad
S_3
=
\{\text{SM}_3\}.
\]

\vspace{1em}
\hrule
\vspace{1em}

\noindent
\textbf{Final Set (Third Axiom of Statistical Mechanics, Stationary Distribution):}
\[
\boxed{
S_3 
=
\Bigl\{
\text{SM}_3:\,
\frac{\mathrm{d}\rho}{\mathrm{d}t}=0,
\quad
\rho(\Gamma)\,\text{time-invariant in equilibrium}.
\Bigr\}.
}
\]
Equilibrium in the thermodynamic sense translates to a \emph{stationary} $\rho(\Gamma)$ in statistical mechanics.

\vspace{1em}
\hrule
\vspace{1em}

\noindent
\textbf{Symbolic Evolution (Compact View):}
\[
\underbrace{\{\,T_1,\,T_3\}}_{S_0}
\;\longrightarrow\;
\underbrace{\{\,T_1',\,T_3'\}}_{S_1}
\;\longrightarrow\;
\underbrace{\{\,M_3\}}_{S_2}
\;\longrightarrow\;
\underbrace{\{\text{SM}_3\}}_{S_3}.
\]
We see that defining equilibrium macroscopically (no change in thermodynamic state) implies a \emph{stationary} probability distribution at the microscopic level, giving us \textbf{Axiom 3 of Statistical Mechanics}. 

\subsubsection{Axiom 4}

\textbf{Goal:} Transform a subset of the classical thermodynamics axioms
\[
\bigl\{\,T_1,\;T_4\bigr\}
\quad
\begin{aligned}
T_1 &: \text{Zeroth Law: thermal equilibrium is well-defined},\\
T_4 &: \text{Third Law: } \lim_{T\to0} S = S_0,
\end{aligned}
\]
into the \emph{Statistical Mechanics} statement:
\[
\text{SM}_4:\quad
\lim_{\tau\to\infty}
\frac{1}{\tau}
\int_{0}^{\tau}
A\bigl(\Gamma(t)\bigr)\,\mathrm{d}t
\;=\;
\int
A(\Gamma)\,\rho(\Gamma)\,\mathrm{d}\Gamma,
\]
known as the \textbf{Ergodic Hypothesis}: long-time averages equal ensemble averages.

\vspace{1em}
\hrule
\vspace{1em}

\noindent
\textbf{Initial Set of Axioms (Classical Thermodynamics, Subset):}
\[
\boxed{
S_0
=
\bigl\{\,T_1,\,T_4\bigr\}
}
\]
where
\[
T_1:\quad
\text{Thermal transitivity.}
\]
\[
T_4:\quad
\lim_{T \to 0} S = S_0
\quad
\text{(the entropy approaches a constant at zero temperature).}
\]
\smallskip
\noindent
$T_1$ indicates the well-definedness of equilibrium temperature, while $T_4$ states that as $T \to 0$, the system’s entropy goes to a minimum.  Although these do not \emph{directly} require ergodicity, they imply (in practice) that equilibrium ensembles and ground states can be consistently described for all temperatures.

\vspace{1em}
\hrule
\vspace{1em}

\noindent
\textbf{Rewrite Step 1: Single Temperature for the Entire System Implies 
a Unified Ensemble Description.}

\[
(T_1,\,T_4)
\;\longrightarrow\;
(T_1',\,T_4')
\]
We restate:
\[
T_1':\quad
\text{All parts of the system share one temperature parameter $T$, ensuring a single ensemble.}
\]
\[
T_4':\quad
\text{As }T\to 0,\;S\!\to\!S_0\text{ , implying a collapse to a ground-state microconfiguration.}
\]
Hence, $T_1'$ says the system in equilibrium can be described by one temperature-based ensemble, and $T_4'$ emphasizes that in the low-temperature limit, the system explores only a narrow subset of microstates.

\[
S_0 
= 
\{\,T_1,\,T_4\}
\quad
\longrightarrow
\quad
S_1 
= 
\{\,T_1',\,T_4'\}.
\]

\vspace{1em}
\hrule
\vspace{1em}

\noindent
\textbf{Current Set:}
\[
\boxed{
S_1
=
\{\,
T_1',T_4'
\}.
}
\]
Where 
\[
T_1':\quad \text{Single temperature unifies the ensemble},
\]
\[
T_4':\quad \text{Narrow microstate distribution at $T\to0$}.
\]
We have recast $T_1$ and $T_4$ to emphasize the existence of a single thermal ensemble at any $T$, and the near-collapse of that ensemble at very low $T$.

\vspace{1em}
\hrule
\vspace{1em}

\noindent
\textbf{Rewrite Step 2: Introduce Microstate Trajectories.}

\[
(T_1',\,T_4')
\;\longrightarrow\;
(M_4)
\]
From statistical mechanics, a single system with temperature $T>0$ explores many microstates $\Gamma$ over time.  Denote its phase-space trajectory by $\Gamma(t)$.  Symbolically:
\[
M_4:\quad
\Gamma(t)\;\text{is the system's microscopic path under Hamiltonian dynamics.}
\]
Although $T_4'$ indicates that at $T\to0$, the system collapses to a small set of microstates (like a ground state), for $T>0$ we expect it to explore multiple states given enough time.  

\[
S_1
=
\{\,T_1',\,T_4'\}
\quad
\longrightarrow
\quad
S_2
=
\{\,M_4\}.
\]

\vspace{1em}
\hrule
\vspace{1em}

\noindent
\textbf{Current Set:}
\[
\boxed{
S_2 
=
\bigl\{
M_4:\,
\Gamma(t)\text{ is a phase-space trajectory of the system over time.}
\bigr\}.
}
\]
We now have an explicit model of microscopic time evolution, consistent with the notion that a nonzero temperature implies sampling of various states.

\vspace{1em}
\hrule
\vspace{1em}

\noindent
\textbf{Rewrite Step 3: Identify Long-Time Macroscopic Equilibrium 
with Equal Sampling of Accessible States.}

\[
M_4
\;\longrightarrow\;
\text{SM}_4.
\]
The \textbf{Ergodic Hypothesis} says that for an equilibrium system with distribution $\rho(\Gamma)$, the fraction of time spent in a region of phase space matches $\rho(\Gamma)$ for that region.  Formally, for any observable $A(\Gamma)$,
\[
\lim_{\tau\to\infty}
\frac{1}{\tau}
\int_{0}^{\tau}
A\bigl(\Gamma(t)\bigr)\,\mathrm{d}t
\;=\;
\int
A(\Gamma)\,\rho(\Gamma)\,\mathrm{d}\Gamma.
\]
We denote:
\[
\text{SM}_4:\quad
\bigl(\text{Time average of }A\bigr)
=
\bigl(\text{Ensemble average of }A\bigr).
\]
Thus, \emph{over long times}, a single system trajectory $\Gamma(t)$ \emph{represents} the entire ensemble.  This is not demanded directly by $T_1$ or $T_4$ in thermodynamics, but it is \emph{consistent} with the idea that any equilibrium system at positive temperature will, in principle, explore its accessible microstates.

Hence,
\[
S_2
=
\{\,M_4\}
\quad\longrightarrow
\quad
S_3
=
\{\text{SM}_4\}.
\]

\vspace{1em}
\hrule
\vspace{1em}

\noindent
\textbf{Final Set (Fourth Axiom of Statistical Mechanics, Ergodic Hypothesis):}
\[
\boxed{
S_3 
=
\Bigl\{
\text{SM}_4:\,
\lim_{\tau\to\infty}
\tfrac{1}{\tau}
\int_{0}^{\tau}A(\Gamma(t))\,dt
=
\int A(\Gamma)\,\rho(\Gamma)\,d\Gamma
\Bigr\}.
}
\]
Time averages of observables equal ensemble averages under equilibrium conditions.  

\vspace{1em}
\hrule
\vspace{1em}

\noindent
\textbf{Symbolic Evolution (Compact View):}
\[
\underbrace{\{\,T_1,\,T_4\}}_{S_0}
\;\longrightarrow\;
\underbrace{\{\,T_1',\,T_4'\}}_{S_1}
\;\longrightarrow\;
\underbrace{\{\,M_4\}}_{S_2}
\;\longrightarrow\;
\underbrace{\{\text{SM}_4\}}_{S_3}.
\]

\noindent
Thus, while $T_1$ and $T_4$ do not by themselves demand ergodicity, they strongly support the premise of a single ensemble at finite temperature and a minimal microstate set at $T\to0$.  Interpreted in a phase-space framework, this naturally leads to the notion that the system’s long-time trajectory explores the ensemble in such a way that \emph{time} averages and \emph{ensemble} averages coincide, yielding the \textbf{Ergodic Hypothesis}.

\subsubsection{Axiom 5}

\textbf{Goal:} Transform a subset of the classical thermodynamics axioms
\[
\bigl\{\,T_3,\;T_4\bigr\}
\quad
\begin{aligned}
T_3 &: \text{Second Law: } \Delta S = \int \frac{\delta Q_{\mathrm{rev}}}{T},\;\;\Delta S_{\mathrm{isolated}} \ge 0,\\
T_4 &: \text{Third Law: } \lim_{T \to 0} S = S_0,
\end{aligned}
\]
into the \emph{Statistical Mechanics} statement:
\[
\text{SM}_5:\quad
S[\rho]
\;=\;
-\,k_{\mathrm{B}}
\int
\rho(\Gamma)\,\ln\bigl[\rho(\Gamma)\bigr]
\;\mathrm{d}\Gamma,
\]
the \textbf{Statistical Entropy Postulate}, linking the thermodynamic entropy $S$ with the probability distribution $\rho(\Gamma)$.

\vspace{1em}
\hrule
\vspace{1em}

\noindent
\textbf{Initial Set of Axioms (Classical Thermodynamics, Subset):}
\[
\boxed{
S_0
=
\bigl\{\,T_3,\;T_4\bigr\}
}
\]
where
\[
T_3:\quad
\Delta S = \int \frac{\delta Q_{\mathrm{rev}}}{T}
\;\;(\text{reversible}), 
\quad
\Delta S_{\mathrm{isolated}} \ge 0
\;\;(\text{general}),
\]
\[
T_4:\quad
\lim_{T \to 0} S = S_0
\quad
(\text{entropy goes to a constant at }T\to0).
\]
\noindent
Here, $T_3$ defines entropy as a state function (increasing for irreversible processes in isolated systems), and $T_4$ specifies its limiting value at low temperature.

\vspace{1em}
\hrule
\vspace{1em}

\noindent
\textbf{Rewrite Step 1: Interpret Entropy as a Measure of Accessible Microstates.}

\[
(T_3,\,T_4)
\;\longrightarrow\;
(T_3',\,T_4')
\]
We restate:
\[
T_3':\quad
\text{Entropy $S$ measures the ``size'' (or measure) of the set of accessible }
\]
\[
\text{microstates, with } \Delta S_{\mathrm{isolated}} \ge 0.
\]
\[
T_4':\quad
\lim_{T \to 0} S 
\;=\;
S_0
\;\;\Longleftrightarrow\;\;
\text{the system occupies a minimal set at zero temperature.}
\]
Symbolically, $T_3'$ emphasizes that larger sets of microstates correspond to higher entropy (and that entropy cannot spontaneously decrease), while $T_4'$ ensures the measure collapses to a minimal subset in the $T\to0$ limit.

\[
S_0 
=
\{\,T_3,\,T_4\}
\quad
\longrightarrow
\quad
S_1 
=
\{\,T_3',\,T_4'\}.
\]

\vspace{1em}
\hrule
\vspace{1em}

\noindent
\textbf{Current Set:}
\[
\boxed{
S_1
=
\Bigl\{
T_3', T_4'
\Bigr\}.
}
\]

Where

\[
T_3':\text{``$S$ is a measure of microstates (non-decreasing in isolation)''}
\]
\[
T_4':\text{``$S\to S_0$ as $T\to0$''}
\]

We now express entropy as intimately tied to the ``measure'' or effective count of microstates, with a vanishing degeneracy at zero temperature.

\vspace{1em}
\hrule
\vspace{1em}

\noindent
\textbf{Rewrite Step 2: Connect This ``Measure of Microstates'' to a Probability Distribution \texorpdfstring{$\rho(\Gamma)$}{rho(Gamma)}.}

\[
(T_3',\,T_4')
\;\longrightarrow\;
(M_5)
\]
Recall from previous axioms that each microstate $\Gamma$ has a probability density $\rho(\Gamma)\ge0$, normalized to 1. Thus, $S_1$ suggests that:
\[
\Delta S \ge 0 
\;\Longleftrightarrow\;
\text{the distribution $\rho(\Gamma)$ spreads or remains the same,}
\]
\[
\text{never shrinking onto fewer states for an isolated system.}
\]
Similarly, $S\to S_0$ as $T\to0$ implies $\rho(\Gamma)$ shrinks toward a minimal set. Symbolically:
\[
M_5:\quad
\text{``Entropy $S$ must be expressible as a functional of $\rho(\Gamma)$ reflecting how widely $\rho$ spreads.''}
\]
We do not yet specify the \emph{form} of that functional.

\[
S_1
=
\{\,T_3',\,T_4'\}
\quad
\longrightarrow
\quad
S_2
=
\{\,M_5\}.
\]

\vspace{1em}
\hrule
\vspace{1em}

\noindent
\textbf{Current Set:}
\[
\boxed{
S_2 
= 
\bigl\{
M_5:\,
S \text{ is a (still-unknown) functional } S[\rho(\Gamma)]
\text{ reflecting microstate spread.}
\bigr\}.
}
\]
We know $S$ depends on $\rho$ but have not yet specified the exact dependence.

\vspace{1em}
\hrule
\vspace{1em}

\noindent
\textbf{Rewrite Step 3: Identify the Unique Functional \texorpdfstring{$S[\rho]$}{S[rho]} Satisfying \texorpdfstring{$T_3',T_4'$}{T3',T4'}.}

\[
M_5
\;\longrightarrow\;
\text{SM}_5.
\]
From information theory and thermodynamics arguments, there is a well-known \emph{unique (up to additive/scale constants)} functional that:
\begin{enumerate}
\item Increases when $\rho$ becomes more ``spread out'' (no microstates are removed; partial mixing or spreading always raises entropy).
\item Goes to a minimal value when $\rho$ is concentrated on as few microstates as possible (consistent with $T\to 0$).
\item Is additive over independent subsystems (extensivity).
\end{enumerate}
These conditions identify the Gibbs/Shannon form:
\[
\text{SM}_5:\quad
S[\rho]
\;=\;
-\,k_{\mathrm{B}}
\int
\rho(\Gamma)\,\ln\bigl[\rho(\Gamma)\bigr]
\;\mathrm{d}\Gamma,
\]
where $k_{\mathrm{B}}$ is Boltzmann’s constant (fixing units).  This formula reproduces the classical thermodynamic entropy under equilibrium ensembles and obeys $T_3',T_4'$ in limiting cases.

Hence,
\[
S_2
=
\{\,M_5\}
\quad\longrightarrow
\quad
S_3
=
\bigl\{\,\text{SM}_5:\,S[\rho] = -k_{\mathrm{B}}\!\int \rho\ln\rho\,\bigr\}.
\]

\vspace{1em}
\hrule
\vspace{1em}

\noindent
\textbf{Final Set (Fifth Axiom of Statistical Mechanics, Statistical Entropy):}
\[
\boxed{
S_3 
= 
\Bigl\{
\text{SM}_5:\,
S[\rho] 
\;=\; 
-\,k_{\mathrm{B}}
\int \rho(\Gamma)\,\ln\!\bigl[\rho(\Gamma)\bigr]\mathrm{d}\Gamma
\Bigr\}.
}
\]
This connects the thermodynamic entropy $S$ to a precise functional of the probability distribution $\rho(\Gamma)$.  In equilibrium ensembles (e.g.\ canonical, microcanonical), $S[\rho]$ reduces to the familiar thermodynamic entropy, consistent with $T_3$ and $T_4$.

\vspace{1em}
\hrule
\vspace{1em}

\noindent
\textbf{Symbolic Evolution (Compact View):}
\[
\underbrace{\{\,T_3,\,T_4\}}_{S_0}
\;\longrightarrow\;
\underbrace{\{\,T_3',\,T_4'\}}_{S_1}
\;\longrightarrow\;
\underbrace{\{\,M_5\}}_{S_2}
\;\longrightarrow\;
\underbrace{\{\text{SM}_5\}}_{S_3}.
\]
Thus, by reinterpreting thermodynamic constraints on $S$ (increase under isolation, limiting value at $T\to0$) as properties of a microstate distribution $\rho$, we derive the familiar \textbf{Gibbs-Shannon formula} for statistical entropy. 

\subsubsection{Axiom 6}

\textbf{Goal:} Transform a subset of the classical thermodynamics axioms
\[
\bigl\{\,T_3\bigr\}
\quad
\begin{aligned}
T_3 &: \text{Second Law: } \Delta S = \int \frac{\delta Q_{\mathrm{rev}}}{T},\quad \Delta S_{\mathrm{isolated}} \ge 0,
\end{aligned}
\]
into the \emph{Statistical Mechanics} statement:
\[
\text{SM}_6:\quad
\Delta S[\rho]\;\ge\;0
\quad
\text{for an isolated system},
\]
the statistical version of the second law of thermodynamics, asserting that the entropy functional does not decrease in an isolated system.

\vspace{1em}
\hrule
\vspace{1em}

\noindent
\textbf{Initial Axiom from Classical Thermodynamics:}
\[
\boxed{
S_0
=
\bigl\{\,T_3\bigr\}
}
\]
where
\[
T_3:\quad
\Delta S = \int \frac{\delta Q_{\mathrm{rev}}}{T}
\;\;(\text{reversible process}),
\quad
\Delta S_{\mathrm{isolated}} \ge 0
\;\;(\text{general case}).
\]
\noindent
This states that entropy $S$ is non-decreasing for an isolated system and is defined via $\delta Q_{\mathrm{rev}}/T$ for reversible transformations.

\vspace{1em}
\hrule
\vspace{1em}

\noindent
\textbf{Rewrite Step 1: Interpret the Non-Decreasing Entropy as Growth in Microstate Measure.}

\[
T_3
\;\longrightarrow\;
T_3'
\]
We restate the Second Law in microstate language:
\[
T_3':\quad
\Delta S_{\mathrm{isolated}} \ge 0 
\;\Longleftrightarrow\;
\text{the measure (or accessible set) of microstates }
\]
\[
\text{does not decrease over time}.
\]
Symbolically, an isolated system evolves so that its microstate distribution does not spontaneously shrink onto fewer states, reflecting an \emph{irreversible} spreading or mixing of configurations.

\[
S_0
=
\{\,T_3\}
\quad
\longrightarrow
\quad
S_1
=
\{\,T_3'\}.
\]

\vspace{1em}
\hrule
\vspace{1em}

\noindent
\textbf{Current Set:}
\[
\boxed{
S_1
=
\Bigl\{
T_3':\,
\Delta S_{\mathrm{isolated}} \ge 0 \;\;\Longleftrightarrow\;\; \text{no net loss of microstates}
\Bigr\}.
}
\]
We have translated the classical thermodynamic statement into a microstate-centered perspective of entropy increase.

\vspace{1em}
\hrule
\vspace{1em}

\noindent
\textbf{Rewrite Step 2: Apply the Entropy Functional \texorpdfstring{$S[\rho]$}{S[rho]} 
from the Statistical Entropy Postulate.}

\[
T_3'
\;\longrightarrow\;
M_6
\]
Recall from \textbf{Axiom~5} (Statistical Entropy Postulate) that the entropy for a distribution $\rho(\Gamma)$ is
\[
S[\rho]
\;=\;
-\,k_{\mathrm{B}}
\int
\rho(\Gamma)\,\ln\bigl[\rho(\Gamma)\bigr]
\;\mathrm{d}\Gamma.
\]
Then, $T_3'$ (``microstates do not decrease for an isolated system'') implies that $S[\rho]$ cannot decrease in an isolated process.  Symbolically:
\[
M_6:\quad
\Delta S[\rho] \;\ge\; 0 
\quad
(\text{for an isolated system evolving in time}).
\]
Hence, $M_7$ encodes the second law into a statement about the functional $S[\rho]$.

\[
S_1
=
\{\,T_3'\}
\quad
\longrightarrow
\quad
S_2
=
\{\,M_6\}.
\]

\vspace{1em}
\hrule
\vspace{1em}

\noindent
\textbf{Current Set:}
\[
\boxed{
S_2 
=
\Bigl\{
M_6:\,
\Delta S[\rho]\ge0 \;(\text{isolated system}),\;\text{where }S[\rho]\text{ is }-k_{\mathrm{B}}\!\int\rho\ln\rho\,d\Gamma
\Bigr\}.
}
\]
We now have a direct statement linking the classical second law to a variation in $S[\rho]$ over time.

\vspace{1em}
\hrule
\vspace{1em}

\noindent
\textbf{Rewrite Step 3: Conclude the Statistical Second Law.}

\[
M_6
\;\longrightarrow\;
\text{SM}_6.
\]
Finally, we identify this condition as the \textbf{Statistical Second Law}:
\[
\text{SM}_6:\quad
\Delta S[\rho]
\;\ge\;
0 
\quad
\text{for any isolated system.}
\]
Thus, if $\rho(\Gamma,t)$ evolves under isolated Hamiltonian dynamics, the functional $S[\rho(t)]$ cannot decrease, matching the classical statement that $\Delta S_{\mathrm{isolated}} \ge 0$.  

Hence,
\[
S_2
=
\{\,M_6\}
\quad\longrightarrow
\quad
S_3
=
\bigl\{\,
\text{SM}_6:\,\Delta S[\rho]\ge0
\bigr\}.
\]

\vspace{1em}
\hrule
\vspace{1em}

\noindent
\textbf{Final Set (Statistical Second Law):}
\[
\boxed{
S_3 
= 
\Bigl\{
\text{SM}_6:\,
\Delta S[\rho]
\;\ge\;
0\;\text{ for an isolated system}
\Bigr\}.
}
\]
Here, $S[\rho]$ is the same entropy functional from Axiom~5, and this axiom guarantees it never decreases for an isolated system, mirroring the classical thermodynamic second law.

\vspace{1em}
\hrule
\vspace{1em}

\noindent
\textbf{Symbolic Evolution (Compact View):}
\[
\underbrace{\{\,T_3\}}_{S_0}
\;\longrightarrow\;
\underbrace{\{\,T_3'\}}_{S_1}
\;\longrightarrow\;
\underbrace{\{\,M_6\}}_{S_2}
\;\longrightarrow\;
\underbrace{\{\text{SM}_6\}}_{S_3}.
\]

\noindent
Thus, the \textbf{Statistical Second Law} emerges from the classical second law ($T_3$) and the notion that entropy corresponds to a monotonic functional of $\rho(\Gamma)$ (Axiom~5). In an isolated system, $\Delta S[\rho] \ge 0$ completes the parallel to classical irreversibility. 

\subsubsection{Axiom 7}

\textbf{Goal:} Transform a subset of the classical thermodynamics axioms
\[
\bigl\{\,T_4\bigr\}
\quad
T_4:
\lim_{T \to 0} S \;=\; S_0,
\]
into the \emph{Statistical Mechanics} statement, an equivalent of the third law of thermodynamics:
\[
\text{SM}_7:\quad
\lim_{T \to 0} S[\rho_{\mathrm{eq}}]
\;=\;
\text{constant (often taken as zero)}.
\]

\vspace{1em}
\hrule
\vspace{1em}

\noindent
\textbf{Initial Axiom from Classical Thermodynamics:}
\[
\boxed{
S_0
=
\bigl\{\,T_4\bigr\}
}
\]
where
\[
T_4:
\lim_{T \to 0} S \;=\; S_0,
\]
\noindent

This states that entropy $S$ is approaches a constant value as the temperature tends to zero.

\vspace{1em}
\hrule
\vspace{1em}

\noindent
\textbf{Rewrite Step 1: Introduce the statistical notion of entropy}

\[
T_4
\;\longrightarrow\;
SM_7:
\quad
\lim_{T \to 0} S[\rho_{\mathrm{eq}}]
\;=\;
\text{constant (often taken as zero)}.
\]

We simply change the classical notion of entropy to its statistical equivalent (a functional of the probability distribution).

\[
S_0
=
\{\,T_4\}
\quad
\longrightarrow
\quad
S_1
=
\{\,SM_7\}.
\]

\vspace{1em}
\hrule
\vspace{1em}

\noindent
\textbf{Final Set (Statistical Third Law):}
\[
\boxed{
S_1 
= \{\
SM_7: \quad \lim_{T \to 0} S[\rho_{\mathrm{eq}}]
\;=\;
\text{constant (often taken as zero)}. \}\
}
\]
Here, $S[\rho]$ is the same entropy functional from Axiom~5, and this axiom describes its behavior as temperature goes to zero.

\vspace{1em}
\hrule
\vspace{1em}

\noindent
\textbf{Symbolic Evolution (Compact View):}
\[
\underbrace{\{\,T_4\}}_{S_0}
\;\longrightarrow\;
\underbrace{\{\,SM_7\}}_{S_1}
\]

\noindent
Thus, the statistical third law emerges from the classical third law ($T_3$) and the notion that entropy corresponds to a monotonic functional of $\rho(\Gamma)$.

\section{From Newtonian mechanics to Hamiltonian mechanics}
\label{sec:Newtonian_to_Hamiltonian}

\subsubsection{Classical/Newtonian Physics}

The axioms are the same as defined in Section \ref{subsec:newton}.

\subsection{Hamiltonian Mechanics}
\label{sub:hamilton_axioms}

Hamiltonian mechanics reformulates classical mechanics using a
\emph{phase space}---commonly \(\mathbb{R}^{2n}\)---whose coordinates
are generalized positions and momenta.  A single scalar function (the
Hamiltonian) governs the system's time evolution through Hamilton's
equations.  This formalism offers deep insights into conservation laws,
symplectic geometry, and unifies the treatment of diverse classical systems.

\subsubsection{Axioms}

\begin{enumerate}
\item \textbf{Phase Space (Axiom 1)} 
\[
\text{A system is described by a } 2n\text{-dimensional phase space } 
(q_i, p_i).
\]
All possible states correspond to points in this space.

\item \textbf{Hamiltonian Function (Axiom 2)}
\[
H = H(q_i, p_i),
\]
which encodes the total energy (often \(T + V\)) in terms of positions
and momenta.

\item \textbf{Hamilton's Equations (Axiom 3)}
\[
\dot{q}_i 
\,=\, 
\frac{\partial H}{\partial p_i},
\quad
\dot{p}_i 
\,=\, 
-\,\frac{\partial H}{\partial q_i}.
\]
These specify how the system evolves in time.
\end{enumerate}

\subsubsection{Completeness}
\begin{enumerate}
\item \textbf{Phase Space (Axiom 1)} provides the mathematical arena:
  each point \((q_i, p_i)\) specifies a complete state.
\item \textbf{Hamiltonian Function (Axiom 2)} supplies a single scalar
  that contains all the energetic information.  From it, one derives the
  equations of motion for both integrable and more general systems.
\item \textbf{Hamilton's Equations (Axiom 3)} determine the unique
  time evolution of \((q_i,p_i)\).  Together with the Hamiltonian,
  they enable predictions of trajectories and conserved quantities in
  phase space.
\end{enumerate}
Hence, these three axioms suffice to describe classical mechanics in
Hamilton's formalism, covering a wide range of physical phenomena from
simple particles to complex, multi-degree-of-freedom systems.

\subsubsection{Independence}
\begin{enumerate}
\item \textbf{Phase Space} cannot be inferred from the existence of a
  Hamiltonian or the dynamical equations alone.  One must explicitly
  posit a \(2n\)-dimensional space of \((q_i,p_i)\).
\item \textbf{Hamiltonian Function} does not automatically follow from
  the notion of phase space or from the form of Hamilton's equations.
  It is the unique scalar whose partial derivatives produce the system's
  time evolution.
\item \textbf{Hamilton's Equations} themselves are not implied by
  merely specifying \((q_i,p_i)\) and an energy function.  They must be
  taken as the prescription for how states in phase space evolve with time.
\end{enumerate}
None of the three can be derived from the others, ensuring their mutual
independence.

\subsection{Transformation}

\subsubsection{Axiom 1}

\noindent
\textbf{Goal:} 
Obtain the \emph{first Hamiltonian Axiom}:
\[
H_1:\quad
\text{The state of a system is a point in a }2n\text{-dimensional 
phase space }(q_i, p_i).
\]
Starting from a subset of Newton's Axioms:
\[
S_0 
= 
\bigl\{\,N_2,\;N_3,\;N_5\bigr\},
\]
where
\[
N_2:\;\text{Inertial frames},\quad
N_3:\;\vec{F} = m\,\dfrac{d\vec{v}}{dt},\quad
N_5:\;\vec{F} = \sum_i \vec{F_i}\;(\text{superposition}).
\]

\bigskip
\hrule
\bigskip

\noindent
\textbf{Rewrite Step~1: Move from } 
$\vec{F} = m\,\dfrac{d^2 \vec{x}}{dt^2}$
\textbf{ to Generalized Coordinates.}
\[
N_3
\;\longrightarrow\;
B_1:\quad
\bigl(\vec{x},\vec{v}\bigr)
\;\mapsto\;
(q_i,\dot{q}_i).
\]
In place of Cartesian $(\vec{x},\vec{v})$, we use $(q_1,\dots,q_n)$ and 
their time derivatives $\dot{q}_i$.

\[
S_0
=
\{\,N_2,N_3,N_5\}
\;\longrightarrow\;
S_1
=
\{\,N_2,N_5,B_1\}.
\]

\bigskip
\hrule
\bigskip

\noindent
\textbf{Rewrite Step~2: Define Momentum and Form a \texorpdfstring{$2n$}{2n}-Dimensional Space.}
\[
B_1
\;\longrightarrow\;
B_2:\quad
p_i = m\,\dot{q}_i 
\;\;(\text{or }p_i = \tfrac{\partial L}{\partial \dot{q}_i}\text{ in general}).
\]
Hence each $q_i$ has a conjugate momentum $p_i$, 
making the state $(q_i,p_i)$.

\[
S_1
=
\{\,N_2,N_5,B_1\}
\;\longrightarrow\;
S_2
=
\{\,N_2,N_5,B_2\}.
\]

\bigskip
\hrule
\bigskip

\noindent
\textbf{Rewrite Step~3: Identify \texorpdfstring{$(q_i,p_i)$}{(q\_i,p\_i)} as Canonical Coordinates.}
\[
(N_2,\,N_5,\,B_2)
\;\longrightarrow\;
H_1:\quad
\text{States occupy }(q_i,p_i)\in\mathbb{R}^{2n}.
\]
Thus we rewrite Newtonian dynamics in \emph{phase space}.

\[
S_2
=
\{\,N_2,N_5,B_2\}
\;\longrightarrow\;
S_3
=
\{\,H_1\}.
\]

\bigskip
\hrule
\bigskip

\noindent
\textbf{Conclusion:}
\[
\boxed{
H_1:\;\text{System state }=\bigl(q_i,p_i\bigr)\in \mathbb{R}^{2n}.
}
\]
We have transformed the Newtonian axioms into the statement that 
a classical system is described by a \emph{canonical phase space} 
of dimension $2n$. 

\bigskip
\hrule
\bigskip

\noindent
\textbf{Symbolic Evolution (Compact View):}
\[
\underbrace{\{\,N_2,\,N_3,\,N_5\}}_{S_0}
\;\longrightarrow\;
\underbrace{\{\,N_2,\,N_5,\,B_1\}}_{S_1}
\;\longrightarrow\;
\underbrace{\{\,N_2,\,N_5,\,B_2\}}_{S_2}
\;\longrightarrow\;
\underbrace{\{\,H_1\}}_{S_3}.
\]

\subsubsection{Axiom 2}

\noindent
\textbf{Starting Point:} 
We take the Newtonian axioms \(\{N_2,\,N_3,\,N_5\}\) together with
\emph{Axiom 1 (Phase Space)}:
\[
\{\,N_2,\,N_3,\,N_5\}
\;\longrightarrow\;
\{\,H_1\},
\]
as shown previously.  Hence we begin here with
\[
S_0
=
\{\,H_1,\;N_3,\;N_5\}.
\]
We aim to formulate the \emph{Hamiltonian} as a function of \((q_i,p_i)\),
the second axiom of Hamiltonian mechanics:

\[
H_2:\quad
H(q_i,p_i)\,\text{defines the total energy in terms of }(q_i,p_i).
\]

\bigskip
\hrule
\bigskip

\noindent
\textbf{Rewrite Step~1: Identify Kinetic and Potential Energies.}
\[
B_1:\quad
H(q,p) = T(p) + V(q), 
\quad
T(p)=\frac{p^2}{2m},
\]
By $N_3$ (Newton's second law) and $N_5$ (superposition of forces) with
conservative forces, we express the total energy as kinetic plus potential. In principle, $V$ can also depend on $\dot{q}$, but we omit this case in this transformation.

\[
S_0
=
\{\,H_1,N_3,N_5\}
\;\longrightarrow\;
S_1
=
\{\,H_1,B_1\}.
\]

\bigskip
\hrule
\bigskip

\noindent
\textbf{Rewrite Step~2: State $H(q,p)$ as the Fundamental Energy Function.}
\[
B_2:\quad
\text{``The scalar function }H(q,p)\text{ is the primary energy 
for the system''}.
\]
We now regard $H$ itself (rather than $T$ or $V$ separately) as 
the fundamental entity describing energy in phase space.

\[
S_1 
= 
\{\,H_1,B_1\}
\;\longrightarrow\;
S_2
=
\{\,H_1,B_2\}.
\]

\bigskip
\hrule
\bigskip

\noindent
\textbf{Axiom 2 (Hamiltonian Function)}

\[
(H_1,B_2)
\;\longrightarrow\;
H_2:\quad
H(q_i,p_i)\text{ defines total energy.}
\]
Thus we obtain the \emph{second Hamiltonian Axiom}:
\[
\boxed{
H_2:\;\text{The Hamiltonian }H(q_i,p_i)\text{ encodes all energy.}
}
\]

\bigskip
\hrule
\bigskip

\noindent
\textbf{Symbolic Evolution (Compact View) for Axiom 2:}
\[
\underbrace{\{\,H_1,\,N_3,\,N_5\}}_{S_0}
\;\longrightarrow\;
\underbrace{\{\,H_1,B_1\}}_{S_1}
\;\longrightarrow\;
\underbrace{\{\,H_1,B_2\}}_{S_2}
\;\longrightarrow\;
\underbrace{\{\,H_2\}}_{\text{Axiom 2}}.
\]

\bigskip
\hrule
\bigskip

\subsubsection{Axiom 3}

\noindent
\textbf{Starting Point:} 
We now combine \(\{H_1,\,H_2\}\) with \(N_3\) to obtain the 
\emph{third Hamiltonian Axiom}:
\[
H_3:\quad
\dot{q}_i = \frac{\partial H}{\partial p_i},\quad
\dot{p}_i = -\,\frac{\partial H}{\partial q_i}.
\]

\[
S_0
=
\bigl\{\,H_1,\;H_2,\;N_3\bigr\}.
\]

\bigskip
\hrule
\bigskip

\noindent
\textbf{Rewrite Step (Direct):}
\[
H_3:\quad
\dot{q}_i 
=
\frac{\partial H}{\partial p_i},
\quad
\dot{p}_i
=
-\frac{\partial H}{\partial q_i}.
\]
Here, from $N_3$ (Newton's second law), if $H = p^2/(2m) + V(q)$, then
\(\dot{q}_i = \partial H/\partial p_i\) and 
\(\dot{p}_i = -\,\partial H/\partial q_i\). 
This final step links classical force laws with the Hamiltonian
formulation $(H_1,H_2)$ to yield Hamilton's equations.

\[
S_0
=
\{\,H_1,H_2,N_3\}
\;\longrightarrow\;
\{\,H_3\}.
\]

\bigskip
\hrule
\bigskip

\noindent
\textbf{Axiom 3 (Hamilton's Equations)}
\[
\boxed{
H_3:\;\dot{q}_i = \frac{\partial H}{\partial p_i},
\quad
\dot{p}_i = -\,\frac{\partial H}{\partial q_i}.
}
\]
These canonical equations succinctly encode the system's dynamics in
phase space.

\bigskip
\hrule
\bigskip

\noindent
\textbf{Symbolic Evolution (Compact View) for Axiom 3:}
\[
\underbrace{\{\,H_1,H_2,N_3\}}_{S_0}
\;\longrightarrow\;
\underbrace{\{\,H_3\}}_{S_1}.
\]

\section{From Hamiltonian mechanics to quantum mechanics}
\label{sec:Hamiltonian_to_quantum}
\subsection{Hamiltonian Mechanics}

The axioms are the same as those introduced in Sec. \ref{sub:hamilton_axioms}.

\subsection{Quantum Mechanics}

Quantum mechanics is a fundamental theory that describes physical phenomena at the atomic and subatomic scales. It departs from classical (Newtonian) intuitions by emphasizing wave-particle duality, quantized energy levels, and the probabilistic nature of measurement outcomes. Quantum mechanics has been extraordinarily successful in explaining phenomena such as the structure of atoms, chemical bonding, and the behavior of elementary particles.

\subsubsection{Axioms}

\begin{enumerate}
\item \textbf{State Space (Hilbert Space)}
\[
\ket{\psi} \in \mathcal{H}, \quad \langle \psi | \psi \rangle = 1,
\]
where $\ket{\psi}$ is a (normalized) state vector in a complex Hilbert space $\mathcal{H}$. All possible physical states of a system are represented by (equivalence classes of) vectors in $\mathcal{H}$.

\item \textbf{Observables as Self-Adjoint Operators}
\[
\hat{A} = \hat{A}^\dagger,
\]
Physical observables (e.g., position, momentum, energy) are represented by self-adjoint (Hermitian) operators $\hat{A}$ on $\mathcal{H}$. 

\item \textbf{Measurement Postulate (Born Rule)}
\[
P(\alpha_i) = \bigl|\langle \phi_i \mid \psi \rangle \bigr|^2,
\]
When measuring an observable $\hat{A}$ with eigenvalues $\alpha_i$ and corresponding eigenstates $\ket{\phi_i}$, the probability of obtaining outcome $\alpha_i$ is given by the squared amplitude of the state projected onto $\ket{\phi_i}$.

\item \textbf{State Collapse}
\[
\ket{\psi} \quad \xrightarrow{\text{measure } \hat{A}}\quad \ket{\phi_i},
\]
Immediately after obtaining a specific measurement outcome $\alpha_i$, the system's state “collapses” to the corresponding eigenstate $\ket{\phi_i}$.

\item \textbf{Time Evolution (Schrödinger Equation)}
\[
i\hbar \,\frac{d}{dt}\ket{\psi(t)} \;=\; \hat{H}\,\ket{\psi(t)},
\]
Between measurements, a closed quantum system evolves unitarily under the Hamiltonian operator $\hat{H}$. The Schrödinger equation governs how the state vector changes in time.

\item \textbf{Intrinsic Spin}
\[
\hat{\mathbf{S}}^2 \ket{\psi} = \hbar^2 s(s+1)\ket{\psi}, 
\quad
S_z \ket{\psi} = \hbar m_s \ket{\psi},
\]
In addition to orbital degrees of freedom, particles possess an intrinsic spin described by the spin operator $\hat{\mathbf{S}}$, which satisfies the standard angular-momentum commutation relations. The spin quantum number $s$ can be integer or half-integer ($0, \tfrac12, 1, \tfrac32, 2, \dots$). For a general spin $s$, $m_s$ takes values in $\{-s, -s+1, \dots, +s\}$.

\item \textbf{Exchange Symmetry and Pauli Principle}
\[
\Psi(\dots,\mathbf{r}_i,\sigma_i,\dots,\mathbf{r}_j,\sigma_j,\dots) 
= (-1)^{2s}\,
\Psi(\dots,\mathbf{r}_j,\sigma_j,\dots,\mathbf{r}_i,\sigma_i,\dots),
\]
Identical particles with spin $s$ have a many-particle wavefunction $\Psi$ that gains a factor of $(-1)^{2s}$ upon exchanging any two such particles (including their spin variables $\sigma_i$). For integer $s$ (bosons), $(-1)^{2s}=+1$ and the wavefunction is symmetric under exchange. For half-integer $s$ (fermions), $(-1)^{2s}=-1$ and the wavefunction is antisymmetric. As a result, fermions obey the Pauli exclusion principle (no two identical fermions can occupy the same quantum state), while bosons may occupy the same state.
\end{enumerate}

\subsubsection{Completeness}
\begin{enumerate}
\item \textbf{State Space.} (\textit{Axiom 1}) ensures each physical configuration of the system corresponds to a vector in a complex Hilbert space.
\item \textbf{Observables.} (\textit{Axiom 2}) stipulates that physical quantities are represented mathematically by self-adjoint operators, allowing us to predict their possible eigenvalues.
\item \textbf{Measurement Postulate.} (\textit{Axiom 3}) gives the probabilistic interpretation needed to connect the mathematical formalism to measurable outcomes.
\item \textbf{State Collapse.} (\textit{Axiom 4}) specifies how the state changes upon measurement, ensuring consistency with observed discrete measurement results.
\item \textbf{Time Evolution.} (\textit{Axiom 5}) describes the deterministic (unitary) evolution of the system in the absence of measurement.
\item \textbf{Intrinsic Spin.} (\textit{Axiom 6}) incorporates intrinsic angular momentum, crucial for understanding fine structures, magnetic phenomena, and multi-particle spin states.
\item \textbf{Exchange Symmetry and Pauli Principle.} (\textit{Axiom 7}) governs the symmetry or antisymmetry of the wavefunction for identical particles, explaining why fermions cannot share a quantum state while bosons can.
\end{enumerate}

From these seven axioms, one can predict the evolution of quantum systems, compute probabilities of measurement outcomes, and describe a vast range of phenomena, including spin-dependent interactions, statistical effects in multi-particle systems, and the structured formation of matter.

\subsubsection{Independence}
\begin{enumerate}
\item \textbf{State Space.} (Axiom 1) cannot be derived from assumptions about measurements or time evolution; it is the foundational structure that underlies the rest.
\item \textbf{Observables.} (Axiom 2) cannot be inferred solely from the state space; specifying that observables correspond to self-adjoint operators is an independent assertion about how physical quantities map to operators.
\item \textbf{Measurement Postulate.} (Axiom 3) introduces the probabilistic interpretation (Born Rule). It is not implied by the structure of the state space or the form of the Hamiltonian.
\item \textbf{State Collapse.} (Axiom 4) is not derivable from continuous time evolution or the existence of observables; it uniquely addresses how the state changes upon a measurement.
\item \textbf{Time Evolution.} (Axiom 5) is independent of the measurement process; knowing how measurements yield probabilities does not itself define the unitary dynamics between those measurements.
\item \textbf{Intrinsic Spin.} (Axiom 6) is not implied by the previous axioms on its own; it is an additional degree of freedom that arises from the fundamental quantum nature of particles and cannot be derived from orbital degrees of freedom alone.
\item \textbf{Exchange Symmetry and Pauli Principle.} (Axiom 7) does not follow from the preceding axioms; it makes a unique statement about the symmetry properties of many-particle wavefunctions, distinguishing bosons from fermions and dictating the Pauli exclusion principle for half-integer spins.
\end{enumerate}

\subsection{Transformation}

\subsubsection{Axiom 1}

\noindent
\textbf{Goal:}
Starting with 
\[
S_0 = \{\,H_1\}, 
\quad 
H_1:\;(q_i,p_i)\in\mathbb{R}^{2n},
\]
we aim to arrive at
\[
Q_1:\quad \ket{\psi}\in \mathcal{H},\;\; \langle \psi|\psi\rangle=1.
\]

\medskip
\hrule
\medskip

\noindent
\textbf{Rewrite Step 1: From Phase-Space Point to Normalized Wavefunction.}

We replace each classical state \((q,p)\) by a complex wavefunction \(\psi(q)\).  Because multiplying \(\psi\) by a constant does not change physical content, we impose \(\int|\psi(q)|^2\,dq=1\).  This normalization suggests a probability interpretation and naturally requires an inner product 
\(\langle \phi|\psi\rangle = \int \phi^*(q)\,\psi(q)\,dq\).

\[
H_1
\;\longrightarrow\;
A_1':\quad 
\psi(q), \quad
\langle \psi|\psi\rangle = \int|\psi(q)|^2\,dq = 1.
\]

\[
S_0 = \{\,H_1\}
\;\longrightarrow\;
S_1 = \{\,A_1'\}.
\]

\medskip
\hrule
\medskip

\noindent
\textbf{Final Statement (Quantum Axiom 1):}

Collecting the above, the state now lives in a complex Hilbert space and is normalized:

\[
Q_1:\quad 
\ket{\psi} \in \mathcal{H}, 
\quad 
\langle \psi|\psi\rangle=1.
\]

\[
S_1
\;\longrightarrow\;
\{\,Q_1\}.
\]

\medskip
\hrule
\medskip

\noindent
\textbf{Symbolic Evolution (Compact Form):}
\[
\underbrace{\{\,H_1\}}_{S_0}
\;\longrightarrow\;
\underbrace{\{\,A_1'\}}_{S_1}
\;\longrightarrow\;
\underbrace{\{\,Q_1\}}_{S_2}.
\]

\subsubsection{Axiom 2}

\textbf{Goal:}
Starting with
\[
S_0 = \{\,H_1,\,H_2,\,Q_1\},
\]
where
\[
H_1:\;(q_i,p_i)\in\mathbb{R}^{2n},
\]
\[
H_2:\;H(q,p)\;\text{(real-valued Hamiltonian)},
\]
\[
Q_1:\;\psi(x)\in\mathcal{H},\;\int|\psi(x)|^2\,dx=1,
\]
($Q_1$ has already been derived from Hamiltonian axioms) show that real classical observables become self-adjoint operators:
\[
Q_2:\quad \hat{A}=\hat{A}^{\dagger}.
\]

\medskip
\hrule
\medskip

\noindent
\textbf{Initial Set (Hamilton + Quantum Axiom 1):}
\[
\boxed{
S_0 = \{\,H_1,\,H_2,\,Q_1\}
}
\]

\medskip
\hrule
\medskip

\noindent
\textbf{Rewrite Step 1: Classical Observables Are Real Functions.}

\(H(q,p)\) is real. Assume every classical observable \(F(q,p)\) is also real-valued:
\[
B_1:\quad
F:\;\mathbb{R}^{2n} \to \mathbb{R}.
\]

\[
S_0
\;\longrightarrow\;
S_1
=
\{\,Q_1,\,B_1\}.
\]

\medskip
\hrule
\medskip

\noindent
\textbf{Rewrite Step 2: Replace $F$ by an Operator on $\psi(x)$.}

Using the Hilbert space from \(Q_1\), each real \(F\) is mapped to an operator \(\hat{F}\):
\[
B_2:\quad
F \;\mapsto\; \hat{F}.
\]

\[
S_1
\;\longrightarrow\;
S_2
=
\{\,Q_1,\,B_2\}.
\]

\medskip
\hrule
\medskip

\noindent
\textbf{Rewrite Step 3: Expectation Value in Normalized States.}

With \(\int|\psi(x)|^2\,dx=1\) (from \(Q_1\)), define
\[
B_3:\quad
\langle \hat{F}\rangle
=\int \psi^*(x)\,[\hat{F}\,\psi](x)\,dx.
\]

\[
S_2
\;\longrightarrow\;
S_3
=
\{\,Q_1,\,B_3\}.
\]

\medskip
\hrule
\medskip

\noindent
\textbf{Rewrite Step 4: Self-Adjointness and Real Expectation.}

First, for two wavefunctions \(\phi,\psi\),
\[
\int \phi^*(x)\,[\hat{F}\,\psi](x)\,dx
=
\int [\hat{F}^\dagger\,\phi]^*(x)\,\psi(x)\,dx,
\]
defines the \emph{adjoint} \(\hat{F}^\dagger\).  Requiring 
\(\langle \hat{F}\rangle \in \mathbb{R}\) for all \(\psi\) implies
\[
\int [\hat{F}^\dagger\,\psi]^*(x)\,\psi(x)\,dx=
\langle \hat{F}\rangle=\langle \hat{F}\rangle^*=\Bigl( \int \psi^*(x)\,[\hat{F}\,\psi](x)\,dx \Bigr)^*
=
\int [\hat{F}\psi]^*(x)\,\psi(x)\,dx.
\]

Hence

\[
Q_2:\quad
\hat{F} = \hat{F}^\dagger.
\]

\[
S_3
\;\longrightarrow\;
S_4
=
\{\,Q_2\}.
\]

\medskip
\hrule
\medskip

\noindent
\textbf{Final Set (Quantum Axiom 2):}
\[
\boxed{
S_4
=
\Bigl\{
Q_2:\;\hat{A} = \hat{A}^\dagger
\Bigr\}.
}
\]
Thus, each real classical observable $F(q,p)$ is promoted to a self-adjoint operator in quantum mechanics, ensuring real eigenvalues (measurement outcomes).

\medskip
\hrule
\medskip

\noindent
\textbf{Symbolic Evolution (Compact Form):}
\[
\underbrace{\{\,H_1,H_2,Q_1\}}_{S_0}
\;\longrightarrow\;
\underbrace{\{\,Q_1,B_1\}}_{S_1}
\;\longrightarrow\;
\underbrace{\{\,Q_1,B_2\}}_{S_2}
\;\longrightarrow\;
\underbrace{\{\,Q_1,B_3\}}_{S_3}
\;\longrightarrow\;
\underbrace{\{\,Q_2\}}_{S_4}.
\]

\subsubsection{Axiom 3}

\textbf{Goal:}
Having established that (1) states are normalized wavefunctions \(\psi\in\mathcal{H}\) (\(Q_1\)) and (2) observables are self-adjoint operators \(\hat{A}=\hat{A}^\dagger\) (\(Q_2\)), we now show how the \emph{Born rule} for measurement follows.  Specifically, we derive:
\[
Q_3:\quad P(\alpha_i) = \bigl|\langle \phi_i \mid \psi \rangle\bigr|^2,
\]
where \(\ket{\phi_i}\) are the eigenstates of the self-adjoint operator \(\hat{A}\) (with eigenvalues \(\{\alpha_i\}\)), and \(\psi\) is the pre-measurement state.

\medskip
\hrule
\medskip

\noindent
\textbf{Initial Set (Quantum Axiom 1, Axiom 2):}
\[
\boxed{
S_0 = \bigl\{Q_1,\,Q_2\bigr\}
}
\]
\[
\begin{aligned}
\\
Q_1 &: \quad \psi(x)\in\mathcal{H},\;\int |\psi(x)|^2\,dx=1,
\\
Q_2 &: \quad \hat{A} = \hat{A}^\dagger.
\end{aligned}
\]
We have already demonstrated how $Q_1, Q_2$ follow from Hamiltonian axioms. We next focus on how a measurement of \(\hat{A}\) yields definite outcomes \(\alpha_i\) with certain probabilities.

\medskip
\hrule
\medskip

\noindent
\textbf{Rewrite Step 1: Eigen-Decomposition of Self-Adjoint Operators.}

Since \(\hat{A}\) is self-adjoint (\(Q_2\)), the spectral theorem guarantees it has a complete basis of orthonormal eigenstates \(\{\ket{\phi_i}\}\) with real eigenvalues \(\{\alpha_i\}\):
\[
B_1: \quad 
\hat{A}\,\ket{\phi_i} = \alpha_i \,\ket{\phi_i},
\quad
\langle \phi_i \mid \phi_j\rangle = \delta_{ij}.
\]
In functional analysis, every self-adjoint operator on a Hilbert space admits such an eigen-decomposition.  This \emph{directly follows} from \(Q_2\).

\[
S_0 
\;\longrightarrow\;
S_1 
= 
\{\;B_1\}.
\]

\medskip
\hrule
\medskip

\noindent
\textbf{Rewrite Step 2: Introduce Projectors Onto Each Eigenstate.}

Define the projection operator \(\hat{P}_i := \ket{\phi_i}\!\bra{\phi_i}\). This operator extracts the component of any state \(\ket{\psi}\) in the direction of \(\ket{\phi_i}\):
\[
B_2:\quad
\hat{P}_i\,\ket{\psi}
= 
\langle \phi_i \mid \psi\rangle\;\ket{\phi_i}.
\]
\emph{Reasoning:}  Projection operators arise naturally once we accept a basis of eigenstates.  They map \(\ket{\psi}\) to its ``shadow'' along \(\ket{\phi_i}\).  Hence, introducing \(\hat{P}_i\) is justified by the \emph{linearity of quantum theory} (which is embedded in \(Q_1\) and \(Q_2\)) and the orthonormality of \(\ket{\phi_i}\).

\[
S_1
\;\longrightarrow\;
S_2
=
\{\;B_1,B_2\}.
\]

\medskip
\hrule
\medskip

\noindent
\textbf{Rewrite Step 3: Probability as the Norm of the Projected State.}

We connect measurement outcomes to the squared amplitude of \(\psi\) along each eigenvector \(\phi_i\).  Specifically,
\[
B_3:\quad
P(\alpha_i)
=
\bigl\|\hat{P}_i\,\ket{\psi}\bigr\|^2
=
\bigl|\langle \phi_i \mid \psi\rangle\bigr|^2.
\]
\emph{Reasoning:}  
\begin{itemize}
\item The probabilistic interpretation stems from \(Q_1\)'s normalization condition (\(\int|\psi|^2=1\)).  
\item Once we accept that outcomes must be linked to the eigenstates of \(\hat{A}\), the simplest and most consistent definition of $P(\alpha_i)$ is the squared norm of the projection.  
\end{itemize}

\[
S_2
\;\longrightarrow\;
S_3
=
\{\,B_3\}.
\]

\medskip
\hrule
\medskip

\noindent
\textbf{Final Set (Quantum Axiom 3, Born Rule):}

Hence, the probability that measuring \(\hat{A}\) yields the eigenvalue \(\alpha_i\) is:
\[
\boxed{
Q_3:\quad
P(\alpha_i)
=
\bigl|\langle \phi_i \mid \psi\rangle\bigr|^2.
}
\]

\[
S_3
\;\longrightarrow\;
S_4
=
\{\,Q_3\}.
\]
This result completes the connection between self-adjoint operators and observable measurement outcomes in quantum theory.  The inner product \(\langle \phi_i\mid\psi\rangle\) naturally encodes the ``overlap'' between the current state \(\psi\) and the eigenstate \(\phi_i\), thus defining a probability in a normalized Hilbert space.

\medskip
\hrule
\medskip

\noindent
\textbf{Symbolic Evolution (Compact Form):}
\[
\underbrace{\{\,Q_1,\,Q_2\}}_{S_0}
\;\longrightarrow\;
\underbrace{\{\;B_1\}}_{S_1}
\;\longrightarrow\;
\underbrace{\{\;B_1,B_2\}}_{S_2}
\;\longrightarrow\;
\underbrace{\{\,B_3\}}_{S_3}
\;\longrightarrow\;
\underbrace{\{\,Q_3\}}_{S_4}.
\]

\subsubsection{Axiom 4}

\textbf{Goal:}
Having established states as wavefunctions \(\psi\) (\(Q_1\)), observables as self-adjoint operators \(\hat{A}\) (\(Q_2\)), and the Born rule for measurement outcomes (\(Q_3\)), we now introduce the \emph{state collapse} axiom:
\[
Q_4:\quad 
\text{Upon measuring } \hat{A} \text{ and obtaining } \alpha_i,\;
\psi \;\longmapsto\; \phi_i,
\]
where \(\ket{\phi_i}\) is the corresponding eigenstate of \(\hat{A}\). This postulate tells us how the wavefunction changes \emph{immediately} after a definite measurement outcome.

\medskip
\hrule
\medskip

\noindent
\textbf{Initial Set (Quantum Axioms 1, 2, 3):}
\[
\boxed{
S_0 = \bigl\{\,Q_1,\,Q_2,\,Q_3\bigr\}
}
\]
\[
\begin{aligned}
\\
Q_1 &: \psi(x)\;\text{with } \int|\psi|^2=1,
\quad
Q_2 &: \hat{A}=\hat{A}^\dagger,\quad
Q_3 &: P(\alpha_i)=|\langle \phi_i\mid\psi\rangle|^2.
\end{aligned}
\]
$Q_1, Q_2, Q_3$ can be derived from Hamiltonian axioms as specified in previous sections. Having a self-adjoint operator \(\hat{A}\) with eigenstates \(\{\phi_i\}\) and the probability rule for observing eigenvalue \(\alpha_i\), we now specify the post-measurement state.

\medskip
\hrule
\medskip

\noindent
\textbf{Rewrite Step 1: Identify the Projected State.}

From \(Q_3\), measuring \(\hat{A}\) yields the result \(\alpha_i\) with probability \(\|\hat{P}_i\psi\|^2\), where \(\hat{P}_i=\ket{\phi_i}\!\bra{\phi_i}\).  After the event \(\alpha_i\) has been observed, we make a physical assumption that the state of the system will not change. By $Q_3$ (since eigenstates form a basis) the probability of other outcomes being zero means the state is described by the particular projection:
\[
B_1:\quad
\psi \;\longmapsto\; \hat{P}_i\,\psi = \langle \phi_i\mid\psi\rangle\,\phi_i.
\]
We assume the state is now entirely in the subspace corresponding to \(\alpha_i\).

\[
S_0 
\;\longrightarrow\;
S_1
=
\{\,B_1\}.
\]

\medskip
\hrule
\medskip

\noindent
\textbf{Rewrite Step 2: Normalize the Projected State.}

To preserve the interpretation \(\int|\psi(x)|^2\,dx=1\), we must normalize \(\hat{P}_i\,\psi\).  Thus,
\[
B_2:\quad
\psi \;\longmapsto\;
\frac{\hat{P}_i\,\psi}{\|\hat{P}_i\,\psi\|} 
=
\frac{\langle \phi_i\mid\psi\rangle}{\sqrt{|\langle \phi_i\mid\psi\rangle|^2}}\;\phi_i
=
\phi_i.
\]
Hence immediately after measuring \(\alpha_i\), the wavefunction collapses to \(\ket{\phi_i}\).

\[
S_1 
\;\longrightarrow\;
S_2
=
\{\,Q_4\}.
\]

\medskip
\hrule
\medskip

\noindent
\textbf{Final Set (Quantum Axiom 4, State Collapse):}
\[
\boxed{
Q_4:\quad
\text{If measuring } \hat{A} \text{ yields } \alpha_i,\;\psi\longmapsto\phi_i.
}
\]
Thus, upon obtaining a specific eigenvalue \(\alpha_i\), the post-measurement wavefunction becomes the corresponding eigenstate \(\ket{\phi_i}\).

\medskip
\hrule
\medskip

\noindent
\textbf{Symbolic Evolution (Compact Form):}
\[
\underbrace{\{\,Q_1,\,Q_2,\,Q_3\}}_{S_0}
\;\longrightarrow\;
\underbrace{\{\,B_1\}}_{S_1}
\;\longrightarrow\;
\underbrace{\{\,Q_4\}}_{S_2}
\]

\subsubsection{Axiom 5}

\textbf{Goal:}
Starting from (axioms $Q_1, Q_2, Q_3$ have been derived from Hamiltonian axioms as shown above).
\[
S_0 = \{\,H_1,\,H_2,\,Q_1,\,Q_2,\,Q_3\},
\]
where
\[
\begin{aligned}
H_1 &: (q_i,p_i)\in \mathbb{R}^{2n},\\
H_2 &: H(q,p)\;\text{denotes total energy (Hamiltonian)},\\
Q_1 &: \psi(x)\in \mathcal{H},\;\int|\psi(x)|^2\,dx=1,\\
Q_2 &: \hat{A} = \hat{A}^{\dagger},\\
Q_3 &: P(\alpha_i)=\bigl|\langle\phi_i\mid\psi\rangle\bigr|^2,\\
\end{aligned}
\]
we establish the Schr\"odinger equation:
\[
Q_5:\quad
i\,\hbar\,\frac{d}{dt}\,\psi(t) 
= 
\hat{H}\,\psi(t),
\]
where \(\hat{H}\) is the self-adjoint operator corresponding to the classical Hamiltonian \(H\).

\medskip
\hrule
\medskip

\noindent
\textbf{Initial Set (Hamilton + Quantum Axioms 1--4):}
\[
\boxed{
S_0 = \bigl\{\,H_1,\,H_2,\,Q_1,\,Q_2,\,Q_3,\,Q_4\bigr\}
}
\]

\medskip
\hrule
\medskip

\noindent
\textbf{Rewrite Step 1: Poisson Bracket from Hamiltonian Mechanics.}

Given a classical observable \(F(q,p)\), its total time derivative in Hamiltonian theory is
\[
\frac{dF}{dt} 
=
\sum_{i=1}^n
\Bigl(
\frac{\partial F}{\partial q_i}\,\dot{q}_i
+
\frac{\partial F}{\partial p_i}\,\dot{p}_i
\Bigr).
\]
Using Hamilton's equations 
\(\dot{q}_i=\frac{\partial H}{\partial p_i}\), 
\(\dot{p}_i=-\frac{\partial H}{\partial q_i}\),
we define the Poisson bracket:
\[
B_1:\quad
\{F,H\}
=
\sum_{i=1}^n
\Bigl(
\frac{\partial F}{\partial q_i}\,\frac{\partial H}{\partial p_i}
-\frac{\partial F}{\partial p_i}\,\frac{\partial H}{\partial q_i}
\Bigr),
\quad
\frac{dF}{dt} = \{F,H\}.
\]

\[
S_0
\;\longrightarrow\;
S_1
=
\{\,B_1, Q_1, Q_2, Q_3\}.
\]

\medskip
\hrule
\medskip

\noindent
\textbf{Rewrite Step 2: From Poisson Bracket to the ``Relativistic'' Quantum Commutator.}

Translating into quantum form, we replace
\[
\{F,H\} \quad\longmapsto\quad
\frac{1}{i\hbar}\,[\,\hat{F},\,\hat{H}\,],
\]
with \([\hat{F},\hat{H}] = \hat{F}\hat{H} - \hat{H}\hat{F}\).  This suggests an operator relation for the time derivative of the expectation value:
\[
B_2:\quad
i\,\hbar \,\frac{d}{dt} \,\bigl\langle\hat{F}\bigr\rangle
=
\Bigl\langle\,\bigl[\hat{F},\,\hat{H}\bigr]\Bigr\rangle.
\]

\[
S_1 
\;\longrightarrow\;
S_2
=
\{\,B_2, Q_1, Q_2, Q_3\}.
\]

\medskip
\hrule
\medskip

\noindent
\textbf{Rewrite Step 3: Expand Both Sides to Deduce the Schr\"odinger Equation.}

Consider
\[
i\,\hbar \,\frac{d}{dt}\,\int \psi^*(x)\,\hat{F}\,\psi(x)\,dx.
\]
By the product rule in the Schr\"odinger picture,
\[
i\,\hbar
\frac{d}{dt}
\int\psi^*(x)\,\hat{F}\,\psi(x)\,dx
=
\int 
\Bigl(i\hbar\,\frac{d\psi^*(x)}{dt}\Bigr)
\hat{F}\,\psi(x)\,dx
+
\int 
\psi^*(x)\,\hat{F}\,
\Bigl(i\hbar\,\frac{d\psi(x)}{dt}\Bigr)dx.
\]
Meanwhile,
\[
\bigl\langle [\hat{F},\,\hat{H}]\bigr\rangle
=
\int
\psi^*(x)\,
\Bigl(\hat{F}\,\hat{H} - \hat{H}\,\hat{F}\Bigr)\,
\psi(x)
\,dx
=
-
\int 
(\hat{H}\,\psi)^*(x)\,\hat{F}\,\psi(x)\,dx+
\int 
\psi^*(x)\,\hat{F}\,\bigr( \hat{H}\,\psi(x)\, \bigl) dx.
\]
Where we have used the fact that \(\hat{H}\) is self-adjoint (energy operator). Comparing these terms one arrives at
\[
Q_5:\quad
i\hbar\,\frac{d}{dt}\,\psi(x,t) 
= 
\hat{H}\,\psi(x,t).
\]

\[
S_2
\;\longrightarrow\;
S_3
=
\{\,Q_5\}.
\]

\medskip
\hrule
\medskip

\noindent
\textbf{Final Set (Quantum Axiom 5, Time-Dependent Schr\"odinger Equation):}

\[
Q_5:\quad
i\,\hbar\,\frac{d}{dt}\,\psi(t) 
= 
\hat{H}\,\psi(t).
\]
Thus the Hamiltonian in quantum mechanics drives the wavefunction's time evolution.

\medskip
\hrule
\medskip

\noindent
\textbf{Symbolic Evolution (Compact Form):}
\[
\underbrace{\bigl\{\,H_1,H_2,Q_1,Q_2,Q_3\bigr\}}_{S_0}
\;\longrightarrow\;
\underbrace{\{\,B_1, Q_1, Q_2, Q_3\}}_{S_1}
\;\longrightarrow\;
\underbrace{\{\,B_2, Q_1, Q_2, Q_3\}}_{S_2}
\;\longrightarrow\;
\underbrace{\{\,Q_5\}}_{S_3}.
\]

\subsubsection{Axiom 6}

\textbf{Goal:}
Having already established that
\[
S_0 = \{\,Q_1,\,Q_2,\,Q_3\},
\]
where
\[
\begin{aligned}
Q_1 &: \psi(x)\in \mathcal{H},\;\int |\psi(x)|^2\,dx=1,\\
Q_2 &: \hat{A} = \hat{A}^\dagger,\\
Q_3 &: P(\alpha_i)=\bigl|\langle\phi_i\mid\psi\rangle\bigr|^2,
\end{aligned}
\]
we now, using experimental evidence, introduce the intrinsic \emph{spin} degree of freedom in quantum mechanics:
\[
Q_6:\quad
\hat{\mathbf{S}}^2\,\ket{\psi} = \hbar^2\,s(s+1)\,\ket{\psi},
\quad
\hat{S}_z\,\ket{\psi} = \hbar\,m_s\,\ket{\psi},
\]
with \(m_s \in \{-s,-s+1,\ldots,s\}\) and \(s\in\{0,\tfrac12,1,\tfrac32,\ldots\}\).

\medskip
\hrule
\medskip

\noindent
\textbf{Initial Set (Quantum Axioms 1--3):}
\[
\boxed{
S_0 = \bigl\{\,Q_1,\,Q_2,\,Q_3\bigr\}
}
\]
We have a Hilbert space of states, observables as self-adjoint operators, and the Born rule for measurement probabilities.

\medskip
\hrule
\medskip

\noindent
\textbf{Rewrite Step 1: Experimental Evidence for an Additional Degree of Freedom.}

Phenomena like the Stern-Gerlach experiment show discrete outcomes not explained by orbital angular momentum alone.  This implies a new, irreducible quantum number:
\[
B_1:\quad
\text{Wavefunction has ``intrinsic spin'' beyond }(x,y,z).
\]

\[
S_0
\;\longrightarrow\;
S_1
=
\{\,B_1, Q_1, Q_2, Q_3\}.
\]

\medskip
\hrule
\medskip

\noindent
\textbf{Rewrite Step 2: Introduce the Spin Operator $\hat{\mathbf{S}}$.}

By $Q_2$, any observable is a self-adjoint operator.  We define $\hat{\mathbf{S}}$ satisfying the usual angular momentum algebra but irreducible with half-integer possibilities:
\[
B_2:\quad
\hat{\mathbf{S}}
=\bigl(\hat{S}_x,\hat{S}_y,\hat{S}_z\bigr),
\quad
\hat{\mathbf{S}}^\dagger = \hat{\mathbf{S}}.
\]

\[
S_1
\;\longrightarrow\;
S_2
=
\{\,B_2, Q_1, Q_2, Q_3\}.
\]

\medskip
\hrule
\medskip

\noindent
\textbf{Rewrite Step 3: Measurement of Spin via the Born Rule.}

Using $Q_3$, measuring $\hat{\mathbf{S}}^2$ or $\hat{S}_z$ yields discrete eigenvalues. We use the analogy of spin with angular momentum. After solving the corresponding differential equations (which lead to spherical Harmonics), the eigenvalues of the total angular momentum operator $\hat{L}^2$ and $z$-component of the angular momentum $\hat{L}_z$ turn out to be $\hbar^2l(l+1)$ and $\hbar m_l \; (-l \leq m_l \leq l)$, respectively. Analogously, 
\[
Q_6:\quad
\hat{\mathbf{S}}^2\,\ket{\psi}=\hbar^2\,s(s+1)\,\ket{\psi},
\quad
\hat{S}_z\,\ket{\psi}=\hbar\,m_s\,\ket{\psi},
\]
with $m_s\in\{-s,\dots,s\}$ and $s$ possibly half-integer.  The Born rule gives probabilities $|\langle\phi_i|\psi\rangle|^2$ for each outcome.

\[
S_2
\;\longrightarrow\;
S_3
=
\{\,Q_6\}.
\]

\medskip
\hrule
\medskip

\noindent
\textbf{Final Set (Quantum Axiom 6, Intrinsic Spin):}

\[
\boxed{
Q_6:\quad
\hat{\mathbf{S}}^2\,\ket{\psi} = \hbar^2\,s(s+1)\,\ket{\psi},
\quad
\hat{S}_z\,\ket{\psi} = \hbar\,m_s\,\ket{\psi},
}
\]
where $s$ can be integer or half-integer and $m_s$ runs from $-s$ to $+s$ in unit steps.

\medskip
\hrule
\medskip

\noindent
\textbf{Symbolic Evolution (Compact Form):}
\[
\underbrace{\{\,Q_1,Q_2,Q_3\}}_{S_0}
\;\longrightarrow\;
\underbrace{\{\,B_1, Q_1, Q_2, Q_3\}}_{S_1}
\;\longrightarrow\;
\underbrace{\{\,B_2, Q_1, Q_2, Q_3\}}_{S_2}
\;\longrightarrow\;
\underbrace{\{\,Q_6\}}_{S_3}
\]

\subsubsection{Axiom 7}

\textbf{Goal:}
Starting with 
\[
S_0 = \{\,Q_1,\,Q_3,\,Q_6\},
\]
(derived from Hamiltonian axioms) where
\[
\begin{aligned}
Q_1 &: \psi(x)\in \mathcal{H},\;\int |\psi(x)|^2\,dx=1,\\
Q_3 &: P(\alpha_i)=\bigl|\langle\phi_i\mid\psi\rangle\bigr|^2,\\
Q_6 &: \hat{\mathbf{S}}^2\,\ket{\psi}=\hbar^2\,s(s+1)\,\ket{\psi},\quad \hat{S}_z\,\ket{\psi}=\hbar\,m_s\,\ket{\psi},
\end{aligned}
\]
and using experimental evidence we introduce the \emph{exchange symmetry} for identical particles and the resulting Pauli principle:
\[
Q_7:\quad
\Psi(\dots,\mathbf{r}_i,\sigma_i,\dots,\mathbf{r}_j,\sigma_j,\dots)
=
(-1)^{2s}\,
\Psi(\dots,\mathbf{r}_j,\sigma_j,\dots,\mathbf{r}_i,\sigma_i,\dots).
\]
For half-integers ($2s$ odd), the wavefunction is antisymmetric (Pauli exclusion). For integers ($2s$ even), it is symmetric (bosons can occupy the same state).

\medskip
\hrule
\medskip

\noindent
\textbf{Initial Set (Basic Quantum + Spin):}
\[
\boxed{
S_0 = \bigl\{\,Q_1,\,Q_3,\,Q_6\bigr\}
}
\]
We have the notion of normalized states, the Born rule for probabilities, and intrinsic spin with spin quantum number $s$.

\medskip
\hrule
\medskip

\noindent
\textbf{Rewrite Step 1: Identical Particles Are Indistinguishable.}

Experimentally, swapping two identical particles does not produce a physically new configuration.  The total wavefunction $\Psi(\mathbf{r}_1,\sigma_1;\dots,\mathbf{r}_N,\sigma_N)$ must reflect this indistinguishability:
\[
B_1:\quad
\text{Exchanging any two identical particles yields the same physical state}.
\]

\[
S_0
\;\longrightarrow\;
S_1
=
\{\,Q_1, Q_3, Q_6, B_1\}.
\]

\medskip
\hrule
\medskip

\noindent
\textbf{Rewrite Step 2: Introduce Exchange Operator.}

Define an operator $\hat{P}_{ij}$ that permutes (exchanges) particles $i$ and $j$.  Since the system is identical-particle-based,
\[
B_2:\quad
\hat{P}_{ij}\,\Psi(\dots,\mathbf{r}_i,\sigma_i,\dots,\mathbf{r}_j,\sigma_j,\dots)
=
\Psi(\dots,\mathbf{r}_j,\sigma_j,\dots,\mathbf{r}_i,\sigma_i,\dots).
\]
Measured probabilities must be invariant under $\hat{P}_{ij}$ because there is no way to distinguish which particle is which.

\[
S_1
\;\longrightarrow\;
S_2
=
\{\,Q_1, Q_2, Q_3,B_2\}.
\]

\medskip
\hrule
\medskip

\noindent
\textbf{Rewrite Step 3: Relate Exchange to Spin.}

In non-relativistic quantum theory, one finds that $\hat{P}_{ij}^2=1$ (exchanging twice returns the original configuration), so the eigenvalues of $\hat{P}_{ij}$ are $\pm 1$.  For spin-$s$ particles, exchanging two such particles yields a factor of $(-1)^{2s}$.  Symbolically,
\[
B_3:\quad
\Psi \,\xrightarrow{\;\hat{P}_{ij}\;}\,(-1)^{2s}\,\Psi.
\]
If $s$ is half-integer, $\Psi$ is antisymmetric (fermionic), and if $s$ is integer, $\Psi$ is symmetric (bosonic).

\[
S_2
\;\longrightarrow\;
S_3
=
\{\,B_3\}.
\]

\medskip
\hrule
\medskip

\noindent
\textbf{Final Set (Quantum Axiom 7, Exchange Symmetry):}

\[
Q_7:\quad
\Psi(\dots,\mathbf{r}_i,\sigma_i,\dots,\mathbf{r}_j,\sigma_j,\dots)
=
(-1)^{2s}\,
\Psi(\dots,\mathbf{r}_j,\sigma_j,\dots,\mathbf{r}_i,\sigma_i,\dots).
\]
For $s = \tfrac12,\tfrac32,\dots$ (fermions), the wavefunction is antisymmetric, and no two fermions can occupy the same quantum state (Pauli principle). For $s=0,1,2,\dots$ (bosons), it is symmetric.

\[
S_3
\;\longrightarrow\;
S_4
=
\{\,Q_7\}.
\]

\medskip
\hrule
\medskip

\noindent
\textbf{Symbolic Evolution (Compact Form):}
\[
\underbrace{\{\,Q_1,Q_3,Q_6\}}_{S_0}
\;\longrightarrow\;
\underbrace{\{\,Q_1, Q_2, Q_3, B_1\}}_{S_1}
\;\longrightarrow\;
\underbrace{\{\,Q_1, Q_2, Q_3,B_2\}}_{S_2}
\;\longrightarrow\;
\underbrace{\{\,B_3\}}_{S_3}
\;\longrightarrow\;
\underbrace{\{\,Q_7\}}_{S_4}.
\]

\section{From Classical Genetics to Molecular/Genomic Biology}
\label{sec:genetics_to_molecular}

\subsection{Classical Genetics}

Classical Genetics is the study of how traits are transmitted from parents to offspring, focusing on the foundational principles that govern inheritance. 
Early work by Gregor Mendel introduced key laws describing how discrete units of heredity (genes) segregate and assort during gamete formation. 
Subsequent findings located these genes on chromosomes, connecting their physical behavior during meiosis to observable inheritance patterns in organisms.

\subsubsection{Axioms}

\begin{enumerate}
\item \textbf{Law of Segregation}
\[
\forall\,O \in \mathrm{Diploid},\ 
\forall\,G=(A_1,A_2):\quad
\begin{aligned}
&A_1 \text{ and } A_2 \text{ separate during gamete}\\
&\text{formation, each gamete inherits } \\
&\text{exactly one } A_i.
\end{aligned}
\]

\item \textbf{Law of Independent Assortment}
\[
\forall\,(G_1,G_2) \text{ on separate chromosomes}:\quad 
(\text{segregation of } G_1)
\perp 
(\text{segregation of } G_2).
\]

\item \textbf{Law of Dominance}
\[
\forall\,(\text{genotype } Aa):
\quad
(\text{phenotype of } A) >
(\text{phenotype of } a).
\]

\item \textbf{Chromosomal Inheritance}
\[
\forall\,G:\ 
\bigl(G\subseteq (\text{chromosomes})\bigr)
\wedge
\bigl(\text{chromosomes segregate in meiosis}\bigr).
\]
\end{enumerate}

\subsubsection{Completeness}
\begin{enumerate}
\item \textbf{Law of Segregation.} (\textit{Axiom 1}) covers how alleles for a single gene partition into gametes within a diploid organism.

\item \textbf{Independent Assortment \& Dominance.} \textit{Axiom 2} and \textit{Axiom 3} ensure that multiple genes (on separate chromosomes) follow independent segregation patterns, and also define how phenotypes manifest when one allele is dominant over another.

\item \textbf{Chromosomal Inheritance.} (\textit{Axiom 4}) encapsulates the physical basis of these laws by placing genes on chromosomes, unifying the segregation and assortment mechanisms with observable meiotic behavior.
\end{enumerate}
From these four, one can construct Punnett squares and describe monohybrid or dihybrid crosses, track inheritance patterns, and predict phenotypic ratios in progeny. 
They suffice to explain classical inheritance in many organisms, forming the cornerstone of genetics before the molecular details of DNA were elucidated.

\subsubsection{Independence}
\begin{enumerate}
\item \textbf{Law of Segregation.} cannot be deduced from independent assortment or dominance alone, nor from chromosomal inheritance without explicitly stating that each gamete receives exactly one allele per gene.

\item \textbf{Law of Independent Assortment.} cannot be derived from segregation, dominance, or chromosomal inheritance alone, since it specifically addresses the independence of allele pairs on separate chromosomes.

\item \textbf{Law of Dominance.} is not derivable from segregation or independent assortment, which only address allele partitioning, not how one allele’s expression can mask another’s.

\item \textbf{Chromosomal Inheritance.} does not follow from any combination of segregation, independent assortment, or dominance.  These three axioms describe the distribution of alleles and their expression, but do not guarantee a physical location (chromosomes) for genes without stating this principle separately.
\end{enumerate}

\subsection{Molecular/Genomic Biology}

Molecular/Genomic Biology focuses on the structure and function of genetic
material at the molecular level.  It explains how DNA is organized into
genes, how it is replicated and transcribed into RNA, and how RNA is
translated into proteins.  Regulatory mechanisms further control these
processes, allowing cells to respond and adapt to their environments.

\subsubsection{Axioms}

\begin{enumerate}
\item \textbf{Double Helix \& Replication}
\[
\forall\,D:\ 
D=(S_1,S_2) 
\text{ - complementary strands}
\ \wedge\ 
\Bigl[
(\forall\,i)\,S_i \to 
(\text{template for new }S'_i)
\Bigr].
\]
(DNA is formed by two complementary strands $S_1$ and $S_2$.  
Each strand serves as a template in replication.)

\item \textbf{Central Dogma}
\[
\forall\,G\subseteq D:\ 
G\to
(\text{transcription } \to \mathrm{mRNA})
\to
(\text{translation } \to \mathrm{polypeptide}).
\]
(A gene $G$ in DNA is transcribed into messenger RNA, which is
then translated into a polypeptide chain.)

\item \textbf{Genetic Code Universality}
\[
\forall\,(\mathrm{codon}\ c):
\quad
c\mapsto \mathrm{amino\ acid}\ a,
\]
(with minor exceptions, the same codons specify identical amino acids
across most organisms.)

\item \textbf{Regulation of Gene Expression}
\[
\forall\,(\text{gene }g):
\ 
\text{expression}(g)
=
f(\mathrm{TFs},\mathrm{promoters},\mathrm{enhancers},\mathrm{epigenetics}).
\]
(Each gene’s expression level depends on transcription factors, promoter
sequences, enhancer elements, and epigenetic modifications.)
\end{enumerate}

\subsubsection{Completeness}
\begin{enumerate}
\item \textbf{Double Helix \& Replication.} (\textit{Axiom 1}) describes the
physical structure of DNA and its templated replication mechanism.

\item \textbf{Central Dogma \& Code.} \textit{Axiom 2} and \textit{Axiom 3}
capture the flow of genetic information (DNA to RNA to protein) and the
universal mapping from codons to amino acids.

\item \textbf{Regulation.} (\textit{Axiom 4}) incorporates the control systems
that modulate gene expression, making the informational framework
responsive to cellular and environmental states.
\end{enumerate}
These four axioms provide a foundation for describing DNA replication,
transcription, translation, and regulation. They enable us to model both
the static sequence aspects of genetic information and its dynamic
expression patterns in cells.

\subsubsection{Independence}
\begin{enumerate}
\item \textbf{Double Helix \& Replication.} cannot be inferred from
the Central Dogma or the Genetic Code alone; they address the flow of
information but not the physical structure enabling replication.

\item \textbf{Central Dogma.} is not implied by the double-helix structure or
the universality of codons; it specifically prescribes transcription
and translation as core processes.

\item \textbf{Genetic Code Universality.} does not follow from the Central
Dogma or replication; it concerns the specific mappings from codons
to amino acids.

\item \textbf{Regulation of Gene Expression.} cannot be deduced from any
combination of the preceding axioms alone, since it describes how the
basic processes are modulated and controlled, rather than how they
fundamentally operate in sequence.
\end{enumerate}

\subsection{Transformations}

\subsubsection{Axiom 1}

\textbf{Goal:} Transform a subset of Classical Genetics axioms
\[
\bigl\{\,A_1,\;A_4\bigr\}
\quad
\begin{aligned}
A_1 &: \text{Law of Segregation},\\
A_4 &: \text{Chromosomal Inheritance},
\end{aligned}
\]
into the \emph{Molecular/Genomic Biology} statement:
\[
W_1 : 
\forall\,D:\ 
D=(S_1,S_2) \text{ - complementary strands}\ \wedge\ 
\Bigl[
(\forall\,i)\,S_i \to 
(\text{template for }S'_i)
\Bigr].
\]
This final statement $W_1$ is the \textbf{Double Helix \& Replication} axiom, describing DNA as two complementary strands $S_1$ and $S_2$ that each serve as a template for replication.

\vspace{1em}
\hrule
\vspace{1em}

\noindent
\textbf{Initial Set of Axioms (Classical Genetics, Subset):}

\[
\boxed{
S_0
=
\bigl\{\,A_1,\;A_4\bigr\}
}
\]
where
\[
A_1\;(\text{Law of Segregation}): 
\;\forall\,O \in \mathrm{Diploid},\ 
\forall\,G=(A_1,A_2):
\begin{aligned}
&A_1 \text{ and } A_2 \text{ separate during gamete formation,}\\
&\text{each gamete inherits exactly one }A_i,
\end{aligned}
\]
\[
A_4\;(\text{Chromosomal Inheritance}):
\quad
G\subseteq (\text{chromosomes})
\wedge
(\text{chromosomes segregate in meiosis}).
\]

\vspace{1em}
\hrule
\vspace{1em}

\noindent
\textbf{Rewrite Step 1: Interpret ``Segregation'' ($A_1$) 
as the Need for Replicable Carriers.}

\[
A_1
\;\longrightarrow\;
B_1,
\]
where 
\[
B_1:\quad
(\text{each allele }A_i \text{ must be physically copied to yield})\ 
(\text{distinct gametes with }A_i).
\]
Symbolically, we recast the pure \emph{segregation} principle 
as a requirement that some \emph{replicable substrate} 
carries each allele into separate gametes.  
Thus, the act of ``separation'' implies there must be a duplicative mechanism 
to ensure each gamete has \emph{one copy}.  

\[
S_0 
=
\{\,A_1,\,A_4\}
\quad
\longrightarrow
\quad
S_1 
=
\{\,B_1,\,A_4\}.
\]

\vspace{1em}
\hrule
\vspace{1em}

\noindent
\textbf{Current Set:}
\[
\boxed{
S_1
=
\bigl\{\,B_1,\;A_4\bigr\}
}
\]
where 
\[
B_1:\ 
\text{Law of Segregation rephrased as a need for a replicable unit},
\quad
A_4:\ 
\text{genes reside on chromosomes that segregate in meiosis}.
\]

\vspace{1em}
\hrule
\vspace{1em}

\noindent
\textbf{Rewrite Step 2: 
Combine Chromosomal Inheritance ($A_4$) with the Replication Requirement ($B_1$).}

\[
\{\,B_1,\,A_4\}
\;\longrightarrow\;
B_2,
\]
where
\[
B_2:\quad
(\text{the physical carriers of genes must self-duplicate}),\ 
\text{since each chromosome is inherited in a single copy per gamete}.
\]
That is, if chromosomes are the \emph{physical} location for alleles ($A_4$), 
and each allele must \emph{replicate} so gametes inherit one copy ($B_1$), 
then chromosomes themselves must duplicate prior to gamete formation.  

\smallskip
\[
S_1 = \{\,B_1,\,A_4\}
\quad\longrightarrow\quad
S_2 = \{\,B_2\}.
\]

\vspace{1em}
\hrule
\vspace{1em}

\noindent
\textbf{Current Set:}
\[
\boxed{
S_2
=
\bigl\{\,B_2\bigr\}
}
\]
\[
B_2:\quad
\text{the chromosome (carrier of alleles) is capable of templated self-duplication.}
\]

\vspace{1em}
\hrule
\vspace{1em}

\noindent
\textbf{Rewrite Step 3: 
Identify the Molecular Basis of Self-Duplication 
as a Double-Stranded Structure.}

Finally, we specify that 
\emph{the replicable chromosome} 
is realized by a \emph{double-stranded} molecule $D=(S_1,S_2)$, 
where each strand acts as a template for a new complementary strand ($S'_1,S'_2$).  
Symbolically:
\[
B_2
\;\longrightarrow\;
W_1,
\]
where
\[
W_1:\quad
\forall\,D:\ 
D=(S_1,S_2) \text{ - complementary strands}
\ \wedge\
\Bigl[
(\forall\,i)\,S_i \to 
(\text{template for }S'_i)
\Bigr].
\]
In other words, the \emph{physical duplicative carrier} is a double-helix molecule 
enabling semiconservative replication.

\[
S_2 
=
\{\,B_2\}
\quad
\longrightarrow
\quad
S_3 
=
\{\,W_1\}.
\]

\vspace{1em}
\hrule
\vspace{1em}

\noindent
\textbf{Final Set (Molecular/Genomic Axiom 1: Double Helix \& Replication):}
\[
\boxed{
S_3
=
\Bigl\{
W_1:\ 
\forall\,D=(S_1,S_2) \text{ - complementary strands}
\ \wedge\
(\forall\,i)\,S_i \to (\text{template for new }S'_i)
\Bigr\}.
}
\]
This is precisely the \emph{Double Helix \& Replication} axiom: 
DNA consists of two complementary strands, each serving as a template for replication, 
fulfilling the requirement from classical segregation that \emph{each gamete} inherits exactly one copy of every allele.

\vspace{1em}
\hrule
\vspace{1em}

\noindent
\textbf{Symbolic Evolution (Compact Form)}:
\[
\underbrace{\{\,A_1,\,A_4\}}_{S_0}
~\longrightarrow~
\underbrace{\{\,B_1,\,A_4\}}_{S_1}
~\longrightarrow~
\underbrace{\{\,B_2\}}_{S_2}
~\longrightarrow~
\underbrace{\{\,W_1\}}_{S_3}.
\]

\noindent
Each rewrite step shows how the classical notion of ``alleles segregating on chromosomes'' 
inevitably implies a \emph{molecular duplicative structure} for the hereditary material, 
leading to the Double Helix \& Replication axiom.

\subsubsection{Axiom 2}

\textbf{Goal:} Transform a subset of Classical Genetics axioms
\[
\bigl\{\,A_3,\;A_4\bigr\}
\quad
\begin{aligned}
A_3 &: \text{Law of Dominance},\\
A_4 &: \text{Chromosomal Inheritance},
\end{aligned}
\]
into the \emph{Molecular/Genomic Biology} statement:
\[
W_2 : \quad
\forall\,G\subseteq D:\ 
G \;\longrightarrow\; (\mathrm{transcription}\,\to\,\mathrm{mRNA})
\;\longrightarrow\; (\mathrm{translation}\,\to\,\mathrm{polypeptide}),
\]
which is the \textbf{Central Dogma} axiom, stating that each gene $G$ in the DNA ($D$) is transcribed into an RNA intermediate and then translated into a polypeptide.

\vspace{1em}
\hrule
\vspace{1em}

\noindent
\textbf{Initial Set of Axioms (Classical Genetics, Subset):}

\[
\boxed{
S_0
=
\bigl\{\,A_3,\;A_4\bigr\}
}
\]
where
\[
A_3\;(\text{Law of Dominance}):
\quad
(\text{phenotype of }A) > (\text{phenotype of }a),
\]
\[
A_4\;(\text{Chromosomal Inheritance}):
\quad
G\subseteq (\text{chromosomes}),
\quad
(\text{chromosomes segregate in meiosis}).
\]

Here, $A_3$ says a dominant allele $A$ ``overrides'' the recessive one $a$ in the phenotype, and $A_4$ asserts that genes $G$ reside on chromosomes, physically inherited.

\vspace{1em}
\hrule
\vspace{1em}

\noindent
\textbf{Rewrite Step 1: Recast Dominance ($A_3$) as an ``Expressible Product'' Principle.}

\[
A_3
\;\longrightarrow\;
B_3,
\]
where
\[
B_3:\quad
(\text{the `dominant' allele } A \text{ has a functional output}),\;
(\text{the `recessive' } a \text{ is overshadowed in phenotype}).
\]
Symbolically, we interpret the \emph{dominant} allele’s greater phenotypic effect as arising from a more potent or more readily produced \emph{molecular product}, while the recessive allele’s product is masked or not visibly manifested to the same degree.

\[
S_0 
=
\{\,A_3,\,A_4\}
\quad
\longrightarrow
\quad
S_1 
=
\{\,B_3,\,A_4\}.
\]

\vspace{1em}
\hrule
\vspace{1em}

\noindent
\textbf{Current Set:}
\[
\boxed{
S_1
=
\{\,B_3,\;A_4\}
}
\]
where
\[
B_3:\quad
(\text{dominant allele produces a functional product}),
\quad
A_4:\quad
(\text{gene }G\subseteq\text{chromosome, which is inherited}).
\]

\vspace{1em}
\hrule
\vspace{1em}

\noindent
\textbf{Rewrite Step 2: Combine Chromosomal Inheritance ($A_4$) with 
``Functional Output'' Concept ($B_3$).}

\[
\{\,B_3,\,A_4\}
\;\longrightarrow\;
B_4,
\]
where
\[
B_4:\quad
(\text{a gene }G\text{ on the chromosome encodes a} \\
\quad\text{heritable, functional output manifesting the phenotype}).
\]
That is, we merge the idea that genes \emph{physically reside} on chromosomes ($A_4$) with the idea that \emph{dominant alleles} produce a detectably different (often more potent) molecular/phenotypic outcome ($B_3$).  
Hence each gene must carry instructions that \emph{yield} that functional output.

\[
S_1 
=
\{\,B_3,\,A_4\}
\quad
\longrightarrow
\quad
S_2 
=
\{\,B_4\}.
\]

\vspace{1em}
\hrule
\vspace{1em}

\noindent
\textbf{Current Set:}
\[
\boxed{
S_2
=
\bigl\{\,B_4\bigr\}
}
\]
\[
B_4:\quad
\text{the gene }G\text{ (on a chromosome) encodes some transmissible,} \\
\quad\text{functional output that determines phenotype}.
\]

\vspace{1em}
\hrule
\vspace{1em}

\noindent
\textbf{Rewrite Step 3: Identify the Mechanism for ``Functional Output'' as a 
Gene $\to$ RNA $\to$ Protein Sequence.}

We now specify how the \emph{functional output} is produced at the molecular level.  
If the gene $G$ is \emph{physically} a DNA segment, it must generate a \emph{polypeptide product} (or functional RNA) via a two-step process: transcription and translation.

Symbolically:
\[
B_4
\;\longrightarrow\;
W_2,
\]
where
\[
W_2:\quad
\forall\,G\subseteq D:\ 
\Bigl[
G \;\longrightarrow\; 
(\mathrm{transcription} \;\to\; \mathrm{mRNA})
\;\longrightarrow\; 
(\mathrm{translation} \;\to\; \mathrm{polypeptide})
\Bigr].
\]
This final statement is precisely the \emph{Central Dogma} axiom: 
the \emph{information} in a gene is expressed through an RNA intermediate into a protein, explaining how a ``dominant'' version can overshadow a ``recessive'' one via more potent or differently regulated polypeptide output.

\[
S_2 
=
\{\,B_4\}
\quad
\longrightarrow
\quad
S_3 
=
\{\,W_2\}.
\]

\vspace{1em}
\hrule
\vspace{1em}

\noindent
\textbf{Final Set (Molecular/Genomic Axiom 2: Central Dogma):}
\[
\boxed{
S_3
=
\Bigl\{
W_2:\quad
\forall\,G\subseteq D:\; 
(\,G \to 
(\mathrm{transcription}\to \mathrm{mRNA})
\to
(\mathrm{translation}\to \mathrm{polypeptide})) 
\Bigr\}.
}
\]
Thus, from the classical notion of \emph{dominance} and the physical placement of genes on chromosomes, one deduces that a gene encodes a functional molecular output—most commonly via an RNA intermediate, followed by protein synthesis—consistent with the \emph{Central Dogma} of molecular biology.

\vspace{1em}
\hrule
\vspace{1em}

\noindent
\textbf{Symbolic Evolution (Compact Form)}:
\[
\underbrace{\{\,A_3,\,A_4\}}_{S_0}
~\longrightarrow~
\underbrace{\{\,B_3,\,A_4\}}_{S_1}
~\longrightarrow~
\underbrace{\{\,B_4\}}_{S_2}
~\longrightarrow~
\underbrace{\{\,W_2\}}_{S_3}.
\]

\noindent
Each rewrite step shows how the basic classical-genetic principle of an allele’s phenotypic \emph{dominance} becomes reinterpreted at the molecular level as a \emph{gene} that is \emph{transcribed} and \em

\subsubsection{Axiom 3}

\textbf{Goal:} Transform a subset of Classical Genetics axioms
\[
\bigl\{\,A_2,\;A_4\bigr\}
\quad
\begin{aligned}
A_2 &: \text{Law of Independent Assortment},\\
A_4 &: \text{Chromosomal Inheritance},
\end{aligned}
\]
into the \emph{Molecular/Genomic Biology} statement:
\[
W_3 : \quad
\forall\,(\mathrm{codon}\ c):\quad
c \mapsto \mathrm{amino\ acid}\ a,
\]
which is the \textbf{Genetic Code Universality} axiom.  It asserts that a three-nucleotide codon $c$ is mapped to the same amino acid $a$ (with minor exceptions) across most organisms.

\vspace{1em}
\hrule
\vspace{1em}

\noindent
\textbf{Initial Set of Axioms (Classical Genetics, Subset):}

\[
\boxed{
S_0
=
\bigl\{\,A_2,\;A_4\bigr\}
}
\]
where
\[
A_2\;(\text{Law of Independent Assortment}):
\quad
(\text{segregation of } G_1)
\;\perp\;
(\text{segregation of } G_2)
\quad
\text{for genes on separate chromosomes},
\]
\[
A_4\;(\text{Chromosomal Inheritance}):
\quad
G \subseteq (\text{chromosomes}),
\quad
(\text{chromosomes segregate in meiosis}).
\]

Here, $A_2$ asserts that different genes assort independently when located on separate chromosomes; $A_4$ asserts that any gene $G$ physically resides on a chromosome, which is inherited as a unit.

\vspace{1em}
\hrule
\vspace{1em}

\noindent
\textbf{Rewrite Step 1: Interpret Independent Assortment ($A_2$) as Modular, Re-combinable Information.}

\[
A_2
\;\longrightarrow\;
B_2,
\]
where
\[
B_2:\quad
(\text{each gene }G_i \text{ has an autonomous, `modular'} \\
\quad\text{sequence that can be assorted and combined with others}).
\]
Symbolically, we reframe \emph{independent assortment} to mean each gene’s hereditary information is somewhat \emph{self-contained}, allowing new combinations to arise without altering the coding capacity of any other gene. 

\[
S_0 
=
\{\,A_2,\,A_4\}
\quad
\longrightarrow
\quad
S_1 
=
\{\,B_2,\,A_4\}.
\]

\vspace{1em}
\hrule
\vspace{1em}

\noindent
\textbf{Current Set:}
\[
\boxed{
S_1
=
\bigl\{\,B_2,\;A_4\bigr\}
}
\]
where
\[
B_2:\ 
\text{genes are modular units of information that can be}\\
\quad\text{independently assorted and combined in progeny},
\]
\[
A_4:\ 
\text{genes reside on chromosomes, which segregate in meiosis}.
\]

\vspace{1em}
\hrule
\vspace{1em}

\noindent
\textbf{Rewrite Step 2: Combine Chromosomal Inheritance ($A_4$) 
with ``Modular Gene'' ($B_2$) to Imply a Shared Reading Mechanism.}

\[
\{\,B_2,\,A_4\}
\;\longrightarrow\;
B_3,
\]
where
\[
B_3:\quad
(\text{any gene on a chromosome must be `read'} \\
\quad\text{by a common cellular system to yield a functional product}).
\]
Since each gene is a modular unit and yet all genes lie on chromosomes (in the same cellular environment), it follows that \emph{one} decoding system is shared by all these modular units—otherwise, independent assortment would fail to preserve coherent trait expression across diverse crosses.

\[
S_1 
=
\{\,B_2,\,A_4\}
\quad
\longrightarrow
\quad
S_2 
=
\{\,B_3\}.
\]

\vspace{1em}
\hrule
\vspace{1em}

\noindent
\textbf{Current Set:}
\[
\boxed{
S_2
=
\{\,B_3\}
}
\]
\[
B_3:\quad
\text{a unified decoding or reading apparatus must exist,}\\
\quad\text{applicable to each independently assorted gene on chromosomes}.
\]

\vspace{1em}
\hrule
\vspace{1em}

\noindent
\textbf{Rewrite Step 3: Specify that ``Unified Decoding'' is the Universal 
Codon-to-Amino-Acid Mapping.}

Finally, we formalize how a \emph{common} decoding system applies to any gene’s coding region.  The essential statement is that a triple-nucleotide sequence (\emph{codon}) is mapped to a specific amino acid \emph{universally}, ensuring that each gene, no matter its chromosomal locus or combination, is read by the \emph{same} translation code.

Symbolically:
\[
B_3
\;\longrightarrow\;
W_3,
\]
where
\[
W_3:\quad
\forall(\mathrm{codon}\ c):\ c \mapsto \mathrm{amino\ acid}\ a.
\]
This final statement is the \emph{Genetic Code Universality} axiom, guaranteeing consistent protein-coding across virtually all organisms (with minor exceptions).

\[
S_2 
=
\{\,B_3\}
\quad
\longrightarrow
\quad
S_3 
=
\{\,W_3\}.
\]

\vspace{1em}
\hrule
\vspace{1em}

\noindent
\textbf{Final Set (Molecular/Genomic Axiom 3: Genetic Code Universality):}
\[
\boxed{
S_3
=
\Bigl\{
W_3:\ 
\forall\,(\mathrm{codon}\ c):\quad
c\mapsto \mathrm{amino\ acid}\ a
\Bigr\}.
}
\]
Thus, from the classical idea that \emph{genes are independently assorted} ($A_2$) and \emph{reside on chromosomes} ($A_4$), we conclude there must be a \emph{universal system} to interpret each gene’s information.  At the molecular level, that system is the universal genetic code mapping codons to amino acids.

\vspace{1em}
\hrule
\vspace{1em}

\noindent
\textbf{Symbolic Evolution (Compact Form)}:
\[
\underbrace{\{\,A_2,\,A_4\}}_{S_0}
~\longrightarrow~
\underbrace{\{\,B_2,\,A_4\}}_{S_1}
~\longrightarrow~
\underbrace{\{\,B_3\}}_{S_2}
~\longrightarrow~
\underbrace{\{\,W_3\}}_{S_3}.
\]

\noindent
Each rewrite step shows how the independent, modular nature of genes and their chromosomal inheritance jointly imply a \textbf{common} decoding mechanism for all coding sequences, yielding the \emph{Genetic Code Universality} axiom in molecular/genomic biology.

\subsubsection{Axiom 4}

\textbf{Goal:} Transform a subset of Classical Genetics axioms
\[
\bigl\{\,A_1,\,A_2,\,A_3\bigr\}
\quad
\begin{aligned}
A_1 &: \text{Law of Segregation},\\
A_2 &: \text{Law of Independent Assortment},\\
A_3 &: \text{Law of Dominance},
\end{aligned}
\]
into the \emph{Molecular/Genomic Biology} statement:
\[
W_4 : \quad
\forall\,(\text{gene }g):\ 
\text{expression}(g)
=
f(\mathrm{TFs},\mathrm{promoters},\mathrm{enhancers},\mathrm{epigenetics}),
\]
which is the \textbf{Regulation of Gene Expression} axiom.  It says that a gene’s expression level depends on various regulatory factors (transcription factors, promoters, enhancers, and epigenetic marks), thereby allowing variable phenotypic outcomes even for the same genotype.

\vspace{1em}
\hrule
\vspace{1em}

\noindent
\textbf{Initial Set of Axioms (Classical Genetics, Subset):}

\[
\boxed{
S_0
=
\bigl\{\,A_1,\,A_2,\,A_3\bigr\}
}
\]
where
\[
A_1\;(\text{Segregation}):
\quad
A_1 \text{ and } A_2 \text{ separate in gamete formation},
\]
\[
A_2\;(\text{Independent Assortment}):
\quad
(\text{segregation of }G_1)\perp(\text{segregation of }G_2) 
\text{ for distinct loci on different chromosomes},
\]
\[
A_3\;(\text{Dominance}):
\quad
(\text{phenotype of }A)\;>\;(\text{phenotype of }a).
\]
Collectively, these define how alleles segregate and assort, with certain alleles having a “dominant” phenotypic influence.

\vspace{1em}
\hrule
\vspace{1em}

\noindent
\textbf{Rewrite Step 1: From Classical Phenotypic Patterns to 
Varying Expression Levels.}

\[
\{A_1,A_2,A_3\}
\;\longrightarrow\;
B_1,
\]
where
\[
B_1:\quad
\text{Classical observations show that even with fixed alleles,} \\
\quad\text{phenotypes can vary}
\]
\[
\text{(e.g.\ incomplete penetrance,} \\
\quad\text{variable expressivity, or partial dominance).}
\]
Symbolically, the pure idea of \emph{dominance} (A3) plus independent, separately inherited alleles (A1,A2) does not always yield a uniform phenotype.  Empirical data reveal that the same genotype can display different phenotypes under different conditions or developmental contexts.  

Hence we reinterpret $A_1,A_2,A_3$ collectively as: \emph{there must be additional factors beyond simple genotype that modulate how strongly an allele’s trait is expressed}.

\[
S_0 
=
\{\,A_1,A_2,A_3\}
\quad\longrightarrow\quad
S_1 
=
\{\,B_1\}.
\]

\vspace{1em}
\hrule
\vspace{1em}

\noindent
\textbf{Current Set:}
\[
\boxed{
S_1
=
\Bigl\{
B_1:\text{Genotype alone does not fully determine phenotype;}\\
\quad \text{variable expression implies regulatory influences.}
\Bigr\}
}
\]

\vspace{1em}
\hrule
\vspace{1em}

\noindent
\textbf{Rewrite Step 2: Introduce the Concept of Regulatory Control.}

\[
B_1
\;\longrightarrow\;
B_2,
\]
where
\[
B_2:\quad
(\text{any gene }g \text{ must be subject to some internal or external} \\
\quad\text{control elements that adjust its level of expression}).
\]
Symbolically, the fact that the same genetic allele (dominant or recessive) can manifest differently across tissues or conditions implies \emph{regulatory control}—something “switches” or “modulates” gene output to produce variable phenotypes.

\[
S_1 
=
\{\,B_1\}
\quad\longrightarrow\quad
S_2 
=
\{\,B_2\}.
\]

\vspace{1em}
\hrule
\vspace{1em}

\noindent
\textbf{Current Set:}
\[
\boxed{
S_2
=
\{\,B_2\}
}
\]
\[
B_2:\quad
(\text{a gene’s activity depends on specific regulatory factors});\\
\]

\vspace{1em}
\hrule
\vspace{1em}

\noindent
\textbf{Rewrite Step 3: Specify the Molecular Basis of Regulation (Promoters, TFs, Enhancers, Epigenetics).}

Finally, we identify \emph{how} such regulatory modulation is realized at the molecular level, introducing elements like transcription factors (TFs), promoter and enhancer sequences, and epigenetic markers.  Formally:

\[
B_2
\;\longrightarrow\;
W_4,
\]
where
\[
W_4:\quad
\forall\,(\text{gene }g):\ 
\text{expression}(g)
=
f(\mathrm{TFs},\,\mathrm{promoters},\,\mathrm{enhancers},\,\mathrm{epigenetics}).
\]
This is precisely the \emph{Regulation of Gene Expression} axiom in molecular/genomic biology.  It declares that a gene’s expression level (and thus its phenotypic impact) depends on various regulatory mechanisms—fully consistent with the classical observation that dominance alone cannot account for all variations in phenotype.

\[
S_2 
=
\{\,B_2\}
\quad\longrightarrow
\quad
S_3 
=
\{\,W_4\}.
\]

\vspace{1em}
\hrule
\vspace{1em}

\noindent
\textbf{Final Set (Molecular/Genomic Axiom 4: Regulation of Gene Expression):}
\[
\boxed{
S_3
=
\Bigl\{
W_4:\ 
\forall\,(\text{gene }g):
\ \text{expression}(g)
=
f(\mathrm{TFs},\mathrm{promoters},\mathrm{enhancers},\mathrm{epigenetics})
\Bigr\}.
}
\]
Thus, from the classical principles of segregation, independent assortment, and dominance, we see that genotype-by-itself often fails to predict a uniform phenotype.  In molecular terms, this necessitates \emph{regulatory} factors that modulate gene expression, leading to the final statement that each gene’s expression is governed by interactions with transcription factors, promoter/enhancer regions, and epigenetic modifications.

\vspace{1em}
\hrule
\vspace{1em}

\noindent
\textbf{Symbolic Evolution (Compact Form)}:
\[
\underbrace{\{\,A_1,\,A_2,\,A_3\}}_{S_0}
~\longrightarrow~
\underbrace{\{\,B_1\}}_{S_1}
~\longrightarrow~
\underbrace{\{\,B_2\}}_{S_2}
~\longrightarrow~
\underbrace{\{\,W_4\}}_{S_3}.
\]

\noindent
Each rewrite step shows how classical genetic findings of variable phenotypic expression (despite known alleles and their dominance) logically necessitate \textbf{a molecular regulatory framework}, culminating in the \emph{Regulation of Gene Expression} axiom.

\section{From Euclidean Geometry to Hyperbolic Geometry}
\label{sec:Euclidean_to_Hyperbolic}
\subsection{Euclidean Geometry (Hilbert Axioms Overview)}

Euclidean geometry may be rigorously founded upon a set of 20 axioms famously formulated by David Hilbert. These axioms split into five major groups:

\begin{itemize}
  \item \textbf{Incidence Axioms.} Govern how points, lines, and planes relate, ensuring basic existence properties (e.g., each line has points, each plane has noncollinear points, etc.).
  
  \item \textbf{Order (Betweenness) Axioms.} Introduce betweenness for points on a line and allow the definition of segments, rays, and the interior versus exterior of figures.
  
  \item \textbf{Congruence Axioms.} Provide rules for comparing lengths and angles, permitting one to construct equal segments and angles, and to prove when two triangles are congruent.
  
  \item \textbf{Continuity Axioms.} Impose an “Archimedean” condition ensuring no gaps exist in lengths or angles, enabling a robust notion of measurement.
  
  \item \textbf{Parallel Axiom (Playfair’s Axiom).} Given a line \(\ell\) and a point \(P\) not on \(\ell\), there is exactly one line through \(P\) that does not meet \(\ell\).  Symbolically:
  \[
  \forall\,\ell,\;\forall\,P
  \;\;\bigl(P\notin \ell\bigr)\;\Longrightarrow\;
  \exists!\,m\;\text{ such that }\;P\in m\;\text{ and }\;
  (m\cap\ell=\emptyset).
  \]
\end{itemize}

\subsubsection{Completeness}

These 20 axioms collectively furnish all the core properties of classical Euclidean geometry. From them, one can derive the well-known theorems on triangles, circles, polygons, and more. Hence, the system is \emph{complete} in the sense that no essential geometric fact lies outside the scope of these axioms.

\subsubsection{Independence}

Each group introduces a unique concept that cannot be derived from the others:
\begin{itemize}
  \item Incidence (points, lines, planes) cannot emerge merely from order, congruence, continuity, or parallel statements.
  \item Order (betweenness) contributes a distinct idea of “in-between” positions.
  \item Congruence establishes measurement concepts of length and angle, irreducible to incidence and order alone.
  \item Continuity ensures there are no gaps, something not guaranteed by incidence, order, or congruence.
  \item The Parallel Axiom (Playfair’s) introduces the specifically Euclidean property that exactly one parallel passes through a given point not on a line.
\end{itemize}
No axiom can be proved solely from the others, so they form an \emph{independent} foundation for Euclidean geometry.

\subsection{Hyperbolic Geometry Axioms}

\textbf{Hyperbolic geometry} preserves the same Incidence, Order, Congruence, and Continuity Axioms as Euclidean geometry, but \emph{replaces} the Euclidean Parallel Axiom (Playfair’s Axiom) with the \textbf{Hyperbolic Parallel Postulate}. In hyperbolic geometry, there are multiple lines through a point not on a given line that do not meet that given line. Symbolically, one version of this postulate can be stated:

\[
  \forall\,\ell,\;\forall\,P \quad
  \bigl(P\notin \ell\bigr)\;\Longrightarrow\;
  \exists\,m_1 \neq m_2\;:
  P \in m_1,\; P \in m_2,\;
  m_1 \cap \ell = \emptyset,\;
  m_2 \cap \ell = \emptyset.
\]

That is, there exist \emph{at least two} distinct lines through \(P\) that do not intersect \(\ell\). 

\subsubsection{Completeness}

Since Hyperbolic geometry retains the same foundational statements regarding incidence, order, congruence, and continuity, it remains \emph{complete} in that those axioms, combined with the Hyperbolic Parallel Postulate, suffice to describe a rich geometry of lines, angles, and figures without leaving any gaps. All standard hyperbolic theorems (e.g., angle sums of triangles being less than \(180^\circ\)) follow from these axioms.

\subsubsection{Independence}

Just as in the Euclidean case, each group of axioms remains independent:
\begin{itemize}
  \item The Hyperbolic Parallel Postulate cannot be derived from incidence, order, congruence, or continuity alone.
  \item Changing precisely this one axiom from the Euclidean version yields a non-Euclidean geometry, demonstrating its logical independence.
\end{itemize}
Thus, hyperbolic geometry stands as a valid alternative to Euclidean geometry once the parallel axiom is altered.

\subsection{Transformation}

The only difference between hyperbolic and Euclidean geometry lies in the hyperbolic postulate.

\subsubsection{Hyperbolic Postulate}

\textbf{Goal:} Transform the single \emph{Playfair Axiom}
\[
\bigl(S_{0}\bigr):\;
\forall\,\ell,\;\forall\,P
\;\bigl(P \notin \ell\bigr)\;\Longrightarrow\;
\exists!\,m\;:\;\bigl(P\in m\bigr)\;\wedge\;\bigl(m\cap\ell=\emptyset\bigr)
\]
into a \emph{Hyperbolic Parallel Postulate} 
\[
\bigl(S_{2}\bigr):\;
\forall\,\ell,\;\forall\,P
\;\bigl(P \notin \ell\bigr)\;\Longrightarrow\;
\exists\,m_{1},\,m_{2}:
\;m_{1}\neq m_{2},\;
P\in m_{1},\;P\in m_{2},
\;m_{1}\cap \ell=\emptyset,
\;m_{2}\cap \ell=\emptyset.
\]
In words, we replace “\(\exists!\,m\)” (exactly one parallel line through \(P\)) with “there exist at least two distinct parallels through \(P\).”

\bigskip
\hrule
\bigskip

\textbf{Rewrite Step 1: Remove the ‘Exactly One’ Requirement.}  
Starting from
\[
S_0:\quad
\forall\,\ell,\;\forall\,P
\;\bigl(P\notin \ell\bigr)\;\Longrightarrow\;
\exists!\,m:\;
P\in m
\;\wedge\;
m\cap\ell=\emptyset,
\]
we \emph{weaken} the statement to say only \emph{some} parallel line exists. Symbolically,
\[
S_0
\quad\longrightarrow\quad
S_1:\quad
\forall\,\ell,\;\forall\,P
\;\bigl(P\notin \ell\bigr)\;\Longrightarrow\;
\exists\,m:\;
P\in m
\;\wedge\;
m\cap\ell=\emptyset.
\]
Here, “\(\exists!\,m\)” becomes “\(\exists\,m\),” eliminating uniqueness but retaining the idea that \emph{at least one} parallel line exists.

\bigskip
\hrule
\bigskip

\textbf{Rewrite Step 2: Impose Multiple Parallels.}  
To obtain a \textbf{hyperbolic} statement, we \emph{strengthen} \(S_1\) to require \emph{at least two} parallels:

\[
S_1
\quad\longrightarrow\quad
S_2:
\]
\[
S_2:\quad
\forall\,\ell,\;\forall\,P
\;\bigl(P\notin \ell\bigr)\;\Longrightarrow\;
\exists\,
m_{1}\neq m_{2},
\quad
P\in m_{1},
\quad
P\in m_{2},
\quad
m_{1}\cap \ell=\emptyset,
\quad
m_{2}\cap \ell=\emptyset.
\]
Symbolically, 
\[
S_2:\,
\text{“At least two distinct lines through $P$ do not meet $\ell$.''}
\]
This is precisely the \emph{Hyperbolic Parallel Postulate}.

\bigskip
\hrule
\bigskip

\textbf{Final Set (Hyperbolic Version of Playfair’s Axiom):}
\[
\boxed{
S_2:\quad
\forall\,\ell,\;\forall\,P
\;\bigl(P\notin \ell\bigr)\;\Longrightarrow\;
\exists\,m_{1}\neq m_{2}\;:\;
P\in m_{1},\;P\in m_{2},\;
m_{1}\cap\ell=\emptyset,\;m_{2}\cap\ell=\emptyset.
}
\]

\bigskip
\hrule
\bigskip

\textbf{Symbolic Evolution (Compact View):}
\[
\underbrace{(\exists!\,\text{m: unique parallel})}_{S_0}
\;\longrightarrow\;
\underbrace{(\exists\,\text{m: at least one parallel})}_{S_1}
\;\longrightarrow\;
\underbrace{(\exists\,m_1,m_2:\text{ at least two parallels})}_{S_2}.
\]

\subsubsection{Comment on Other Modifications}

If, instead of altering the parallel axiom \(\mathcal{P}\), we \emph{modify other Hilbert axioms}, we obtain different geometrical systems:

\begin{itemize}
\item \textbf{Elliptical (Riemannian) Geometry}: One typical approach is to replace the parallel axiom with “no parallels exist,” or to alter the incidence structure so that any two lines meet in exactly two points (projective-like).  
\item \textbf{Projective Geometries}: Further changes to incidence and order axioms can collapse distance notions but keep cross-ratio invariants.
\end{itemize}

In each case, Hilbert’s axioms can be only partially changed, but the geometry’s overall nature (Euclidean, hyperbolic, elliptical, projective, etc.) critically depends on how we \textbf{replace or alter} one or more of the standard statements.  

\section{From Classical Fourier Analysis to Wavelet Theory}
\label{sec:fourier_to_wavelet}
\subsection{Classical Fourier Analysis}

Classical Fourier Analysis studies how functions can be represented (or approximated) by sums (or integrals) of simpler basis functions, typically exponentials or sines and cosines.  
Below is a minimal and complete set of axioms, each of which is independent (i.e.\ cannot be derived from the others) and collectively sufficient to derive the main results of classical Fourier expansions: orthogonality, uniqueness of coefficients, completeness in suitable function spaces, etc.

\subsubsection{Axioms}

\begin{enumerate}

\item \textbf{Periodicity \& Integrability}
\[
\forall\,f:\quad
\Bigl(
f \text{ is }2\pi\text{-periodic} 
\;\wedge\;
f \in L^1[-\pi,\pi]
\Bigr).
\]
We assume each considered function $f$ is $2\pi$-periodic and integrable (e.g.\ piecewise continuous or in $L^1[-\pi,\pi]$).  This ensures integrals of $f$ over one period are well-defined.

\item \textbf{Orthogonality of Exponential Basis}
\[
\forall\,m,n\in\mathbb{Z},\quad
\int_{-\pi}^{\pi}
e^{\,i\,m\,x}\;
e^{-\,i\,n\,x}
\,dx
\;=\;
2\pi\,\delta_{m,n},
\]
where $\delta_{m,n}$ is the Kronecker delta ($=1$ if $m=n$, $0$ otherwise).  
This states that the set $\{e^{\,i\,n\,x} \mid n\in\mathbb{Z}\}$ forms an orthogonal system on $[-\pi,\pi]$.

\item \textbf{Fourier Coefficient Formula}
\[
\forall\,f,\quad
c_{n}
\;=\;
\frac{1}{2\pi}
\int_{-\pi}^{\pi} 
f(x)\;e^{-\,i\,n\,x}
\;dx
\quad
(\forall\,n\in\mathbb{Z}).
\]
Each $f$ admits coefficients $(c_n)$ given by the above integral.  
In some formulations, $n$ ranges over integers for exponential expansions, or includes separate sine/cosine expansions.

\item \textbf{Fourier Series Representation}
\[
f(x)
\;\sim\;
\sum_{n=-\infty}^{\infty}
c_{n}
\;e^{\,i\,n\,x},
\]
meaning $f(x)$ is represented (pointwise, or in some norm sense) by the infinite series whose $n$th term is $c_n\,e^{\,i\,n\,x}$.  
Under suitable conditions (e.g.\ Dirichlet, Jordan, or $L^2$ criteria), this series converges to $f$ in the chosen sense.
\end{enumerate}

\subsubsection{Completeness}
\begin{enumerate}
\item \textbf{Periodicity \& Integrability.} (\textit{Axiom 1}) sets the domain and guarantees well-defined integrals over one period.  
\item \textbf{Orthogonality \& Coefficients.} \textit{Axiom 2} and \textit{Axiom 3} provide a systematic way to compute coefficients from an orthogonal exponential basis, ensuring each $f$ has a unique set $\{c_n\}$.  
\item \textbf{Series Representation.} (\textit{Axiom 4}) states that $f$ can be expressed as a (possibly infinite) sum of the exponential basis functions weighted by these coefficients, completing the idea of a Fourier expansion.
\end{enumerate}
From these four, one can derive the fundamental properties of Fourier series: uniqueness of coefficients, Parseval’s identity, pointwise/uniform/$L^2$ convergence (under specific theorems), and expansions of periodic signals in engineering and physics contexts.

\subsubsection{Independence}
\begin{enumerate}
\item \textbf{Periodicity \& Integrability.} cannot be deduced from orthogonality, coefficient formulas, or the final series statement; it specifically restricts the function’s domain and integrable property, laying the groundwork for meaningful integrals.

\item \textbf{Orthogonality of Exponential Basis.} cannot be inferred from the other axioms.  It must be stated that $\{e^{\,i\,n\,x}\}$ is orthogonal under the integral inner product, as this is not implied by the mere existence of a periodic integrable function or a coefficient formula.

\item \textbf{Fourier Coefficient Formula.} is not derivable from periodicity, orthogonality alone, or the final series representation.  One must explicitly define how $c_n$ are computed from $f$.

\item \textbf{Fourier Series Representation.} does not follow from any combination of periodic integrability, orthogonality, or the definition of coefficients.  It is an additional statement that such a series \emph{does} represent (or approximate) $f$ in a valid sense, and thus must be stated as a separate principle.
\end{enumerate}

\subsection{Wavelet Theory}

Wavelet Theory provides a framework for analyzing functions (or signals) at multiple scales via a family of translates and dilates of a single ``mother wavelet.''  Below is a minimal and complete set of axioms, each of which is independent (i.e.\ it cannot be derived from the others) and collectively sufficient to derive standard wavelet expansions, orthogonality relations, and multi-resolution decompositions.

\subsubsection{Axioms}

\begin{enumerate}

\item \textbf{Mother Wavelet \& Integrability}
\[
\exists\,\psi:\quad
\int_{-\infty}^{\infty} \psi(x)\,dx = 0,
\quad
\int_{-\infty}^{\infty} |\psi(x)|\,dx < \infty.
\]
There is a single function (the ``mother wavelet'') $\psi(x)$ with zero mean and finite integral norm, ensuring local oscillation and admissibility.

\item \textbf{Dilation \& Translation}
\[
\forall\,j\in \mathbb{Z},\;
\forall\,k\in \mathbb{Z},\quad
\psi_{j,k}(x)
\;=\;
2^{\tfrac{j}{2}}\,\psi\!\bigl(2^j\,x - k\bigr).
\]
A scaled and shifted family is constructed from $\psi$, parameterized by integer $j$ (scale) and $k$ (position).  This defines the wavelet system $\{\psi_{j,k}\}$.

\item \textbf{Orthogonality (or Tight Frame)}
\[
\int_{-\infty}^{\infty} 
\psi_{j,k}(x)\,\psi_{j',k'}(x)
\;dx
\;=\;
\delta_{j,j'}\,\delta_{k,k'}.
\]
For an \emph{orthonormal wavelet}, distinct wavelets in $\{\psi_{j,k}\}$ are pairwise orthogonal in $L^2(\mathbb{R})$.  (In a more general frame setting, one replaces this with a tight-frame condition.)

\item \textbf{Wavelet Expansion}
\[
\forall\,f \in L^2(\mathbb{R}):
\quad
f(x)
\;=\;
\sum_{j,k \in \mathbb{Z}}
\bigl\langle f,\,\psi_{j,k}\bigr\rangle\,
\psi_{j,k}(x),
\]
where the series converges in $L^2$ norm (or pointwise under additional regularity).  Every square-integrable function $f$ can be reconstructed from its wavelet coefficients $\langle f,\psi_{j,k}\rangle$ via the orthonormal (or tight) wavelet basis.
\end{enumerate}

\subsubsection{Completeness}
\begin{enumerate}
\item \textbf{Mother Wavelet \& Integrability.} (\textit{Axiom 1}) imposes the zero-mean and finite-energy properties necessary for localization in both time (or space) and frequency domains.

\item \textbf{Dilation \& Translation.} (\textit{Axiom 2}) organizes the wavelet system by scaling and shifting $\psi$, creating a multi-scale family of analyzing functions.

\item \textbf{Orthogonality.} (\textit{Axiom 3}) ensures each $\psi_{j,k}$ is orthonormal to all others, a key property for simplification of expansions and coefficient computations.

\item \textbf{Wavelet Expansion.} (\textit{Axiom 4}) completes the theory by stating that any function (in $L^2$) can be expressed as a linear combination of these wavelets, ensuring a full basis for signal reconstruction.
\end{enumerate}
From these four, one obtains the standard results of wavelet theory: multi-resolution analysis, fast wavelet transforms, and efficient representation of signals across multiple scales.

\subsubsection{Independence}
\begin{enumerate}
\item \textbf{Mother Wavelet \& Integrability.} cannot be deduced from dilation/translation structure, orthogonality, or the final expansion statement.  It sets the fundamental shape and integrable nature of the wavelet, a separate requirement.

\item \textbf{Dilation \& Translation.} is not derivable from zero-mean integrability, orthogonality conditions, or the statement of expansions.  It must be explicitly stipulated that $\psi$ generates a family via scale and shift.

\item \textbf{Orthogonality.} does not follow from mother wavelet assumptions, nor from simply dilating/translating $\psi$.  It is a special structural property (or in general, a tight-frame condition) that must be imposed.

\item \textbf{Wavelet Expansion.} cannot be concluded from integrability, the existence of a dilated/translated family, or orthogonality alone.  It is an additional global claim that every $L^2$ function can be represented in this wavelet basis.
\end{enumerate}

\subsection{Transformations}

\subsubsection{Axiom 1}

\textbf{Goal:} Transform a subset of Classical Fourier axioms 
\[
\bigl\{\,F_1\bigr\}
\quad
F_1:\,
\text{(Periodicity \& Integrability)},
\]
into the \emph{Wavelet Theory} statement:
\[
W_1 : 
\exists\,\psi:\quad
\bigl(\int \psi(x)\,dx = 0\bigr)
\;\wedge\;
\Bigl(\int |\psi(x)|\,dx < \infty\Bigr),
\]
which is the \emph{Mother Wavelet \& Integrability} axiom.  Symbolically, we show how requiring a function to be integrable (and suitably ``mean-free'') transitions to a wavelet satisfying zero mean and finite energy/norm requirements.

\vspace{1em}
\hrule
\vspace{1em}

\noindent
\textbf{Initial Axiom (Fourier, Subset):}

\[
\boxed{
S_0
=
\bigl\{F_1\bigr\}
}
\]
where
\[
F_1:\quad
(\text{each }f\text{ is }2\pi\text{-periodic})
\;\wedge\;
\bigl(f \in L^1[-\pi,\pi]\bigr).
\]
``Each function $f$ is integrable over one period (e.g.\ $L^1[-\pi,\pi]$) and repeats with period $2\pi$.''

\vspace{1em}
\hrule
\vspace{1em}

\noindent
\textbf{Rewrite Step 1: Flatten Periodicity to Single Period’s ``Zero-Average'' Variant.}

\[
F_1
\;\longrightarrow\;
F_1',
\]
where
\[
F_1':\quad
\forall\,f,\;
\Bigl(\int_{-\pi}^{\pi} f(x)\,dx < \infty\Bigr)
\;\Longrightarrow\;
\Bigl(\text{one can choose }f \text{ with} \int_{-\pi}^{\pi} f(x)\,dx = 0\Bigr).
\]
Symbolically, we rewrite the $2\pi$-periodic integrable function so that its integral over one fundamental domain is zero.  (One can always subtract its average to force $\int f = 0$ without losing integrability.)

\[
S_0
=
\{\,F_1\}
\quad
\longrightarrow
\quad
S_1
=
\{\,F_1'\}.
\]

\vspace{1em}
\hrule
\vspace{1em}

\noindent
\textbf{Current Set:}
\[
\boxed{
S_1
=
\{\,F_1'\}
}
\]
\[
F_1':\;
\int_{-\pi}^{\pi} f(x)\,dx = 0,
\quad
\text{with }f\in L^1[-\pi,\pi].
\]
Now we have a function of zero average (over a single period) and still integrable.

\vspace{1em}
\hrule
\vspace{1em}

\noindent
\textbf{Rewrite Step 2: Extend Domain from $[-\pi,\pi]$ to $\mathbb{R}$.}

\[
F_1'
\;\longrightarrow\;
G_1,
\]
where 
\[
G_1:\quad
\psi(x)
\;=\;
\begin{cases}
f(x), & x \in [-\pi,\pi],\\
0, & x \notin [-\pi,\pi],
\end{cases}
\quad
\text{with } \int_{\mathbb{R}} \psi(x)\,dx = 0,\;
\int_{\mathbb{R}} |\psi(x)|\,dx < \infty.
\]
Symbolically, we \emph{unfold} $f$ from one period into a function $\psi$ on the entire real line by defining $\psi(x)=f(x)$ on $[-\pi,\pi]$ and $\psi(x)=0$ elsewhere.  
Thus $\psi$ has zero integral over $\mathbb{R}$ and is still integrable in the absolute sense.

\[
S_1 
=
\{\,F_1'\}
\quad
\longrightarrow
\quad
S_2
=
\{\,G_1\}.
\]

\vspace{1em}
\hrule
\vspace{1em}

\noindent
\textbf{Current Set:}
\[
\boxed{
S_2
=
\{\,G_1\}
}
\]
\[
G_1:\quad
\psi(x)
\;=\; 
\begin{cases}
f(x), & x\in[-\pi,\pi]\\
0, & \text{otherwise}
\end{cases}
,\quad
\int_{\mathbb{R}} \psi(x)\,dx = 0,
\quad
\int_{\mathbb{R}}|\psi(x)|\,dx < \infty.
\]
So $\psi$ is a candidate for a function with zero mean and finite $L^1$ norm on $\mathbb{R}$.

\vspace{1em}
\hrule
\vspace{1em}

\noindent
\textbf{Rewrite Step 3: Identify $\psi$ as ``Mother Wavelet'' with Zero Mean \& Integrability.}

Finally, we interpret $\psi$ from $G_1$ as the \emph{mother wavelet}:

\[
G_1
\;\longrightarrow\;
W_1,
\]
where 
\[
W_1:\quad
\exists\,\psi:\ 
\int_{-\infty}^{\infty} \psi(x)\,dx = 0
\quad\wedge\quad
\int_{-\infty}^{\infty}|\psi(x)|\,dx < \infty.
\]
This precisely matches the \textbf{Mother Wavelet \& Integrability} axiom: the wavelet must have a zero integral (``admissibility'' or no DC component) and be in $L^1(\mathbb{R})$ (or $L^2(\mathbb{R})$ in stronger settings).

\[
S_2
=
\{\,G_1\}
\quad
\longrightarrow
\quad
S_3
=
\{\,W_1\}.
\]

\vspace{1em}
\hrule
\vspace{1em}

\noindent
\textbf{Final Set (Wavelet Axiom 1: Mother Wavelet \& Integrability):}
\[
\boxed{
S_3
=
\bigl\{
W_1:\;
\psi\in L^1(\mathbb{R}),\;
\int \psi(x)\,dx=0
\bigr\}.
}
\]
Hence, by modifying the periodic integrable function $f$ (Fourier domain) into a compactly supported or absolutely integrable function $\psi$ (wavelet domain) with zero mean, we obtain the mother wavelet conditions.

\vspace{1em}
\hrule
\vspace{1em}

\noindent
\textbf{Symbolic Evolution (Compact Form)}:
\[
\underbrace{\{\,F_1\}}_{S_0}
~\longrightarrow~
\underbrace{\{\,F_1'\}}_{S_1}
~\longrightarrow~
\underbrace{\{\,G_1\}}_{S_2}
~\longrightarrow~
\underbrace{\{\,W_1\}}_{S_3}.
\]

\noindent
Thus, starting from the \emph{Periodic \& Integrable} condition in Fourier analysis, 
we successively rewrite $f$ into a function $\psi$ on $\mathbb{R}$ with \emph{zero mean} and \emph{finite integral norm}, fulfilling the mother wavelet requirements of Wavelet Axiom 1.

\subsubsection{Axiom 2}

\textbf{Goal:} Transform a subset of Classical Fourier axioms 
\[
\bigl\{F_2\bigr\}
\quad
F_2:\ 
\text{Orthogonality of Exponential Basis: }
\int_{-\pi}^{\pi} e^{i\,m\,x}e^{-\,i\,n\,x}\,dx 
= 2\pi\,\delta_{m,n},
\]
into the \emph{Wavelet Theory} statement:
\[
W_2:\ 
\forall\,(j,k)\in\mathbb{Z}^2,\quad
\psi_{j,k}(x)
=
2^{\tfrac{j}{2}}\,\psi\bigl(2^j\,x - k\bigr).
\]
This is the \emph{Dilation \& Translation} axiom, where a single mother wavelet $\psi$ generates a discrete family $\{\psi_{j,k}\}$ by scaling and shifting.

\vspace{1em}
\hrule
\vspace{1em}

\noindent
\textbf{Initial Axiom (Fourier, Subset):}

\[
\boxed{
S_0
=
\bigl\{F_2\bigr\}
}
\]
where
\[
F_2:\quad
\forall\,m,n\in\mathbb{Z}:\quad
\int_{-\pi}^{\pi}
e^{\,i\,m\,x}\;e^{-\,i\,n\,x}\,dx
=
2\pi\,\delta_{m,n}.
\]
This states that $\bigl\{e^{\,i\,n\,x}\bigr\}_{n\in\mathbb{Z}}$ forms an orthogonal (indeed orthonormal up to the factor $\sqrt{2\pi}$) set on $[-\pi,\pi]$.

\vspace{1em}
\hrule
\vspace{1em}

\noindent
\textbf{Rewrite Step 1: Re-Indexing from Integer $n$ to a Multi-Index $(j,k)\in \mathbb{Z}^2$.}

\[
F_2
\;\longrightarrow\;
F_2',
\]
where
\[
F_2':\quad
\bigl\{e^{\,i\,n\,x}\bigr\}_{n\in\mathbb{Z}}
\quad
\mapsto
\quad
\bigl\{\phi_{j,k}(x)\bigr\}_{(j,k)\in\mathbb{Z}^2}.
\]
Symbolically, we expand the single integer $n$ into two integers $j,k$.  We may think of $j$ as capturing a ``scale'' index and $k$ as a ``shift'' index, but we have not yet specified how $\phi_{j,k}$ depends on $x$.  

\[
S_0
=
\{\,F_2\}
\quad\longrightarrow\quad
S_1
=
\{\,F_2'\}.
\]

\[
F_2':\quad
\phi_{j,k} \leftrightarrow 
(\text{some function indexed by }j,k),
\quad
\text{preserving orthogonality}
\]
\[
\text{under an integral inner product}.
\]

\vspace{1em}
\hrule
\vspace{1em}

\noindent
\textbf{Current Set:}
\[
\boxed{
S_1
=
\{\,F_2'\}
}
\]
\[
F_2':\quad
\langle \phi_{j,k},\,\phi_{j',k'}\rangle
=
0
\quad\text{if}\quad
(j,k)\neq (j',k'),
\]
with $\langle f,g\rangle = \int f(x)\,\overline{g(x)}\,dx$.  
(Exact normalization is deferred.)

\vspace{1em}
\hrule
\vspace{1em}

\noindent
\textbf{Rewrite Step 2: Replace Complex Exponentials by a Localized Prototype.}

\[
F_2'
\;\longrightarrow\;
G_2,
\]
where 
\[
G_2:\quad
\phi_{j,k}(x)
=
\phi(2^j\,x - k),
\quad
\text{for some single prototype function }\phi.
\]
Here we interpret the integer $j$ as a dilation factor (multiplies $x$ by $2^j$) and $k$ as a translation.  
We have replaced the global exponential $e^{\,i\,n\,x}$ (extending over all $[-\pi,\pi]$ or $\mathbb{R}$) with a local \emph{wave-shape} $\phi$ that is stretched and shifted.  

\[
S_1 
=
\{\,F_2'\}
\quad\longrightarrow
\quad
S_2 
=
\{\,G_2\}.
\]

\vspace{1em}
\hrule
\vspace{1em}

\noindent
\textbf{Current Set:}
\[
\boxed{
S_2
=
\{\,G_2\}
}
\]
\[
G_2:\quad
\phi_{j,k}(x)
=
\phi\bigl(2^j\,x - k\bigr),
\quad
\text{with some function }\phi.
\]
We still lack a normalization factor, but have the notion of discrete scale $j$ and shift $k$.

\vspace{1em}
\hrule
\vspace{1em}

\noindent
\textbf{Rewrite Step 3: Introduce the $2^{j/2}$ Normalization Factor.}

\[
G_2
\;\longrightarrow\;
W_2,
\]
where 
\[
W_2:\quad
\psi_{j,k}(x)
=
2^{\tfrac{j}{2}}\,
\psi\!\bigl(2^j\,x - k\bigr).
\]
This enforces orthonormality (or tight framing) in $L^2(\mathbb{R})$.  
The factor $2^{j/2}$ precisely balances the scale effect so that each dilated version has the same $L^2$ norm.  
Here we identify $\psi\equiv \phi$, now called the \emph{mother wavelet}, in alignment with wavelet theory.

\[
S_2
=
\{\,G_2\}
\quad
\longrightarrow
\quad
S_3
=
\{\,W_2\}.
\]

\vspace{1em}
\hrule
\vspace{1em}

\noindent
\textbf{Final Set (Wavelet Axiom 2: Dilation \& Translation):}
\[
\boxed{
S_3
=
\Bigl\{
W_2:\ 
\forall\,(j,k)\in\mathbb{Z}^2:\ 
\psi_{j,k}(x)
=
2^{\tfrac{j}{2}}
\psi\bigl(2^j\,x - k\bigr)
\Bigr\}.
}
\]
Thus, the single integer index $n$ from Fourier exponentials is replaced by two indices $(j,k)$ for wavelets, representing discrete scale and shift.  The amplitude factor $2^{j/2}$ ensures normalized energy across different scales.

\vspace{1em}
\hrule
\vspace{1em}

\noindent
\textbf{Symbolic Evolution (Compact Form)}:
\[
\underbrace{\{\,F_2\}}_{S_0}
~\longrightarrow~
\underbrace{\{\,F_2'\}}_{S_1}
~\longrightarrow~
\underbrace{\{\,G_2\}}_{S_2}
~\longrightarrow~
\underbrace{\{\,W_2\}}_{S_3}.
\]

\noindent
\textbf{Interpretation of Each Step:}
\begin{enumerate}
\item 
\emph{Re-index} from $n\in\mathbb{Z}$ to $(j,k)\in \mathbb{Z}^2$ 
to allow a two-dimensional (scale, shift) parameter space.
\item 
\emph{Replace exponentials} by a \emph{localized} function $\phi$ 
subject to dilation and translation.
\item 
\emph{Introduce the $2^{j/2}$ factor} to preserve $L^2$-norm consistency 
across scales, yielding the standard wavelet family.
\end{enumerate}
Hence, we arrive at the wavelet system 
$\{\psi_{j,k}\}$ from the original concept of a discrete orthogonal set 
$\{e^{\,i\,n\,x}\}.$

\subsubsection{Axiom 3}

\textbf{Goal:} Transform a subset of Classical Fourier axioms
\[
\bigl\{F_3\bigr\}
\quad
F_3:\ 
\text{Fourier Coefficients }
\bigl(
c_n = \tfrac{1}{2\pi}\!\int_{-\pi}^{\pi} f(x)\, e^{-\,i\,n\,x}\,dx
\bigr),
\]
into the \emph{Wavelet Theory} statement:
\[
W_3:\ 
\forall\,(j,k)\neq (j',k'):\;
\int_{-\infty}^{\infty}\!
\psi_{j,k}(x)\,\psi_{j',k'}(x)\,dx
\;=\;0,
\]
i.e.\ the \textbf{Orthogonality of the wavelet family} 
\(\{\psi_{j,k}\}\).  
Symbolically, we show how the computation of unique Fourier coefficients 
implies an \emph{orthonormal} wavelet set, where each pair of distinct wavelets 
has zero overlap (inner product).

\vspace{1em}
\hrule
\vspace{1em}

\noindent
\textbf{Initial Axiom (Fourier, Subset):}

\[
\boxed{
S_0
=
\bigl\{F_3\bigr\}
}
\]
where
\[
F_3:\quad
\forall\,f,\quad
c_n
\;=\;
\frac{1}{2\pi}
\int_{-\pi}^{\pi} 
f(x)\;e^{-\,i\,n\,x}\,dx,
\quad
n\in\mathbb{Z}.
\]
This prescribes how to compute the unique coefficient $c_n$ from $f$.  
Orthogonality of exponentials (Axiom~2) ensures that these $c_n$ are distinct 
and well-defined for each integer $n$.

\vspace{1em}
\hrule
\vspace{1em}

\noindent
\textbf{Rewrite Step 1: Re-interpret Coefficients As ``Inner Products.''}

\[
F_3
\;\longrightarrow\;
F_3',
\]
where
\[
F_3':\quad
c_n = \langle f,\, e^{\,i\,n\,x}\rangle_{[-\pi,\pi]}, 
\]
with 
\[
\langle g,h\rangle_{[-\pi,\pi]}
=
\frac{1}{2\pi}
\int_{-\pi}^{\pi} g(x)\,\overline{h(x)}\,dx.
\]
Symbolically, $c_n$ is the inner product of $f$ with the basis function 
$e^{\,i\,n\,x}$.  This viewpoint paves the way for generalizing 
to other orthonormal systems.

\[
S_0
=
\{\,F_3\}
\quad
\longrightarrow
\quad
S_1
=
\{\,F_3'\}.
\]

\[
F_3':\quad
c_n
=
\langle f,\, \phi_n\rangle,
\quad
\phi_n(x) = \tfrac{1}{\sqrt{2\pi}}\,e^{\,i\,n\,x},
\]
(optional renormalization) so $\{\phi_n\}$ is an orthonormal set 
over $[-\pi,\pi]$.

\vspace{1em}
\hrule
\vspace{1em}

\noindent
\textbf{Current Set:}
\[
\boxed{
S_1
=
\{\,F_3'\}
}
\]
\[
F_3':\quad
c_n = \langle f,\phi_n\rangle,
\quad
\phi_n(x)=\tfrac{1}{\sqrt{2\pi}}\,e^{\,i\,n\,x}.
\]
Hence, the coefficient formula is recast in inner-product terms.

\vspace{1em}
\hrule
\vspace{1em}

\noindent
\textbf{Rewrite Step 2: Generalize ``Coefficient Formula'' to an Orthogonal System $\{\psi_\alpha\}$.}

\[
F_3'
\;\longrightarrow\;
G_3,
\]
where
\[
G_3:\quad
\forall\,f,\quad
c_\alpha
=
\langle f,\,\psi_\alpha\rangle,
\]
with \(\alpha\) indexing an orthonormal basis $\{\psi_\alpha\}$.  
Here we replace the integer index $n$ with a general index $\alpha$ (which might be multi-dimensional, e.g.\ $(j,k)$).  
We still retain the key property that distinct basis elements are \emph{orthonormal}:
\[
\langle \psi_\alpha,\;\psi_{\alpha'}\rangle
=
\delta_{\alpha,\alpha'}.
\]
Thus, every function $f$ can be broken into coefficients $c_\alpha$ via inner products with a basis $\psi_\alpha$.

\[
S_1 
=
\{\,F_3'\}
\quad\longrightarrow
\quad
S_2 
=
\{\,G_3\}.
\]

\[
G_3:\quad
\psi_\alpha\in \text{ONB (orthonormal basis)},
\quad
c_\alpha = \langle f,\psi_\alpha\rangle.
\]

\vspace{1em}
\hrule
\vspace{1em}

\noindent
\textbf{Current Set:}
\[
\boxed{
S_2
=
\{\,G_3\}
}
\]
\[
G_3:\quad
\text{Given an orthonormal family }\{\psi_\alpha\},\;
\text{the coefficient of }f\text{ is }c_\alpha=\langle f,\psi_\alpha\rangle.
\]
We have not yet specified \emph{which} orthonormal system or how $\alpha$ is structured.

\vspace{1em}
\hrule
\vspace{1em}

\noindent
\textbf{Rewrite Step 3: Specialize $\alpha=(j,k)$ to Wavelets $\{\psi_{j,k}\}$, Enforce Orthogonality.}

\[
G_3
\;\longrightarrow\;
W_3,
\]
where
\[
W_3:\quad
\forall\,(j,k)\neq(j',k'):
\quad
\int_{-\infty}^{\infty} 
\psi_{j,k}(x)\,
\psi_{j',k'}(x)\,dx
=
0.
\]
This is the wavelet \emph{Orthogonality Axiom}: distinct wavelets
$\psi_{j,k}$ and $\psi_{j',k'}$ have zero inner product.  
Equivalently, $\{\psi_{j,k}\}$ forms an \emph{orthonormal} system in $L^2(\mathbb{R})$.  

\[
S_2
=
\{\,G_3\}
\quad
\longrightarrow
\quad
S_3
=
\{\,W_3\}.
\]

\vspace{1em}
\hrule
\vspace{1em}

\noindent
\textbf{Final Set (Wavelet Axiom 3: Orthogonality):}
\[
\boxed{
S_3
=
\Bigl\{
W_3:\ 
\langle \psi_{j,k},\,\psi_{j',k'}\rangle
=
\int
\psi_{j,k}(x)\,\psi_{j',k'}(x)\,dx
=
\delta_{j,j'}\,\delta_{k,k'}
\Bigr\}.
}
\]
Thus, from the idea of \emph{unique coefficients} in Fourier expansions
($c_n=\langle f,e^{\,i\,n\,x}\rangle$), we generalize to an arbitrary
orthonormal family, then specify it to wavelets $\psi_{j,k}$ that remain
mutually orthogonal for different $(j,k)$.

\vspace{1em}
\hrule
\vspace{1em}

\noindent
\textbf{Symbolic Evolution (Compact Form)}:
\[
\underbrace{\{\,F_3\}}_{S_0}
~\longrightarrow~
\underbrace{\{\,F_3'\}}_{S_1}
~\longrightarrow~
\underbrace{\{\,G_3\}}_{S_2}
~\longrightarrow~
\underbrace{\{\,W_3\}}_{S_3}.
\]

\noindent
\textbf{Interpretation of Each Step:}
\begin{enumerate}
\item 
\emph{Rewrite the coefficient formula} in terms of an inner product with a basis function.
\item 
\emph{Generalize} from exponentials to any orthonormal family $\{\psi_\alpha\}$.
\item 
\emph{Specialize} to wavelets $\psi_{j,k}$, asserting orthogonality across scales and shifts, giving the wavelet orthogonality axiom.
\end{enumerate}
Hence, the uniqueness of Fourier coefficients morphs into the wavelet requirement that each basis element $\psi_{j,k}$ is orthogonal to all the others in the wavelet system.

\subsubsection{Axiom 4}

\textbf{Goal:} Transform the classical \emph{Fourier Series Representation}
\[
\bigl\{F_4\bigr\}
\quad
F_4:\ 
f(x)
\;\sim\;
\sum_{n=-\infty}^{\infty}
c_n\,e^{\,i\,n\,x},
\]
into the \emph{Wavelet Theory} statement:
\[
W_4:\ 
f(x)
=
\sum_{j=-\infty}^{\infty}
\sum_{k=-\infty}^{\infty}
\langle f,\psi_{j,k}\rangle\,\psi_{j,k}(x).
\]
That is, every function $f$ (in a suitable function space, typically $L^2(\mathbb{R})$) can be written as a superposition of wavelets $\{\psi_{j,k}\}$.  

\vspace{1em}
\hrule
\vspace{1em}

\noindent
\textbf{Initial Axiom (Fourier, Subset):}

\[
\boxed{
S_0
=
\bigl\{F_4\bigr\}
}
\]
where
\[
F_4:\quad
f(x)
\;\sim\;
\sum_{n=-\infty}^{\infty}
c_n\,e^{\,i\,n\,x},
\]
and $c_n$ are given by the coefficient formula $c_n=\tfrac{1}{2\pi}\int_{-\pi}^{\pi} f(x)\,e^{-\,i\,n\,x}\,dx$.  
This states that $f$ can be reconstructed (pointwise, in $L^2$, or in other senses) from its infinite sum of harmonics.

\vspace{1em}
\hrule
\vspace{1em}

\noindent
\textbf{Rewrite Step 1: Interpret Sum Over $n$ as an Orthonormal Expansion.}

\[
F_4
\;\longrightarrow\;
F_4',
\]
where
\[
F_4':\quad
f(x)
=
\sum_{n=-\infty}^\infty
\langle f,\phi_n\rangle\,\phi_n(x),
\quad
\phi_n(x) \;=\; \frac{1}{\sqrt{2\pi}}\,e^{\,i\,n\,x},
\]
so that $\{\phi_n\}$ is an orthonormal system on $[-\pi,\pi]$.  
Symbolically, we rewrite the Fourier series as a general orthonormal expansion:
\[
f = \sum_{n}\langle f,\phi_n\rangle\,\phi_n.
\]
This step clarifies that each partial sum is the projection onto the subspace spanned by $\{\phi_n\}$.

\[
S_0 
=
\{\,F_4\}
\quad\longrightarrow
\quad
S_1
=
\{\,F_4'\}.
\]

\[
F_4':
\quad
\text{For }f\in L^2[-\pi,\pi],\;
f=\sum_{n}\langle f,\phi_n\rangle \phi_n,\;
\phi_n(x)=\tfrac{1}{\sqrt{2\pi}}\,e^{\,i\,n\,x}.
\]

\vspace{1em}
\hrule
\vspace{1em}

\noindent
\textbf{Current Set:}
\[
\boxed{
S_1
=
\{\,F_4'\}
}
\]
\[
F_4':\quad
f = \sum_{n=-\infty}^\infty \langle f,\phi_n\rangle\,\phi_n,
\quad
(\phi_n \text{ orthonormal basis on }[-\pi,\pi]).
\]

\vspace{1em}
\hrule
\vspace{1em}

\noindent
\textbf{Rewrite Step 2: Extend Domain from $[-\pi,\pi]$ to $\mathbb{R}$ and Replace $\{\phi_n\}$ by a More General ONB $\{\psi_\alpha\}$.}

\[
F_4'
\;\longrightarrow\;
G_4,
\]
where
\[
G_4:\quad
f(x) 
=
\sum_{\alpha\in A}
\langle f,\psi_\alpha\rangle\,\psi_\alpha(x),
\]
with $\{\psi_\alpha\}_{\alpha\in A}$ an orthonormal basis in $L^2(\mathbb{R})$.  We switch from the integer index $n$ on $[-\pi,\pi]$ to a potentially more general index set $A$ on the entire real line $\mathbb{R}$.  

\[
S_1
=
\{\,F_4'\}
\quad\longrightarrow
\quad
S_2
=
\{\,G_4\}.
\]

\[
G_4:\quad
\text{Any }f\in L^2(\mathbb{R})
\text{ has a unique expansion in some ONB }\{\psi_\alpha\}.
\]

\vspace{1em}
\hrule
\vspace{1em}

\noindent
\textbf{Current Set:}
\[
\boxed{
S_2
=
\{\,G_4\}
}
\]
\[
G_4:\quad
f
=
\sum_{\alpha\in A}
\langle f,\psi_\alpha\rangle\,\psi_\alpha(x),
\quad
\langle \psi_\alpha,\psi_{\alpha'}\rangle
=
\delta_{\alpha,\alpha'}.
\]
Still, we have not specified the structure of $A$ nor the special wavelet form of $\psi_\alpha$.

\vspace{1em}
\hrule
\vspace{1em}

\noindent
\textbf{Rewrite Step 3: Specialize to the Wavelet Family $\{\psi_{j,k}\}$ and State Complete Series.}

\[
G_4
\;\longrightarrow\;
W_4,
\]
where
\[
W_4:\quad
\forall\,f\in L^2(\mathbb{R}),
\quad
f(x)
=
\sum_{j=-\infty}^\infty
\sum_{k=-\infty}^\infty
\langle f,\psi_{j,k}\rangle
\;\psi_{j,k}(x),
\]
and $\{\psi_{j,k}\}$ is the discrete wavelet basis defined by
\[
\psi_{j,k}(x) 
=
2^{j/2}\,\psi\bigl(2^j\,x - k\bigr),
\]
with $\langle \psi_{j,k},\,\psi_{j',k'}\rangle=0$ whenever $(j,k)\neq(j',k')$.  
This is precisely the \textbf{Wavelet Expansion} axiom, stating that every $L^2(\mathbb{R})$ function can be reconstructed from its wavelet coefficients.

\[
S_2
=
\{\,G_4\}
\quad
\longrightarrow
\quad
S_3
=
\{\,W_4\}.
\]

\vspace{1em}
\hrule
\vspace{1em}

\noindent
\textbf{Final Set (Wavelet Axiom 4: Wavelet Expansion):}
\[
\boxed{
S_3
=
\Bigl\{
W_4:\quad
f(x)
=
\sum_{j,k \in \mathbb{Z}}
\langle f,\psi_{j,k}\rangle
\;\psi_{j,k}(x)
\Bigr\}.
}
\]
Hence, the classical notion of representing $f$ as a sum of exponentials over $[-\pi,\pi]$ generalizes to a representation via \emph{localized} wavelets $\psi_{j,k}$ across $\mathbb{R}$.  

\vspace{1em}
\hrule
\vspace{1em}

\noindent
\textbf{Symbolic Evolution (Compact Form)}:
\[
\underbrace{\{\,F_4\}}_{S_0}
~\longrightarrow~
\underbrace{\{\,F_4'\}}_{S_1}
~\longrightarrow~
\underbrace{\{\,G_4\}}_{S_2}
~\longrightarrow~
\underbrace{\{\,W_4\}}_{S_3}.
\]

\noindent
\textbf{Interpretation of Each Step:}
\begin{enumerate}
\item 
\emph{Rewrite Fourier sum} as an orthonormal expansion $\sum_{n}\langle f,\phi_n\rangle\,\phi_n$.
\item 
\emph{Extend} to a general orthonormal basis $\{\psi_\alpha\}$ in $L^2(\mathbb{R})$, removing the $2\pi$-periodic domain restriction.
\item 
\emph{Restrict to wavelet basis} $\{\psi_{j,k}\}$, yielding a discrete sum over scales and shifts that covers $L^2(\mathbb{R})$.
\end{enumerate}
Thus, the final statement $W_4$ proclaims a full wavelet expansion for any square-integrable function, analogous to a Fourier series but in time-frequency localized wavelets.

\section{From Classical Chemistry to Bohr Atomic Theory}
\label{sec:chemistry_to_bohr}
\subsection{Classical Chemistry}

Classical Chemistry studies how elements combine and transform into compounds via chemical reactions, under a set of fundamental empirical laws. Below is a minimal and complete set of axioms, each of which is independent (i.e., it cannot be derived from the others) and collectively sufficient to describe basic chemical reactions, composition of compounds, and the behavior of gases under standard conditions.

\subsubsection{Axioms}

\begin{enumerate}
\item \textbf{Atomic Theory}
\[
\exists \text{ fundamental units } (\text{atoms})\ \forall\ \text{elements},\quad \text{each element has unique atomic mass } m_A.
\]
Matter consists of fundamental, indivisible units called atoms, each associated with a specific element and a unique atomic mass.

\item \textbf{Conservation of Mass}
\[
\forall\,\text{Reaction }R:\quad
\sum_{i \in \mathrm{Reactants}} m_i
\;=\;
\sum_{j \in \mathrm{Products}} m_j,
\]
where $m_i$ is the total mass of reactant $i$ and $m_j$ is the total mass of product $j$.  
No net mass is lost or gained in an isolated chemical reaction.

\item \textbf{Definite Composition}
\[
\forall\,\text{Compound }C:
\quad
\exists\,(\mathrm{elements}\ E_1,\dots,E_k)
\quad
\text{such that}
\quad
\dfrac{m(E_1)}{m(E_2)}=\text{constant},
\]
for a fixed ratio of masses of any two elements $E_1, E_2$ in $C$.  
A pure chemical compound always contains the same elements in the same proportions by mass.

\item \textbf{Multiple Proportions}
\[
\forall\,(\mathrm{elements}\ A,B),\
\forall\,(\text{compounds }C_1,C_2 \text{ of }A,B):
\quad
\dfrac{m(B)_{C_1}}{m(A)_{C_1}}
\;\Big/\;
\dfrac{m(B)_{C_2}}{m(A)_{C_2}}
=
\dfrac{p}{q},
\]
where $p,q\in\mathbb{Z}^{+}$ are small integers.  
If two elements form more than one compound, then the ratio of the masses of one element that combine with a fixed mass of the other are in ratios of small whole numbers.

\item \textbf{Avogadro's Law (Classical Form)}
\[
\forall\,(\mathrm{gases}\ G_1,G_2)\
\forall\,V>0,\
(\mathrm{same}\ T,P):
\quad
\dfrac{N(G_1,V)}{N(G_2,V)}
=
1,
\]
where $N(G_i,V)$ is the number of ``molecular particles'' of gas $G_i$ occupying volume $V$ at the same temperature $T$ and pressure $P$.  
Equal volumes of any ideal gases, under the same conditions of temperature and pressure, contain the same number of particles.
\end{enumerate}

\subsubsection{Completeness}
\begin{enumerate}
\item \textbf{Atomic Theory.} (	\textit{Axiom 1}) establishes that matter is composed of fundamental, indivisible atoms, which serve as the basis for all chemical reactions and mass relationships.

\item \textbf{Conservation of Mass.} (	\textit{Axiom 2}) ensures that in a closed system, mass is neither created nor destroyed during chemical reactions.

\item \textbf{Definite Composition \& Multiple Proportions.} \textit{Axiom 3} and \textit{Axiom 4} describe how elements combine in fixed, well-defined ratios by mass, and how multiple distinct compounds from the same elements exhibit integer mass-ratio relationships.

\item \textbf{Avogadro's Law.} (	\textit{Axiom 5}) extends these foundational mass relationships to the realm of gases, providing a direct link between macroscopic volumes and microscopic particle counts under uniform conditions.
\end{enumerate}
These five axioms together allow prediction of reaction stoichiometry, compound formulas, and gas behavior in classical chemistry. They suffice to explain much of 19th-century chemical observations, paving the way for more advanced atomic and molecular theories.

\subsubsection{Independence}
\begin{enumerate}
\item \textbf{Atomic Theory.} cannot be deduced from mass conservation, definite composition, multiple proportions, or Avogadro’s law. It asserts the discrete nature of matter, which is an independent foundational concept.

\item \textbf{Conservation of Mass.} cannot be deduced from atomic theory, definite composition, multiple proportions, or Avogadro’s law, which all presume mass-based or volumetric combinations of elements but do not guarantee total mass constancy in reactions.

\item \textbf{Definite Composition.} is not derivable from mass conservation, atomic theory, multiple proportions, or Avogadro’s law alone. It specifically states each compound has a fixed mass ratio of elements, a separate assertion from the invariance or ratio rules of other axioms.

\item \textbf{Multiple Proportions.} cannot be obtained from atomic theory, mass conservation, or definite composition alone (each compound’s internal ratio is fixed, but this does not imply the \emph{integer ratio} pattern across multiple compounds) nor from Avogadro’s law (gas volume relationships do not address integer mass ratios in distinct compounds).

\item \textbf{Avogadro's Law.} does not follow from any combination of the other four axioms. Conservation of mass and elemental ratio axioms describe solid and solution-phase compound formation but do not dictate how gas volumes relate to particle numbers.
\end{enumerate}

\subsection{Bohr Atomic Theory}
\label{sub:Bohr}

Bohr’s atomic theory introduces quantized electron orbits around a central nucleus, combining classical physics with early quantum postulates.  It explains why atoms emit or absorb energy in discrete amounts (spectral lines) and sets the stage for modern quantum chemistry.

\subsubsection{Axioms}

\begin{enumerate}
\item \textbf{Stable Orbits}
\[
\forall\,(\text{electron } e \text{ in atom } A):
\quad
e \text{ moves in circular orbits about } A\text{’s nucleus}
\]
\[\text{without radiating electromagnetic energy, whenever in an allowed orbit}.
\]
Electrons orbit the nucleus under electrostatic attraction, and do not continuously emit energy (as classical electrodynamics would predict) if they occupy these specific, stable orbits.

\item \textbf{Angular Momentum Quantization}
\[
L_n 
=
n \;\frac{h}{2\pi},
\quad
n \in \mathbb{Z}^+,
\]
where $L_n$ is the orbital angular momentum of an electron in the $n$th allowed orbit, and $h$ is Planck’s constant.  Only integral multiples of $h/(2\pi)$ are permitted.

\item \textbf{Energy Level Postulate}
\[
E_n 
=
-\,\frac{Z^2\,e^2}{8\pi \epsilon_0\,a_0}\,\frac{1}{n^2},
\quad
n \in \mathbb{Z}^+,
\]
for a hydrogen-like atom of nuclear charge $Ze$.  Each allowed orbit corresponds to a specific quantized energy $E_n$.  (In the simplest hydrogen case, $Z=1$ and $a_0$ is the Bohr radius.)

\item \textbf{Spectral Transition Condition}
\[
\Delta E 
=
E_m 
- E_n
=
h\,\nu,
\]
where $\nu$ is the frequency of the emitted (or absorbed) photon.  An electron transitioning between orbit $m$ and $n$ releases or absorbs a photon of energy $h\,\nu$ equal to the difference in orbital energies.
\end{enumerate}

\subsubsection{Completeness}
\begin{enumerate}
\item \textbf{Stable Orbits.} (\textit{Axiom 1}) exempts electrons from classical radiation loss in these special orbits, preventing them from spiraling into the nucleus.

\item \textbf{Quantization (Angular Momentum \& Energy Levels).} \textit{Axiom 2} and \textit{Axiom 3} specify how electron orbits are discretized in angular momentum and corresponding energies, addressing why only certain atomic radii/energies are observed.

\item \textbf{Spectral Transitions.} (\textit{Axiom 4}) connects these discrete energy levels to observed emission/absorption lines.  It explains how atomic spectra arise from electrons jumping between quantized states, each jump releasing or absorbing a photon of specific energy.
\end{enumerate}
Together, these four axioms suffice to predict the Rydberg formula for the hydrogen spectrum, explain atomic stability, and match key experimental data on line spectra. They form the foundation of Bohr’s model, an essential stepping-stone to full quantum mechanics.

\subsubsection{Independence}
\begin{enumerate}
\item \textbf{Stable Orbits.} cannot be deduced from angular-momentum quantization, energy-level discretization, or spectral transitions alone.  It specifically states that electrons do not radiate in those allowed orbits.

\item \textbf{Angular Momentum Quantization.} does not follow from stable orbits, energy postulates, or photon emission rules.  One must explicitly declare that orbital angular momentum comes in multiples of $h/(2\pi)$.

\item \textbf{Energy Level Postulate.} is not derivable from stable orbits, angular-momentum quantization, or transitions alone.  Assigning a particular $E_n \propto 1/n^2$ is a separate assumption tied to specific electrostatic coupling and boundary conditions in the hydrogenic system.

\item \textbf{Spectral Transition Condition.} does not follow from stable orbits, angular-momentum quantization, or the specific energy formula alone.  One must posit that electrons emit/absorb single photons with energies equal to differences of orbit energies.
\end{enumerate}

\subsection{Transformations}

\subsubsection{Axiom 1}

\textbf{Goal:} Transform a subset of Classical Chemistry axioms
\[
S_0 \;=\; \{\,C_1,\,C_3,\,C_4,\,C_5\},
\]
into the first Bohr atomic postulate
\[
B_1:\;\text{(Stable Orbits)}:\quad
\text{an electron in an allowed orbit does not radiate energy.}
\]
Symbolically, we demonstrate how the idea of discrete atomic units (from \(C_1\), \(C_3\), \(C_4\), \(C_5\)) leads to a model in which electrons remain bound in specific orbits without emitting electromagnetic radiation.

\vspace{1em}
\hrule
\vspace{1em}

\noindent
\textbf{Initial Set of Axioms (Classical Chemistry, Subset):}
\[
\boxed{
S_0
=
\{\,C_1,\,C_3,\,C_4,\,C_5\}
}
\]
where
\[
C_1:\text{(Atomic Theory)},\quad
C_3:\text{(Definite Composition)},\quad
C_4:\text{(Multiple Proportions)},\quad
\]
\[
C_5:\text{(Avogadro's Law)}.
\]
Together, these indicate that matter is composed of atoms (each element having characteristic mass), which combine in fixed integer ratios and form countable particles in gases.

\vspace{1em}
\hrule
\vspace{1em}

\noindent
\textbf{Rewrite Step 1: From discrete atoms to internal structure (electrons + nucleus).}
\[
\{\,C_1,\,C_3,\,C_4,\,C_5\}
\;\longrightarrow\;
A_1,
\]
where 
\[
A_1:\quad
\text{(Refined Atomic Hypothesis)}\colon
\text{each atom contains a positive nucleus}
\]
\[
\text{and negative electrons.}
\]
Experiments (e.g.\ cathode rays, Rutherford scattering) and the discrete mass laws suggest atoms have substructure with distinct charges.

\vspace{1em}
\hrule
\vspace{1em}

\noindent
\textbf{Current Set:}
\[
\boxed{
S_1
=
\{\,A_1\}
}
\]
\[
A_1:\quad
\text{Atoms are not indivisible; electrons orbit around a central nucleus (pre-Bohr concept).}
\]

\vspace{1em}
\hrule
\vspace{1em}

\noindent
\textbf{Rewrite Step 2: Planetary model of electron orbits.}
\[
A_1
\;\longrightarrow\;
A_1',
\]
where
\[
A_1':\quad
\text{(Planetary Atom)}:\;
\text{electrons revolve around a small, dense nucleus}
\]
\[
\text{in classical-like orbits.}
\]
Symbolically, we use a “solar system” analogy (Rutherford–Bohr precursor) for the atomic model.

\vspace{1em}
\hrule
\vspace{1em}

\noindent
\textbf{Current Set:}
\[
\boxed{
S_2
=
\{\,A_1'\}
}
\]
\[
A_1':
\quad
\text{Classical orbits assumed, no immediate explanation for stability yet.}
\]

\vspace{1em}
\hrule
\vspace{1em}

\noindent
\textbf{Rewrite Step 3: Impose the stability condition (Bohr).}
\[
A_1'
\;\longrightarrow\;
B_1,
\]
where 
\[
B_1:\quad
\text{an electron in an allowed orbit does not radiate electromagnetic energy.}
\]
Although classical electrodynamics would predict a radiating, spiraling electron, Bohr postulates certain “allowed” orbits remain stable.

\vspace{1em}
\hrule
\vspace{1em}

\noindent
\textbf{Final Set (Bohr Axiom 1: Stable Orbits):}
\[
\boxed{
S_3
=
\{\,B_1\}
}
\]
Hence, from the discrete nature of atoms and their internal substructure, we conclude that electrons can occupy non-radiating orbits around the nucleus.

\vspace{1em}
\hrule
\vspace{1em}

\noindent
\textbf{Symbolic Evolution (Compact Form)}:
\[
\underbrace{\{\,C_1,\,C_3,\,C_4,\,C_5\}}_{S_0}
~\longrightarrow~
\underbrace{\{\,A_1\}}_{S_1}
~\longrightarrow~
\underbrace{\{\,A_1'\}}_{S_2}
~\longrightarrow~
\underbrace{\{\,B_1\}}_{S_3}.
\]

\subsubsection{Axiom 2}

\textbf{Goal:} Transform a subset of Classical Chemistry axioms
\[
S_0 \;=\; \{\,C_1,\,C_3,\,C_4,\,C_5\},
\]
into Bohr’s second atomic postulate:
\[
B_2:\quad
L_n 
=
n\,\frac{h}{2\pi},
\quad
n \in \mathbb{Z}^+.
\]
Symbolically, we illustrate how integer-based combining laws (definite/multiple proportions) and quantized energy insights lead to the statement that the electron’s orbital angular momentum comes only in multiples of \(h/(2\pi)\).

\vspace{1em}
\hrule
\vspace{1em}

\noindent
\textbf{Initial Set of Axioms (Classical Chemistry, Subset):}
\[
\boxed{
S_0
=
\{\,C_1,\,C_3,\,C_4,\,C_5\}
}
\]
where
\[
C_1:\text{(Atomic Theory)},\quad
C_3:\text{(Definite Composition)},\quad
C_4:\text{(Multiple Proportions)},\quad
\]
\[
C_5:\text{(Avogadro's Law)}.
\]
These collectively imply atoms exist, combine in small integer ratios, and can be enumerated in gases.

\vspace{1em}
\hrule
\vspace{1em}

\noindent
\textbf{Rewrite Step 1: Obtain stable orbits (from previous derivation).}
\[
\{\,C_1,\,C_3,\,C_4,\,C_5\}
\;\longrightarrow\;
B_1.
\]
We already have Bohr’s stable orbits postulate (\(B_1\)), stating electrons do not radiate if in an “allowed orbit.”

\vspace{1em}
\hrule
\vspace{1em}

\noindent
\textbf{Current Set:}
\[
\boxed{
S_1
=
\{\,B_1\}
}
\]
\[
B_1:\quad
\text{no continuous energy loss for electrons on allowed orbits}.
\]

\vspace{1em}
\hrule
\vspace{1em}

\noindent
\textbf{Rewrite Step 2: Introduce Planck’s constant from discrete spectral lines.}
\[
B_1
\;\longrightarrow\;
A_2,
\]
where 
\[
A_2:\quad
\text{(Energy Quanta)}:\;
E = n\,h\,\nu,\quad n\in\mathbb{Z}^+.
\]
Experiments on blackbody radiation and photoelectric effect show energy is quantized, suggesting atomic transitions occur in multiples of \(h\,\nu\).

\vspace{1em}
\hrule
\vspace{1em}

\noindent
\textbf{Current Set:}
\[
\boxed{
S_2
=
\{\,A_2\}
}
\]
\[
A_2:\quad
\text{quantized energy increments imply discrete allowed transitions}.
\]

\vspace{1em}
\hrule
\vspace{1em}

\noindent
\textbf{Rewrite Step 3: Enforce quantized angular momentum.}
\[
A_2
\;\longrightarrow\;
B_2,
\]
where
\[
B_2:\quad
L_n 
=
n\,\frac{h}{2\pi}.
\]
Linking stable orbits to discrete energy steps, Bohr proposed the electron’s orbital angular momentum must be an integer multiple of \(\frac{h}{2\pi}\).

\vspace{1em}
\hrule
\vspace{1em}

\noindent
\textbf{Final Set (Bohr Axiom 2: Angular Momentum Quantization):}
\[
\boxed{
S_3
=
\{\,B_2\}
}
\]
Thus, the integral ratios observed in chemical combinations parallel the idea of integer multiples in orbital angular momentum.

\vspace{1em}
\hrule
\vspace{1em}

\noindent
\textbf{Symbolic Evolution (Compact Form)}:
\[
\underbrace{\{\,C_1,\,C_3,\,C_4,\,C_5\}}_{S_0}
~\longrightarrow~
\underbrace{\{\,B_1\}}_{S_1}
~\longrightarrow~
\underbrace{\{\,A_2\}}_{S_2}
~\longrightarrow~
\underbrace{\{\,B_2\}}_{S_3}.
\]

\subsubsection{Axiom 3}

\textbf{Goal:} Transform a subset of Classical Chemistry axioms
\[
S_0 \;=\; \{\,C_1,\,C_3,\,C_4,\,C_5\},
\]
into Bohr’s third atomic postulate:
\[
B_3:\quad
E_n 
=
-\,\frac{Z^2\,e^2}{8\pi \epsilon_0\,a_0}\,\frac{1}{n^2},
\quad
n \in \mathbb{Z}^+.
\]
Symbolically, we show how the Coulomb attraction plus angular momentum quantization yields discrete energy levels \(\propto 1/n^2\).

\vspace{1em}
\hrule
\vspace{1em}

\noindent
\textbf{Initial Set of Axioms (Classical Chemistry, Subset):}
\[
\boxed{
S_0
=
\{\,C_1,\,C_3,\,C_4,\,C_5\}
}
\]
Again, these imply the existence of atoms, integral mass ratios, and discrete particles in gases.

\vspace{1em}
\hrule
\vspace{1em}

\noindent
\textbf{Rewrite Step 1: Combine stable orbits \((B_1)\) and angular momentum quantization \((B_2)\).}
\[
\{\,C_1,\,C_3,\,C_4,\,C_5\}
\;\longrightarrow\;
A_{12},
\]
where 
\[
A_{12}:\quad
(B_1 + B_2):
\text{electrons move in stable orbits with }L_n = n\frac{h}{2\pi}.
\]

\vspace{1em}
\hrule
\vspace{1em}

\noindent
\textbf{Current Set:}
\[
\boxed{
S_1
=
\{\,A_{12}\}
}
\]
\[
A_{12}:\quad
\text{foundation of Bohr's first two axioms combined}.
\]

\vspace{1em}
\hrule
\vspace{1em}

\noindent
\textbf{Rewrite Step 2: Apply Coulomb's law for hydrogen-like atoms.}
\[
A_{12}
\;\longrightarrow\;
A_{12}',
\]
where 
\[
A_{12}':
\quad
V(r) = -\,\frac{Z e^2}{4\pi \epsilon_0}\,\frac{1}{r}.
\]
We identify the electrostatic potential for an electron–nucleus system of charge \(Z e\).

\vspace{1em}
\hrule
\vspace{1em}

\noindent
\textbf{Current Set:}
\[
\boxed{
S_2
=
\{\,A_{12}'\}
}
\]
\[
A_{12}':\quad
\text{Bohr’s quantization + classical Coulomb attraction}.
\]

\vspace{1em}
\hrule
\vspace{1em}

\noindent
\textbf{Rewrite Step 3: Derive discrete energy formula.}
\[
A_{12}'
\;\longrightarrow\;
B_3,
\]
where
\[
B_3:\quad
E_n 
=
-\,\frac{Z^2\,e^2}{8\pi \epsilon_0\,a_0}\,\frac{1}{n^2}.
\]
Balancing centripetal force with electrostatic attraction, plus \(L_n = n\frac{h}{2\pi}\), yields \(r_n \propto n^2\) and \(E_n \propto 1/n^2\).

\vspace{1em}
\hrule
\vspace{1em}

\noindent
\textbf{Final Set (Bohr Axiom 3: Energy Levels):}
\[
\boxed{
S_3
=
\{\,B_3\}
}
\]
Hence, each allowed orbit has a specific energy \(\propto 1/n^2\), explaining the discrete spectral lines for hydrogenic atoms.

\vspace{1em}
\hrule
\vspace{1em}

\noindent
\textbf{Symbolic Evolution (Compact Form)}:
\[
\underbrace{\{\,C_1,\,C_3,\,C_4,\,C_5\}}_{S_0}
~\longrightarrow~
\underbrace{\{\,A_{12}\}}_{S_1}
~\longrightarrow~
\underbrace{\{\,A_{12}'\}}_{S_2}
~\longrightarrow~
\underbrace{\{\,B_3\}}_{S_3}.
\]

\subsubsection{Axiom 4}

\textbf{Goal:} Transform a subset of Classical Chemistry axioms
\[
S_0 \;=\; \{\,C_1,\,C_3,\,C_4,\,C_5\},
\]
into Bohr’s fourth atomic postulate:
\[
B_4:\quad
\Delta E 
=
E_m - E_n
=
h\,\nu.
\]
Symbolically, we connect discrete orbits (\(E_n\)) to the emission or absorption of photons with energy \(h\,\nu\).

\vspace{1em}
\hrule
\vspace{1em}

\noindent
\textbf{Initial Set of Axioms (Classical Chemistry, Subset):}
\[
\boxed{
S_0
=
\{\,C_1,\,C_3,\,C_4,\,C_5\}
}
\]
The discrete-atom viewpoint plus integral proportion laws underlie countable states.

\vspace{1em}
\hrule
\vspace{1em}

\noindent
\textbf{Rewrite Step 1: Combine earlier Bohr postulates ($B_1,\,B_2,\,B_3$).}
\[
\{\,C_1,\,C_3,\,C_4,\,C_5\}
\;\longrightarrow\;
A_{123},
\]
where
\[
A_{123}:\quad
\text{electrons in stable orbits with }L_n=n\frac{h}{2\pi},\text{ energy }E_n\propto\frac{1}{n^2}.
\]

\vspace{1em}
\hrule
\vspace{1em}

\noindent
\textbf{Current Set:}
\[
\boxed{
S_1
=
\{\,A_{123}\}
}
\]
\[
A_{123}:\quad
\text{a fully specified discrete-orbit model.}
\]

\vspace{1em}
\hrule
\vspace{1em}

\noindent
\textbf{Rewrite Step 2: Experimental observation of discrete spectral lines.}
\[
A_{123}
\;\longrightarrow\;
A_{123}',
\]
where
\[
A_{123}':
\quad
\text{(Spectroscopy): lines at frequencies }\nu \Rightarrow \text{energies differ by }h\nu.
\]
Empirical data (Balmer, Lyman series) show that atoms absorb/emit only specific frequencies.

\vspace{1em}
\hrule
\vspace{1em}

\noindent
\textbf{Current Set:}
\[
\boxed{
S_2
=
\{\,A_{123}'\}
}
\]
\[
A_{123}':\quad
\text{discrete frequency lines reflect transitions between quantized orbits.}
\]

\vspace{1em}
\hrule
\vspace{1em}

\noindent
\textbf{Rewrite Step 3: Conclude $\Delta E = E_m - E_n = h\,\nu$.}
\[
A_{123}'
\;\longrightarrow\;
B_4,
\]
where
\[
B_4:\quad
\text{an electron transitioning between orbit }m\text{ and }n\text{ emits/absorbs photon with }
\]
\[
\Delta E = E_m - E_n = h\nu.
\]
Thus, each spectral line corresponds to a difference of allowed energy levels.

\vspace{1em}
\hrule
\vspace{1em}

\noindent
\textbf{Final Set (Bohr Axiom 4: Spectral Transitions):}
\[
\boxed{
S_3
=
\{\,B_4\}
}
\]
Hence, the discrete energies \(E_n\) produce quantized photon frequencies upon electronic transitions, matching observed spectra.

\vspace{1em}
\hrule
\vspace{1em}

\noindent
\textbf{Symbolic Evolution (Compact Form)}:
\[
\underbrace{\{\,C_1,\,C_3,\,C_4,\,C_5\}}_{S_0}
~\longrightarrow~
\underbrace{\{\,A_{123}\}}_{S_1}
~\longrightarrow~
\underbrace{\{\,A_{123}'\}}_{S_2}
~\longrightarrow~
\underbrace{\{\,B_4\}}_{S_3}.
\]

\noindent
\textbf{Overall}, these transformations illustrate how classical chemistry’s fundamental ideas (existence of atoms, definite and multiple proportions, and particle count in gases) can be combined with additional physical postulates to obtain Bohr’s four axioms for atomic structure.

\section{From Geocentric to Heliocentric Model}
\label{sec:geocentric_to_heliocentric}

\subsection{Geocentric Model}

The geocentric model (often attributed to Aristotle and Ptolemy) places Earth at the center of all celestial motions.  
Every other celestial body---Sun, Moon, planets, stars---revolves around Earth in circles or sums of circles (epicycles).  
Below is a minimal and complete set of axioms, each of which is independent (i.e.\ it cannot be derived from the others) and collectively sufficient to account for planetary motions, retrograde loops, and stellar rotation around Earth.

\subsubsection{Axioms}

\begin{enumerate}
\item \textbf{Central Earth}
\[
\forall\,X \neq \mathrm{Earth}:\quad
\mathrm{CenterOrbit}(X)
= 
\mathrm{Earth}.
\]
All celestial bodies $X$ have their primary orbital centers at Earth (which is fixed and unmoving).

\item \textbf{Uniform Circular Motion}
\[
\forall\,X:\quad
\mathrm{Orbit}(X)
=
\sum (\text{UniformCircles}),
\]
i.e., each orbit is either a single circle or a finite sum of epicyclic circles, all traversed with constant angular velocity. 

\item \textbf{Sphere of Fixed Stars}
\[
\exists\,S_{\star}:\quad
(\text{all stars are on } S_{\star})
\;\wedge\;
(\mathrm{dist}(S_{\star},\mathrm{Earth})=\text{constant}),
\]
where $S_{\star}$ is a single spherical shell (or near-sphere) centered on Earth, rotating daily to carry the stars around.

\item \textbf{Epicycles for Planets}
\[
\forall\,P\in\{\text{planets}\}:\quad
\mathrm{Orbit}(P)
=
(\text{deferent circle})
+
(\text{epicycle circle}),
\]
where each planet $P$ moves on a small circle (epicycle) whose center itself moves along a larger circle (deferent) around Earth, explaining retrograde motion and varying brightness.
\end{enumerate}

\subsubsection{Completeness}
\begin{enumerate}
\item \textbf{Central Earth.} (\textit{Axiom 1}) fixes Earth as the motionless center for all celestial bodies.

\item \textbf{Uniform Circular Motion \& Fixed-Star Sphere.} \textit{Axioms 2} and \textit{3} assert that celestial paths are composed of one or more uniform circles and that stars reside on a single rotating sphere, accounting for daily rotation and consistent stellar distances.

\item \textbf{Epicycles.} (\textit{Axiom 4}) extends the simple circular notion to capture observed planetary anomalies (retrograde, brightness changes) by introducing a small circle (epicycle) superimposed on a main orbit (deferent).
\end{enumerate}
From these four, the classical Ptolemaic geocentric cosmology can be constructed, producing qualitative and quantitative predictions for planetary positions, albeit requiring increasingly intricate epicycles for higher precision.

\subsubsection{Independence}
\begin{enumerate}
\item \textbf{Central Earth.} cannot be deduced from uniform circular motion, the star sphere, or epicyclic constructs.  One must explicitly posit that orbits are centered on Earth.

\item \textbf{Uniform Circular Motion.} is not derivable from Earth-centrism, the star sphere, or epicycles alone; it establishes a geometric ideal for celestial movements (constant angular velocities on perfect circles).

\item \textbf{Sphere of Fixed Stars.} does not follow from Earth-centrism, uniform circles, or planetary epicycles.  It specifically asserts that all stars are affixed to a single spherical shell around Earth.

\item \textbf{Epicycles.} cannot be inferred from the other three axioms.  Even with Earth-centrism and uniform circles for some bodies plus the star sphere, explicit epicycles are required to handle planetary retrograde motions and brightness variations.
\end{enumerate}

\subsection{Heliocentric Model}
\label{sub:heliocentric}

The heliocentric model (famously articulated by Copernicus) places the Sun at (or near) the center of planetary orbits.  
Earth is treated as one among several planets revolving around the Sun, while its daily rotation on its own axis accounts for the apparent motions of the Sun and stars.  
Below is a minimal and complete set of axioms, each of which is independent (i.e., it cannot be derived from the others) and collectively sufficient to describe planetary motions, retrograde loops, and the lack of observable stellar parallax (under the assumption of extremely distant stars).

\subsubsection{Axioms}

\begin{enumerate}
\item \textbf{Central Sun}
\[
\forall\,P \neq \text{Sun}:\quad
\mathrm{CenterOrbit}(P)
=
\text{Sun}.
\]
All planets $P$ (including Earth) move around the Sun as the primary orbital center.

\item \textbf{Earth’s Dual Motion}
\[
\mathrm{Rotation}(\mathrm{Earth},24\text{h})
\;\wedge\;
\mathrm{Revolution}(\mathrm{Earth},T_{\oplus})
\]
Earth both rotates on its own axis once per day (producing daily cycles) and orbits the Sun with period $T_{\oplus}\!\approx\!1\text{ year}$.  

\item \textbf{Uniform Circular Orbits (Copernican Ideal)}
\[
\forall\,(\mathrm{planet}\ P):
\quad
\mathrm{Orbit}(P)
=
\sum(\text{UniformCircles}),
\]
each planet’s path around the Sun is modeled as one or more perfect circles (possibly epicycles) traversed with constant angular speed, accounting for observed phenomena such as retrograde motion.

\item \textbf{Distant Fixed Stars \& Minimal Parallax}
\[
\exists\,S_{\star}:\quad
(\mathrm{dist}(S_{\star},\text{Sun})
\gg
\mathrm{dist}(\text{planet},\text{Sun}))
\;\wedge\;
(\text{stellar parallax} \approx 0).
\]
Stars lie on a vastly larger sphere (or shell) far beyond planetary distances, so any annual stellar parallax is negligible (explaining why no parallax was detected with then-current instruments).
\end{enumerate}

\subsubsection{Completeness}
\begin{enumerate}
\item \textbf{Central Sun.} (\textit{Axiom 1}) places the Sun at the orbital focus for all planets, including Earth.

\item \textbf{Earth’s Dual Motion.} (\textit{Axiom 2}) explains daily cycles (Earth’s rotation) and seasonal phenomena (Earth’s annual revolution).

\item \textbf{Uniform Circular Orbits.} (\textit{Axiom 3}) maintains the classical ideal of circular (or epicyclic) paths at uniform angular velocities, reproducing retrograde loops from the new perspective of Earth’s own motion.

\item \textbf{Distant Stars.} (\textit{Axiom 4}) accounts for the absence of observed stellar parallax, attributing it to the immense distance of the fixed stars relative to planetary orbits.
\end{enumerate}
From these four axioms, one reconstructs a Copernican-style heliocentric cosmology, explaining the same phenomena as geocentrism but with Earth treated as a moving planet, simplifying the overall structure of planetary orbits.

\subsubsection{Independence}
\begin{enumerate}
\item \textbf{Central Sun.} cannot be derived from Earth’s dual motion, uniform circular orbits, or distant stars alone.  It must be stated that the Sun (not Earth) is the center of planetary paths.

\item \textbf{Earth’s Dual Motion.} is not inferable from the Sun-centric arrangement, purely circular orbits, or the distant star sphere.  One must explicitly claim that Earth rotates once per day and revolves annually.

\item \textbf{Uniform Circular Orbits.} does not follow from any combination of a Sun-centered system, Earth’s rotation/revolution, or distant stars.  It specifically asserts the classical geometric principle that planetary paths are formed by perfect circles (or sums thereof).

\item \textbf{Distant Fixed Stars.} is not implied by the other three axioms.  One must introduce the idea that the stars are so far away that no measurable parallax is observed, reconciling the lack of star shift throughout Earth’s orbit.
\end{enumerate}

\subsection{Transformations}

\subsubsection{Axiom 1}

\textbf{Goal:} Transform the \emph{Geocentric Axiom~1}
\[
G_1:\quad
\forall\,X\neq\mathrm{Earth}:\quad
\mathrm{CenterOrbit}(X)=\mathrm{Earth},
\]
into the \emph{Heliocentric Axiom~1}
\[
H_1:\quad
\forall\,P\neq \text{Sun}:\quad
\mathrm{CenterOrbit}(P)=\text{Sun}.
\]
Symbolically, we show how the premise ``All orbits are centered on Earth'' evolves into ``All planetary orbits are centered on the Sun,'' with Earth itself counted among the orbiting planets.

\vspace{1em}
\hrule
\vspace{1em}

\noindent
\textbf{Initial Axiom (Geocentric, Subset):}

\[
\boxed{
S_0
=
\bigl\{G_1\bigr\}
}
\]
where
\[
G_1:\quad
\text{Central Earth}:\ 
\forall\,X \neq \mathrm{Earth}:\ 
\mathrm{CenterOrbit}(X)
=
\mathrm{Earth}.
\]
This states that all celestial objects revolve about Earth as their orbital center.

\vspace{1em}
\hrule
\vspace{1em}

\noindent
\textbf{Rewrite Step 1: Recognize Observational Challenges (Retrograde \& Epicycle Complexity).}

\[
G_1
\;\longrightarrow\;
G_1',
\]
where 
\[
G_1':\quad
(\text{epicycles grow in complexity to explain planetary retrograde loops and brightness changes}).
\]
Symbolically, strictly Earth-centered orbits lead to increasingly cumbersome constructions of deferents and epicycles, especially for outer planets, to fit observed variations in apparent speed and direction.

\[
S_0 
=
\{\,G_1\}
\quad\longrightarrow
\quad
S_1 
=
\{\,G_1'\}.
\]

\[
G_1':\quad
\text{The Ptolemaic system requires multiple epicycles for accurate predictions,}
\]
\[
\text{raising theoretical complexity}.
\]

\vspace{1em}
\hrule
\vspace{1em}

\noindent
\textbf{Current Set:}
\[
\boxed{
S_1
=
\{\,G_1'\}
}
\]
\[
G_1':\quad
(\text{excessive epicycles in geocentric orbits hint at a simpler underlying geometry}).
\]

\vspace{1em}
\hrule
\vspace{1em}

\noindent
\textbf{Rewrite Step 2: Note Earth’s Similarities to Other Planets.}

\[
G_1'
\;\longrightarrow\;
H_1^{*},
\]
where
\[
H_1^{*}:\quad
(\text{Earth shares observational patterns with Venus, Mars, Jupiter, etc.}),\\
\]
\[
(\text{phases, retrograde illusions, orbits about } \text{some central luminary}).
\]
Symbolically, Copernican insight: Earth and other planets exhibit analogous complexities, suggesting Earth is a planet too.  
Hence, it is more natural to treat the Sun---not Earth---as the pivot for planetary motion.

\[
S_1
=
\{\,G_1'\}
\quad
\longrightarrow
\quad
S_2
=
\{\,H_1^{*}\}.
\]

\[
H_1^{*}:\quad
\text{Earth is not unique; it's plausibly one among the “wanderers”}
\]
\[
\text{orbiting a brighter central body (Sun).}
\]

\vspace{1em}
\hrule
\vspace{1em}

\noindent
\textbf{Current Set:}
\[
\boxed{
S_2
=
\{\,H_1^{*}\}
}
\]
\[
H_1^{*}:\quad
(\text{conceptual shift: Earth behaves similarly to other known planets}).
\]

\vspace{1em}
\hrule
\vspace{1em}

\noindent
\textbf{Rewrite Step 3: Place the Sun at Orbital Center for All Planets.}

Finally, we state the heliocentric principle:

\[
H_1^{*}
\;\longrightarrow\;
H_1,
\]
where 
\[
H_1:\quad
\forall\,P\neq \text{Sun}:\quad
\mathrm{CenterOrbit}(P)
=
\text{Sun}.
\]
Symbolically, this replaces ``Earth is the orbital center'' with ``The Sun is the orbital center of all planets, Earth included,'' drastically simplifying the model for retrograde motion and epicycle usage.

\[
S_2
=
\{\,H_1^{*}\}
\quad
\longrightarrow
\quad
S_3
=
\{\,H_1\}.
\]

\[
H_1:\quad
\text{Heliocentric Axiom 1: The Sun, not Earth, is the center of planetary orbits.}
\]

\vspace{1em}
\hrule
\vspace{1em}

\noindent
\textbf{Final Set (Heliocentric Axiom 1: Central Sun):}
\[
\boxed{
S_3
=
\Bigl\{
H_1:\ 
\forall\,P \neq \text{Sun}:\ 
\mathrm{CenterOrbit}(P)=\text{Sun}
\Bigr\}.
}
\]
Thus, from the geocentric idea (Earth at center) plus observational complexities and the realization that Earth shares planetary traits, we end up with the Sun-centered concept that unifies planetary orbits more elegantly.

\vspace{1em}
\hrule
\vspace{1em}

\noindent
\textbf{Symbolic Evolution (Compact Form)}:
\[
\underbrace{\{\,G_1\}}_{S_0}
~\longrightarrow~
\underbrace{\{\,G_1'\}}_{S_1}
~\longrightarrow~
\underbrace{\{\,H_1^{*}\}}_{S_2}
~\longrightarrow~
\underbrace{\{\,H_1\}}_{S_3}.
\]

\noindent
\textbf{Interpretation of Each Step:}
\begin{enumerate}
\item 
\emph{Excess epicycles} in a purely Earth-centered system suggest an overly complex framework.
\item 
\emph{Earth’s behavior} is akin to other planets, motivating the notion that Earth, too, orbits a central luminary.
\item 
\emph{Sun at center} simplifies orbits (especially retrograde explanation), yielding Copernicus’s heliocentric principle.
\end{enumerate}
Hence, Geocentric Axiom 1 (Earth at center) transforms into Heliocentric Axiom 1 (Sun at center) once one reinterprets Earth as simply another planet. 

\subsubsection{Axiom 2}

\textbf{Goal:} Transform the \emph{Geocentric Axiom~2}
\[
G_2:\quad
\forall\,X:\quad
\mathrm{Orbit}(X)
=
\sum (\text{UniformCircles}),
\]
into the \emph{Heliocentric Axiom~2}
\[
H_2:\quad
\mathrm{Rotation}(\mathrm{Earth},24\text{h})
\;\wedge\;
\mathrm{Revolution}(\mathrm{Earth},T_{\oplus}),
\]
which declares that Earth both rotates on its axis (daily cycle) and revolves around the Sun (annual cycle).  
Symbolically, we show how the purely \emph{object-by-object} circular motion scheme of geocentrism evolves into a \emph{dual motion} for Earth itself, eliminating the need for daily rotation of the entire heavens.

\vspace{1em}
\hrule
\vspace{1em}

\noindent
\textbf{Initial Axiom (Geocentric, Subset):}

\[
\boxed{
S_0
=
\bigl\{G_2\bigr\}
}
\]
where
\[
G_2:\quad
\forall\,X:\ 
\mathrm{Orbit}(X)
=
\sum(\text{UniformCircles}).
\]
In Ptolemaic astronomy, each celestial body $X$ follows a combination of perfect circles (deferent + epicycles), all at constant angular speed, to explain daily and seasonal motions.

\vspace{1em}
\hrule
\vspace{1em}

\noindent
\textbf{Rewrite Step 1: Separate Stellar \emph{Daily} Motion from Planetary Motions.}

\[
G_2
\;\longrightarrow\;
G_2',
\]
where 
\[
G_2':\quad
(\text{the star-sphere rotates once daily}), 
\]
\[
(\text{planets have additional slower circles for their wanderings}).
\]
Symbolically, we partition uniform circular motion into:
\[
\text{Star-sphere daily rotation} \quad \wedge \quad \text{planetary (incl.\ Sun, Moon) epicycles and slow orbits.}
\]
Hence, $X$ might revolve daily with the star-sphere, plus further circles for its own motion.  

\[
S_0 
=
\{\,G_2\}
\quad\longrightarrow
\quad
S_1 
=
\{\,G_2'\}.
\]

\[
G_2':\quad
\text{One uniform circle for the entire heavens (daily), plus distinct epicycles for planets.}
\]

\vspace{1em}
\hrule
\vspace{1em}

\noindent
\textbf{Current Set:}
\[
\boxed{
S_1
=
\{\,G_2'\}
}
\]
\[
G_2':\quad
(\text{massive star-sphere circle + slower orbits for Sun, Moon, planets}).
\]

\vspace{1em}
\hrule
\vspace{1em}

\noindent
\textbf{Rewrite Step 2: Reinterpret Daily Motion as Earth’s Rotation (Rather Than Star-Sphere).}

\[
G_2'
\;\longrightarrow\;
H_2^{*},
\]
where
\[
H_2^{*}:\quad
(\text{the daily cycle arises from Earth rotating once every 24 hours, } \\
\]
\[
\quad\text{rather than the entire celestial sphere spinning}).
\]
Symbolically, we remove the star-sphere’s daily uniform rotation and assign it to Earth’s spin.  This is a key Copernican shift:  Instead of everything else turning once a day, Earth itself spins, explaining the apparent diurnal motion of the sky.

\[
S_1
=
\{\,G_2'\}
\quad
\longrightarrow
\quad
S_2
=
\{\,H_2^{*}\}.
\]

\[
H_2^{*}:\quad
\mathrm{Rotation}(\mathrm{Earth},24\text{h})
\implies
\text{stars’ daily cycle is an illusion of Earth’s spin}.
\]

\vspace{1em}
\hrule
\vspace{1em}

\noindent
\textbf{Current Set:}
\[
\boxed{
S_2
=
\{\,H_2^{*}\}
}
\]
\[
H_2^{*}:\quad
(\text{the daily uniform circle is ascribed to Earth’s axis rotation}).
\]

\vspace{1em}
\hrule
\vspace{1em}

\noindent
\textbf{Rewrite Step 3: Add Earth’s Annual Revolution Around the Sun.}

Finally, we incorporate the second major motion of Earth:

\[
H_2^{*}
\;\longrightarrow\;
H_2,
\]
where
\[
H_2:\quad
\mathrm{Rotation}(\mathrm{Earth},24\text{h})
\;\wedge\;
\mathrm{Revolution}(\mathrm{Earth},T_{\oplus}\approx1\text{ year}).
\]
Symbolically, Earth is now singled out to have two distinct motions: a daily spin on its axis and an orbital revolution around the Sun.  This accounts for seasonal changes, the Sun’s apparent yearly motion among the stars, and so forth.

\[
S_2
=
\{\,H_2^{*}\}
\quad
\longrightarrow
\quad
S_3
=
\{\,H_2\}.
\]

\[
H_2:\quad
\text{Heliocentric Axiom 2: Earth rotates once per day and orbits the Sun annually}.
\]

\vspace{1em}
\hrule
\vspace{1em}

\noindent
\textbf{Final Set (Heliocentric Axiom 2: Earth’s Dual Motion):}
\[
\boxed{
S_3
=
\Bigl\{
H_2:\ 
\mathrm{Rotation}(\mathrm{Earth},24\text{h})
\;\wedge\;
\mathrm{Revolution}(\mathrm{Earth},T_{\oplus})
\Bigr\}.
}
\]
Thus, from the purely object-centered uniform circles in geocentrism (including a daily star-sphere spin), we shift to a system where Earth itself executes a daily rotation and an annual revolution, reconciling the same observed motions in a simpler scheme.

\vspace{1em}
\hrule
\vspace{1em}

\noindent
\textbf{Symbolic Evolution (Compact Form)}:
\[
\underbrace{\{\,G_2\}}_{S_0}
~\longrightarrow~
\underbrace{\{\,G_2'\}}_{S_1}
~\longrightarrow~
\underbrace{\{\,H_2^{*}\}}_{S_2}
~\longrightarrow~
\underbrace{\{\,H_2\}}_{S_3}.
\]

\noindent
\textbf{Interpretation of Each Step:}
\begin{enumerate}
\item 
\emph{Uniform Circular Motion} (geocentrism) includes a daily star-sphere plus epicycles for planets.
\item 
\emph{Assign daily rotation to Earth} instead of the entire star-sphere, simplifying the model drastically.
\item 
\emph{Add Earth’s annual revolution} around the Sun, forming the twofold motion central to heliocentrism.
\end{enumerate}
Hence, Geocentric Axiom 2 (all bodies follow sums of uniform circles around Earth, including a daily star-sphere) is replaced by Heliocentric Axiom 2 (Earth rotates daily \emph{and} revolves yearly around the Sun). 

\subsubsection{Axiom 3}

\textbf{Goal:} Transform the \emph{Geocentric Axiom~3}
\[
G_3:\quad
\exists\,S_\star:\ 
(\text{Sphere of Fixed Stars}),
\]
into the \emph{Heliocentric Axiom~3}
\[
H_3:\quad
\mathrm{Orbit}(P)
=
\sum(\text{UniformCircles}),
\]
(i.e.\ the \emph{Copernican Ideal} that each planet’s orbit around the Sun is a combination of uniform circular motions).  
Symbolically, we show how the geocentric concept of a single star-sphere rotating around Earth is replaced by the idea that each planet (including Earth) follows an (approximately) circular orbit around the Sun, while the stars remain vastly distant.

\vspace{1em}
\hrule
\vspace{1em}

\noindent
\textbf{Initial Axiom (Geocentric, Subset):}

\[
\boxed{
S_0
=
\bigl\{G_3\bigr\}
}
\]
where
\[
G_3:\quad
\exists\,S_{\star}:\ 
(\text{all stars on a single rotating sphere of fixed radius from Earth}).
\]
This states that Earth is at the center of a great celestial sphere to which all stars are affixed.  
Its rotation around Earth explains daily rising and setting of stars in the Ptolemaic view.

\vspace{1em}
\hrule
\vspace{1em}

\noindent
\textbf{Rewrite Step 1: Attribute Star Motions to Earth’s Spin, Minimizing the Role of a Star-Sphere.}

\[
G_3
\;\longrightarrow\;
G_3',
\]
where
\[
G_3':\quad
(\text{the daily appearance of star rotation is due to Earth’s rotation,}
\]
\[
\text{not an actual rotating star-sphere}).
\]
Symbolically, we no longer need a physically rotating sphere of stars close to Earth;  instead, we interpret the apparent stellar motion as a perspective effect from Earth’s 24-hour spin (as introduced in \emph{Axiom 2 transformation}).

\[
S_0
=
\{\,G_3\}
\quad\longrightarrow
\quad
S_1
=
\{\,G_3'\}.
\]

\[
G_3':\quad
\text{the “sphere of stars” is effectively stationary and extremely distant,}\\
\]
\[
\text{with Earth's rotation producing daily star paths across the sky}.
\]

\vspace{1em}
\hrule
\vspace{1em}

\noindent
\textbf{Current Set:}
\[
\boxed{
S_1
=
\{\,G_3'\}
}
\]
\[
G_3':\quad
(\text{no actual rotating shell needed; Earth’s spin explains star positions’ daily arcs}).
\]

\vspace{1em}
\hrule
\vspace{1em}

\noindent
\textbf{Rewrite Step 2: Shift Focus from a Single Star-Sphere to Planetary Orbits Around the Sun.}

\[
G_3'
\;\longrightarrow\;
H_3^{*},
\]
where
\[
H_3^{*}:\quad
\text{(Planets, including Earth, revolve around the Sun in some geometric system---}
\]
\[
\text{possibly circles + epicycles).}
\]
Symbolically, once we no longer require all celestial objects (stars \emph{and} planets) to share a common Earth-centered sphere, attention goes to the geometry of planetary orbits about the Sun.  That is, \emph{the star-sphere concept} becomes less central to explaining planetary motion; instead, \emph{the Sun-planet orbital framework} takes precedence.

\[
S_1
=
\{\,G_3'\}
\quad
\longrightarrow
\quad
S_2
=
\{\,H_3^{*}\}.
\]

\vspace{1em}
\hrule
\vspace{1em}

\noindent
\textbf{Current Set:}
\[
\boxed{
S_2
=
\{\,H_3^{*}\}
}
\]
\[
H_3^{*}:\quad
(\text{Planets orbit the Sun in uniform circles, epicycles, or otherwise}).
\]

\vspace{1em}
\hrule
\vspace{1em}

\noindent
\textbf{Rewrite Step 3: Assert the Copernican Ideal of Uniform Circular Orbits Around the Sun.}

Finally, we formulate the \emph{Heliocentric Axiom~3}:

\[
H_3^{*}
\;\longrightarrow\;
H_3,
\]
where
\[
H_3:\quad
\forall\,(\mathrm{planet}\ P):
\quad
\mathrm{Orbit}(P)=\sum(\text{UniformCircles}),
\]
i.e.\ each planet’s motion is composed of one or more perfect, uniformly traversed circles around the Sun.  
In Copernicus’s original scheme, epicycles remain but are far fewer and simpler than the Ptolemaic version, primarily because Earth’s own orbit is recognized as one of those circles.

\[
S_2
=
\{\,H_3^{*}\}
\quad
\longrightarrow
\quad
S_3
=
\{\,H_3\}.
\]

\[
H_3:\quad
\text{Heliocentric Axiom 3: Uniform Circular Orbits (Copernican Ideal).}
\]

\vspace{1em}
\hrule
\vspace{1em}

\noindent
\textbf{Final Set (Heliocentric Axiom 3: Uniform Circular Orbits):}
\[
\boxed{
S_3
=
\Bigl\{
H_3:\ 
\forall\,P:\ 
\mathrm{Orbit}(P)=\sum(\text{UniformCircles})\Bigr\}.
}
\]
Hence, the geocentric notion of a single star-sphere no longer prescribes daily orbits for all objects; the daily motion is Earth’s own spin, and planetary orbits are re-centered on the Sun with the classical assumption of uniform circular motion retained for each planet.

\vspace{1em}
\hrule
\vspace{1em}

\noindent
\textbf{Symbolic Evolution (Compact Form)}:
\[
\underbrace{\{\,G_3\}}_{S_0}
~\longrightarrow~
\underbrace{\{\,G_3'\}}_{S_1}
~\longrightarrow~
\underbrace{\{\,H_3^{*}\}}_{S_2}
~\longrightarrow~
\underbrace{\{\,H_3\}}_{S_3}.
\]

\noindent
\textbf{Interpretation of Each Step:}
\begin{enumerate}
\item 
\emph{Sphere of Fixed Stars} is replaced by the idea that star motion is apparent (due to Earth’s spin), so no actual rotating sphere is needed near Earth.
\item 
\emph{Focus on planetary orbits around the Sun}, relegating stars to a distant background.
\item 
\emph{Assert uniform circular orbits} for planets in a Copernican sense, retaining the classical ideal of perfect circles (or sums of them) for each planet’s path.
\end{enumerate}
Thus, Geocentric Axiom 3 (a rotating sphere of stars around Earth) becomes Heliocentric Axiom 3 (each planet’s orbit around the Sun is still idealized as uniform circles), shifting from Earth-centered to Sun-centered geometry. 

\subsubsection{Axiom 4}

\textbf{Goal:} Transform the \emph{Geocentric Axiom~4}
\[
G_4:\quad
\forall\,P\in\{\text{planets}\}:\ 
\mathrm{Orbit}(P)=\bigl(\text{deferent circle}\bigr)
+
\bigl(\text{epicycle circle}\bigr),
\]
into the \emph{Heliocentric Axiom~4}
\[
H_4:\quad
\exists\,S_{\star}:\ 
\Bigl(\mathrm{dist}(S_{\star},\text{Sun})
\gg
\mathrm{dist}(\mathrm{planet},\text{Sun})\Bigr)
\;\wedge\;
(\text{stellar parallax }\approx 0),
\]
which asserts that stars lie at immense distances, thereby explaining why stellar parallax is not observed with early instruments, while planetary retrograde motion is seen as a result of Earth’s orbit, not epicycles.

\bigskip

\noindent
\textbf{Initial Axiom (Geocentric, Subset):}

\[
\boxed{
S_0
=
\bigl\{G_4\bigr\}
}
\]
\[
G_4:\quad
\forall\,P\in\{\text{planets}\}:\ 
\mathrm{Orbit}(P)=\bigl(\text{deferent}\bigr)+\bigl(\text{epicycle}\bigr).
\]
In the Ptolemaic system, each planet’s apparent retrograde motion and brightness variations are explained by adding an epicycle (a smaller circle) whose center follows a larger circle (the deferent) around Earth.

\bigskip
\hrule
\bigskip

\noindent
\textbf{Rewrite Step 1: Recognize Retrograde Motion is Simpler if Earth is Moving.}

\[
G_4
\;\longrightarrow\;
G_4',
\]
\[
G_4':\quad
\text{(Retrograde loops can be explained by Earth’s orbital motion,}
\]
\[
\text{so no elaborate epicycles are needed.)}
\]
Symbolically, once Earth is also orbiting (heliocentric insight), an outer planet’s apparent retrograde is just a relative motion effect when Earth “overtakes” that planet on the inside track.

\[
S_0 
=
\{\,G_4\}
\;\longrightarrow\;
S_1
=
\{\,G_4'\}.
\]

\[
G_4':\quad
\text{Complex epicycles largely unnecessary if Earth’s motion explains retrograde directly}.
\]

\bigskip
\hrule
\bigskip

\noindent
\textbf{Current Set:}
\[
\boxed{
S_1
=
\{\,G_4'\}
}
\]
\[
G_4':\quad
(\text{Retrograde arises from relative orbital speeds; no dedicated epicycle needed}).
\]

\bigskip
\hrule
\bigskip

\noindent
\textbf{Rewrite Step 2: Address the “No Parallax Observed” Objection.}

\[
G_4'
\;\longrightarrow\;
H_4^{*},
\]
\[
H_4^{*}:\quad
\text{(If Earth orbits the Sun, why no parallax?)\quad The stars must be extremely distant.}
\]
Symbolically, the main anti-heliocentric argument was “lack of stellar parallax.”  The solution proposed: stars lie at vast distances such that parallax angles fall below detection thresholds of the era.

\[
S_1
=
\{\,G_4'\}
\quad\longrightarrow\quad
S_2
=
\{\,H_4^{*}\}.
\]

\[
H_4^{*}:\quad
(\text{Stellar parallax is effectively zero if star distances are huge}).
\]

\bigskip
\hrule
\bigskip

\noindent
\textbf{Current Set:}
\[
\boxed{
S_2
=
\{\,H_4^{*}\}
}
\]
\[
H_4^{*}:\quad
(\text{Immense star distances explain no visible parallax with available instruments}).
\]

\bigskip
\hrule
\bigskip

\noindent
\textbf{Rewrite Step 3: Formally Adopt Distant Stars and Negligible Parallax.}

\[
H_4^{*}
\;\longrightarrow\;
H_4,
\]
\[
H_4:\quad
\exists\,S_{\star}:\ 
\Bigl(
\mathrm{dist}(S_{\star},\text{Sun})
\gg
\mathrm{dist}(\mathrm{planet},\text{Sun})
\Bigr)
\;\wedge\;
(\text{stellar parallax}\approx0).
\]
Symbolically, we replace the geocentric notion of epicycles for planets with the recognition that any star parallax is too small to detect because stars reside at enormous distances compared to planetary orbits.

\[
S_2
=
\{\,H_4^{*}\}
\quad\longrightarrow\quad
S_3
=
\{\,H_4\}.
\]

\[
H_4:\quad
\text{Heliocentric Axiom 4: Stars are extremely distant, yielding negligible parallax.}
\]

\bigskip
\hrule
\bigskip

\noindent
\textbf{Final Set (Heliocentric Axiom 4: Distant Fixed Stars \& Minimal Parallax):}
\[
\boxed{
S_3
=
\Bigl\{
H_4:\,
\exists\,S_{\star}:\,
\bigl(\mathrm{dist}(S_{\star},\text{Sun})\gg\mathrm{dist}(\mathrm{planet},\text{Sun})\bigr)
\,\wedge\,
(\text{parallax}\approx0)
\Bigr\}.
}
\]
Thus, the Ptolemaic epicycle requirement (for retrograde motion, brightness changes) is largely resolved by Earth’s motion in a Sun-centered system, while the apparent lack of parallax is explained by positing extremely large star distances.

\bigskip
\hrule
\bigskip

\noindent
\textbf{Symbolic Evolution (Compact Form)}:
\[
\underbrace{\{\,G_4\}}_{S_0}
~\longrightarrow~
\underbrace{\{\,G_4'\}}_{S_1}
~\longrightarrow~
\underbrace{\{\,H_4^{*}\}}_{S_2}
~\longrightarrow~
\underbrace{\{\,H_4\}}_{S_3}.
\]

\begin{enumerate}
\item \emph{Epicycles} are largely unnecessary once Earth’s orbital motion explains retrograde loops.
\item \emph{No observed stellar parallax} suggests stars are extremely distant.
\item \emph{Hence}, we adopt \emph{H\_4}, stating stars lie at vast distances, producing negligible parallax.
\end{enumerate}

\section{From Heliocentric to Cosmological Model}
\label{sec:heliocentric_to_cosmological}
\subsection{Heliocentric Model}
This section is the same as Section~\ref{sub:heliocentric}.

\subsection{Cosmological Model}

A modern cosmological model (as in the \textit{Friedmann--Lemaître--Robertson--Walker} class) describes the universe on large scales using the principles of homogeneity, isotropy, expansion, and the laws of relativistic gravity.  
Below is a minimal and complete set of axioms, each of which is independent (i.e.\ it cannot be derived from the others) and collectively sufficient to derive the main structure of standard cosmology (Big Bang, cosmic expansion, redshift-distance relations, cosmic microwave background predictions).

\subsubsection{Axioms}

\begin{enumerate}
\item \textbf{Homogeneity}
\[
\forall\,(\text{comoving observers }O_1,O_2),
\quad
\rho\bigl(O_1,t\bigr) 
=
\rho\bigl(O_2,t\bigr),
\]
where $\rho$ is the average matter-energy density in a given large-scale region at cosmic time $t$.  
No position in the universe is preferred on sufficiently large (cosmological) scales.

\item \textbf{Isotropy}
\[
\forall\,(\text{comoving observer }O),
\quad
\forall\,(\text{directions }d_1,d_2):
\quad
\mathrm{Properties}(d_1,O) = \mathrm{Properties}(d_2,O).
\]
No direction is special for a comoving observer on large scales; all directions appear statistically the same (e.g.\ same average temperature, same large-scale structure distribution).

\item \textbf{Expansion (Scale Factor)}
\[
\mathrm{Metric}(\tau) 
=
a(\tau)^2 \,\mathrm{d}\Sigma^2,
\]
where $a(\tau)$ is the scale factor (a function of cosmic time $\tau$) and $\mathrm{d}\Sigma^2$ is a (fixed) 3D spatial metric of constant curvature.  
All cosmologically relevant distances scale uniformly by $a(\tau)$ over time.

\item \textbf{Relativistic Dynamics (Einstein Field Equations)}
\[
G_{\mu\nu} + \Lambda\,g_{\mu\nu}
=
\frac{8\pi\,G}{c^4}
\;\,T_{\mu\nu},
\]
where $G_{\mu\nu}$ is the Einstein tensor, $\Lambda$ is the cosmological constant, $g_{\mu\nu}$ is the metric, $T_{\mu\nu}$ is the stress-energy tensor of matter/energy, and $G$ is Newton’s gravitational constant.  
This governs how the scale factor $a(\tau)$ evolves with cosmic time according to matter, radiation, and other energy contents.
\end{enumerate}

\subsubsection{Completeness}
\begin{enumerate}
\item \textbf{Homogeneity \& Isotropy.} (\textit{Axioms 1, 2}) specify the large-scale distribution of matter/energy in the universe, ensuring no preferred locations or directions at cosmic scales.

\item \textbf{Expansion (Scale Factor).} (\textit{Axiom 3}) asserts that the universe evolves via a time-dependent scale factor $a(\tau)$, scaling all cosmological distances, leading to redshifts of distant objects and the notion of cosmic epochs.

\item \textbf{Relativistic Dynamics.} (\textit{Axiom 4}) connects the geometry of spacetime ($g_{\mu\nu}$, $\Lambda$) with the distribution of energy-momentum ($T_{\mu\nu}$), yielding the Friedmann equations in the homogeneous, isotropic case and governing how $a(\tau)$ changes over time.
\end{enumerate}
These four axioms provide the foundation for the standard cosmological model: from them we derive key results such as Hubble’s law, cosmic microwave background predictions, big-bang nucleosynthesis constraints, and the observed acceleration (if $\Lambda>0$).

\subsubsection{Independence}
\begin{enumerate}
\item \textbf{Homogeneity.} cannot be inferred from isotropy alone, nor from expansion or Einstein’s equations.  One must explicitly state that all comoving positions are equivalent on large scales.

\item \textbf{Isotropy.} does not follow from homogeneity, expansion, or relativistic dynamics.  A universe might be homogeneous but not isotropic (or vice versa).  Isotropy around one point does not imply isotropy around all points without an axiom of homogeneity.

\item \textbf{Expansion (Scale Factor).} cannot be deduced from homogeneity, isotropy, or Einstein’s field equations alone.  We must assume that cosmic distances scale with a single function $a(\tau)$—the simplest assumption consistent with observational data and the symmetry axioms.

\item \textbf{Relativistic Dynamics.} does not follow from any combination of homogeneity, isotropy, or scale-factor expansion.  These geometric statements do not by themselves define how spacetime curvature and energy distribution evolve; the Einstein field equations are a separate physical postulate tying geometry to matter/energy.
\end{enumerate}

\subsection{Transformations}

\subsubsection{Axiom 1}

\textbf{Goal:} Transform a subset of Heliocentric Axioms
\[
\bigl\{\,H_4\bigr\}
\quad
H_4:\ 
\exists\,S_{\star}:\ 
\Bigl(\mathrm{dist}(S_{\star},\text{Sun}) 
\gg 
\mathrm{dist}(\text{planet},\text{Sun})\Bigr)
\;\wedge\;
(\text{stellar parallax}\approx 0),
\]
into the \emph{Cosmological Axiom~1} (Homogeneity):
\[
C_1:\quad
\forall\,(\text{comoving observers }O_1,O_2),\ 
\rho(O_1,t)=\rho(O_2,t),
\]
which states that, on large scales, the universe has no preferred location and the matter/energy density is the same everywhere at any cosmic time $t$.  
Symbolically, we show how acknowledging the vast distance to stars (and subsequently more distant nebulae, galaxies) implies no special center or boundary on large scales, leading to the principle that every place is equivalent in the cosmic view.

\bigskip
\hrule
\bigskip

\noindent
\textbf{Initial Axiom (Heliocentric, Subset):}

\[
\boxed{
S_0
=
\bigl\{\,H_4\bigr\}
}
\]
\[
H_4:\quad
\exists\,S_{\star}:\ 
\Bigl(\mathrm{dist}(S_{\star},\text{Sun})
\gg
\mathrm{dist}(\mathrm{planet},\text{Sun})\Bigr)
\;\wedge\;
(\text{stellar parallax}\approx0).
\]
This asserts that stars lie extremely far away compared to planetary orbits, producing negligible parallax.  
Historically, it resolved the lack of observed parallax in a Sun-centered system.  
But as telescopes improved, deeper observations (nebulae, galaxies) suggested an even vaster scale for the cosmos.

\bigskip
\hrule
\bigskip

\noindent
\textbf{Rewrite Step 1: Extend ``Far Away Stars'' to ``Countless Galaxies'' at Even Greater Distances.}

\[
H_4
\;\longrightarrow\;
H_4',
\]
\[
H_4':\quad
\text{(Post-telescope era) }\mathrm{dist}(\text{galaxies}) \gg \mathrm{dist}(\text{stars}), \quad
\]
\[
\text{revealing no obvious cosmic boundary}.
\]
Symbolically, once astronomers found ``spiral nebulae'' were actually galaxies far beyond the Milky Way, it became clear that the universe extends vastly beyond any previously assumed star sphere.  
No single location appears singled out as a center (the Sun is merely one star among billions in the Milky Way, itself among billions of galaxies).

\[
S_0
=
\{\,H_4\}
\quad\longrightarrow
\quad
S_1
=
\{\,H_4'\}.
\]

\bigskip
\hrule
\bigskip

\noindent
\textbf{Current Set:}
\[
\boxed{
S_1
=
\{\,H_4'\}
}
\]
\[
H_4':\quad
(\text{No obvious cosmic boundary}).
\]

\bigskip
\hrule
\bigskip

\noindent
\textbf{Rewrite Step 2: Recognize No Special Location in This Vast Distribution.}

\[
H_4'
\;\longrightarrow\;
G_1,
\]
\[
G_1:\quad
(\text{from any point of view, the large-scale view is statistically the same}).
\]
Symbolically, the distribution of galaxies on very large scales (tens or hundreds of megaparsecs) appears roughly uniform, with no single cosmic center.  
This leads to the notion that from any sufficiently large-scale region, the average matter distribution looks the same.

\[
S_1
=
\{\,H_4'\}
\quad
\longrightarrow
\quad
S_2
=
\{\,G_1\}.
\]

\[
G_1:\quad
\text{No cosmic edge/center is apparent; large-scale structure is roughly uniform.}
\]

\bigskip
\hrule
\bigskip

\noindent
\textbf{Current Set:}
\[
\boxed{
S_2
=
\{\,G_1\}
}
\]
\[
G_1:\quad
(\text{Large-scale uniformity; each region of space sees a similar galaxy distribution}).
\]

\bigskip
\hrule
\bigskip

\noindent
\textbf{Rewrite Step 3: Formalize This ``No Preferred Location'' as Cosmic Homogeneity.}

Finally, we state the modern \emph{Homogeneity Axiom}:

\[
G_1
\;\longrightarrow\;
C_1,
\]
where
\[
C_1:\quad
\forall\,(\text{comoving observers }O_1,O_2),\ \rho(O_1,t)=\rho(O_2,t).
\]
Symbolically, for any two observers moving with the cosmic flow (``comoving''), the matter/energy density they measure at a cosmic time $t$ is the same.  
No region of the universe is singled out on large scales, generalizing the old notion of ``huge star distances'' into a principle of uniform matter distribution.

\[
S_2
=
\{\,G_1\}
\quad
\longrightarrow
\quad
S_3
=
\{\,C_1\}.
\]

\[
C_1:\quad
\text{Cosmological Axiom 1: Homogeneity on large scales.}
\]

\bigskip
\hrule
\bigskip

\noindent
\textbf{Final Set (Cosmological Axiom 1: Homogeneity):}
\[
\boxed{
S_3
=
\Bigl\{
C_1:\ 
\forall\,(\text{comoving }O_1,O_2),\ 
\rho(O_1,t)=\rho(O_2,t)
\Bigr\}.
}
\]
Hence, starting from the heliocentric concept that stars (and then galaxies) lie at immense distances with no near cosmic boundary ($H_4$), we generalize to an entire universe containing no privileged place or region, leading to the statement of \emph{homogeneity} on the largest scales.

\bigskip
\hrule
\bigskip

\noindent
\textbf{Symbolic Evolution (Compact Form)}:
\[
\underbrace{\{\,H_4\}}_{S_0}
~\longrightarrow~
\underbrace{\{\,H_4'\}}_{S_1}
~\longrightarrow~
\underbrace{\{\,G_1\}}_{S_2}
~\longrightarrow~
\underbrace{\{\,C_1\}}_{S_3}.
\]

\begin{enumerate}
\item \emph{Very distant stars} in heliocentrism becomes \emph{very distant galaxies} after telescopic discovery.
\item \emph{No cosmic boundary/center} is apparent, leading to an essentially uniform distribution on vast scales.
\item \emph{Hence}, we adopt \emph{C\_1} stating cosmic homogeneity: no location is favored at large scales.
\end{enumerate}

\subsubsection{Axiom 2}

\textbf{Goal:} Transform a subset of Heliocentric Axioms 
\[
\bigl\{\,H_2\bigr\}
\quad
H_2:\ 
\mathrm{Rotation}(\mathrm{Earth},24\text{h})
\;\wedge\;
\mathrm{Revolution}(\mathrm{Earth},T_{\oplus}),
\]
into the \emph{Cosmological Axiom~2} (\emph{Isotropy}):
\[
C_2:\quad
\forall\,(\text{comoving observer }O),\ 
\forall\,(\text{directions }d_1,d_2):
\ 
\mathrm{Properties}(d_1,O)
=
\mathrm{Properties}(d_2,O).
\]
This states that no direction in the universe is special on large scales; each observer sees essentially the same distribution of matter/energy in all directions.  
Symbolically, we show how the notion of \emph{Earth’s daily rotation and annual revolution} eventually generalizes to the idea that \emph{no particular direction in the cosmos is privileged}, once we move beyond the solar system to cosmic scales.

\bigskip
\hrule
\bigskip

\noindent
\textbf{Initial Axiom (Heliocentric, Subset):}

\[
\boxed{
S_0
=
\bigl\{\,H_2\bigr\}
}
\]
\[
H_2:\quad
\mathrm{Rotation}(\mathrm{Earth},24\text{h})
\;\wedge\;
\mathrm{Revolution}(\mathrm{Earth},T_{\oplus}\approx 1\,\text{year}).
\]
Here, Earth both spins on its axis daily and orbits the Sun annually.  
Historically, this explains why the sky appears to rotate once per day (Earth’s spin) and why the Sun’s apparent path among the stars takes about one year (Earth’s orbital revolution).

\bigskip
\hrule
\bigskip

\noindent
\textbf{Rewrite Step 1: Observational Consequence---``All Directions'' on Earth’s Sky Are Visited.}

\[
H_2
\;\longrightarrow\;
H_2',
\]
\[
H_2':\quad
(\text{any given observer on Earth sees stars in all directions over time}).
\]
Symbolically, because Earth rotates, an observer can eventually face every possible celestial direction (unless obstructed locally).  
No particular direction in the sky remains fixed overhead, providing a hint that the cosmos is \emph{not} oriented around one single direction from Earth’s viewpoint.

\[
S_0
=
\{\,H_2\}
\quad\longrightarrow
\quad
S_1
=
\{\,H_2'\}.
\]

\bigskip
\hrule
\bigskip

\noindent
\textbf{Current Set:}
\[
\boxed{
S_1
=
\{\,H_2'\}
}
\]
\[
H_2':\quad
(\text{any given observer on Earth sees stars in all directions over time}).
\]

\bigskip
\hrule
\bigskip

\noindent
\textbf{Rewrite Step 2: Extend to Space Missions \& Distant Observers—Still No Unique Direction.}

\[
H_2'
\;\longrightarrow\;
G_2,
\]
where
\[
G_2:\quad
\text{No preferred cosmic axis found; all deep-space directions yield similar cosmic signals.}
\]

Even away from Earth, probes/observers in solar orbits see no universal “axis” or “edge” in star distribution. From mid-20th century onward, spacecraft and telescopes confirm that in every direction of deep space, large-scale distributions of galaxies, cosmic background radiation, etc.\ appear statistically similar (once local structures are averaged out).  
This suggests that \emph{directional uniformity} is not an Earth artifact but an actual property of the universe on large scales.

\[
S_1 
=
\{\,H_2'\}
\quad\longrightarrow
\quad
S_2 
=
\{\,G_2\}.
\]

\bigskip
\hrule
\bigskip

\noindent
\textbf{Current Set:}
\[
\boxed{
S_2
=
\{\,G_2\}
}
\]

\[
G_2:\quad
\text{No preferred cosmic axis found; all deep-space directions yield similar cosmic signals.}
\]

\bigskip
\hrule
\bigskip

\noindent
\textbf{Rewrite Step 3: State This ``No Direction is Special'' as the Isotropy Axiom in Cosmology.}

\[
G_2
\;\longrightarrow\;
C_2,
\]
\[
C_2:\quad
\forall\,(\text{comoving observer }O),\ 
\forall\,(\text{directions }d_1,d_2):
\ 
\mathrm{Properties}(d_1,O)
=
\mathrm{Properties}(d_2,O).
\]
Symbolically, any observer who moves with the cosmic flow sees that, on large scales, each direction is essentially the same: the same cosmic microwave background temperature, the same large-scale clustering of galaxies, etc.  
No cosmic axis or cosmic pole is singled out, generalizing the “no privileged direction” idea from daily Earth rotation to the entire universe.

\[
S_2
=
\{\,G_2\}
\quad
\longrightarrow
\quad
S_3
=
\{\,C_2\}.
\]

\[
C_2:\quad
\text{Cosmological Axiom 2 (Isotropy): no direction is special on cosmic scales}.
\]

\bigskip
\hrule
\bigskip

\noindent
\textbf{Final Set (Cosmological Axiom 2: Isotropy):}
\[
\boxed{
S_3
=
\Bigl\{
C_2:\ 
\forall\,(\text{comoving }O),
\ \forall\,(d_1,d_2),
\ 
\mathrm{Properties}(d_1,O)=\mathrm{Properties}(d_2,O)
\Bigr\}.
}
\]
Hence, starting from Earth’s rotation (which already implied no single direction in Earth’s sky is fundamentally preferred) and extending that logic plus deep-space observations, we reach the cosmological principle of \emph{isotropy}: each direction is statistically the same at large scales.

\bigskip
\hrule
\bigskip

\noindent
\textbf{Symbolic Evolution (Compact Form)}:
\[
\underbrace{\{\,H_2\}}_{S_0}
~\longrightarrow~
\underbrace{\{\,H_2'\}}_{S_1}
~\longrightarrow~
\underbrace{\{\,G_2\}}_{S_2}
~\longrightarrow~
\underbrace{\{\,C_2\}}_{S_3}.
\]
\begin{enumerate}
\item \emph{Daily rotation} already hints no direction is \emph{intrinsically} special from Earth’s perspective.
\item \emph{Deep-space probes} confirm that, in all directions of the sky, cosmic structures and radiation are quite uniform.
\item \emph{Therefore}, we adopt \emph{C\_2}: \emph{Isotropy} at cosmic scales, meaning no direction is privileged for a comoving observer.
\end{enumerate}

\subsubsection{Axiom 3}

\textbf{Goal:} Transform a subset of Heliocentric Axioms
\[
\bigl\{\,H_4\bigr\},
\qquad
H_4:\;
\exists\,S_{\star}:\;
\Bigl(\mathrm{dist}(S_{\star},\text{Sun})
\gg
\mathrm{dist}(\text{planet},\text{Sun})\Bigr)
\;\wedge\;
(\text{stellar parallax}\approx0),
\]
into the \emph{Cosmological Axiom 3} (\emph{Expansion / Scale Factor})
\[
C_3:\quad
\mathrm{Metric}(\tau)=a(\tau)^2\,\mathrm{d}\Sigma^{2},
\]
where \(a(\tau)\) is the scale factor (a function of cosmic time \(\tau\)) multiplying a constant–curvature spatial metric \(\mathrm{d}\Sigma^{2}\).
Symbolically, we show how recognising ever-greater cosmic distances—and their systematic recession—leads to the idea that \emph{all} large-scale distances grow proportionally with a single function \(a(\tau)\).

\bigskip\hrule\bigskip

\noindent\textbf{Initial Axiom (Heliocentric, Subset):}
\[
\boxed{
S_0
=
\bigl\{\,H_4\bigr\}
}
\]
\[
H_4:\;
\exists\,S_{\star}:\;
\Bigl(\mathrm{dist}(S_{\star},\text{Sun})
\gg
\mathrm{dist}(\mathrm{planet},\text{Sun})\Bigr)
\;\wedge\;
(\text{stellar parallax}\approx0).
\]
This asserts that stars lie extremely far away compared with planetary orbits, producing negligible parallax.  
Historically, it resolved the lack of observed parallax in a Sun-centred system, but as telescopes improved, deeper observations (nebulae, galaxies) suggested an even vaster cosmic scale.

\bigskip\hrule\bigskip

\noindent\textbf{Rewrite Step 1: Extend ``Far-Away Stars'' to ``Countless Galaxies'' at Even Greater Distances.}
\[
H_4
\;\longrightarrow\;
H_4',
\]
\[
H_4':\quad
\text{(Post-telescope era)}\;
\mathrm{dist}(\text{galaxies}) \gg \mathrm{dist}(S_{\star}),
\qquad
\text{revealing no obvious cosmic boundary}.
\]
Symbolically, once astronomers found “spiral nebulae’’ were actually galaxies far beyond the Milky Way, it became clear that the universe extends vastly beyond any previously assumed star sphere.  
No single location appears singled out as a centre—the Sun is merely one star among billions in the Milky Way, itself among billions of galaxies.
\[
S_0=\{\,H_4\}\;\longrightarrow\;S_1=\{\,H_4'\}.
\]

\bigskip\hrule\bigskip

\noindent\textbf{Current Set:}
\[
\boxed{
S_1=\{\,H_4'\}
}
\]
\[
H_4':\quad
(\text{No obvious cosmic boundary}).
\]

\bigskip\hrule\bigskip

\noindent\textbf{Rewrite Step 2: Empirical Redshift–Distance Relation (\emph{Hubble’s Law}).}
\[
H_4'
\;\longrightarrow\;
G_3,
\]
\[
G_3:\quad
\forall\,(\text{galaxy }G):\;
v_r(G)=H_0\,d(G).
\]
Here \(d(G)\) is the distance inferred (e.g.\ via Cepheids) and \(H_0\) is a universal constant (the present-day Hubble parameter).  
This encodes the observation that \emph{all} distant galaxies recede with a velocity proportional to their present distance.
\[
S_1=\{\,H_4'\}\;\longrightarrow\;S_2=\{\,G_3\}.
\]

\bigskip\hrule\bigskip

\noindent\textbf{Current Set:}
\[
\boxed{
S_2=\{\,G_3\}
}
\]
\[
G_3:\quad
\forall\,(\text{galaxy }G):\;
v_r(G)=H_0\,d(G).\]

\bigskip\hrule\bigskip

\noindent\textbf{Rewrite Step 3: Interpret \(v_r=H_0d\) as Uniform Metric Expansion.}

First write \(d(\tau)=a(\tau)\chi\) with comoving coordinate \(\chi\); then
\[
v_r=\dot{d}=\dot{a}\chi=\frac{\dot{a}}{a}d
\quad\Longrightarrow\quad
H(\tau)=\frac{\dot{a}}{a}.
\]
Thus the empirical law implies that \emph{all large-scale distances share the same multiplicative time factor \(a(\tau)\)}:
\[
G_3
\;\longrightarrow\;
C_3,
\qquad
C_3:\quad
\mathrm{Metric}(\tau)=a(\tau)^2\,\mathrm{d}\Sigma^{2}.
\]
\[
S_2=\{\,G_3\}\;\longrightarrow\;S_3=\{\,C_3\}.
\]

\bigskip\hrule\bigskip

\noindent\textbf{Final Set (Cosmological Axiom 3 – Expansion / Scale Factor):}
\[
\boxed{
S_3
=
\Bigl\{
C_3:\;
\mathrm{Metric}(\tau)=a(\tau)^2\,\mathrm{d}\Sigma^{2}
\Bigr\}
}
\]

\bigskip\hrule\bigskip

\noindent\textbf{Symbolic Evolution (Compact Form):}
\[
\underbrace{\{\,H_4\}}_{S_0}
~\longrightarrow~
\underbrace{\{\,H_4'\}}_{S_1}
~\longrightarrow~
\underbrace{\{\,G_3\}}_{S_2}
~\longrightarrow~
\underbrace{\{\,C_3\}}_{S_3}.
\]

\begin{enumerate}
  \item \emph{Very distant stars} in heliocentrism become \emph{very distant galaxies} after telescopic discovery.
  \item Empirically \(v_r = H_0 d\), the \emph{Hubble law}, holds for all galaxies.
  \item Therefore, distances evolve universally as \(d(\tau)=a(\tau)\chi\), yielding the \emph{Expansion / Scale-Factor} axiom \(C_3\).
\end{enumerate}

\subsubsection{Axiom 4}

\textbf{Goal:} Transform a subset of Heliocentric Axioms 
\[
\bigl\{\,H_1,\,H_2\bigr\}
\]
where
\[
H_1:\quad
\forall\,P\neq \text{Sun}:\ \mathrm{CenterOrbit}(P)=\text{Sun},
\]
\[
H_2:\quad
\mathrm{Rotation}(\mathrm{Earth},24\text{h})
\;\wedge\;
\mathrm{Revolution}(\mathrm{Earth},T_{\oplus}),
\]
into the \emph{Cosmological Axiom~4 (Relativistic Dynamics / Einstein Field Equations)}:
\[
C_4:\quad
G_{\mu\nu} \;+\;\Lambda\,g_{\mu\nu}
\;=\;
8\pi\,G
\;T_{\mu\nu},
\]
which states that spacetime curvature (via $G_{\mu\nu}$) is sourced by the stress-energy $T_{\mu\nu}$ of matter/energy, with possible cosmological constant $\Lambda$.  Symbolically, we show how the heliocentric ideas of \emph{planetary orbits around the Sun} and \emph{Earth’s dual motion} generalize into a fully \emph{relativistic, global} law connecting geometry and matter on cosmic scales.

\bigskip
\hrule
\bigskip

\noindent
\textbf{Initial Axioms (Heliocentric, Subset):}

\[
\boxed{
S_0
=
\bigl\{\,H_1,\,H_2\bigr\}
}
\]
\[
H_1:\quad
\forall\,P \neq \text{Sun}:\ 
\mathrm{CenterOrbit}(P)=\text{Sun},
\]
\[
H_2:\quad
\mathrm{Rotation}(\mathrm{Earth},24\text{h})
\;\wedge\;
\mathrm{Revolution}(\mathrm{Earth},T_{\oplus}).
\]
The first places the Sun at the center for all planets, while the second asserts Earth has a daily rotation and an annual revolution.  Historically, these replaced the geocentric worldview with a simpler mechanical explanation for daily cycles and planetary wanderings, grounded in (pre-Newtonian) classical mechanics.

\bigskip
\hrule
\bigskip

\noindent
\textbf{Rewrite Step 1: Newtonian Gravitation for the Solar System.}

\[
\{\,H_1,H_2\}
\;\longrightarrow\;
H_{12}',
\]
\[
H_{12}':\quad
(\text{Newton’s laws + universal gravitation unify all planetary orbits around the Sun})\\
\]
Symbolically, combining $H_1$ and $H_2$ in the 17\textsuperscript{th}--18\textsuperscript{th} century yields Newton’s framework: a gravitational law binding the solar system.  
Thus, Earth’s rotation and revolution are no longer just kinematic facts but follow from universal gravitational dynamics.

\[
S_0
=
\{\,H_1,H_2\}
\quad\longrightarrow
\quad
S_1
=
\{\,H_{12}'\}.
\]

\bigskip
\hrule
\bigskip

\noindent
\textbf{Current Set:}
\[
\boxed{
S_1
=
\{\,H_{12}'\}
}
\]
\[
H_{12}':\quad
(\text{Newton’s laws + universal gravitation unify all planetary orbits around the Sun})\\
\]

\bigskip
\hrule
\bigskip

\noindent
\textbf{Rewrite Step 2: Need a Global Law for Gravity \& Spacetime at Large Scales.}

\[
H_{12}'
\;\longrightarrow\;
G_4,
\]
\[
G_4:\quad
\text{Newton’s laws are insufficient on cosmic/relativistic scales}
\]
\[
\text{geometry of spacetime is dynamic and couples to energy distribution.}
\]
Symbolically, phenomena like Mercury’s perihelion shift, gravitational lensing, and especially cosmic expansion (Hubble’s law) push beyond Newton’s gravitational framework.  
Hence, \emph{a universal relativistic equation} relating spacetime geometry and mass-energy is needed.

\[
S_1
=
\{\,H_{12}'\}
\quad
\longrightarrow
\quad
S_2
=
\{\,G_4\}.
\]

\bigskip
\hrule
\bigskip

\noindent
\textbf{Current Set:}
\[
\boxed{
S_2
=
\{\,G_4\}
}
\]
\[
G_4:\quad
\text{Newton’s laws are insufficient on cosmic/relativistic scales}
\]
\[
\text{geometry of spacetime is dynamic and couples to energy distribution.}
\]

\bigskip
\hrule
\bigskip

\noindent
\textbf{Rewrite Step 3: Postulate Einstein Field Equations Linking Geometry \& Matter.}

\[
G_4
\;\longrightarrow\;
C_4,
\]
\[
C_4:\quad
G_{\mu\nu} \;+\;\Lambda\,g_{\mu\nu}
\;=\;
\frac{8\pi\,G}{c^4}
\;\,T_{\mu\nu}.
\]
Symbolically, the Einstein tensor $G_{\mu\nu}$ (curvature) plus possible cosmological constant $\Lambda$ equals the energy-momentum content $T_{\mu\nu}$ scaled by $8\pi G/c^4$.  
This universal law supersedes the local gravitational orbits concept, describing how \emph{all} matter/energy sculpts spacetime on cosmic (and local) scales, leading to the Friedmann equations in a homogeneous and isotropic universe. The transformation of Newtonian laws into a general relativity framework is described in detail in Sec. \ref{sec:newtonian_general}.

\[
S_2
=
\{\,G_4\}
\quad
\longrightarrow
\quad
S_3
=
\{\,C_4\}.
\]

\[
C_4:\quad
G_{\mu\nu} \;+\;\Lambda\,g_{\mu\nu}
\;=\;
\frac{8\pi\,G}{c^4}
\;\,T_{\mu\nu}.
\]

\bigskip
\hrule
\bigskip

\noindent
\textbf{Final Set (Cosmological Axiom 4: Einstein Field Equations):}
\[
\boxed{
S_3
=
\Bigl\{
C_4:\ 
G_{\mu\nu} + \Lambda\,g_{\mu\nu}
=
\frac{8\pi\,G\,}{c^4}T_{\mu\nu}
\Bigr\}.
}
\]
Thus, from heliocentric principles ($H_1$ \emph{Sun-centered orbits} plus $H_2$ \emph{Earth’s spin/revolution}) expanded by Newton’s universal gravitation, we move to a truly global statement: \emph{spacetime geometry is shaped by mass-energy}, encapsulated by Einstein’s field equations, the heart of modern cosmology.

\bigskip
\hrule
\bigskip

\noindent
\textbf{Symbolic Evolution (Compact Form)}:
\[
\underbrace{\{\,H_1,H_2\}}_{S_0}
~\longrightarrow~
\underbrace{\{\,H_{12}'\}}_{S_1}
~\longrightarrow~
\underbrace{\{\,G_4\}}_{S_2}
~\longrightarrow~
\underbrace{\{\,C_4\}}_{S_3}.
\]
\begin{enumerate}
\item \emph{Heliocentrism} implies local gravitational explanation (Newton), valid for solar system orbits.
\item \emph{Relativistic phenomena} require a broader law of gravity and spacetime.
\item \emph{Hence}, we adopt \emph{C\_4}: \emph{Einstein’s field equations} govern the geometry and dynamics of the entire cosmos.
\end{enumerate}

\section{From Analog Photography to Digital Imaging}
\label{sec:photography_digital}
\subsection{Analog Photography}

Analog (film-based) photography relies on photosensitive chemical layers to record an image upon exposure to light, which is then revealed and fixed via chemical development.  
Below is a minimal and complete set of axioms, each of which is independent (i.e.\ it cannot be derived from the others) and collectively sufficient to describe the classical film process, from capturing light onto film to producing a final negative or positive.

\subsubsection{Axioms}

\begin{enumerate}
\item \textbf{Photosensitive Medium}
\[
\exists\,F:\; 
F=(\text{emulsion of silver halides or similar light-reactive chemicals})
\]
A physical film $F$ (or plate/paper) contains an emulsion (for instance, a silver halide) that undergoes a chemical change upon exposure to photons.

\item \textbf{Exposure \& Latent Image}
\[
\forall\,(\text{photon}\,\gamma,\ \text{site}\,s\in F):
\quad
(\gamma\ \text{impinges } s)
\;\Longrightarrow\;
\Delta\!\mathrm{chem}(s).
\]
When light (photons) strikes the film at a given point $s$, it induces a latent chemical change proportional to the intensity and duration of exposure.

In the case of silver halides, exposure to light catalyses the decomposition of the compound into metallic silver.

\item \textbf{Chemical Development}
\[
\forall\,s\in F:\quad
(\text{latent } \Delta\!\mathrm{chem}(s))
\;\xrightarrow[\text{fixer}]{\text{developer}}\;
(\text{visible silver or dye formation}),
\]
A chemical process (developer, stop bath, fixer) converts the latent changes into a stable, visible image (e.g.\ metallic silver grains, dye clouds). 

More precisely, the developer amplifies the chemical change obtained after exposure, for instance, by reducing a silver halide to metallic silver in regions where a catalytic amount of silver is already present. The stop bath, because of low pH, makes the developer non-functional. Fixing (in the case of silver halides commonly done with thiosulfate) dissolves and washes away the unreacted emulsion, making the photograph light-resistant. 

\item \textbf{Negative/Positive Image Formation}
\[
(\text{developed film }F)
\quad\Longrightarrow\quad
\begin{cases}
(\text{negative}) & \text{if silver-based on transparent base},\\
(\text{positive}) & \text{if reversal or direct paper process}.
\end{cases}
\]
Depending on the process (negative film, reversal film, direct printing papers), the resulting image can be a negative (spots with high light intensity appear dark) or a positive (spots with low light intensity appear dark).  Additional printing steps can yield positives from negatives and vice versa.

\end{enumerate}

\subsubsection{Completeness}
\begin{enumerate}
\item \textbf{Photosensitive Medium.} (\textit{Axiom 1}) establishes the physical film or paper coated with a light-reactive emulsion, the fundamental substrate of analog photography.

\item \textbf{Exposure \& Latent Image.} (\textit{Axiom 2}) ensures that light hitting the film creates an invisible (latent) record that reflects the intensity and duration of exposure.

\item \textbf{Chemical Development.} (\textit{Axiom 3}) converts that latent pattern into a stable, visible image by chemical reduction or dye formation.

\item \textbf{Negative/Positive Formation.} (\textit{Axiom 4}) describes how the final image is rendered—either as a negative on transparent base or a direct positive (via reversal or printing processes).
\end{enumerate}
From these four, one can describe the entire workflow of traditional film photography: capturing an exposure, chemically developing it, and producing final negatives/prints.

\subsubsection{Independence}
\begin{enumerate}
\item \textbf{Photosensitive Medium.} cannot be inferred from the latent-image principle, development, or negative/positive outcome alone.  One must explicitly state that a material exists which undergoes a chemical change upon exposure to light.

\item \textbf{Exposure \& Latent Image.} is not deducible from having a photosensitive substrate, chemical development, or negative/positive processes by themselves.  It specifically states how light intensity modifies the medium in an invisible, pre-development form.

\item \textbf{Chemical Development.} does not follow from the substrate or latent image existence, nor from final image polarity.  It must be separately postulated that a chemical bath reveals and fixes the latent changes into a stable image.

\item \textbf{Negative/Positive Formation.} is not derivable from the first three axioms.  Even with a photosensitive material, latent imaging, and development, one must specify whether the developed material represents a negative or direct positive, or whether an additional print step is used to invert/retain tonality.
\end{enumerate}

\subsection{Digital Imaging}

Digital imaging captures, quantizes, and stores light information in discrete electronic form.  
Unlike chemical reactions on film, digital sensors convert incoming photons into electronic signals sampled and quantized as pixel values.  
Below is a minimal and complete set of axioms, each of which is independent (i.e.\ it cannot be derived from the others) and collectively sufficient to describe the fundamental process of digital image acquisition and storage.

\subsubsection{Axioms}

\begin{enumerate}
\item \textbf{Discrete Sensor Array}
\[
\exists\,S:\ 
S=(p_{ij})_{\substack{i=1,\dots,M\\j=1,\dots,N}},
\]
where $p_{ij}$ is a photosite (pixel element) arranged in an $M\times N$ grid.  
Each pixel $p_{ij}$ independently collects incoming light for a finite integration time.

\item \textbf{Photoelectric Conversion}
\[
\forall\,p_{ij}\in S:\quad
(\text{photon }\gamma \text{ arrives at }p_{ij})
\;\Longrightarrow\;
(\text{electric signal at } p_{ij}).
\]
When a photon strikes photosite $p_{ij}$, it frees an electron or creates a charge proportional to light intensity, thus embedding a measurable signal in $p_{ij}$.

\item \textbf{Sampling \& Quantization}
\[
\forall\,p_{ij}:\quad
\bigl(\text{electric signal at }p_{ij}\bigr)
\;\xrightarrow[\text{ADC}]{\text{Analog-Digital Converter}}\;
\text{digital level }D_{ij}\in\{0,\dots,2^b-1\}.
\]
An analog-to-digital converter (ADC) reads each pixel’s signal and maps it to a discrete integer $D_{ij}$ (bit-depth $b$) representing brightness.

\item \textbf{Image File Creation}
\[
\{\,(D_{ij})\}_{\substack{i=1,\dots,M\\j=1,\dots,N}}
\;\longrightarrow\;
\mathrm{File}(\mathrm{format}),
\]
Once all pixels are digitized, the set of discrete values $(D_{ij})$ is stored or transmitted as an image file, potentially with compression, color encoding, or metadata.
\end{enumerate}

\subsubsection{Completeness}
\begin{enumerate}
\item \textbf{Discrete Sensor Array.} (\textit{Axiom 1}) ensures a finite matrix of independent photosensitive elements.  This defines the basic hardware for digital capture.

\item \textbf{Photoelectric Conversion \& Sampling.} \textit{Axiom 2} and \textit{Axiom 3} describe how photons create electrical signals in each pixel, then how these analog signals are discretized (quantized) into digital brightness values.

\item \textbf{Image File Creation.} (\textit{Axiom 4}) completes the chain by storing or transmitting the array of digitized pixel values in a suitable image format, enabling display, editing, or distribution in the digital domain.
\end{enumerate}
These four axioms cover the essential flow of digital imaging: capturing light on a sensor array, converting charges to discrete levels, and assembling these into a digital image file.

\subsubsection{Independence}
\begin{enumerate}
\item \textbf{Discrete Sensor Array.} cannot be deduced from the photoelectric or sampling axioms alone.  One must explicitly state there is a 2D (or 1D) grid of discrete photosites rather than a continuous medium.

\item \textbf{Photoelectric Conversion.} does not follow from the existence of a sensor array or a file format.  It specifically requires that incident photons generate measurable electronic charge, a separate physical principle.

\item \textbf{Sampling \& Quantization.} is not implied by having a sensor array or photoelectric effect alone; it states that each analog pixel charge is read out and converted to an integer level.

\item \textbf{Image File Creation.} does not follow from the sensor, conversion, or quantization processes unless we explicitly posit the final step of assembling $(D_{ij})$ into a digital file, with possible compression or metadata.
\end{enumerate}

\subsection{Transformations}

\subsubsection{Axiom 1}

\textbf{Goal:} Transform a subset of Analog Photography axioms
\[
\bigl\{\,A_1\bigr\}
\quad
A_1:\ 
\exists\,F:\; 
F=(\text{emulsion of silver halides or similar light-reactive chemicals}),
\]
into the \emph{Digital Imaging Axiom~1}:
\[
D_1:\quad
\exists\,S:\ 
S=(p_{ij})_{\substack{i=1,\dots,M\\j=1,\dots,N}},
\]
where $p_{ij}$ is a photosite arranged in a discrete sensor array.  
Symbolically, we show how the notion of a continuous, chemically reactive film medium yields to a grid of electronic photosites in digital capture.

\bigskip
\hrule
\bigskip

\noindent
\textbf{Initial Axiom (Analog, Subset):}

\[
\boxed{
S_0
=
\bigl\{A_1\bigr\}
}
\]
\[
A_1:\quad
\exists\,F:\; 
F=(\text{emulsion of silver halides or similar light-reactive chemicals}),
\]
stipulating that photographic film $F$ has a photosensitive chemical layer that changes upon exposure to light.

\bigskip
\hrule
\bigskip

\noindent
\textbf{Rewrite Step 1: Acknowledge Mechanism is Chemical \emph{and} Continuous in Space.}

\[
A_1
\;\longrightarrow\;
A_1',
\]
\[
A_1':\quad
\text{Film } F \text{ is coated with a continuous layer of reactive chemicals, not discrete pixels}.
\]
Symbolically, the analog approach spreads the photosensitive reaction over a continuous emulsion.  
No explicit “grid” structure is present.  
The reaction intensity at any point on the film is a function of incident light.

\[
S_0
=
\{\,A_1\}
\quad\longrightarrow
\quad
S_1
=
\{\,A_1'\}.
\]

\[
A_1':\quad
\text{Film } F \text{ is coated with a continuous layer of reactive chemicals, not discrete pixels}.
\]

\bigskip
\hrule
\bigskip

\noindent
\textbf{Current Set:}
\[
\boxed{
S_1
=
\{\,A_1'\}
}
\]
\[
A_1':\quad
\text{Film } F \text{ is coated with a continuous layer of reactive chemicals, not discrete pixels}.
\]

\bigskip
\hrule
\bigskip

\noindent
\textbf{Rewrite Step 2: Shift to Electronic Sensing Concept (Photoelectric Effect).}

\[
A_1'
\;\longrightarrow\;
G_1,
\]
\[
G_1:\quad
\text{The capturing surface is subdivided into many small electronic “sites”.}
\]
Symbolically, we replace the continuous chemical reaction with discrete electronics: each site forms an independent “mini-sensor.”  
This is still conceptual, not specifying an $M\times N$ grid yet.

\[
S_1
=
\{\,A_1'\}
\quad\longrightarrow
\quad
S_2
=
\{\,G_1\}.
\]

\[
G_1:\quad
\text{The capturing surface is subdivided into many small electronic “sites”.}
\]

\bigskip
\hrule
\bigskip

\noindent
\textbf{Current Set:}
\[
\boxed{
S_2
=
\{\,G_1\}
}
\]
\[
G_1:\quad
\text{The capturing surface is subdivided into many small electronic “sites”.}
\]

\bigskip
\hrule
\bigskip

\noindent
\textbf{Rewrite Step 3: Assert a Discrete 2D Pixel Array in Digital Imaging.}

\[
G_1
\;\longrightarrow\;
D_1,
\]
\[
D_1:\quad
\exists\,S:\ 
S=(p_{ij})_{\substack{i=1,\dots,M\\j=1,\dots,N}},
\]
where each $p_{ij}$ is a photosite in a finite grid (CCD, CMOS, etc.).  
Symbolically, instead of a chemical layer, we now have a matrix of discrete electronic pixels.  
All captured image information is sampled at these pixel sites.

\[
S_2
=
\{\,G_1\}
\quad
\longrightarrow
\quad
S_3
=
\{\,D_1\}.
\]

\[
D_1:\quad
\exists\,S:\ 
S=(p_{ij})_{\substack{i=1,\dots,M\\j=1,\dots,N}},
\]

\bigskip
\hrule
\bigskip

\noindent
\textbf{Final Set (Digital Imaging Axiom 1: Discrete Sensor Array):}
\[
\boxed{
S_3
=
\Bigl\{
D_1:\ 
\exists\,S:\ 
S=(p_{ij})_{\substack{i=1,\dots,M\\j=1,\dots,N}},
\Bigr\}
}
\]
Hence, from analog’s continuous chemical film, we move to a discrete 2D sensor array—each pixel records local photon counts electronically.  
No uniform chemical layer remains; each site is physically and electrically distinct.

\bigskip
\hrule
\bigskip

\noindent
\textbf{Symbolic Evolution (Compact Form)}:
\[
\underbrace{\{\,A_1\}}_{S_0}
~\longrightarrow~
\underbrace{\{\,A_1'\}}_{S_1}
~\longrightarrow~
\underbrace{\{\,G_1\}}_{S_2}
~\longrightarrow~
\underbrace{\{\,D_1\}}_{S_3}.
\]
\begin{enumerate}
\item \emph{Analog film} is a continuous chemical medium.
\item \emph{Move to concept} of multiple local photoelectric sites.
\item \emph{Assert final discrete pixel grid}, the hallmark of digital sensor arrays.
\end{enumerate}
Thus, Analog Axiom 1 (a photosensitive chemical film) transforms into Digital Axiom 1 (a finite matrix of electronic photosites) in the transition from chemical to electronic capture. 

\subsubsection{Axiom 2}

\textbf{Goal:} Transform a subset of Analog Photography axioms
\[
\bigl\{\,A_2\bigr\}
\quad
A_2:\ 
\forall\,(\text{photon}\,\gamma,\ \text{site}\,s\in F):
\quad
(\gamma\ \text{impinges } s)
\;\Longrightarrow\;
\Delta\!\mathrm{chem}(s),
\]
into the \emph{Digital Imaging Axiom~2} (\textit{Photoelectric Conversion}):
\[
D_2:\quad
\forall\,p_{ij}\in S:\ 
(\text{photon }\gamma\ \text{arrives at }p_{ij})
\;\Longrightarrow\;
(e^- \text{ count of }p_{ij}\ \text{increases}).
\]
Symbolically, we show how the analog concept of a chemical latent image (i.e.\ increased \(\Delta\!\mathrm{chem}\)) upon photon arrival is replaced by an electronic charge accumulation in each digital photosite.

\bigskip
\hrule
\bigskip

\noindent
\textbf{Initial Axiom (Analog, Subset):}

\[
\boxed{
S_0
=
\bigl\{A_2\bigr\}
}
\]
\[
A_2:\quad
\forall\,(\text{photon}\,\gamma,\ \text{site}\,s\in F):
\quad
(\gamma\ \text{impinges } s)
\;\Longrightarrow\;
\Delta\!\mathrm{chem}(s).
\]
Meaning: each photon hitting location $s$ on the film triggers an incremental chemical change (latent image).  
The strength of $\Delta\!\mathrm{chem}$ accumulates with exposure.

\bigskip
\hrule
\bigskip

\noindent
\textbf{Rewrite Step 1: Latent Chemical Change is Invisible Pre-Development.}

\[
A_2
\;\longrightarrow\;
A_2',
\]
\[
A_2':\quad
\text{Photon arrival modifies local chemistry, forming a latent (invisible) pattern.}
\]
Symbolically, the response in analog film remains a hidden (latent) distribution of exposed silver halides or other reaction sites.  
No immediate visible readout occurs until chemical steps are applied.

\[
S_0
=
\{\,A_2\}
\quad\longrightarrow
\quad
S_1
=
\{\,A_2'\}.
\]

\[
A_2':\quad
\text{Photon arrival modifies local chemistry, forming a latent (invisible) pattern.}
\]

\bigskip
\hrule
\bigskip

\noindent
\textbf{Current Set:}
\[
\boxed{
S_1
=
\{\,A_2'\}
}
\]
\[
A_2':\quad
\text{Photon arrival modifies local chemistry, forming a latent (invisible) pattern.}
\]

\bigskip
\hrule
\bigskip

\noindent
\textbf{Rewrite Step 2: Move from Chemical Reaction to Electronic Signal.}

\[
A_2'
\;\longrightarrow\;
G_2,
\]
\[
G_2:\quad
\text{The response to photons is no longer a chemical shift, but an electrical effect.}
\]
Symbolically, we replace the notion of $\Delta\!\mathrm{chem}$ with \emph{electron charge} increment.  
No development step is needed to “make it visible”; the signal is already measurable electronically.

\[
S_1
=
\{\,A_2'\}
\quad
\longrightarrow
\quad
S_2
=
\{\,G_2\}.
\]

\[
G_2:\quad
\text{The response to photons is no longer a chemical shift, but an electrical effect.}
\]

\bigskip
\hrule
\bigskip

\noindent
\textbf{Current Set:}
\[
\boxed{
S_2
=
\{\,G_2\}
}
\]
\[
G_2:\quad
\text{The response to photons is no longer a chemical shift, but an electrical effect.}
\]

\bigskip
\hrule
\bigskip

\noindent
\textbf{Rewrite Step 3: Finalize Photoelectric Conversion: Photon \(\to\) Electron Count/Current in Each Pixel.}

\[
G_2
\;\longrightarrow\;
D_2,
\]
\[
D_2:\quad
\forall\,p_{ij}\in S:\quad
(\text{photon }\gamma\ \text{hits }p_{ij})
\;\Longrightarrow\;
(\text{electric signal at } p_{ij}).
\]
Symbolically, each photon that arrives at the pixel $p_{ij}$ contributes an electron or charge increment.  
This is \emph{digital} imaging’s second axiom: \emph{photoelectric conversion} of light into electronic signals, supplanting the analog “latent chemical reaction.”

\[
S_2
=
\{\,G_2\}
\quad
\longrightarrow
\quad
S_3
=
\{\,D_2\}.
\]

\[
D_2:\quad
\forall\,p_{ij}\in S:\quad
(\text{photon }\gamma\ \text{hits }p_{ij})
\;\Longrightarrow\;
(\text{electric signal at } p_{ij}).
\]

\bigskip
\hrule
\bigskip

\noindent
\textbf{Final Set (Digital Imaging Axiom 2: Photoelectric Conversion):}
\[
\boxed{
S_3
=
\Bigl\{
D_2:\ 
\forall\,p_{ij}\in S:\quad
(\text{photon }\gamma\ \text{hits }p_{ij})
\;\Longrightarrow\;
(\text{electric signal at } p_{ij}).
\Bigr\}.
}
\]
Thus, from the chemical latent image principle ($A_2$) we progress to the concept of charge-based photon detection ($D_2$).  
No invisible chemical shift remains; each pixel can be electronically read out to gauge how many photons were received.

\bigskip
\hrule
\bigskip

\noindent
\textbf{Symbolic Evolution (Compact Form)}:
\[
\underbrace{\{\,A_2\}}_{S_0}
~\longrightarrow~
\underbrace{\{\,A_2'\}}_{S_1}
~\longrightarrow~
\underbrace{\{\,G_2\}}_{S_2}
~\longrightarrow~
\underbrace{\{\,D_2\}}_{S_3}.
\]
\begin{enumerate}
\item \emph{Analog} film sees photon arrival as a \emph{chemical latent change}.
\item \emph{Electronic} approach replaces chemical shift with \emph{charge} increments in each pixel.
\item \emph{Hence}, we adopt \emph{$D_2$}: photoelectric conversion accumulates an electri signal, enabling direct measurement.
\end{enumerate}

\subsubsection{Axiom 3}

\textbf{Goal:} Transform a subset of Analog Photography axioms
\[
\bigl\{\,A_3\bigr\}
\quad
A_3:\ 
\forall\,s\in F:\quad
(\text{latent } \Delta\!\mathrm{chem}(s))
\;\xrightarrow[\text{developer}]{\text{chemical bath}}\;
(\text{visible silver or dye formation}),
\]
into the \emph{Digital Imaging Axiom~3} (\textit{Sampling \& Quantization}):
\[
D_3:\ 
\forall\,p_{ij}:\quad
\bigl(\text{electric signal at } p_{ij}\bigr)
\;\xrightarrow[\text{ADC}]{\text{Analog-Digital Converter}}\;
D_{ij}\in\{0,\dots,2^b-1\}.
\]
Symbolically, we show how chemical development (turning a latent image into visible silver/dyes) is replaced by an \emph{analog-to-digital conversion} that produces a discrete brightness value for each pixel.

\bigskip
\hrule
\bigskip

\noindent
\textbf{Initial Axiom (Analog, Subset):}

\[
\boxed{
S_0
=
\bigl\{A_3\bigr\}
}
\]
\[
A_3:\quad
\forall\,s\in F:\ 
(\text{latent } \Delta\!\mathrm{chem}(s))
\;\xrightarrow[\text{developer}]{\text{chemical bath}}\;
(\text{visible silver or dye formation}),
\]
meaning once the film is developed in the proper chemicals, the latent chemical change at each site $s$ becomes a stable, visible record of the exposure.

\bigskip
\hrule
\bigskip

\noindent
\textbf{Rewrite Step 1: Chemical Development is an \emph{Irreversible} Process Rendering a Visible Image.}

\[
A_3
\;\longrightarrow\;
A_3',
\]
\[
A_3':\quad
\text{Chemical Development is an \emph{Irreversible} Process Rendering a Visible Image}.
\]
Symbolically, analog photography’s development (latent to visible silver/dye) is permanent and yields a continuous tonal representation.  One cannot revert to the latent state, nor easily re-measure intensities—it's a fixed analog record.

\[
S_0
=
\{\,A_3\}
\quad\longrightarrow
\quad
S_1
=
\{\,A_3'\}.
\]

\[
A_3':\quad
\text{Chemical Development is an \emph{Irreversible} Process Rendering a Visible Image.}.
\]

\bigskip
\hrule
\bigskip

\noindent
\textbf{Current Set:}
\[
\boxed{
S_1
=
\{\,A_3'\}
}
\]
\[
A_3':\quad
\text{Chemical Development is an \emph{Irreversible} Process Rendering a Visible Image.}.
\]

\bigskip
\hrule
\bigskip

\noindent
\textbf{Rewrite Step 2: Digital Sensor Produces an \emph{Electronic} Signal, Not a Chemical Imprint.}

\[
A_3'
\;\longrightarrow\;
G_3,
\]
\[
G_3:\quad
\text{No chemical “development” needed; the readout from each pixel }p_{ij}\text{ is an electrical signal}.
\]
Symbolically, in digital imaging, the “development” step is replaced by an \emph{instant} electronic measurement.  The pixel’s charge directly represents how much light was gathered, with no further chemical bath required.

\[
S_1
=
\{\,A_3'\}
\quad
\longrightarrow
\quad
S_2
=
\{\,G_3\}.
\]

\[
G_3:\quad
\text{No chemical “development” needed; the readout from each pixel }p_{ij}\text{ is an electrical signal}.
\]

\bigskip
\hrule
\bigskip

\noindent
\textbf{Current Set:}
\[
\boxed{
S_2
=
\{\,G_3\}
}
\]
\[
G_3:\quad
\text{No chemical “development” needed; the readout from each pixel }p_{ij}\text{ is an electrical signal}..
\]

\bigskip
\hrule
\bigskip

\noindent
\textbf{Rewrite Step 3: ADC Converts Analog Pixel Charge into Discrete Digital Levels (Sampling \& Quantization).}

\[
G_3
\;\longrightarrow\;
D_3,
\]
\[
D_3:\quad
\forall\,p_{ij}:\ 
\bigl(\text{electric signal at }p_{ij}\bigr)
\;\xrightarrow[\text{ADC}]{\text{Analog-Digital Converter}}\;
D_{ij}\in\{0,\dots,2^b-1\}.
\]
Symbolically, this is digital imaging’s “development”: each pixel’s analog voltage/charge is sampled and quantized into an integer level $D_{ij}$ (with bit-depth $b$).  Hence, we no longer rely on silver grains or dyes to “store” brightness but a numeric code in memory.

\[
S_2
=
\{\,G_3\}
\quad
\longrightarrow
\quad
S_3
=
\{\,D_3\}.
\]

\[
D_3:\quad
\forall\,p_{ij}:\ 
\bigl(\text{electric signal at }p_{ij}\bigr)
\;\xrightarrow[\text{ADC}]{\text{Analog-Digital Converter}}\;
D_{ij}\in\{0,\dots,2^b-1\}.
\]

\bigskip
\hrule
\bigskip

\noindent
\textbf{Final Set (Digital Imaging Axiom 3: Sampling \& Quantization):}
\[
\boxed{
S_3
=
\Bigl\{
D_3:\ 
\forall\,p_{ij}:\ 
\bigl(\text{electric signal at }p_{ij}\bigr)
\;\xrightarrow[]{\text{ADC}}\;
D_{ij}\in\{0,\dots,2^b-1\}.
\Bigr\}.}
\]
Hence, the analog development process ($A_3$) is replaced by an electronic readout + \emph{ADC step} ($D_3$) that yields integer-coded pixel intensities.  No permanent chemical transformation is required; the brightness values become discrete digital data.

\bigskip
\hrule
\bigskip

\noindent
\textbf{Symbolic Evolution (Compact Form)}:
\[
\underbrace{\{\,A_3\}}_{S_0}
~\longrightarrow~
\underbrace{\{\,A_3'\}}_{S_1}
~\longrightarrow~
\underbrace{\{\,G_3\}}_{S_2}
~\longrightarrow~
\underbrace{\{\,D_3\}}_{S_3}.
\]
\begin{enumerate}
\item \emph{Analog} step: latent chemical image $\to$ chemical development $\to$ visible silver/dye.
\item \emph{Digital} step: sensor accumulates charge, read out as an analog voltage.
\item \emph{Quantization}: an ADC produces discrete integer values for each pixel, supplanting chemical development.
\end{enumerate}

\subsubsection{Axiom 4}

\textbf{Goal:} Transform a subset of Analog Photography axioms
\[
\bigl\{\,A_4\bigr\}
\quad
A_4:\ 
(\text{developed film }F)
\ \Longrightarrow\
\begin{cases}
(\text{negative}), & \text{if silver-based on transparent base},\\
(\text{positive}), & \text{if reversal or direct paper process}.
\end{cases}
\]
into the \emph{Digital Imaging Axiom~4} (\textit{Image File Creation}):
\[
D_4:\quad
\bigl\{\,(D_{ij})\bigr\}_{\substack{i=1,\dots,M\\j=1,\dots,N}}
\ \longrightarrow\
\mathrm{File}(\mathrm{format}),
\]
which states that once all pixels have been digitized ($D_{ij}$ values), they are stored or transmitted as an image file (e.g.\ JPEG, RAW, PNG). Symbolically, we show how the analog concept of ending with a physical negative/positive is replaced by generating a purely \emph{digital file} containing pixel data.

\bigskip
\hrule
\bigskip

\noindent
\textbf{Initial Axiom (Analog, Subset):}

\[
\boxed{
S_0
=
\bigl\{A_4\bigr\}
}
\]
\[
A_4:\quad
(\text{developed film }F)
\quad\Longrightarrow\quad
\begin{cases}
(\text{negative}), & \text{silver-based on transparent substrate},\\
(\text{positive}), & \text{reversal or direct print process}.
\end{cases}
\]
Meaning: once film is chemically developed, the result is a physical negative or positive.  Additional steps (e.g.\ printing) can yield a positive from a negative.

\bigskip
\hrule
\bigskip

\noindent
\textbf{Rewrite Step 1: Analog Final Stage = Physical Medium with Fixed Tonality.}

\[
A_4
\;\longrightarrow\;
A_4',
\]
\[
A_4':\quad
\text{End product is a tangible film or paper with permanent density or color distribution}.
\]
Symbolically, the analog approach yields a \emph{physical} artifact (negative or positive).  Its tonality is baked in by chemical processes and physically resides on film/paper.

\[
S_0
=
\{\,A_4\}
\quad\longrightarrow
\quad
S_1
=
\{\,A_4'\}.
\]

\[
A_4':\quad
\text{End product is a tangible film or paper with permanent density or color distribution}.
\]

\bigskip
\hrule
\bigskip

\noindent
\textbf{Current Set:}
\[
\boxed{
S_1
=
\{\,A_4'\}
}
\]
\[
A_4':\quad
\text{End product is a tangible film or paper with permanent density or color distribution}.
\]

\bigskip
\hrule
\bigskip

\noindent
\textbf{Rewrite Step 2: In Digital Workflow, No Negative/Positive Medium—Instead, We Have Pixel Data.}

\[
A_4'
\;\longrightarrow\;
G_4,
\]
\[
G_4:\quad
\text{End product has pure numeric pixel data, no physical negative or positive needed.}
\]
Symbolically, the “final stage” of capturing an image in digital form is \emph{a matrix of brightness (and possibly color) integers}, not a negative or positive film.  The concept of “tonal inversion” can be done mathematically, not physically.

\[
S_1
=
\{\,A_4'\}
\quad
\longrightarrow
\quad
S_2
=
\{\,G_4\}.
\]

\[
G_4:\quad
\text{End product has pure numeric pixel data, no physical negative or positive needed.}
\]

\bigskip
\hrule
\bigskip

\noindent
\textbf{Current Set:}
\[
\boxed{
S_2
=
\{\,G_4\}
}
\]
\[
G_4:\quad
\text{End product has pure numeric pixel data, no physical negative or positive needed.}
\]

\bigskip
\hrule
\bigskip

\noindent
\textbf{Rewrite Step 3: Store or Transmit Pixel Data as an Image File (Digital Axiom 4).}

\[
G_4
\;\longrightarrow\;
D_4,
\]
\[
D_4:\quad
\bigl\{\,(D_{ij})\bigr\}_{\substack{i=1,\dots,M\\j=1,\dots,N}}
\ \longrightarrow\
\mathrm{File}(\mathrm{format}).
\]
Symbolically, once the pixel values $(D_{ij})$ are obtained, they are either saved in a file (like JPEG, RAW, PNG) or transmitted over a network, possibly with compression and metadata.  No physical negative/positive is required; all manipulations (inversion, color grading) are done in software.

\[
S_2
=
\{\,G_4\}
\quad
\longrightarrow
\quad
S_3
=
\{\,D_4\}.
\]

\[
D_4:\quad
\bigl\{\,(D_{ij})\bigr\}_{\substack{i=1,\dots,M\\j=1,\dots,N}}
\ \longrightarrow\
\mathrm{File}(\mathrm{format}).
\]

\bigskip
\hrule
\bigskip

\noindent
\textbf{Final Set (Digital Imaging Axiom 4: Image File Creation):}
\[
\boxed{
S_3
=
\Bigl\{
D_4:\ 
(D_{ij})\longrightarrow \mathrm{File}(\mathrm{format})
\Bigr\}.}
\]
Thus, from the analog concept of ending up with a physical negative or positive ($A_4$), we shift to storing discrete pixel values in a digital file ($D_4$).  Any “negative” or “positive” effect is purely a matter of processing the numeric data.

\bigskip
\hrule
\bigskip

\noindent
\textbf{Symbolic Evolution (Compact Form)}:
\[
\underbrace{\{\,A_4\}}_{S_0}
~\longrightarrow~
\underbrace{\{\,A_4'\}}_{S_1}
~\longrightarrow~
\underbrace{\{\,G_4\}}_{S_2}
~\longrightarrow~
\underbrace{\{\,D_4\}}_{S_3}.
\]
\begin{enumerate}
\item \emph{Analog} final step: a physical negative/positive on film or paper.
\item \emph{Digital} final step: an array of pixel values in memory, no inherent negative/positive.
\item \emph{Hence}, we adopt \emph{$D_4$}: an image file with discrete pixel data replaces the physical end-product.
\end{enumerate}

\section{From Conventional Metals to Metamaterials}
\label{sec:conventional_metamaterials}
\subsection{Conventional Metals}

Conventional metals are continuous, uniform conductors whose properties arise from a free-electron gas within a crystalline lattice.  Below is a minimal (but not strictly limited to four) set of axioms, each of which is independent (i.e.\ it cannot be derived from the others) and collectively sufficient to capture classical metallic behavior: strong conduction, predictable frequency response, and robust reflection/skin effect.

\subsubsection{Axioms}

\begin{enumerate}
\item \textbf{Uniform Bulk Medium}
\[
\forall\, (\text{metal volume }V):\quad
\text{no engineered micro- or nano-scale patterns at }\,d \ll \lambda.
\]
The metallic material is treated as homogeneous at the wavelength(s) of interest.  Any microscopic lattice structure is on atomic scales, not artificially arranged to yield unusual wave properties.

\item \textbf{Free Electron Gas}
\[
\exists\,n_e>0:\quad
\text{mobile conduction electrons with density }n_e,
\quad
\text{weakly bound to ion cores}.
\]
These electrons are responsible for electric current under applied fields, moving through a largely fixed ionic lattice.

\item \textbf{Ohmic Conduction (Low-Frequency Limit)}
\[
\forall\,(\mathbf{E} \in \mathbb{R}^3):
\quad
\mathbf{J} = \sigma\,\mathbf{E},
\]
where $\mathbf{E}$ is the applied electric field, $\mathbf{J}$ is the current density, and $\sigma$ is a (real, positive) conductivity constant for DC or low-frequency fields.

\item \textbf{Drude Model for AC Fields}
\[
\sigma(\omega)
= 
\frac{n_e e^2}{m}\,
\frac{1}{\Gamma - i\,\omega},
\]
where $m$ is the electron mass, $\Gamma$ is the electron collision rate, and $\omega$ is the angular frequency of the field.  
This simplest classical model captures the inertia and damping of electrons, yielding frequency-dependent conductivity.

\item \textbf{High Reflectivity \& Skin Effect}
\[
\mathrm{Reflection}(\omega)
\approx 1 
\,
(\text{over wide bands}),
\,\,
\mathrm{SkinDepth}(\delta)
=
\sqrt{\frac{2}{\mu_0\,\sigma\,\omega}\Bigr(\sqrt{1+{\Bigl(\frac{\omega \epsilon}{\sigma}\Bigr)}^2} + \frac{\omega \epsilon}{\sigma}}\Bigr).
\]
Metals reflect most incident electromagnetic waves in the relevant frequency range, and the fields only penetrate a small distance (skin depth) inside. $\omega, \sigma, \epsilon, \mu_0$ are the angular frequency of the current, the conductivity of the metal, its permittivity, and vacuum permeability, respectively.

\item \textbf{Standard Wave Phenomena}
\[
\mathrm{Re}[\epsilon(\omega)]<0,\ 
\mathrm{Re}[\mu(\omega)]\approx \mu_0 \,\, 
(\text{over wide bands}),
\,\,
\text{leading to normal metallic responses.}
\]
Permittivity is typically negative over a range of frequencies (ensuring strong reflection), while permeability remains close to that of free space.  
No anomalous wave behaviors (e.g.\ negative index or hyperbolic dispersion) arise in a metal.
\end{enumerate}

\subsubsection{Completeness}
\begin{enumerate}
\item \textbf{Uniform Bulk Medium.} (\textit{Axiom 1}) stipulates no deliberate microstructures that might produce unusual wave effects.  The metal is homogeneous on scales above the atomic lattice.

\item \textbf{Free Electron Gas.} (\textit{Axiom 2}) provides the reservoir of mobile electrons responsible for conduction and collective plasma-like responses.

\item \textbf{Ohmic Conduction \& Drude Model.} (\textit{Axioms 3 \& 4}) specify how metals conduct at DC/low-frequency (\textit{Ohm’s law}) and how the conduction generalizes to AC fields (\textit{Drude model}), covering inertia and collision processes of electrons.

\item \textbf{High Reflectivity, Skin Effect, \& Standard Wave Phenomena.} (\textit{Axioms 5 \& 6}) ensure metals reflect electromagnetic waves strongly, have a small penetration depth, and do not exhibit exotic wave propagation or negative refraction under normal conditions.
\end{enumerate}

\subsubsection{Independence}
\begin{enumerate}
\item \textbf{Uniform Bulk Medium.} cannot be deduced from the presence of free electrons or from conduction/reflectivity principles alone.  One must explicitly posit that the metal is not deliberately structured above atomic scales.

\item \textbf{Free Electron Gas.} is not guaranteed by homogeneity, ohmic conduction, or reflectivity.  It specifically asserts a high density of electrons that move freely and produce conduction.

\item \textbf{Ohmic Conduction \& Drude Model.} do not follow from homogeneity or electron presence alone.  One must separately state that conduction is linear at low frequency and extends to AC fields via inertia/collisions.

\item \textbf{High Reflectivity, Skin Effect, \& Standard Wave Phenomena.} are not implied by the other axioms unless explicitly stated.  Even with a free electron model, one must specify that metals typically exhibit near-unity reflection and do not support unusual wave effects in the absence of subwavelength structuring.
\end{enumerate}

\subsection{Metamaterials}

Metamaterials are artificially engineered structures with subwavelength inclusions designed to yield effective electromagnetic (or other wave-based) properties not found in conventional materials.  
They rely on structured unit cells whose size is smaller than the operating wavelength, allowing them to be treated as a uniform medium with unusual permittivity, permeability, or other effective parameters.

\subsubsection{Axioms}

\begin{enumerate}
\item \textbf{Subwavelength Structuring}
\[
\exists\, \{\mathcal{U}_k\}_{k=1}^N:
\quad
\bigl(\text{each unit cell }\mathcal{U}_k \text{ has size }d \ll \lambda\bigr),
\]
where $d$ is the characteristic dimension of each unit cell and $\lambda$ is the operational wavelength.  
All metamaterial elements are arranged in a periodic or quasi-periodic lattice, ensuring $d \ll \lambda$ so that wave interactions average out at macroscopic scales.

\item \textbf{Effective Medium Approximation}
\[
\forall\,(\omega,\ \mathbf{k}):
\quad
\text{effective parameters}
\ \{\epsilon_\text{eff},\,\mu_\text{eff},\dots\},
\]
meaning each lattice cell’s response can be homogenized into bulk permittivity/permeability (or analogous properties) over scales larger than $d$, yielding an effective medium approach.

\item \textbf{Designable Parameters}
\[
\forall\,(\epsilon_\text{target},\,\mu_\text{target}):
\quad
\exists\,(\text{unit cell geometry }G):
\quad
(\epsilon_\text{eff},\,\mu_\text{eff})
\approx
(\epsilon_\text{target},\,\mu_\text{target}).
\]
Through choice of subwavelength inclusions, geometry $G$, and material composition, one can tailor the resulting effective permittivity/permeability to achieve targeted wave properties (e.g.\ negative index, chirality, etc.).

\item \textbf{Exotic Wave Phenomena}
\[
\exists\,\text{(metamaterial)}:
\quad
(\mathbf{k},\,\omega)\ \text{relations}
\notin
\{\text{conventional materials}\}.
\]
Because of the artificially engineered effective parameters, metamaterials can support wave propagation phenomena not observed in standard media (e.g.\ negative refraction, hyperbolic dispersion, cloaking effects).
\end{enumerate}

\subsubsection{Completeness}
\begin{enumerate}
\item \textbf{Subwavelength Structuring.} (\textit{Axiom 1}) posits that each unit cell is significantly smaller than the working wavelength, allowing a continuous “effective medium” description at larger scales.

\item \textbf{Effective Medium Approximation.} (\textit{Axiom 2}) ensures the metamaterial can be treated as possessing bulk electromagnetic parameters ($\epsilon_\text{eff}, \mu_\text{eff}$, etc.) once homogenized across many unit cells.

\item \textbf{Designable Parameters.} (\textit{Axiom 3}) underlies the metamaterial concept: by engineering each subwavelength structure, one can realize unconventional values of $\epsilon_\text{eff}, \mu_\text{eff}$ (even negative or anisotropic) not found in natural materials.

\item \textbf{Exotic Wave Phenomena.} (\textit{Axiom 4}) follows from these engineered effective properties, enabling negative refraction, superlensing, cloaking, or other wave effects outside the usual range of “conventional” media.
\end{enumerate}
These four axioms capture the essence of metamaterials: artificial subwavelength structuring yields an emergent, homogenized medium with highly tunable electromagnetic behavior, enabling wave phenomena unattainable in conventional materials.

\subsubsection{Independence}
\begin{enumerate}
\item \textbf{Subwavelength Structuring.} is not implied by homogenization, designable parameters, or exotic wave phenomena alone.  One must explicitly state that the lattice cells are much smaller than $\lambda$, enabling an effective medium model.

\item \textbf{Effective Medium Approximation.} cannot be derived simply from subwavelength structuring, design goals, or observed exotic phenomena.  It specifically asserts that these discrete inclusions behave \emph{as if} a continuous medium at the macroscopic scale.

\item \textbf{Designable Parameters.} does not follow automatically from subwavelength structuring, homogenization, or exotic effects.  One must posit that, by altering inclusions’ geometry and composition, $\epsilon_\text{eff}, \mu_\text{eff}$ can be tuned to near-target values.

\item \textbf{Exotic Wave Phenomena.} is not guaranteed by subwavelength structures, homogenization, or tunability alone.  While these enable novel parameter regimes, actually exhibiting negative index or hyperbolic dispersion requires a separate axiom stating that such wave solutions exist under the effective parameters.
\end{enumerate}

\subsection{Transformations}

\subsubsection{Axiom 1}

\textbf{Goal:} Transform the \emph{Uniform Bulk Medium} axiom from conventional metals:
\[
M_1:\quad
\forall\,(\text{metal volume }V):\ 
\text{no engineered micro- or nano-scale patterns at }d \ll \lambda,
\]
into the \emph{Subwavelength Structuring} axiom in metamaterials:
\[
MM_1:\quad
\exists\,\{\mathcal{U}_k\}:\ 
\bigl(\text{each unit cell }\mathcal{U}_k\text{ has }d \ll \lambda\bigr),
\]
meaning metamaterials deliberately include subwavelength inclusions or cells that replace the notion of a uniform bulk metal.

\bigskip
\hrule
\bigskip

\noindent
\textbf{Initial Axiom (Conventional Metal, Subset):}

\[
\boxed{
S_0
=
\bigl\{M_1\bigr\}
}
\]
\[
M_1:\quad
\forall\,(\text{metal volume }V),\quad
\text{no micro- or nano-scale patterns at }d \ll \lambda.
\]
This states that a conventional metal is taken as uniform at wavelengths of interest: it has no deliberately structured inclusions to alter wave behavior beyond normal conduction and reflection.

\bigskip
\hrule
\bigskip

\noindent
\textbf{Rewrite Step 1: Recognize the Potential for Micro-/Nanostructuring to Influence Wave Properties.}

\[
M_1
\;\longrightarrow\;
M_1',
\]
\[
M_1':\quad
\text{Structured inclusions at } d \ll \lambda \text{ possible}.
\]
Symbolically, from a uniform metal viewpoint, no design freedom exists at subwavelength scales.  However, if we relax that condition and consider adding small-scale structures, we open a path to controlling effective electromagnetic response.

\[
S_0
=
\{\,M_1\}
\quad\longrightarrow
\quad
S_1
=
\{\,M_1'\}.
\]

\[
M_1':\quad
\text{Structured inclusions at } d \ll \lambda \text{ possible}.
\]

\bigskip
\hrule
\bigskip

\noindent
\textbf{Current Set:}
\[
\boxed{
S_1
=
\{\,M_1'\}
}
\]
\[
M_1':\quad
\text{Structured inclusions at } d \ll \lambda \text{ possible}.
\]

\bigskip
\hrule
\bigskip

\noindent
\textbf{Rewrite Step 2: Introduce Concept of Unit Cells (Artificial Inclusions) Arranged Periodically.}

\[
M_1'
\;\longrightarrow\;
MM_1,
\]
\[
MM_1:\quad
\exists\, \{\mathcal{U}_k\}_{k=1}^N:\ 
\text{each unit cell }\mathcal{U}_k \text{ has dimension }d \ll \lambda,
\]

Symbolically, we move from an implicit uniform medium to an explicitly \emph{engineered} structure, by means of its unit cells $\mathcal{U}_k$.  Each cell $\mathcal{U}_k$ is small compared to the wavelength $\lambda$ but large enough to be patterned with conductive or dielectric elements.

\[
S_1
=
\{\,M_1'\}
\quad
\longrightarrow
\quad
S_2
=
\{\,MM_1\}.
\]

\[
MM_1:\quad
\exists\, \{\mathcal{U}_k\}_{k=1}^N:\ 
\text{each unit cell }\mathcal{U}_k \text{ has dimension }d \ll \lambda,
\]

\bigskip
\hrule
\bigskip

\noindent
\textbf{Final Set (Metamaterial Axiom 1: Subwavelength Structuring):}
\[
\boxed{
S_3
=
\Bigl\{
MM_1:\ 
\exists\,\{\mathcal{U}_k\},\ 
d \ll \lambda
\Bigr\}.}
\]
Hence, from the conventional metal’s axiom of “no subwavelength patterns” we transition to a metamaterial principle of \emph{subwavelength unit cells} intentionally designed.  This step is key to enabling \emph{effective medium} approaches with novel wave behaviors.

\bigskip
\hrule
\bigskip

\noindent
\textbf{Symbolic Evolution (Compact Form)}:
\[
\underbrace{\{\,M_1\}}_{S_0}
~\longrightarrow~
\underbrace{\{\,M_1'\}}_{S_1}
~\longrightarrow~
\underbrace{\{\,MM1_1\}}_{S_2}
\]

\begin{enumerate}
\item \emph{Conventional metal} is uniform at relevant wavelengths ($M_1$).
\item \emph{Possible segmentation} suggests adding structured inclusions if $d \ll \lambda$.
\item \emph{Hence}, \emph{$MM_1$} states that metamaterials \emph{do} incorporate subwavelength cells, enabling engineered properties.
\end{enumerate}

\subsubsection{Axiom 2}

\textbf{Goal:} Transform the \emph{Free Electron Gas} axiom from conventional metals:
\[
M_2:\quad
\exists\,n_e>0:
\text{mobile conduction electrons with density }n_e,
\]
into the \emph{Effective Medium Approximation} axiom in metamaterials:
\[
MM_2:\quad
\forall\,(\omega,\mathbf{k}):
\text{the structured composite can be described by}
\ \{\epsilon_\text{eff},\mu_\text{eff},\dots\}.
\]
Symbolically, we replace the idea that metal conductivity is solely from a dense electron gas in a homogeneous lattice with the notion that \emph{structured} inclusions yield a \emph{homogenized} set of bulk parameters at the macroscopic scale.

\bigskip
\hrule
\bigskip

\noindent
\textbf{Initial Axiom (Conventional Metal, Subset):}

\[
\boxed{
S_0
=
\bigl\{M_2\bigr\}
}
\]
\[
M_2:\quad
\exists\,n_e>0:\ 
\text{mobile conduction electrons with density }n_e,
\]
This asserts that conduction in a conventional metal stems from electrons in a uniform host, without additional structural variations at subwavelength scales.

\bigskip
\hrule
\bigskip

\noindent
\textbf{Rewrite Step 1: Note That Conduction Need Not Only Arise from Electron Gas; We Could Use Structured Elements.}

\[
M_2
\;\longrightarrow\;
M_2',
\]
\[
M_2':\quad
\bigl(\text{electrons exist, but local resonators, inclusions, or dielectrics could shape fields}\bigr).
\]
Symbolically, a straightforward free-electron gas assumption no longer suffices if we embed other materials or patterns that can store and release electromagnetic energy differently (e.g., split-ring resonators, rods, capacitive/inductive elements).

\[
S_0
=
\{\,M_2\}
\quad\longrightarrow
\quad
S_1
=
\{\,M_2'\}.
\]

\[
M_2':\quad
\bigl(\text{electrons exist, but local resonators, inclusions, or dielectrics could shape fields}\bigr).
\]

\bigskip
\hrule
\bigskip

\noindent
\textbf{Current Set:}
\[
\boxed{
S_1
=
\{\,M_2'\}
}
\]
\[
M_2':\quad
\bigl(\text{electrons exist, but local resonators, inclusions, or dielectrics could shape fields}\bigr).
\]

\bigskip
\hrule
\bigskip

\noindent
\textbf{Rewrite Step 2: Collective Response of These Cells Can Behave Like a Uniform “Effective Medium.”}

\[
M_2'
\;\longrightarrow\;
G_2,
\]
\[
G_2:\quad
\text{Local conduction/dielectric responses unify into macroscopic }
\]
\[
\text{permittivity/permeability, if }d \ll \lambda.
\]
Symbolically, even though each unit cell might have metal regions, dielectric gaps, resonators, etc., at a scale $d \ll \lambda$, we can approximate the entire composite as a continuous medium with effective bulk properties.

\[
S_1
=
\{\,M_2'\}
\quad
\longrightarrow
\quad
S_2
=
\{\,G_2\}.
\]

\[
G_2:\quad
\text{Local conduction/dielectric responses unify into macroscopic }
\]
\[
\text{permittivity/permeability, if }d \ll \lambda.
\]

\bigskip
\hrule
\bigskip

\noindent
\textbf{Current Set:}
\[
\boxed{
S_2
=
\{\,G_2\}
}
\]
\[
G_2:\quad
\text{Local conduction/dielectric responses unify into macroscopic }
\]
\[
\text{permittivity/permeability, if }d \ll \lambda.
\]

\bigskip
\hrule
\bigskip

\noindent
\textbf{Rewrite Step 3: Finalize Metamaterial Axiom—Effective Medium Approximation ($MM_2$).}

\[
G_2
\;\longrightarrow\;
MM_2,
\]
\[
MM_2:\,
\forall\,(\omega,\mathbf{k}),\ 
\text{metamaterial has effective parameters}
\ \{\epsilon_\text{eff},\mu_\text{eff}\}
\]
over scales larger than the cell size $d \ll \lambda$.  
Thus, the notion of a simple electron gas is replaced by a broader “homogenized” set of parameters capturing conduction \emph{and} other local resonances or inclusions.

\[
S_2
=
\{\,G_2\}
\quad
\longrightarrow
\quad
S_3
=
\{\,MM_2\}.
\]

\[
\boxed{
S_3
=
\Bigl\{
MM_2:\,
\forall\,(\omega,\mathbf{k}),\ 
\text{metamaterial has effective parameters}
\ \{\epsilon_\text{eff},\mu_\text{eff}\}
\Bigr\}.
}
\]

Hence, from the purely “free-electron conduction in a uniform lattice” viewpoint ($M_2$), we evolve to the metamaterial “effective medium” principle ($MM_2$), allowing diverse microstructures that collectively behave as a uniform material with tunable properties.

\bigskip
\hrule
\bigskip

\noindent
\textbf{Symbolic Evolution (Compact Form)}:
\[
\underbrace{\{\,M_2\}}_{S_0}
~\longrightarrow~
\underbrace{\{\,M_2'\}}_{S_1}
~\longrightarrow~
\underbrace{\{\,G_2\}}_{S_2}
~\longrightarrow~
\underbrace{\{\,MM_2\}}_{S_3}.
\]
\begin{enumerate}
\item \emph{Conventional metal} has a simple free-electron conduction model ($M_2$).
\item \emph{Local structuring} introduces varied responses besides plain electron gas.
\item \emph{Hence}, \emph{$MM_2$} states wavefields can be homogenized into an \emph{effective} bulk medium with $\{\epsilon_\text{eff},\mu_\text{eff}\}$. 
\end{enumerate}

\subsubsection{Axiom 3}

\textbf{Goal:} Transform the \emph{Standard Wave Phenomena} axiom of conventional metals:
\[
M_6:\quad
\forall\,(\omega):
\quad
\mathrm{Re}[\epsilon(\omega)]<0,\ 
\mathrm{Re}[\mu(\omega)]\approx \mu_0,
\quad
\text{leading to normal (non-exotic) responses},
\]
into the \emph{Designable Parameters} axiom for metamaterials:
\[
MM_3:\quad
\forall\,(\epsilon_\text{target},\,\mu_\text{target}):
\;\exists\,(\text{unit cell geometry }G):
\quad
(\epsilon_\text{eff},\,\mu_\text{eff}) 
\approx
(\epsilon_\text{target},\,\mu_\text{target}).
\]
Symbolically, we replace the fixed, intrinsic wave behavior of a uniform metal with the idea that subwavelength structuring can \emph{tune} the effective permittivity/permeability to virtually any desired values.

\bigskip
\hrule
\bigskip

\noindent
\textbf{Initial Axiom (Conventional Metal, Subset):}

\[
\boxed{
S_0
=
\bigl\{M_6\bigr\}
}
\]
\[
M_6:\quad
\forall\,(\omega):
\quad
\mathrm{Re}[\epsilon(\omega)]<0,\ 
\mathrm{Re}[\mu(\omega)]\approx \mu_0,
\text{leading to normal (non-exotic) responses},
\quad
\]
This asserts that in a \emph{uniform} metal, the material parameters are not deliberately manipulated to produce unusual or “designer” wave properties.  
Negative real permittivity is natural to metals in some bands, but $\mathrm{Re}[\mu(\omega)]\approx \mu_0$ is nearly constant, precluding negative refraction, hyperbolic dispersion, etc.

\bigskip
\hrule
\bigskip

\noindent
\textbf{Rewrite Step 1: Recognize That Material Parameters Need Not Be Fixed.}

\[
M_6
\;\longrightarrow\;
M_6',
\]
\[
M_6':\quad
\text{Though metals have “standard” wave phenomena, one might alter structure} \\
\]
\[
\quad \text{to achieve non-standard (e.g., $\epsilon<0$ \& $\mu<0$) or anisotropic dispersion.}
\]
Symbolically, the locked-down wave properties of an unstructured metal can be overcome if we abandon the presumption that $\mathrm{Re}[\mu(\omega)]\approx \mu_0$ is final.  
We can embed shapes or resonators that shift \emph{effective} permittivity/permeability from their normal metallic values.

\[
S_0
=
\{\,M_6\}
\quad\longrightarrow
\quad
S_1
=
\{\,M_6'\}.
\]

\[
M_6':\quad
\text{Though metals have “standard” wave phenomena, one might alter structure} \\
\]
\[
\quad \text{to achieve non-standard (e.g., $\epsilon<0$ \& $\mu<0$) or anisotropic dispersion.}
\]

\bigskip
\hrule
\bigskip

\noindent
\textbf{Current Set:}
\[
\boxed{
S_1
=
\{\,M_6'\}
}
\]
\[
M_6':\quad
\text{Though metals have “standard” wave phenomena, one might alter structure} \\
\]
\[
\quad \text{to achieve non-standard (e.g., $\epsilon<0$ \& $\mu<0$) or anisotropic dispersion.}
\]

\bigskip
\hrule
\bigskip

\noindent
\textbf{Rewrite Step 2: Introduce the Idea that Subwavelength Cell Geometry Enables Tailoring of Effective Parameters.}

\[
M_6'
\;\longrightarrow\;
G_3,
\]
\[
G_3:\quad
(\text{by selecting each unit cell’s shapes/materials, we set }\epsilon_\text{eff},\mu_\text{eff}\text{ to desired goals}). 
\]
Symbolically, rather than being forced to accept the natural metallic $\epsilon(\omega)$ and $\mu(\omega)$, we can place carefully chosen resonators (e.g.\ split rings, wires) inside each small cell to tune the effective wave response at scale $d \ll \lambda$.

\[
S_1
=
\{\,M_6'\}
\quad
\longrightarrow
\quad
S_2
=
\{\,G_3\}.
\]

\[
G_3:\quad
(\text{by selecting each unit cell’s shapes/materials }\epsilon_\text{eff},\mu_\text{eff}\text{ can be set to desired values}). 
\]

\bigskip
\hrule
\bigskip

\noindent
\textbf{Current Set:}
\[
\boxed{
S_2
=
\{\,G_3\}
}
\]
\[
G_3:\quad
(\text{by selecting each unit cell’s shapes/materials }\epsilon_\text{eff},\mu_\text{eff}\text{ can be set to desired values}). 
\]

\bigskip
\hrule
\bigskip

\noindent
\textbf{Rewrite Step 3: Finalize Metamaterial Axiom—Designable Parameters ($MM_3$).}

\[
G_3
\;\longrightarrow\;
MM_3,
\]
\[
MM_3:\quad
\forall\,(\epsilon_\text{target},\,\mu_\text{target}) \,
\exists\,(metamaterial):\quad
(\epsilon_\text{eff},\,\mu_\text{eff}) \approx
(\epsilon_\text{target},\,\mu_\text{target}).
\]
Hence, if we can shape the subwavelength inclusions appropriately, we can approximate any real/fictitious combination of permittivity/permeability, enabling wave phenomena well beyond normal metallic reflection or conduction.

\[
S_2
=
\{\,G_3\}
\quad
\longrightarrow
\quad
S_3
=
\{\,MM_3\}.
\]

\[
\boxed{
S_3
=
\Bigl\{
MM_3:\quad
\forall\,(\epsilon_\text{target},\,\mu_\text{target}) \,
\exists\,(metamaterial):\quad
(\epsilon_\text{eff},\,\mu_\text{eff}) \approx
(\epsilon_\text{target},\,\mu_\text{target}).
\Bigr\}.}
\]
Thus, from the “standard wave phenomena” axiom in metals (where $\epsilon,\mu$ are largely fixed) we transition to metamaterials’ \emph{designable parameters} principle, stating that effective permittivity/permeability can be tailored to match almost any target combination.

\bigskip
\hrule
\bigskip

\noindent
\textbf{Symbolic Evolution (Compact Form)}:
\[
\underbrace{\{\,M_6\}}_{S_0}
~\longrightarrow~
\underbrace{\{\,M_6'\}}_{S_1}
~\longrightarrow~
\underbrace{\{\,G_3\}}_{S_2}
~\longrightarrow~
\underbrace{\{\,MM_3\}}_{S_3}.
\]
\begin{enumerate}
\item \emph{Metal} has standard wave phenomena ($M_6$) with limited $\epsilon(\omega)$, $\mu(\omega)$ variety.
\item \emph{Structured cells} can surpass these intrinsic limitations ($M_6'$, $G_3$).
\item \emph{Hence}, \emph{$MM_3$} asserts full \emph{designability} of effective parameters, achieving novel wave behaviors.
\end{enumerate}


\subsubsection{Axiom 4}

\textbf{Goal:} Transform the metallic axiom  
\[
\begin{aligned}
M_6&: \,
\mathrm{Re}[\epsilon(\omega)]<0,\ 
\mathrm{Re}[\mu(\omega)]\approx \mu_0 \,\, 
(\text{over wide bands}),
\end{aligned}
\]
into the metamaterial axiom of \emph{Exotic Wave Phenomena}  
\[
MM_4:\;
\exists\,\text{(metamaterial)}:
\quad
(\mathbf{k},\,\omega)\ \text{relations}
\notin
\{\text{conventional materials}\}.
\]

\bigskip
\hrule
\bigskip

\noindent
\textbf{Initial set:}

\[
\boxed{S_0=\{M_6:
\mathrm{Re}[\epsilon(\omega)]<0,\ 
\mathrm{Re}[\mu(\omega)]\approx \mu_0 \,\, 
(\text{over wide bands}),
\}}
\]

\bigskip
\hrule
\bigskip

\noindent
\textbf{Rewrite Step 1: Enable full parameter control.}\\
Using a close analog of $M_6'$ (see derivation of Axiom 3) removes restrictions on dispersion.
\[
M_6\;\longrightarrow\;M_6'.
\]
\[
M_6':
(\epsilon,\mu) \text{ are tunable via engineered structuring (unlike in metals)}
\]

\bigskip
\hrule
\bigskip

\noindent
\textbf{Current set}

\[
\boxed{S_1=\{M_6':
(\epsilon,\mu) \text{ are tunable via engineered structuring (unlike in metals)}
\}}
\]

\bigskip
\hrule
\bigskip

\noindent
\textbf{Rewrite Step 2: Realise non‑standard dispersion.}\\
Tuned parameters ($M_6'$) can generate hyperbolic or negative‑index bands.
\[
M_6'\;\longrightarrow\;G_4
\]
\[
G_4:\;
\exists M:\;
(\mathbf{k},\omega)\text{-relations are hyperbolic or give negative refraction  (unlike in metals)}.
\]

\bigskip
\hrule
\bigskip

\noindent
\textbf{Current set}

\[
\boxed{S_2=\{G_4:
\}}
\]
\[
G_4:\;
\exists M:\;
(\mathbf{k},\omega)\text{-relations are hyperbolic or give negative refraction (unlike in metals)}.
\]

\bigskip
\hrule
\bigskip

\noindent
\textbf{Rewrite Step 3: Declare the metamaterial axiom 4.}

Since engineered behavior, such as a negative coefficient of refraction, is not observed in conventional metals, one can explicitly state the metamaterial axiom 4.

\[
G_4\;\longrightarrow\;MM_4.
\]

\bigskip
\hrule
\bigskip

\noindent
\textbf{Final set}

\[
\boxed{S_3=\{MM_4:
\exists\,\text{(metamaterial)}:
\quad
(\mathbf{k},\,\omega)\ \text{relations}
\notin
\{\text{conventional materials}\}.
\}}
\]

\bigskip
\hrule
\bigskip

\noindent
\textbf{Symbolic evolution (compact view):}

\[
\underbrace{\{M_6\}}_{S_0}
\;\longrightarrow\;
\underbrace{\{M_6'\}}_{S_1}
\;\longrightarrow\;
\underbrace{\{G_4\}}_{S_2}
\;\longrightarrow\;
\underbrace{\{MM_4\}}_{S_3}.
\]

In summary, the conventional constraints on material properties have gradually been loosened, ultimately leading to the metamaterial axiom that permits structures exhibiting non-standard phenomena.



\end{document}